\documentclass[11pt]{cernrep}
\usepackage{graphicx}
\usepackage{here}
\usepackage{epsf}
\usepackage{epsfig}
\usepackage{subfigure}
\usepackage{rotate}

\renewcommand{\thesection}{\arabic{section}}
\renewcommand{\thesubsection}{\arabic{section}.\arabic{subsection}}

\begin{document}

\pagestyle{empty}

\title{B DECAYS AT THE LHC\\[-2.3cm]}
\author{{\bf Convenors}: P.~Ball, R.~Fleischer, G.F.~Tartarelli, P.~Vikas,
G.~Wilkinson\\
  {\bf Contributing authors}: 
J.~Baines, S.P.~Baranov, P.~Bartalini, M.~Beneke, E.~Bouhova,
G.~Buchalla, I.~Caprini, F.~Charles, J.~Charles, Y.~Coadou,
P.~Colangelo, P.~Colrain, J.~Damet, F.~De Fazio, A.~Dighe,
H.~Dijkstra, P.~Eerola, N.~Ellis, B.~Epp, S.~Gadomski, P.~Galumian,
I.~Gavrilenko, S.~George, V.M.~Ghete, V.~Gibson, L.~Guy, Y.~Hasegawa, P.~Iengo,
A.~Jacholkowska, R.~Jones, A.~Khodjamirian, E.~Kneringer, P.~Koppenburg,
H.~Korsmo, N.~Labanca, L.~Lellouch, M.~Lehto, Y.~Lemoigne, J.~Libby,
J.~Matias, S.~Mele, M.~Misiak, A.M.~Nairz, T.~Nakada, A.~Nikitenko, N.~Nikitin,
A.~Nisati, F.~Palla, E.~Polycarpo, J.~Rademacker, F.~Rizatdinova,
S.~Robins, D.~Rousseau, W.~Ruckstuhl, M.A.~Sanchis, O.~Schneider, M.~Shapiro,
C.~Shepherd-Themistocleous, P.~Sherwood, L.~Smirnova, M.~Smizanska, 
A.~Starodumov, N.~Stepanov, Z.~Xie, N.~Zaitsev\\
{\bf Acknowledgements}: Thanks to M.~Ciuchini, A.~Deandrea, G.~Eigen,
A.~Lenz, L.~Moroni, N.~Neufeld and D.~Wyler who also contributed to
the workshop.}
\maketitle

\begin{flushright}
~\\[-8cm]CERN--TH/2000--101\\[6.6cm]
\end{flushright} 

\begin{abstract}
We review the prospects for $B$-decay physics at the LHC as discussed
in the 1999 workshop on Standard Model physics at the LHC.\\[-0.8cm]
\end{abstract}

\tableofcontents

\setcounter{page}{0}
\newpage

\pagestyle{plain}

\setcounter{equation}{0}
\section[THEORETICAL INTRODUCTION]{THEORETICAL 
INTRODUCTION\protect\footnote{Section coordinators: 
P. Ball and R. Fleischer, with help from G. Buchalla and L. Lellouch.}}

The exploration of physics with $b$-flavoured hadrons offers a very
fertile testing ground for the Standard Model (SM) description of
electroweak interactions. One of the key problems to be studied is the
phenomenon of CP violation, which, although already discovered in 1964
by Christenson, Cronin, Fitch and Turlay in the neutral kaon system 
\cite{CP-discovery}, is still one of the experimentally least
constrained phenomena. Another main topic is the study of rare $b$
decays induced by flavour changing neutral current (FCNC) transitions
$b\to s,d$, which are loop-suppressed in the SM and thus very
sensitive to new-physics effects.

During the last few years, $B$ physics has received a lot of attention,
both from theorists and experimentalists, and we are presently at
the beginning of the $B$-factory era in particle physics. The BaBar 
(SLAC), BELLE (KEK) and  HERA-B (DESY) detectors have already seen their 
first events,
and  CLEO-III (Cornell), CDF-II and D0-II (Fermilab) will start 
taking data in the near future (see \cite{paula} for a recent experimental
overview). Although the physics potential of these 
experiments is very promising, it may well be that the ``definite'' answer 
in the search for new physics in $B$ decays will be left for 
second-generation $B$ experiments at hadron machines. In the following, 
we will give an overview of the $B$-physics potential of the LHC experiments 
ATLAS, CMS and LHCb, with the main focus on SM physics.

\subsection{CP Violation in the B System}
Among the most interesting aspects and unsolved
 mysteries\label{p:mystery} of modern 
particle physics is the violation of CP symmetry. Studies of CP 
violation are particularly exciting, as they may open a window to the 
physics beyond the SM. There are many interesting ways
to explore CP violation, for instance through certain rare $K$- or $D$-meson
decays (a very recent comprehensive description of all aspects of
CP symmetry and its
violation can be found in Ref.~\cite{book}). 
However, for testing the SM description of CP 
violation in a quantitative way, the $B$ system appears to be most 
promising \cite{CP-revs1,CP-revs2,BaBar}.

\subsubsection{The SM Description of CP Violation}
Within the framework of the SM, CP violation is closely related 
to the Cabibbo--Kobayashi--Maskawa (CKM) matrix \cite{cab,KM}, connecting 
the electroweak eigenstates $(d',s',b')$ of the down, strange and bottom 
quarks with their mass eigenstates $(d,s,b)$ through the following unitary 
transformation:
\begin{equation}\label{ckm}
\left(\begin{array}{c}
d'\\
s'\\
b'
\end{array}\right)=\left(\begin{array}{ccc}
V_{ud}&V_{us}&V_{ub}\\
V_{cd}&V_{cs}&V_{cb}\\
V_{td}&V_{ts}&V_{tb}
\end{array}\right)\cdot
\left(\begin{array}{c}
d\\
s\\
b
\end{array}\right)\equiv\hat V_{\mbox{{\scriptsize CKM}}}\cdot
\left(\begin{array}{c}
d\\
s\\
b
\end{array}\right).
\end{equation}
The elements of the CKM matrix describe charged-current couplings, as
can be seen easily by expressing the nonleptonic charged-current 
interaction Lagrangian in terms of the electroweak eigenstates (\ref{ckm}):
\begin{equation}\label{cc-lag2}
{\cal L}_{\mbox{{\scriptsize int}}}^{\mbox{{\scriptsize CC}}}=
-\frac{g_2}{\sqrt{2}}\left(\begin{array}{ccc}\bar
u_{\mbox{{\scriptsize L}}},& \bar c_{\mbox{{\scriptsize L}}},&
\bar t_{\mbox{{\scriptsize L}}}\end{array}\right)\gamma^\mu\,\hat
V_{\mbox{{\scriptsize CKM}}}
\left(
\begin{array}{c}
d_{\mbox{{\scriptsize L}}}\\
s_{\mbox{{\scriptsize L}}}\\
b_{\mbox{{\scriptsize L}}}
\end{array}\right)W_\mu^\dagger\quad+\quad h.c.,
\end{equation}
where the gauge coupling $g_2$ is related to the gauge group 
SU$_{\rm L}$(2) and the $W_\mu^{(\dagger)}$ fields 
describe the charged $W$-bosons.

In the case of three generations, three generalized Cabibbo-type angles 
\cite{cab} and a single {\it complex phase} \cite{KM} are needed in order 
to parametrize the CKM matrix. This complex phase allows one to accommodate
CP violation in the SM, as was pointed out by Kobayashi and
Maskawa in 1973~\cite{KM}. A closer look shows that CP-violating observables 
are proportional to the following combination of CKM matrix elements
\cite{jarlskog}:
\begin{equation}
J_{\rm CP}=\pm\,\mbox{Im} \left(V_{ik}V_{jl}V_{il}^\ast 
V_{jk}^\ast\right)\quad(i\not=j,\,l\not=k)\,,
\end{equation}
which represents a measure of the ``strength'' of CP violation in the 
SM. Since $J_{\rm CP}={\cal O}(10^{-5})$, CP 
violation is a small effect. However, in scenarios of new physics 
\cite{new-phys}, typically several additional complex couplings are 
present, leading to new sources of CP violation.

As far as phenomenological applications are concerned, the following 
parametrization of the CKM matrix, the ``Wolfenstein parametrization'' 
\cite{wolf}, which corresponds to a phenomenological expansion in powers 
of the small quantity $\lambda\equiv|V_{us}|=\sin\theta_{\rm C}
\approx0.22$, turns out to be very useful: 
\begin{equation}\label{wolf2}
\hat V_{\mbox{{\scriptsize CKM}}} =\left(\begin{array}{ccc}
1-\frac{1}{2}\lambda^2 & \lambda & A\lambda^3 (\rho-i\eta) \\
-\lambda & 1-\frac{1}{2}\lambda^2 & A\lambda^2\\
A\lambda^3(1-\rho-i\eta) & -A\lambda^2 & 1
\end{array}\right)+\,{\cal O}(\lambda^4).
\end{equation}
The terms of ${\cal O}(\lambda^4)$ can be taken into account systematically 
\cite{BLO}, and will play an important r\^{o}le below. 

\subsubsection{The Unitarity Triangle(s) of the CKM Matrix}
\begin{figure}
\begin{tabular}{lr}
 \vspace{-0.3cm}
   \epsfysize=4.5cm
   \epsffile{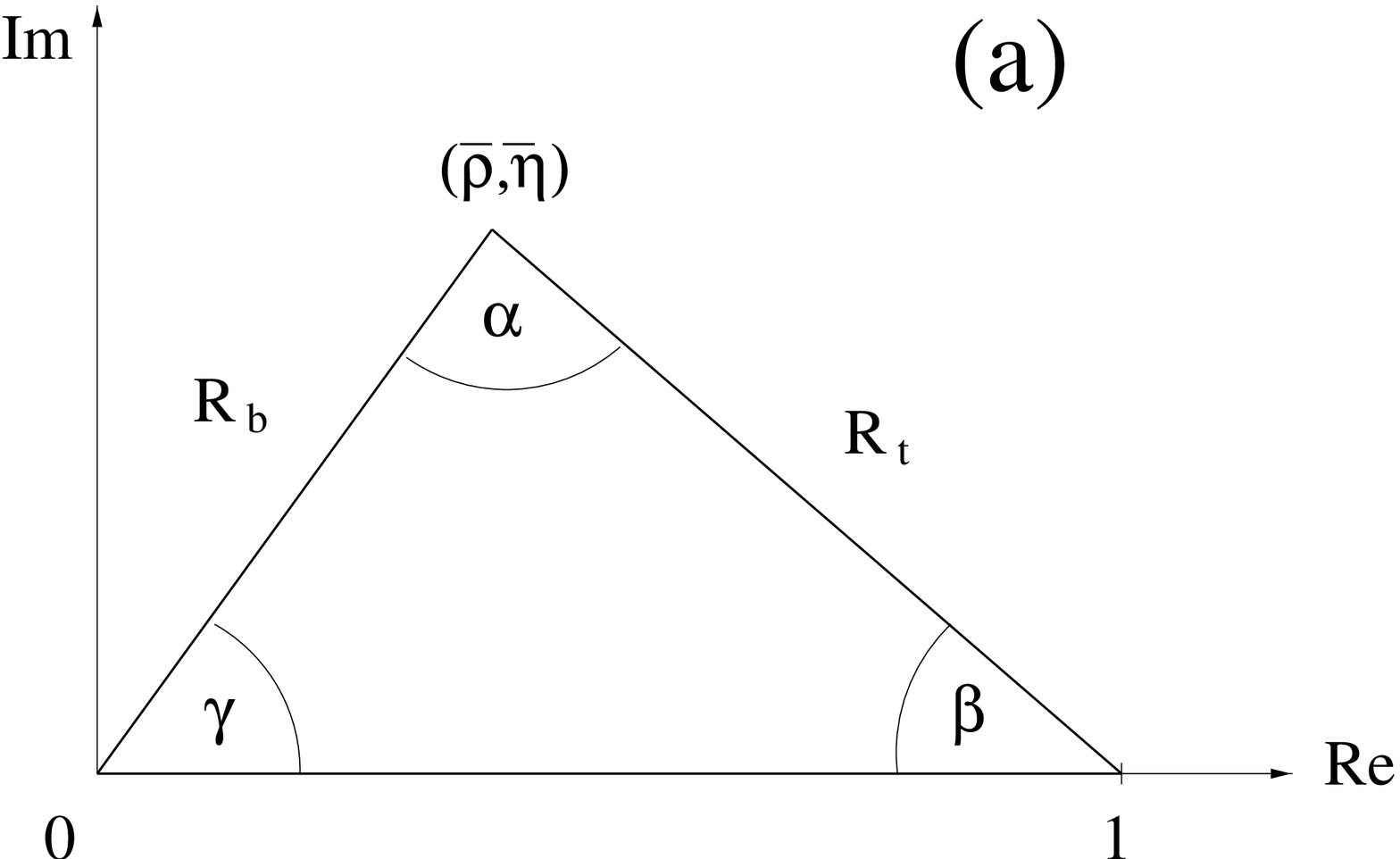}
&
   \epsfysize=4.5cm
   \epsffile{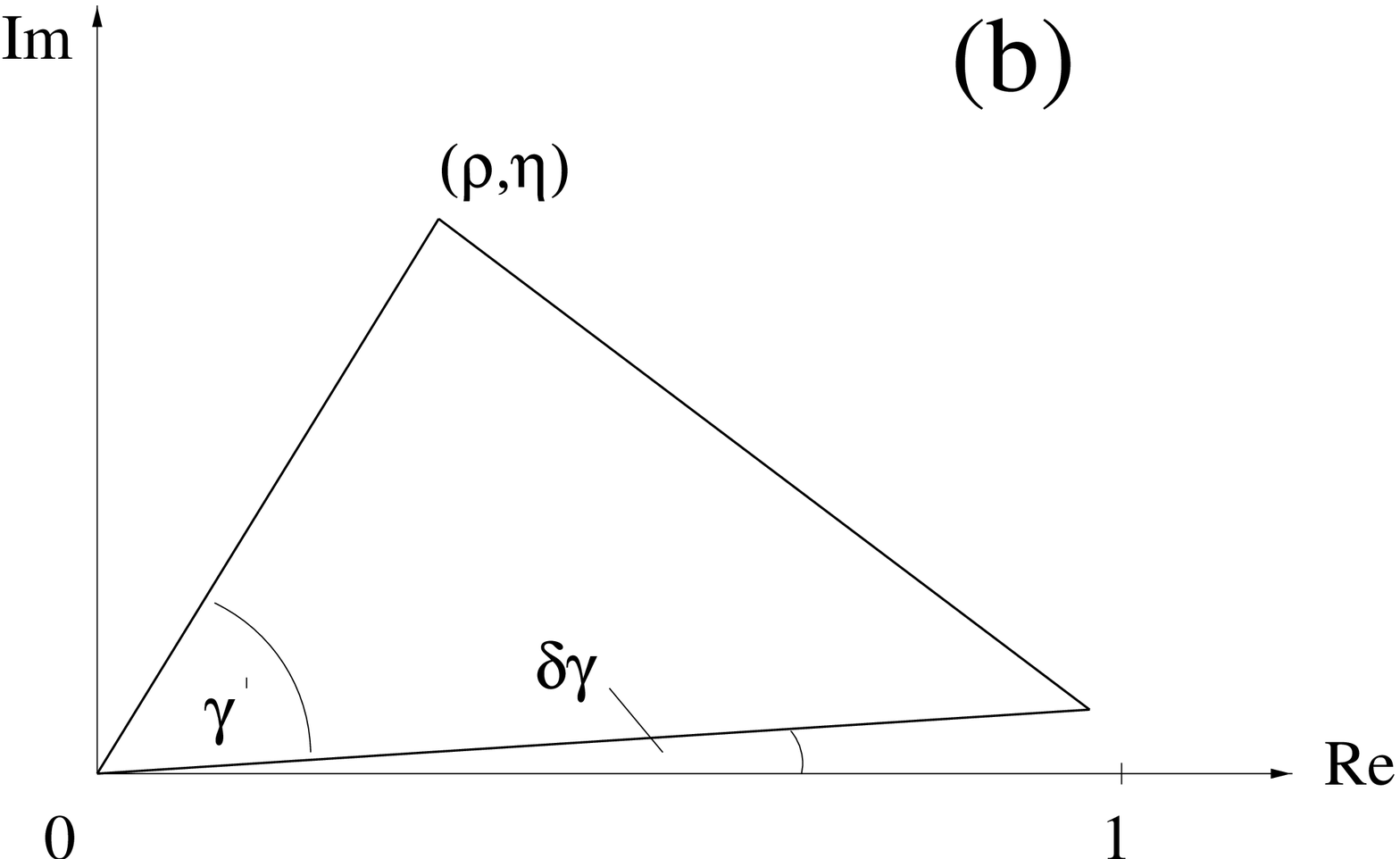}
\end{tabular}
\caption[]{The two non-squashed unitarity triangles of the CKM matrix: 
(a) and (b) correspond to the orthogonality relations (\ref{UT1}) and 
(\ref{UT2}), respectively.}
\label{fig:UT}
\end{figure}

Concerning tests of the CKM picture of CP violation, the central targets 
are the {\it unitarity triangles} of the CKM matrix. The unitarity of the 
CKM matrix, which is described by
\begin{equation}
\hat V_{\mbox{{\scriptsize CKM}}}^{\,\,\dagger}\cdot\hat 
V_{\mbox{{\scriptsize CKM}}}=
\hat 1=\hat V_{\mbox{{\scriptsize CKM}}}\cdot\hat V_{\mbox{{\scriptsize 
CKM}}}^{\,\,\dagger},
\end{equation}
leads to a set of 12 equations, consisting of 6 normalization 
and 6 orthogonality relations. The latter can be represented as 6 triangles
in the complex plane, which all have 
the same area \cite{AKL}. However, in only 
two of them, all three sides are of comparable magnitude 
${\cal O}(\lambda^3)$, while in the remaining ones, one side is suppressed 
relative to the others by ${\cal O}(\lambda^2)$ or ${\cal O}(\lambda^4)$.
The orthogonality relations describing the non-squashed triangles are 
given as follows:
\begin{eqnarray}
V_{ud}\,V_{ub}^\ast+V_{cd}\,V_{cb}^\ast+V_{td}\,V_{tb}^\ast&=&0\label{UT1}\\
V_{ud}^\ast\, V_{td}+V_{us}^\ast\, V_{ts}+V_{ub}^\ast\, V_{tb}&=&0.\label{UT2}
\end{eqnarray}
The two non-squashed triangles agree at leading order in the Wolfenstein 
expansion, i.e.\ at ${\cal O}(\lambda^3)$, so that we actually have to 
deal with a single triangle at this order, which is usually referred to as 
``the'' unitarity triangle of the CKM matrix \cite{ut}. However, in the 
LHC era, the experimental accuracy will be so tremendous that we will have
to take into account the next-to-leading order terms of the Wolfenstein 
expansion, and distinguish between the unitarity triangles 
described by (\ref{UT1}) and (\ref{UT2}), which are illustrated in 
Fig.~\ref{fig:UT}. Here, $\overline{\rho}$ and $\overline{\eta}$ are 
related to the Wolfenstein parameters $\rho$ and $\eta$ through \cite{BLO}
\begin{equation}
\overline{\rho}\equiv\left(1-\lambda^2/2\right)\rho,\quad
\overline{\eta}\equiv\left(1-\lambda^2/2\right)\eta.
\end{equation}
Note
the angles of the triangles,  in particular those designated by
$\alpha$, $\beta$, $\gamma$ and $\delta \gamma$.  These will be
referred to frequently throughout this report.
The sides $R_b$ and $R_t$ of the unitarity triangle shown in 
Fig.~\ref{fig:UT}(a) are given as follows:
\begin{eqnarray}
R_b&=&\left(1-\frac{\lambda^2}{2}\right)\frac{1}{\lambda}\left|
\frac{V_{ub}}{V_{cb}}\right|\,=\,\sqrt{\overline{\rho}^2+\overline{\eta}^2}
\,=\,0.41\pm0.07,\label{Rb-intro}\\
R_t&=&\frac{1}{\lambda}\left|\frac{V_{td}}{V_{cb}}\right|\,=\,
\sqrt{(1-\overline{\rho})^2+\overline{\eta}^2}\,=\,{\cal O}(1),
\end{eqnarray}
and will also appear in the following discussion.

\subsubsection{Nonleptonic $B$ Decays and Low-Energy Effective Hamiltonians}
With respect to testing the SM description of CP violation, 
the major r\^ole is played by nonleptonic $B$ decays, which can be divided 
into three decay classes: decays receiving both ``tree'' and ``penguin'' 
contributions, pure ``tree'' decays, and pure ``penguin'' decays. There are 
two types of penguin topologies: gluonic (QCD) and electroweak (EW) penguins, 
which are related to strong and electroweak interactions, respectively. 
Because of the large top-quark mass, also the latter operators play an 
important r\^ole in several processes \cite{RF-EWP1,RF-EWP2,RF-rev}. 

In order to analyse nonleptonic $B$ decays theoretically, one uses 
low-energy effective Hamiltonians, which are calculated by making use 
of the operator product expansion, yielding transition matrix elements 
of the following structure:
\begin{equation}\label{ee2}
\langle f|{\cal H}_{\mbox{\scriptsize eff}}|i\rangle\propto\sum_k C_k(\mu)
\langle f|Q_k(\mu)|i\rangle\,.
\end{equation}
The operator product expansion allows one to separate the short-distance
contributions to this transition amplitude from the long-distance 
ones, which are described by perturbative Wilson coefficient 
functions $C_k(\mu)$ and non-perturbative hadronic matrix elements 
$\langle f|Q_k(\mu)|i\rangle$, respectively. As usual, $\mu$ denotes an 
appropriate renormalization scale. 

In the case of $|\Delta B|=1$, $\Delta C=\Delta U=0$ transitions, which 
will be of particular interest for the exploration of CP violation in the
$B$ system, we have
\begin{equation}\label{e3}
{\cal H}_{\mbox{{\scriptsize eff}}}={\cal H}_{\mbox{{\scriptsize 
eff}}}(\Delta B=-1)+{\cal H}_{\mbox{{\scriptsize eff}}}(\Delta B=-1)^\dagger,
\end{equation}
where 
\begin{equation}\label{e4}
{\cal H}_{\mbox{{\scriptsize eff}}}(\Delta B=-1)=\frac{G_{\mbox{{\scriptsize 
F}}}}{\sqrt{2}}\left[\sum\limits_{j=u,c}V_{jq}^\ast V_{jb}\left\{\sum
\limits_{k=1}^2Q_k^{jq}\,C_k(\mu)+\sum\limits_{k=3}^{10}Q_k^{q}\,C_k(\mu)
\right\}\right].
\end{equation}
Here $\mu={\cal O}(m_b)$ is a renormalization scale, the $Q_k^{jq}$ are 
four-quark operators, the label $q\in\{d,s\}$ corresponds to $b\to d$ and 
$b\to s$ transitions, and $k$ distinguishes between current--current 
$(k\in\{1,2\})$, QCD $(k\in\{3,\ldots,6\})$ and EW $(k\in\{7,\ldots,10\})$ 
penguin operators. The calculation of such low-energy effective Hamiltonians 
has been reviewed in \cite{BBL-rev}, where the four-quark operators 
$Q_k^{jq}$ are given explicitly, and where also numerical values for their 
Wilson coefficient functions can be found.

\subsubsection{$B$--$\bar B$ Mixing}

The eigenstates of flavour, $B_q=(\bar bq)$ and $\bar B_q=(b\bar q)$
($q=d$, $s$), degenerate in pure QCD, mix on account of weak
interactions. The quantum mechanics of the two-state system,
with basis \{$|1\rangle$, $|2\rangle$\} $\equiv$
\{$|B_q\rangle$, $|\bar B_q\rangle$\},
is described by a complex, $2\times 2$ Hamiltonian matrix
\begin{equation}\label{hmg}
{\bf H}={\bf M}-\frac{i}{2}{\bf\Gamma}=
\left(
\begin{array}{cc}
M & M_{12}\\
M^*_{12} & M \\
\end{array}\right)
-\frac{i}{2}
\left(
\begin{array}{cc}
\Gamma & \Gamma_{12}\\
\Gamma^*_{12} & \Gamma \\
\end{array}\right)
\end{equation}
with Hermitian matrices ${\bf M}$ and ${\bf\Gamma}$.
The off-diagonal elements in (\ref{hmg}) arise from $\Delta B=2$
flavour-changing transitions with virtual ($M_{12}$) or real
intermediate states ($\Gamma_{12}$), in the latter case corresponding
to decay channels common to $B$ and $\bar B$.

\begin{figure}[t]
   \vspace{-0.5cm}
   \epsfxsize=10cm
   \centerline{\epsffile{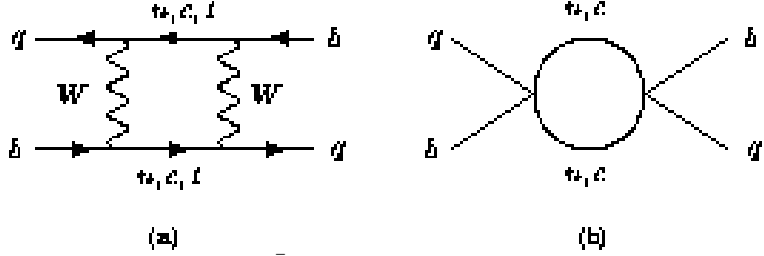}}
   \vspace*{-0.5cm}
\caption{\label{fig:boxb2} (a): General box diagram describing
$B$--$\bar B$ mixing. (b): Special case of diagram (a) with
internal $u$ and $c$, whose absorptive part determines $\Gamma_{12}$.}
\end{figure}

Diagonalizing (\ref{hmg}), one obtains the physical eigenstates
$B_H$ (`heavy'), $B_L$ (`light') and the corresponding eigenvalues
$M_{H,L}-\frac{i}{2}\Gamma_{H,L}$. The mass and width differences
read
\begin{equation}\label{delm}
\Delta M_q \equiv M_H^{(q)}-M_L^{(q)} =2|M^{(q)}_{12}|,\qquad
\Delta\Gamma_q \equiv \Gamma_H^{(q)}-\Gamma_L^{(q)} = 
\frac{2{\rm Re}(M^{(q)*}_{12}\Gamma^{(q)}_{12})}{|M^{(q)}_{12}|}\,.
\end{equation}
$\Delta M$ is positive by definition, $\Delta\Gamma$
is defined in such a way that a negative value\footnote{Note that also
  the opposite sign convention for 
$\Delta\Gamma$ is used in the literature.} is obtained in
the SM for the case of $B_s$, where a sizable width
difference is expected.
In the SM, the off-diagonal elements $M_{12}$ and $\Gamma_{12}$
inducing B mixing are described by the
box diagrams in Fig.~\ref{fig:boxb2}.

Detailed numerical results will be given Sec.~\ref{sec:mix}; here we only
summarize a few important general characteristics of 
$\Delta B=2$ second-order weak processes.
The relative size of the various contributions is controlled
by CKM quantities and quark masses.
With $\lambda^{(q)}_i=V^*_{iq}V_{ib}$, and denoting the magnitude
in powers of the Wolfenstein parameter $\lambda$, we have
$\lambda^{(d)}_u\sim\lambda^{(d)}_c\sim\lambda^{(d)}_t\sim\lambda^3$
for $B_d$, and
$\lambda^{(s)}_u\sim\lambda^4$,
$\lambda^{(s)}_c\sim\lambda^{(s)}_t\sim\lambda^2$
for $B_s$.
Because the box amplitude strongly grows with large
($\gg m_b$) internal quark masses $m_i$, proportional
to $m^2_i$ for $m_i\gg M_W$, it is clear, considering the
above CKM hierarchy, that the top-quark contribution completely
dominates the {\it dispersive part\/} $M_{12}$.
The remaining contributions ($i=u$, $c$) are safely negligible
for both the $B_d$ and $B_s$ system.
Since $m_t$, $M_W\gg m_b$, $M_{12}$ is described by an effectively
local interaction already at scales far above $m_b$.
External mass scales can thus be neglected and the resulting
$\Delta B=2$ effective Hamiltonian is governed by a single
operator. It acquires the simple form
\begin{equation}\label{hbf2}
{\cal H}^{\Delta B=2}_{\mbox{\scriptsize eff}}=(V^*_{tq}V_{tb})^2 
C\left(x_t\right)
(\bar qb)_{V-A}(\bar qb)_{V-A}
\end{equation}
with $C$ the short-distance Wilson-coefficient and $x_t=m_t^2/m_W^2$,
whence $M_{12}$ is obtained as
\begin{equation}\label{m12h}
M_{12}=\frac{1}{2M_B}\langle B|
{\cal H}^{\Delta B=2}_{\mbox{\scriptsize eff}}|\bar B\rangle.
\end{equation}

For the {\it absorptive part\/} $\Gamma_{12}$ the situation
is more complicated.
First of all, the top contribution, dominant for $M_{12}$,
cannot contribute to $\Gamma_{12}$, since top quarks
are kinematically forbidden as on-shell final states in $B$ decays.
$\Gamma_{12}$ is then determined by the (absorptive parts of)
box diagrams with up and charm quarks. Both up and charm
are important for $B_d$ because
$\lambda^{(d)}_u\sim\lambda^{(d)}_c$.
In the case of $B_s$, the up-quark sector is negligible
($\lambda^{(s)}_u\ll\lambda^{(s)}_c$).
In calculating $\Gamma_{12}$, the
heavy $W$-boson lines can be contracted to form two local
$\Delta B=1$ four-quark interactions (Fig.~\ref{fig:boxb2}(b)). 
By contrast,
$u$ and $c$ are lighter than the relevant scale of the process
($\sim m_b$) and cannot be integrated out, unlike the top quark in
$M_{12}$.
Consequently, $\Gamma_{12}$ is given as the matrix element of
a non-local (or `bi-local') product of two
local $\Delta B=1$ Hamiltonian operators 
${\cal H}^{\Delta B=1}_{\mbox{\scriptsize eff}}$,
the usual effective weak Hamiltonian describing $B$ decays:
\begin{equation}\label{g12h}
\Gamma_{12}=\frac{1}{2M_B}\langle B|{\rm Im}\,
i\int d^4x\, T\,{\cal H}^{\Delta B=1}_{\mbox{\scriptsize eff}}(x) 
{\cal H}^{\Delta B=1}_{\mbox{\scriptsize eff}}(0)|\bar B\rangle\,.
\end{equation}
To lowest order in strong interactions, (\ref{g12h})
corresponds to the absorptive part (Im) of the diagram in
Fig.~\ref{fig:boxb2}(b).
Taking the absorptive part inside the formal
expression (\ref{g12h}), the T-product is transformed into an
ordinary product of the two factors 
${\cal H}^{\Delta B=1}_{\mbox{\scriptsize eff}}$.
Inserting a complete set of hadronic final states $f$ gives
\begin{equation}
\Gamma_{12}^{\mbox{\scriptsize (hadron)}} = \sum_f 
\langle B|{\cal H}^{\Delta B=1}_{\mbox{\scriptsize eff}}|f\rangle
\langle f|{\cal H}^{\Delta B=1}_{\mbox{\scriptsize eff}}|\bar B\rangle,
\end{equation}
where one recognizes the usual expression for a decay rate,
generalized here to the off-diagonal entry $\Gamma_{12}$.
This connection, which allows one to write $\Gamma_{12}$ in
(\ref{g12h}) as the absorptive part of the
$\bar B\to B$ forward scattering amplitude, is known as the
optical theorem. $\Gamma_{12}^{\mbox{\scriptsize (hadron)}}$ does,
however, escape
direct calculation, which instead starts from (\ref{g12h}):
 taking advantage
of the large momentum $\sim m_b\gg\Lambda_{\rm QCD}$ flowing through
the internal $u$ and $c$ quark lines of the box diagram, one expands
the operator product into a series of local operators
\cite{BIG1}:
\begin{equation}\label{hqe}
\Gamma^{\mbox{\scriptsize (quark)}}_{12} = \frac{1}{2M_B}\,\langle B \mid
{\rm Im}\, i\int d^4x\, T\,{\cal H}^{\Delta B=1}_{\mbox{\scriptsize eff}}(x) 
{\cal H}^{\Delta B=1}_{\mbox{\scriptsize eff}}(0) \mid \bar B\rangle =
\frac{1}{2M_B}\,\sum_n \frac{C_n}{m^n_b} 
\langle B\mid Q^{\Delta B=2}_n \mid \bar B\rangle\,.
\end{equation}
The identification of the exact
$\Gamma^{\mbox{\scriptsize (hadron)}}_{12}$
with the approximation
$\Gamma^{\mbox{\scriptsize (quark)}}_{12}$ based on the
heavy quark expansion is equivalent to the assumption of local
quark-hadron duality (`local' in this context refers to the fact
that the large energy scale $m_b$ is, in practice, a fixed number,
rather than a variable allowing for the consideration of some
(`global') averaging procedure).
When viewed as a function of $m_b$, $\Gamma^{\mbox{\scriptsize (hadron)}}_{12}$
is expected to include terms of the form 
$\exp(-(m_b/\Lambda_{\rm QCD})^k)\,\sin((m_b/\Lambda_{\rm QCD})^k)$.
Such oscillating and exponentially suppressed terms are related to
the opening of new decay channels as $m_b$ is increased.
They are however completely missed in $\Gamma^{\mbox{\scriptsize
    (quark)}}_{12}$ 
to any
finite order in the heavy quark expansion, which is just a
power series in $\Lambda_{\rm QCD}/m_b$. Of course, for
asymptotically large $m_b/\Lambda_{\rm QCD}\to\infty$, these terms
vanish much faster than power corrections. In any case
$\Gamma^{\mbox{\scriptsize (quark)}}_{12}\to\Gamma^{\mbox{\scriptsize 
(hadron)}}_{12}$ in the strict limit $m_b\to\infty$.
Nevertheless, for realistic values of $m_b$ those terms may introduce
a deviation of $\Gamma^{\mbox{\scriptsize (quark)}}_{12}$ from 
the correct $\Gamma^{\mbox{\scriptsize (hadron)}}_{12}$ (beyond the omission of
higher power corrections). This error is referred to as a violation
of local duality. Theoretical knowledge from first principles about 
duality violating contributions is so far rather limited.
Interesting general discussions and further information can be 
found in Refs.~\cite{BSZ,BIG2,GL}.

\subsubsection{CP Violation in Neutral $B$ Decays into CP 
Eigenstates}\label{intro-neutCP}
The description of CP violation in terms of weak phases becomes
particularly simple for decays of neutral $B_q$-mesons ($q\in\{d,s\}$) into CP 
self-conjugate final states $|f\rangle$, satisfying the relation 
\begin{equation}
({\cal CP})|f\rangle=\pm\,|f\rangle. 
\end{equation}
In this case, the corresponding time-dependent CP asymmetry can be 
expressed as
\begin{eqnarray}
{\cal A}_{\rm CP}(t)&\equiv&\frac{\Gamma(B^0_q(t)\to f)-
\Gamma(\overline{B^0_q}(t)\to f)}{\Gamma(B^0_q(t)\to f)+
\Gamma(\overline{B^0_q}(t)\to f)}\nonumber\\
&=&2\,e^{-\Gamma_q t}\left[\frac{{\cal A}_{\rm CP}^{\rm dir}(B_q\to f)
\cos(\Delta M_q t)+{\cal A}_{\rm CP}^{\rm mix}(B_q\to f)\sin(\Delta M_q t)}{
e^{-\Gamma_{\rm H}^{(q)}t}+e^{-\Gamma_{\rm L}^{(q)}t}+
{\cal A}_{\rm \Delta\Gamma}(B_q\to f)\left(e^{-\Gamma_{\rm H}^{(q)}t}-
e^{-\Gamma_{\rm L}^{(q)}t}\right)} \right],\label{ee6}
\end{eqnarray}
where $\Delta M_q\equiv M_{\rm H}^{(q)}-M_{\rm L}^{(q)}$ denotes the
mass difference between the $B_q$ mass eigenstates, and 
$\Gamma_{\rm H,L}^{(q)}$ are the corresponding decay widths, with 
\begin{equation}
\Gamma_q\equiv\frac{\Gamma_{\rm H}^{(q)}+\Gamma_{\rm L}^{(q)}}{2}\,.
\end{equation} 
In Eq.\ (\ref{ee6}), we have separated the ``direct'' from the 
``mixing-induced'' CP-violating contributions, which are described by
\begin{equation}\label{ee7}
{\cal A}^{\mbox{{\scriptsize dir}}}_{\mbox{{\scriptsize CP}}}(B_q\to f)\equiv
\frac{1-\bigl|\xi_f^{(q)}\bigr|^2}{1+\bigl|\xi_f^{(q)}\bigr|^2}\quad
\mbox{and}\quad
{\cal A}^{\mbox{{\scriptsize mix}}}_{\mbox{{\scriptsize
CP}}}(B_q\to f)\equiv\frac{2\,\mbox{Im}\left\{\xi^{(q)}_f\right\}}{1+
\bigl|\xi^{(q)}_f\bigr|^2}\,,
\end{equation}
respectively. Here direct CP violation refers to CP-violating effects
arising directly in the corresponding decay amplitudes, whereas 
mixing-induced CP violation is due to interference effects between 
$B_q^0$--$\overline{B_q^0}$ mixing and decay processes. Whereas the
width difference $\Delta\Gamma_q$
is negligible in the $B_d$ system, it may be sizeable in the 
$B_s$ system \cite{DGamma-cal1,DGamma-cal2}, thereby providing the observable 
\begin{equation}\label{ADGam}
{\cal A}_{\rm \Delta\Gamma}(B_q\to f)\equiv
\frac{2\,\mbox{Re}\left\{\xi^{(q)}_f\right\}}{1+\bigl|\xi^{(q)}_f
\bigr|^2},
\end{equation}
which is not independent of ${\cal A}^{\mbox{{\scriptsize 
dir}}}_{\mbox{{\scriptsize CP}}}(B_q\to f)$ and 
${\cal A}^{\mbox{{\scriptsize mix}}}_{\mbox{{\scriptsize CP}}}(B_q\to f)$:
\begin{equation}\label{Obs-rel}
\Bigl[{\cal A}_{\rm CP}^{\rm dir}(B_s\to f)\Bigr]^2+
\Bigl[{\cal A}_{\rm CP}^{\rm mix}(B_s\to f)\Bigr]^2+
\Bigl[{\cal A}_{\Delta\Gamma}(B_s\to f)\Bigr]^2=1.
\end{equation}

Essentially all the information needed to evaluate the CP asymmetry
(\ref{ee6}) is included in the following quantity:
\begin{equation}\label{xi-expr}
\xi_f^{(q)}=\mp\,e^{-i\phi_q}\,
\frac{\sum\limits_{j=u,c}V_{jr}^\ast V_{jb}\bigl\langle f\bigl|{\cal Q}^{jr}
\bigr|\overline{B^0_q}\bigr\rangle}{\sum\limits_{j=u,c}V_{jr}
V_{jb}^\ast\bigl\langle f\bigl|{\cal Q}^{jr}
\bigr|\overline{B^0_q}\bigr\rangle}\,,
\end{equation}
where 
\begin{equation}
{\cal Q}^{jr}\equiv\sum\limits_{k=1}^2Q_k^{jr}C_k(\mu)+
\sum\limits_{k=3}^{10}Q_k^{jr}C_k(\mu),
\end{equation}
$r\in\{d,s\}$ distinguishes between $\bar b\to\bar d$ and $\bar b\to\bar s$ 
transitions, and 
\begin{equation}\label{eq:grummelbrummel}
\phi_q=\left\{\begin{array}{cr}
+2\beta&\mbox{($q=d$)}\\
-2\delta\gamma&\mbox{($q=s$)}\end{array}\right.
\end{equation}
is the weak $B_q^0$--$\overline{B_q^0}$ mixing phase.
In general, the 
observable $\xi_f^{(q)}$ suffers from hadronic uncertainties, which are 
introduced by the hadronic matrix elements in Eq.~(\ref{xi-expr}).
However, if the decay $B_q\to f$ is dominated by a single CKM amplitude, 
the corresponding matrix elements cancel, and $\xi_f^{(q)}$ takes the 
simple form
\begin{equation}\label{ee10}
\xi_f^{(q)}=\mp\exp\left[-i\left(\phi_q-\phi_{\mbox{{\scriptsize 
D}}}^{(f)}\right)\right],
\end{equation}
where $\phi_{\mbox{{\scriptsize D}}}^{(f)}$ is a weak decay phase, 
which is given as follows:
\begin{equation}
\phi_{\mbox{{\scriptsize D}}}^{(f)}=\left\{\begin{array}{cc}
-2\gamma&\mbox{for dominant 
$\bar b\to\bar u\,u\,\bar r$ CKM amplitudes,}\\
0&\,\mbox{for dominant $\bar b\to\bar c\,c\,\bar r\,$ CKM 
amplitudes.}
\end{array}\right.
\end{equation}

This simple formalism has several interesting applications, probably the 
most important one is the extraction of the CKM angle $\beta$ from
CP-violating effects in the ``gold-plated'' mode $B_d\to J/\psi\,K_{\rm S}$.
In addition to the CP-violating effects in neutral $B$ decays into CP 
eigenstates discussed above, also certain modes into non-CP eigenstates, 
for example $B_d\to D^{(\ast)\pm} \pi^\mp$ and $B_s\to D_s^\pm K^\mp$, play 
an outstanding r\^{o}le to extract CKM phases. These decays will be
discussed in
more detail in a later part of this report.

\subsubsection{The ``El Dorado'' for the LHC: the $B_s$ System}
The $e^+$--\,$e^-$ $B$-factories operating at the $\Upsilon(4S)$ resonance 
will not be in a position to explore the $B_s$ system. Since it is, 
moreover, very desirable to have large data samples available to study
$B_s$ decays, they are of particular interest for hadron machines and were
one of the central targets of this LHC workshop. There are important 
differences between the $B_d$ and $B_s$ systems:
\begin{itemize}
\item Within the SM, a large $B^0_s$--$\overline{B^0_s}$ 
mixing parameter 
$
x_s\equiv\Delta M_s/\Gamma_s={\cal O}(20) 
$
is expected, whereas the mixing phase $\phi_s=-2\lambda^2\eta$ is expected 
to be very small.

\item There may be a sizeable width difference 
$\Delta\Gamma_s/\Gamma_s={\cal O}(15\%)$, whereas $\Delta\Gamma_d$ is
negligible.
\end{itemize}

\noindent The mass difference $\Delta M_s$ plays an important r\^ole
to constrain the apex of the unitarity triangle shown in 
Fig.~\ref{fig:UT}(a), and the non-vanishing width difference 
$\Delta\Gamma_s$ may allow studies of CP-violating effects in 
``untagged'' $B_s$ rates, \cite{dun}--\cite{RF-Bs}, which are defined as 
follows:
\begin{equation}\label{untag-def}
\Gamma_s[f(t)]\equiv\Gamma(B^0_s(t)\to f)+\Gamma(\overline{B^0_s}(t)
\to f)=\mbox{PhSp}\times\left[R_{\rm H}\,e^{-\Gamma_{\rm H}^{(s)}t}+
R_{\rm L}\,e^{-\Gamma_{\rm L}^{(s)}t}\right],
\end{equation}
where  ``PhSp'' denotes an appropriate, straightforwardly calculable 
phase-space factor. Interestingly, there are no rapid oscillatory 
$\Delta M_st$ terms present in this expression. Although it should be 
no problem to resolve these $B^0_s$--$\overline{B^0_s}$ oscillations 
at the LHC, studies of such untagged rates, which allow the extraction of 
the observable ${\cal A}_{\Delta\Gamma}$ introduced in (\ref{ADGam}) 
as
\begin{equation}
{\cal A}_{\Delta\Gamma}=\frac{R_{\rm H}-R_{\rm L}}{R_{\rm H}+R_{\rm L}},
\end{equation}
are interesting in terms of efficiency, acceptance and purity.

\subsubsection{CP Violation in Charged $B$ Decays}\label{subsubcharged}
Since there are no mixing effects present in the charged $B$-meson system, 
non-vanishing CP asymmetries of the kind 
\begin{equation}\label{CP-charged}
{\cal A}_{\mbox{{\scriptsize CP}}}(B^+\to\overline{f})
\equiv\frac{\Gamma(B^+\to\overline{f})-
\Gamma(B^-\to f)}{\Gamma(B^+\to\overline{f})+\Gamma(B^-\to f)}
\end{equation}
would give us unambiguous evidence for ``direct'' CP violation in the 
$B$ system, which has recently been demonstrated in the kaon system by 
the new experimental results of the KTeV (Fermilab) and NA48 (CERN) 
collaborations for $\mbox{Re}(\varepsilon'/\varepsilon)$ \cite{epsprime}. 

The CP asymmetries (\ref{CP-charged}) arise from the interference between 
decay amplitudes with both different CP-violating weak and different 
CP-conserving strong phases. In the SM, the weak phases are 
related to the phases of the CKM matrix elements, whereas the strong phases 
are induced by final-state-interaction processes. In general, the strong 
phases introduce severe theoretical uncertainties into the calculation of 
${\cal A}_{\mbox{{\scriptsize CP}}}(B^+\to\overline{f})$, thereby destroying 
the clean relation to the CP-violating weak phases. However, there is an 
important tool to overcome these problems, which is provided by {\it amplitude 
relations} between certain nonleptonic $B$ decays. There are two kinds 
of such relations:
\begin{itemize}
\item Exact relations, which involve $B\to DK$ decays (pioneered by 
Gronau and Wyler \cite{gw}).
\item Approximate relations, which rely on the flavour symmetries of 
strong interactions and certain plausible dynamical assumptions, and
involve $B\to \pi K$, $\pi\pi$, $K\overline{K}$ decays (pioneered by 
Gronau, Hern\'andez, London and Rosner \cite{GRL,GHLR}).
\end{itemize}
Unfortunately, the $B\to DK$ approach, which allows a {\it theoretically 
clean} determination of $\gamma$, makes use of certain amplitude triangles 
that are expected to be very squashed ones. Moreover, there are
additional experimental problems \cite{ads}, so that this approach is 
very challenging from a practical point of view. The flavour-symmetry 
relations between the $B\to \pi K$, $\pi\pi$, $K\overline{K}$ decay 
amplitudes have received considerable attention in the literature during 
the last couple of years and led to interesting strategies to probe the 
CKM angle $\gamma$.

\subsubsection{Outline of the CP Violation Part}
The outline of the part of this Chapter dealing with aspects related to
CP violation and the determination of the angles of the unitarity triangles 
is as follows: after an overview of the experimental aspects in 
Sec.~\ref{sec:expover}, we have a closer look at the benchmark modes
to explore CP violation in Sec.~\ref{sec:benchmark}, where we will
discuss the extraction of $\beta$ from the ``gold-plated'' decay
$B_d\to J/\psi\,K_{\rm S}$, the prospects to probe $\alpha$ with  
$B_d\to\pi^+\pi^-$ and $B\to\rho\pi$ modes, as well as extractions of 
$\gamma$ from $B_d\to D^{\ast\pm}\pi^\mp$ and $B_s\to D_s^\pm K^\mp$ decays. 
Finally, we will also give a discussion of $\gamma$ determinations from 
$B\to DK$ modes. 

Section~\ref{sec:Bspsiphi} is devoted to a detailed analysis of another 
CP benchmark mode, $B_s\to J/\psi\,\phi$, which is particularly promising 
for the LHC experiments because of its favourable experimental signature and 
its rich physics potential, allowing one to extract the 
$B^0_s$--$\overline{B^0_s}$ mixing parameters $\Delta M_s$ and 
$\Delta\Gamma_s$, as well as the corresponding CP-violating weak mixing 
phase $\phi_s$. Since the CP-violating effects in $B_s\to J/\psi\,\phi$
are tiny in the SM, this channel offers an important tool to
search for new physics. 

In Sec.~\ref{sec:newstrat}, we focus on strategies to extract CKM phases 
that were not considered for the LHC experiments so far, and on new methods, 
which were developed during this workshop \cite{new-over}. We discuss 
extractions of the angle $\gamma$ from $B\to\pi K$ decays, which received 
a lot of attention in the literature during the last couple of years. 
Moreover, we discuss extractions of $\gamma$ that are provided by 
$B_{s(d)}\to J/\psi\,K_{\rm S}$ and $B_{d(s)}\to D^+_{d(s)}D^-_{d(s)}$
decays, and a simultaneous determination of $\beta$ and $\gamma$ from
a combined analysis of the decays $B_d\to \pi^+\pi^-$ and $B_s\to K^+K^-$.

Systematic error considerations in CP measurements are discussed in
Sec.~\ref{sec:syst}, and the reach for the $B^0_s$--$\overline{B^0_s}$ 
mixing parameters $\Delta M_s$ and $\Delta\Gamma_s$ is presented in
Sec.~\ref{sec:mix}.

\subsection{Rare B Decays}

By rare $B$ decays, one commonly understands heavily
Cabibbo-suppressed $b\to u$ transitions or flavour-changing neutral
currents (FCNC) $b\to s$ or $b\to d$ that in the SM
 are forbidden at
tree-level. Rare decays are an important testing ground of the
SM and offer a strategy in the search for new physics 
complementary to that of direct searches
by probing the indirect effects of new interactions in
higher order processes. Assuming the validity of the SM,
rare FCNC decays allow the measurement of the CKM matrix elements
$|V_{ts}|$ and $|V_{td}|$ and thus complement their
determination from $B^0$--$\bar{B}^0$ mixing. Any significant deviation
between these two determinations 
would hint at new physics. With the large statistics available at
the LHC, also decay spectra will be accessible, which will allow a
direct measurement of virtual new physics effects: 
in some contrast to the investigation of CP violation,
 we are in the lucky situation that the impact of new physics on
FCNC processes can be defined in a {\it model-independent}
way\footnote{Barring the possibility that new physics induces new
  operators not present in the SM, like e.g.\ a left-right symmetric
  model would do.}: at
quark-level, $b\to q$, $q=(d,s)$, transitions can be described in
terms of an effective Hamiltonian obtained by integrating out virtual
effects of heavy particles (top quark and $W$ boson in the SM):
\begin{equation}
{\cal H}_{\mbox{\scriptsize eff}}(b\to q) = -4\,\frac{G_F}{\sqrt{2}}\, V_{tb}
V_{tq}^*\, \sum\limits_{i=1}^{10} C_i(\mu) O_i(\mu).
\end{equation}
The relevant operators will be specified in Sec.~\ref{sec:rare}; 
here we would like to stress that the short-distance
coefficients $C_i(\mu)$ encode both perturbative QCD evolution between
the hadronic scale $\mu\sim O(m_b)$ and the scale of heavy particles $M_H$
and information on the physics at that scale itself, contained in
 $C_i(M_H)$. A
measurement of these coefficients that significantly deviates from the SM
expectation thus would constitute immediate and unambiguous evidence
for new physics beyond the SM. 

In these proceedings we concentrate on decays that have a favourable
experimental signature at the LHC 
and for which experimental studies exist at the
time of writing: the exclusive decays 
$B_{d,s}\to\mu^+\mu^-$, $B_d\to K^*\gamma$ and
$B_d\to K^*\mu^+\mu^-$. Although it is generally believed that theoretical
uncertainties due to nonperturbative QCD effects are larger for
exclusive than for inclusive decays, the experimental environment of a
hadronic machine renders it exceedingly difficult to perform inclusive
measurements. 
There has, however, been recent progress in the calculation of exclusive
hadronic matrix elements \cite{BB}, which narrows down the theoretical
uncertainty, and as we shall elaborate on in Sec.~\ref{sec:rare}, 
one can define
experimental observables in which a large fraction of theoretical
uncertainties cancels.

\subsection{Other B Physics Topics}

The B physics potential of the LHC is by far not exhausted by the
programme sketched above. Possible further lines of investigation
include physics with $b$-flavoured baryons (lifetime measurements,
spectra, decay dynamics etc.), physics of $b$-flavoured mesons other
than $B_{u,d,s}$ (radial and orbital excitations, $B_c$), and the
study of purely leptonic or semileptonic decays, $B_q\to e\nu$,
$B_q\to M e\nu$, where $M$ stands for a meson. {}From the theory point
of view, one major topic whose relevance goes beyond the LHC is the
calculation of nonleptonic decay amplitudes from first principles:
whereas the discussion in Secs.~3 to 5 promotes a very pragmatic
approach which aims at eliminating (``controlling'') the effects of
strong interactions by measuring a large number of observables that
are related by certain approximate symmetry principles, it remains a
challenge for theory to provide {\it predictions} for
nonleptonic decay amplitudes, both in factorization approximation and beyond.

Only a limited number of such topics were discussed during the
workshop, and so we restrict ourselves to the presentation of selected
aspects and review the present status of the theory of nonleptonic
decays in Sec.~\ref{sec:nonlept}, relevant for the prediction of decay rates in
general and the
extraction of weak phases from CP asymmetries in theoretically
``dirty'' channels in particular; in Sec.~10, we give an overview of
the physics opportunities and predicted decay rates in $B_c$ decays.

\setcounter{equation}{0}
\section[EXPERIMENTAL OVERVIEW]{EXPERIMENTAL OVERVIEW\protect\footnote{Section
    coordinator: G.F.\ Tartarelli, with help from Y. Lemoigne and
    C. Shepherd-Themistocleous.}}\label{sec:expover}

The LHC will represent a unique opportunity for B physics studies.
At a centre-of-mass energy of $\sqrt{s}=14$ TeV the production cross-section
for $b \overline{b}$ pairs will be very high. While current
theoretical predictions of the absolute value are rather uncertain, 
it is expected that it will be
about a factor of five higher than the one obtainable at the Tevatron, 
running at $\sqrt{s}=2$ TeV. 
Naturally,  therefore,  B physics has been an important consideration
in the optimisation of the LHC experimental programme.   
The two multi-purpose experiments, ATLAS~\cite{ATLASTP,ATLASPTDR}  
and CMS~\cite{CMSPTDR} have the capabilities to realise a rich
and competitive programme and a dedicated experiment, LHCb~\cite{LHCbTP}, 
will have the sole task of exploiting as wide a range of 
B physics topics as possible.

\subsection{Introduction}

\begin{figure}
 \begin{center}
   \includegraphics[width=0.46\textwidth,
     clip]{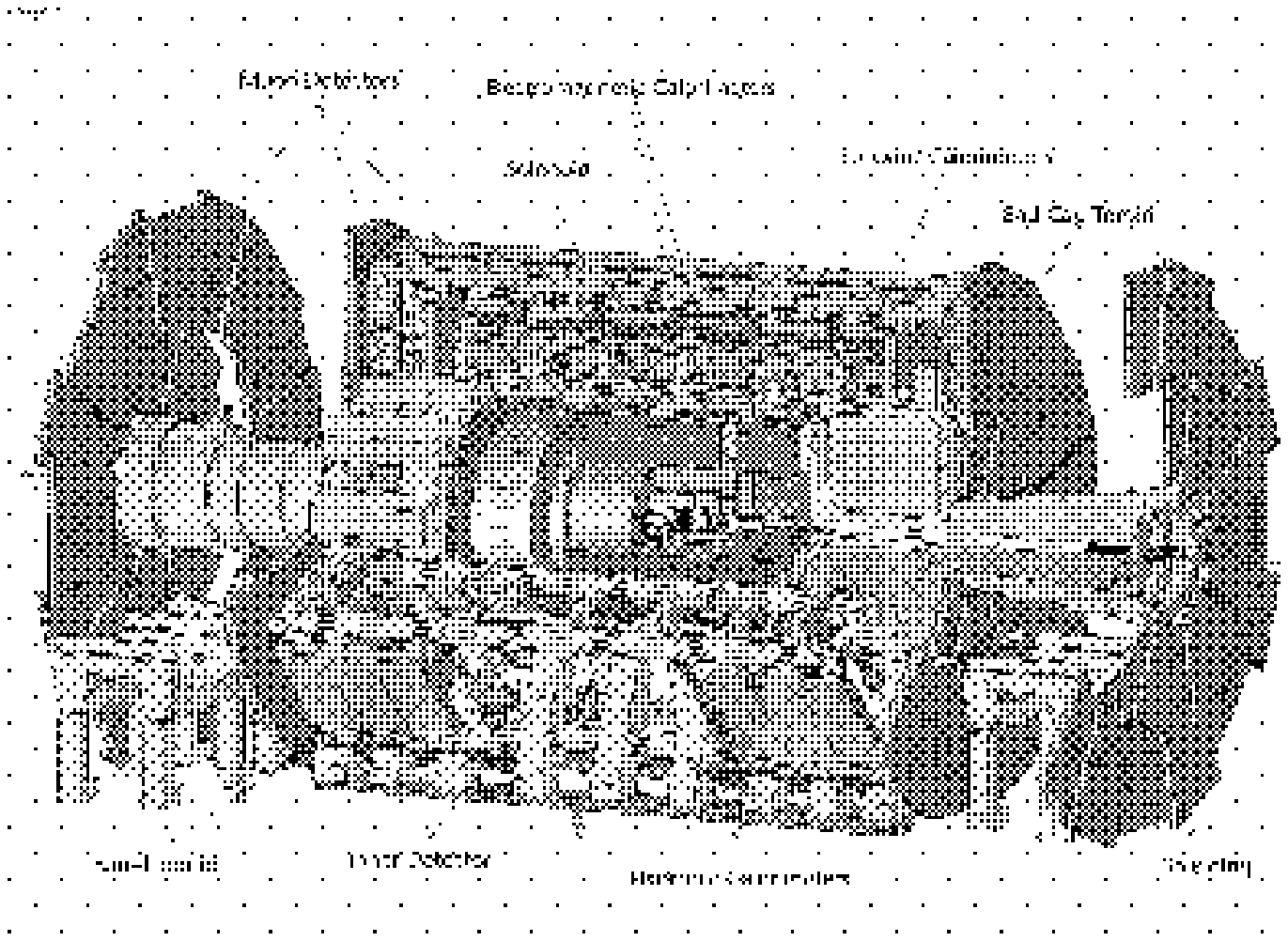}
   \includegraphics[width=0.46\textwidth,
     ]{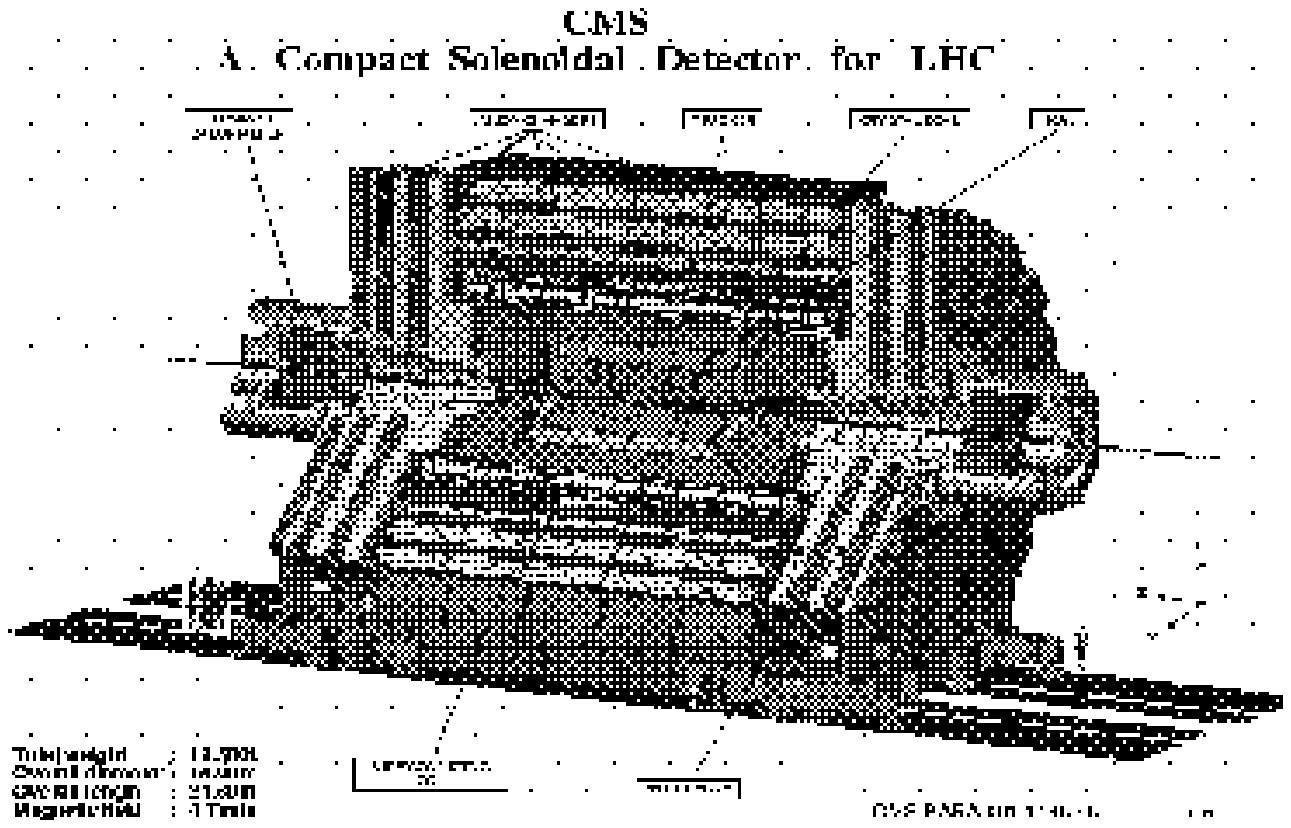}
\end{center}
\vspace*{-0.5cm}
     \caption[]{Pictorial 3D-views of the two central multi-purpose LHC
      detectors: ATLAS (left) and CMS (right).}
     \label{fig:DETECT}
\end{figure}

The ATLAS and CMS detectors (see Fig.~\ref{fig:DETECT}) 
have been designed primarily
to search for new particles, such as the Higgs boson. The detectors therefore
should be able to operate at the highest LHC luminosity and be sensitive 
to the highest mass scale. However, specific features required for 
B-hadron reconstruction
have been accommodated in the design. Both experiments have also put
large emphasis on `$b$ tagging' (discrimination between $b$ jets and jets from
light quarks, which is used in a variety of physics analyses), 
but this is not discussed in this chapter.

Both the ATLAS and CMS detectors cover
the central region of the $pp$ interaction point and have forward-backward 
and azimuthal symmetry. Inside a superconducting solenoid (generating a 
2$\,$T magnetic field in ATLAS and a 4$\,$T one in CMS, parallel to
the beam line), a multi-layer tracking 
system (ATLAS~\cite{ATtrk}, CMS~\cite{CMtrk}) covering the $|\eta|<2.5$ 
region is located. The system has 
higher granularity detector layers at small radii (silicon pixel and 
microstrip detectors) for good impact parameter 
resolution and track separation and extends to large radii to improve 
the transverse momentum 
resolution (in ATLAS the tracking system has also additional electron/pion 
separation as explained Sec.~\ref{exp_id}). In both experiments, the tracking 
system is surrounded by 
electromagnetic and hadronic calorimetry (ATLAS~\cite{ATcalo}, 
CMS~\cite{CMcalo}) which extends up to about 
$|\eta|=5.0$. Finally, outside 
the calorimeters there are high-precision muon chambers 
 (in the region 
$|\eta|<2.7$ in ATLAS~\cite{ATmu} and $|\eta|<2.4$ 
in CMS~\cite{CMmu}) and muon trigger chambers in 
a smaller pseudorapidity range ($|\eta|<2.4$ in both ATLAS and CMS).

\begin{figure}
\begin{center}
\epsfysize=7cm
\epsffile{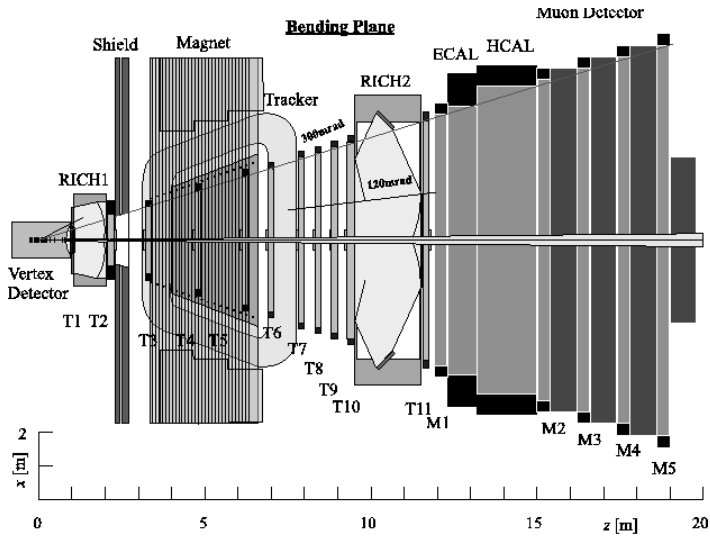}
\end{center}
\vspace*{-0.8cm}
\caption{Schematic 2D view of the LHCb detector in the bending plane.  The 
interaction point is at $z=0$.}
\label{fig:lhcbschem}
\end{figure}

The LHCb detector is a single-arm spectrometer covering the forward region 
of the $pp$ interactions. 
A schematic view is shown in Fig.~\ref{fig:lhcbschem}.
The detector covers the angular region from 
10$\,$mrad up to 300$\,$mrad in the horizontal plane (the 
{\em bending plane}) and from 10$\,$mrad up 
to 250$\,$mrad in the vertical plane (the {\em non-bending plane}),
corresponding to the approximate range $2.1<\eta<5.3$ in terms of 
pseudorapidity. 
Starting from 
the interaction point, it consists of a silicon vertex detector, a RICH 
detector and a tracking system; the tracking system is followed by a 
second RICH detector, electromagnetic and hadron calorimeters and by muon 
detectors. The vertex detector, which is located 
inside the beam pipe, also includes a pile-up veto counter to reject events 
with multiple $pp$ interactions. The tracking system is partly included in 
a dipole magnet field having a maximum value of 1.1$\,$T in the vertical 
direction. The calorimetry system extends from 30$\,$mrad to 
300$\,$(250)$\,$mrad
in the horizontal (vertical) direction. Muon coverage is assured in the
angular range 25$\,$(15)$\,$mrad to 294$\,$(245)$\,$mrad in the 
horizontal (vertical) direction.

\subsection{Luminosity}      

The LHC is being built to run at  
a design luminosity of $10^{34}\,$cm$^{-2}$ s$^{-1}$ to maximise the
potential for discovering new, heavy particles. {}From the point of view 
of B-hadron reconstruction, multiple interactions and pile-up effects 
in the detectors are a complication both at trigger level and in the 
reconstruction of relatively low-$p_{T}$ particles. Moreover, the high
luminosity will deteriorate the performance (both in terms of radiation
damage and occupancy) of the innermost tracking layer when the 
reconstruction of the B meson vertex position is needed. 

It is expected, however, that the LHC will reach design luminosity
only gradually in time, starting at $10^{33}\,$cm$^{-2}$ s$^{-1}$
and taking three years to reach $10^{34}\,$cm$^{-2}$ s$^{-1}$.
ATLAS and CMS will take advantage of this so-called
{\em low-luminosity} period in order to carry out most of their
B physics programme. 
At this luminosity, each crossing will have an average of 2 to 3 pile-up 
events in the tracking detectors which, however, have been shown not
to affect significantly the detector performances.
It is under current investigation if it is possible to continue
certain studies
at higher luminosity: for some critical channels, like very rare decays
(see Sec.~\ref{sec:rare}), this has been already
demonstrated to be feasible (both at trigger and reconstruction level). 

In order to have a clean environment, well suited to B physics,
the luminosity at LHCb will be locally controlled to have a mean value of
$2 \times 10^{32}\,$cm$^{-2}$ s$^{-1}$,  even when the machine is operating
at design luminosity.
This value is chosen to optimise the number of single interaction
bunch crossings, which will make up $\sim 75\%$ of crossings within an
interaction, and to ensure that radiation damage and occupancy
problems are not too severe.

In this report we will present estimates of the potential of the three
experiments at various integrated luminosities. When a simple comparison
among the potential of the three experiments is needed, the results will be 
normalized to one year of running: this corresponds to 
$2 \times 10^3\,$pb$^{-1}$ for LHCb and to $10^4\,$pb$^{-1}$ for ATLAS and 
CMS running at low luminosity. 
More often the full potential of each experiment is presented:
here we take 5 years of running for LHCb 
and  3~years at low luminosity for 
ATLAS and CMS (unless the study can be extended into the high-luminosity 
running period). Whenever possible, the results of the three experiments 
have been statistically combined to estimate the {\em ultimate} LHC potential.

\subsection{Monte Carlo Generators, Simulation Methods 
and Assumed Cross-Sections}

For the performance studies presented in this Chapter, large samples of
B hadron events have been produced using the 
PYTHIA 5.7/JETSET 7.4~\cite{PYTHIA} event
generator. In the ATLAS Monte Carlo flavour-creation, flavour-excitation and
gluon splitting
production processes were included. In CMS, flavour-creation and gluon
splitting were included (see also discussion in the "Bottom
production" Chapter of this report \cite{bcprod}).
The LHCb Monte Carlo production was based on flavour-creation and 
flavour-excitation
processes,  with additional samples including gluon-splitting.
The CTEQ2L~\cite{CTEQ} set of parton-distribution 
functions has been
chosen. The Peterson function (with $\epsilon_b = 0.007$) has been used
to fragment $b$ quarks to B hadrons. Other PYTHIA physics parameters
are the default ones. The agreement between PYTHIA predictions and
theoretical calculations is discussed elsewhere in this report.

The response of the detectors to the generated particles is simulated with
programs based on the GEANT~\cite{GEANT} package. Then the event is 
reconstructed
in the sub-detectors relevant to each particular analysis; event 
reconstruction
includes full pattern recognition in the tracking detectors, vertexing
and particle identification (muons and electrons and $\pi / K$ separation,
if available). 

The procedure detailed above is called {\em full simulation} and has
been used for the majority of the analyses presented. In some cases, a 
{\em fast simulation} which makes no use of GEANT, but of a simple 
parametrization of the detector response has been used.   

The results have been normalized assuming a total inelastic cross-section of 
$80\,$mb and a $b\overline{b}$ cross-section of $500\,\mu$b.

\subsection{Proper Time Resolution}\label{exp_tres}

Different detector layouts used by the three 
experiments lead to differences in the impact parameter and in the proper 
decay time resolutions. 

\begin{figure}
\begin{center}
\includegraphics[width=0.33\textwidth,
bbllx=5,bblly=15,bburx=215,bbury=215,clip]{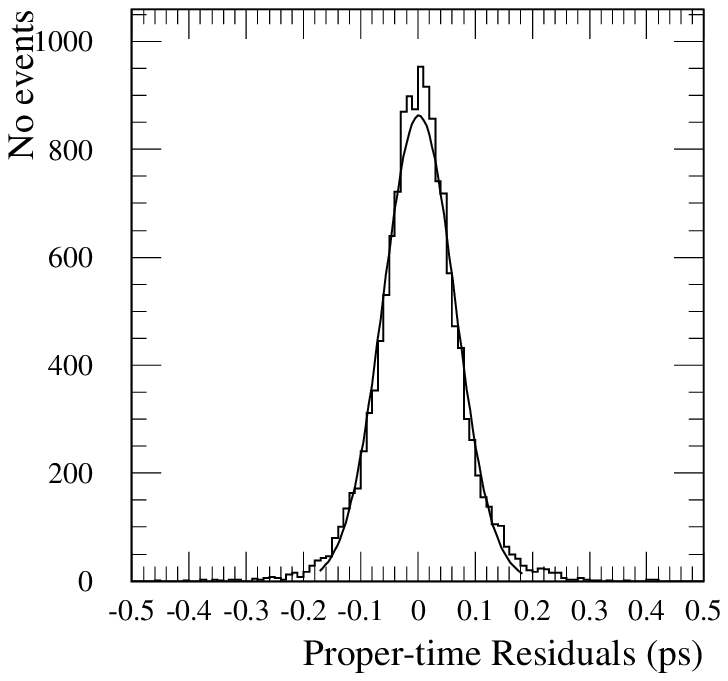}
\epsfxsize=0.3\textwidth\epsffile{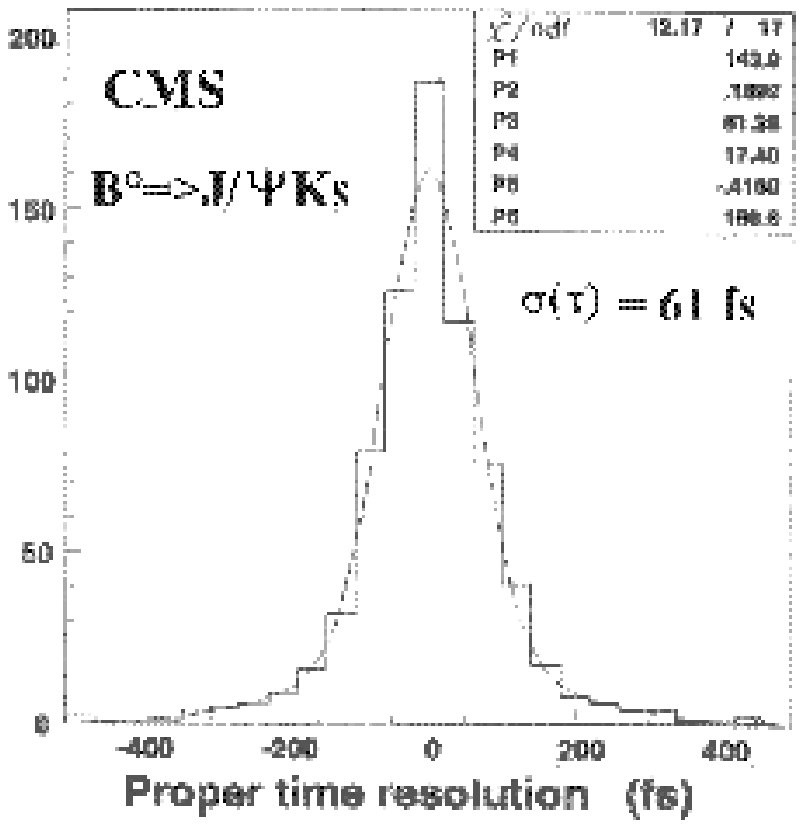}
\includegraphics[width=0.3\textwidth,
bbllx=10,bblly=10,bburx=515,bbury=515,clip]{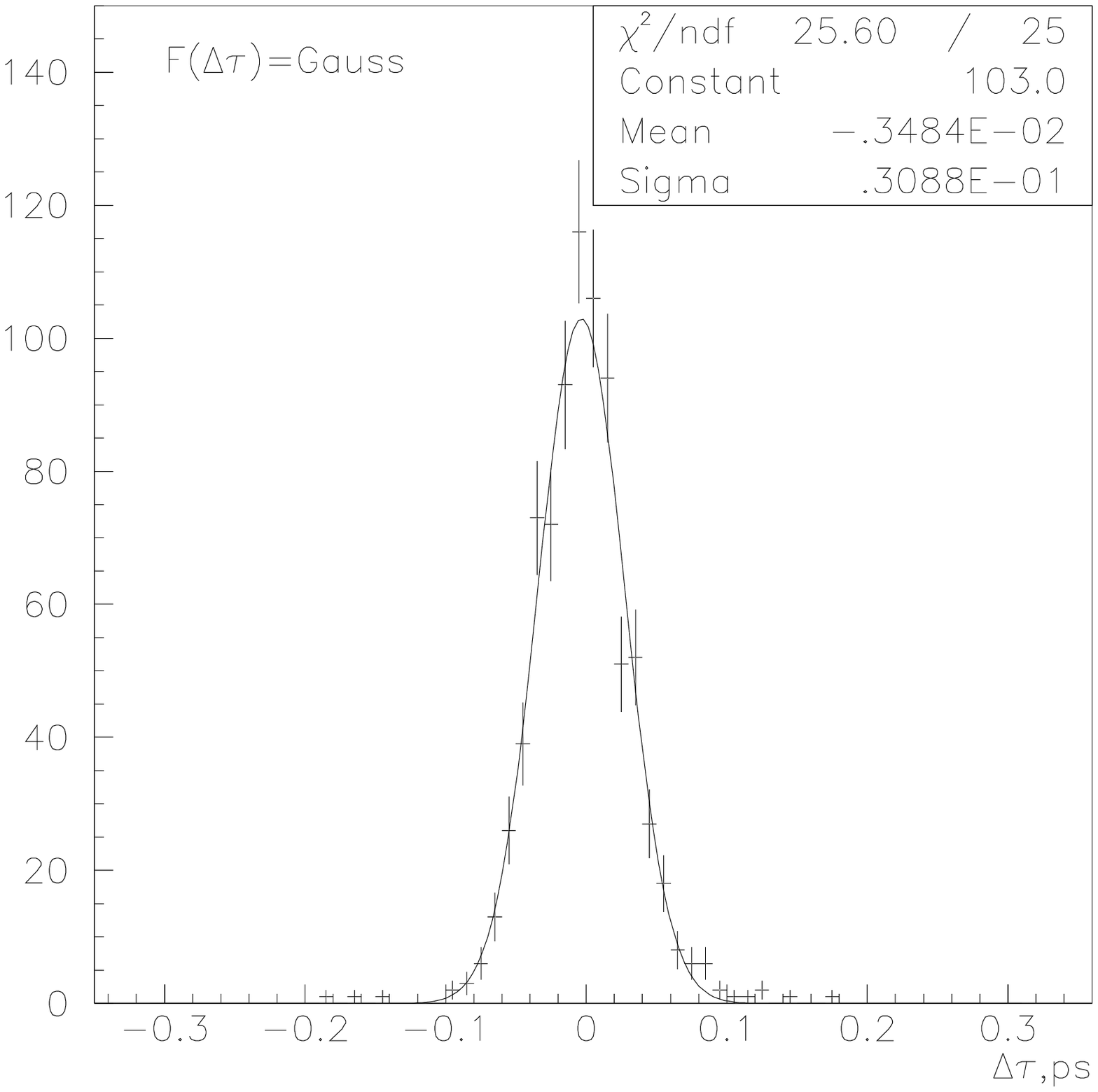}
\end{center}
\vspace*{-0.7cm} 
\caption[]{Typical proper time residual distributions for ATLAS (left), CMS 
 (center) amd LHCb (right). The plots refer
      to reconstructed $B^0_s \rightarrow J/\psi \phi$  (ATLAS, LHCb)
       and $B^0_d \rightarrow J/\psi K^0_s$ (CMS) decays.}
      \label{fig:resolb}
\end{figure}

In LHCb the impact parameter is measured in the $R$--$z$ plane: 
the resolution
increases with transverse momentum and reaches an asymptotic value
of about 40 $\mu$m already for tracks with transverse 
momenta $p_T > 3\,$GeV \cite{LHCbTP}. Particles coming from 
B decays are mostly above this threshold and so LHCb can achieve a 
proper time resolution (for fully reconstructed exclusive decays) of about 
$0.031\,$ps (see Fig.\ref{fig:resolb}).

The ATLAS and CMS experiments measure precisely the projection of the track
impact parameter in the $R$--$\phi$ plane \cite{ATLASPTDR,CMSPTDR}.
The plateau value (for high-$p_T$ tracks) of the transverse impact parameter
resolution is about 11 $\mu$m (for comparison,
the asymptotic value for the impact parameter in the $R$--$z$ plane is 
about 90 $\mu$m); however, most of the
tracks from B decays concentrate in 
the low-$p_T$ region where the resolution degrades due to multiple scattering.
The proper time resolutions in ATLAS and CMS for typical fully reconstructed
B decays are 
characterized by a width of a Gaussian distribution of about $0.060\,$ps
(see Fig.\ref{fig:resolb}).
 
The proper time resolution estimates summarized
in this section refer either to 
the $B^0_s \rightarrow J/\psi \phi$ decay analysis discussed in 
Sec.~\ref{sec:Bspsiphi}  or to the $B^0_d \rightarrow J/\psi K^0_s$ sample
(see Sec.~\ref{subsec:BdpsiKS}). Slightly different values are estimated 
according to the B decay channel under study.

\subsection{Particle Identification}\label{exp_id}

Particle identification is a very important tool in many B physics channels. 
In 
particular, $\pi / K$ separation plays a key r\^{o}le in hadronic B decays 
(see Secs.~\ref{sec:benchmark} and \ref{sec:newstrat}), allowing the 
separation of the decays of interest from similar, and indeed identical,
topologies  that would 
otherwise have overlapping (and in some cases overwhelming) spectra.
Moreover, $\pi / K$ separation is crucial for one of the techniques
({\em kaon} tagging) used to identify the flavour of the $b$ hadron  
at production (see Sec.~\ref{exp_ft} for a short review of flavour tagging
methods).
 
\begin{figure}
\begin{minipage}[t]{0.3\textwidth}
\epsfxsize=0.95\textwidth
\centerline{\epsffile{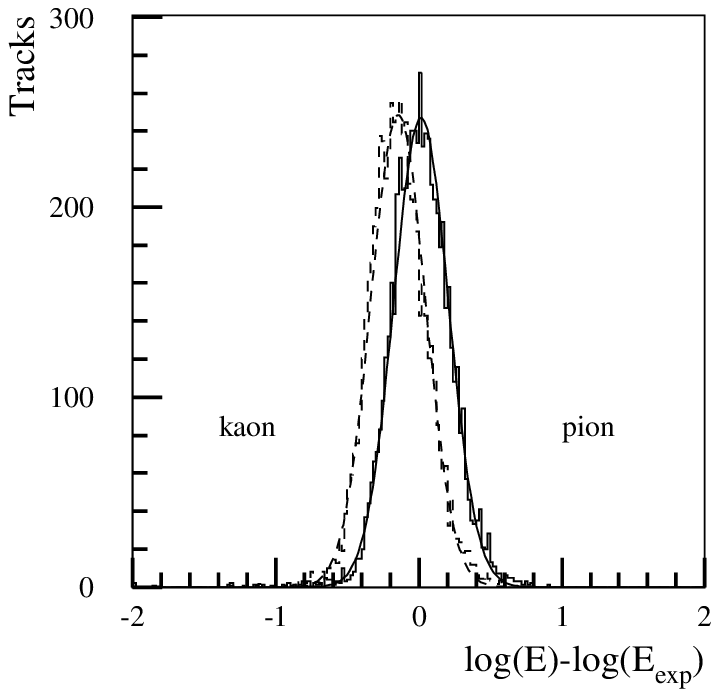}}
\vspace*{-0.4cm}
\caption{Logarithm of
the energy deposited in the ATLAS straw tracker normalised to the expected
energy deposit of a pion, for $p_T=5\,$GeV  pions (solid line) and kaons
(dashed line). 
}\label{fig:atlaspiksep}
\end{minipage}
\hfill
\begin{minipage}[t]{0.3\textwidth}
\includegraphics[width=1.05\textwidth,clip]{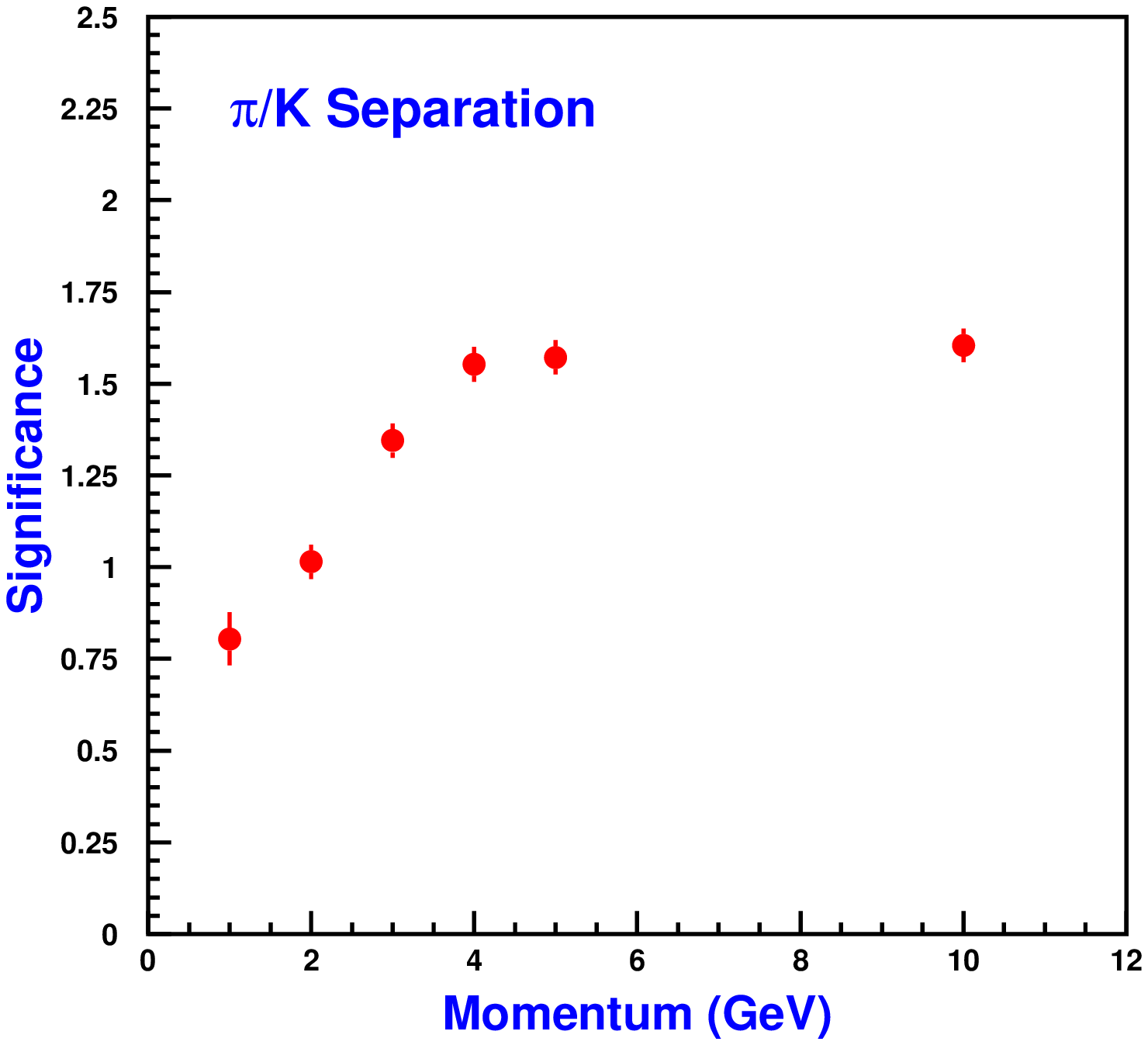}      
\vspace*{-0.9cm}
\caption{CMS $dE/dx$ pion/kaon separation, plotted in units of
$\sigma$ versus track momentum.}
\label{fig:cmspiksep}
\end{minipage}
\hfill
\begin{minipage}[t]{0.33\textwidth}
\epsfxsize=0.855\textwidth
\centerline{\epsffile{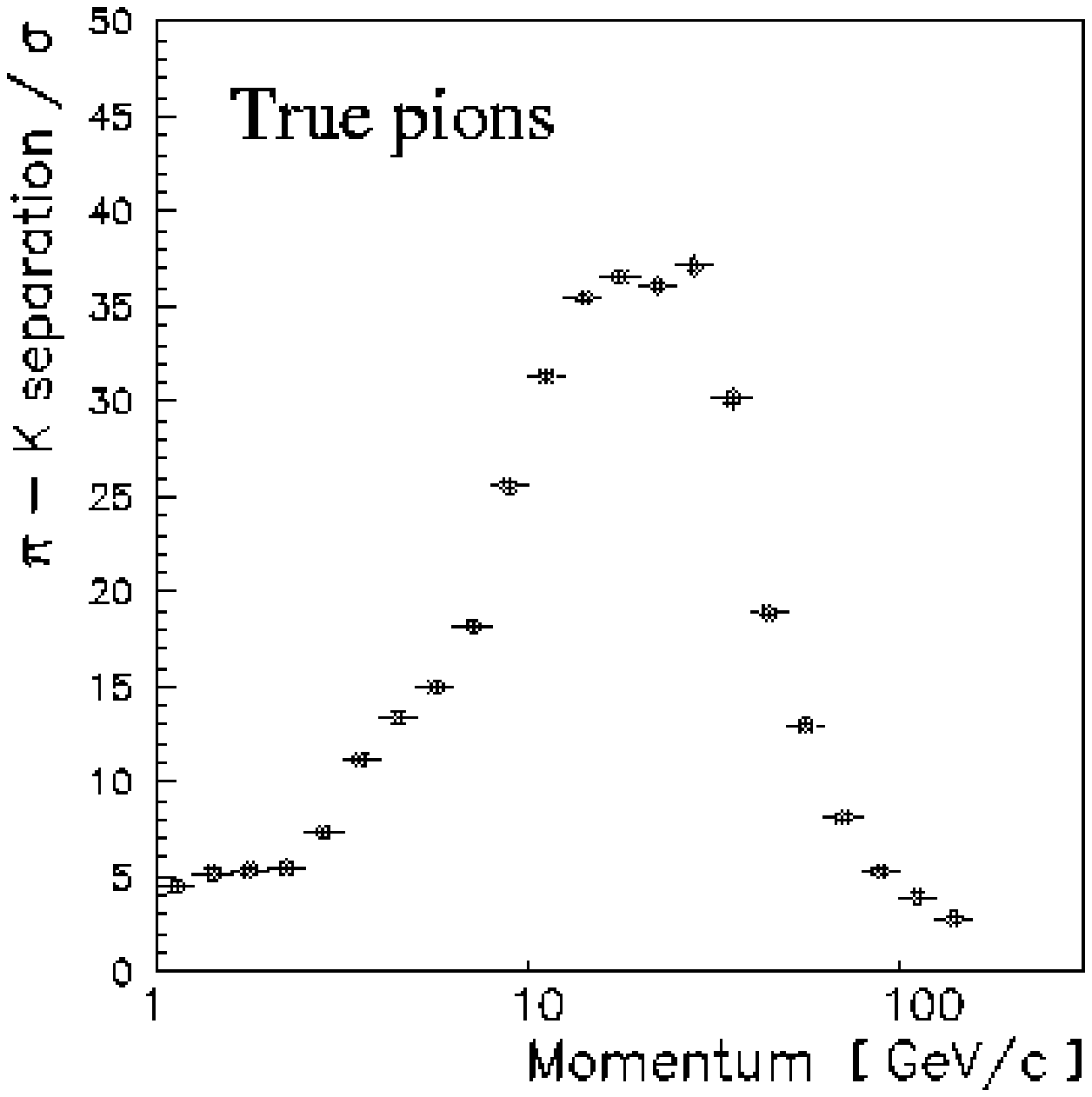}}
\vspace*{-0.4cm}
\caption{LHCb RICH pion/kaon separation, after 
pattern recognition, plotted in units of
$\sigma$ against momentum for true pions.}
\label{fig:lhcbpiksep}
\end{minipage}
\end{figure}

For this purpose, 
the LHCb detector has a dedicated system composed of two RICH detectors.
The first system, RICH1, located upstream of the magnet, uses silica aerogel
and C$_4$F$_{10}$ as radiators: this detector is intended to identify
low-momentum particles over the full angular acceptance. The RICH2 detector,
which uses CF$_4$, is located downstream of the magnet and covers a smaller
solid angle. The purpose of this detector is to complement RICH1 by 
covering the high-end of the momentum spectrum.  
The performance of LHCb's RICH is shown in Fig.~\ref{fig:lhcbpiksep}.
There is  significant $\pi / K$ separation over $1 < p < 150$ GeV, 
exceeding 10$\,\sigma$ for most of this range. Efficiencies and purities 
are expected to be in excess of 90\%.

In the absence of dedicated detectors for particle identification, ATLAS 
and CMS have studied other methods to obtain some level of pion/kaon 
separation, although with reduced performance. The CMS silicon tracker 
has analogue read-out electronics so that the pulse height information is
preserved and can be used to estimate $dE/dx$. Preliminary
results have been obtained~\cite{CMSpid} using a full GEANT simulation of the 
CMS tracker system described in~\cite{CMtrk}. This study estimates the
asymptotic performance of the detector: a number of effects that can 
influence the $dE/dx$ resolution have not been simulated and will be the 
subject of future investigations when test-beam data will be
available. The estimated $\pi/K$ separation, shown in 
Fig.~\ref{fig:cmspiksep} as a function of the
particle momentum, has been used to obtain some of the CMS results 
presented in Sec.~\ref{sec:benchmark}.

The ATLAS outer tracking system, which uses 
drift tubes (or {\em straws}) to provide an average of 36 hits per 
track, has electron/pion separation capability. The space between the 
straws is filled with radiator material and transition-radiation photons,
created by electrons traversing it, are detected by using a xenon-based
gas mixture in the straws and a double-threshold read-out electronics.
This detector can provide some $\pi / K$ separation
using $dE/dx$, although the pulse-height is not 
measured~\cite{ATLASPTDR}. 
Information about the deposited energy is extracted from the offset
and accuracy of the measured drift distance, the fraction of
high--threshold hits and the fraction of missing low--threshold hits.
A preliminary study has concluded that, by combining all this information,
a $\pi / K$ separation of
 0.8$\,\sigma$ for tracks with $p_T \sim 5\,$GeV can be obtained.
The expected performance of this method is shown in
Fig.~\ref{fig:atlaspiksep}.
This separation power is not enough to identify pions and kaons, but can be
used on a statistical basis.
A more recent study, incorporating some changes to the readout format of the
straw-tracker data, which provide a measurement of time-over-threshold for
low-threshold hits, improves significantly this separation.

\subsection{Triggers}\label{exp:trig}

Triggering is the key issue for B physics studies at the LHC. Careful trigger 
strategies are needed to extract interesting channels from  
inelastic collisions. Different trigger strategies to approach this
problem will be used by 
ATLAS~\cite{atl0} and CMS~\cite{cms0}, on the one side, and LHCb (whose trigger
is entirely dedicated to B decays), on the other side.  
For robustness and flexibility, all three experiments 
will use multi-level trigger systems with the ATLAS and CMS triggers being 
divided into three levels and the LHCb trigger into four levels.

The lowest trigger level 
of ATLAS~\cite{atl1} and CMS~\cite{cms1}, called Level-1, which operates 
at the 40$\,$MHz machine 
bunch-crossing frequency, uses reduced-granularity data from the muon 
trigger chambers and from the calorimeters. B physics is accommodated in 
these triggers by pushing the lepton transverse-momentum thresholds down to 
the minimum possible, still keeping the output trigger rate compatible with 
the acceptance rate of the next trigger level, Level-2. In ATLAS this is 
achieved by requiring a single muon with $p_T>6$ GeV in
$|\eta|<2.4$. The possibility of using a Level-1 dimuon trigger with 
$\eta$-dependent thresholds is under study 
as a means of increasing statistics. However,
all ATLAS studies reported in this Chapter have been obtained requiring 
at least one muon with $p_T>6\,$GeV.  
In CMS, lower transverse momentum thresholds can be achieved, by adding
to the 
single lepton trigger ($p_T>7\,$GeV for muons and $p_T>12$ GeV for electrons)
also double-lepton triggers ($\mu \mu$, $\mu e$ and $ee$) with thresholds 
which vary with pseudorapidity and can go down to 2 or 4$\,$GeV for the 
two-muon case and to 5$\,$GeV for the two-electron case. 

The lowest trigger level in LHCb, called Level-0, works at 40$\,$Mhz and is 
based on the identification 
of single leptons, hadrons and photons with high-$p_T$ in calorimeters and 
muon  chambers. Because of the forward geometry,  and high output rate,  
the `high'-$p_T$ threshold can be as low as 1 GeV.
The hadron trigger allows the collection of large event samples
in rare decay channels without leptons.
The Level-0 trigger is combined with 
the pile-up veto to reject bunch crossings likely to contain more than 
one $pp$ 
interaction. After the pile-up veto, the rate is reduced to about 9 MHz 
already, so that the high-$p_T$ trigger has to provide only an additional
reduction factor of about 10 to match the design Level-0 output rate of
about 1 MHz.    The allocation of bandwidth between the trigger components
and the assignment of thresholds is adjustable to match running conditions
and physics requirements.   At present the nominal thresholds for the single
particle triggers are 1 GeV for muons, 2.3 GeV for electrons, 2.4 GeV for 
hadrons and 4 GeV for photons.    

In ATLAS, the Level-2 trigger~\cite{atl2} uses full-granularity data from 
the muon system, 
the calorimeters and from the tracking system. The Level-2 trigger will 
confirm and refine the Level-1 information and then look for specific 
final states according to the physics channel to be studied. Fast algorithms 
will be used to reconstruct tracks in the tracking system to allow $p_T$ and
mass cuts. The second-muon
trigger threshold will be set to $p_T=3$ GeV. The dimuon trigger covers 
both some
rare B decays and channels with $J/\psi$'s in the final state. 
Triggers with $J/\psi \rightarrow e e$, with the $p_T$ threshold on the two 
electrons as low as 0.5 GeV, will also be available. Hadronic triggers will 
be available for selected channels. 
The maximum total Level-2 output rate is limited to about 1$\,$kHz.
CMS will follow a similar strategy.

In LHCb, the next trigger-level after Level-0, called Level-1, uses
information from the {\em vertex} detector. This trigger is meant to
complement the Level-0 information by exploiting the displacement of
$b$ decay vertices. The vertex trigger will 
first reconstruct the event primary vertex and then look for track pairs
with significant impact parameters with respect to the primary vertex, which 
are close in space.   This signature provides high efficiency in all
B decay modes.
The total output rate is about 40 kHz. Successively, 
the Level-2 trigger will refine the vertex trigger by adding momentum 
information to the tracks forming the secondary vertices and reduce the
data rate to 5 kHz. 

For the three experiments, the final trigger decision will be taken by a 
Level-3 trigger which feeds
full event data from all detectors to an offline-like algorithm to reconstruct
specific final states. Selected events will be stored for offline analysis.
  
The trigger performance of the experiments
will be summarized 
elsewhere in this report, for certain important decay modes.  It will
become clear that the enormous rate of B production
at the LHC can indeed be properly exploited.

\subsection{Flavour Tagging}\label{exp_ft}

An important issue of many CP-violation and $B^0$ mixing studies is the 
determination of the flavour of a $b$ hadron at production. The LHC
experiments have already successfully investigated
several tagging strategies, but the studies are not yet completed 
(ATLAS~\cite{ATtag}, CMS~\cite{CMtag}, LHCB~\cite{LHCbTP}). 

Tagging algorithms can
be divided into two broad categories: {\em Opposite Side} (OS)
and {\em Same Side} (SS) algorithms, according whether one studies the
$b$ or the $\overline{b}$ quark in the event. The 
nomenclature, OS and SS, used to distinguish between the $b$ and 
$\overline{b}$ quarks, is used for historical reasons 
(it is derived from the
LEP experiments), but it does not imply that
the two quarks are produced in separate hemispheres. Indeed,
for the LHCb experiment there is {\em no} other side 
and both the $b$ and the $\overline{b}$ quarks are produced predominantly 
in the same forward-cone. Moreover, for the LHC experiments, the importance   
of the gluon splitting mechanism for producing $b\overline{b}$ 
pairs implies that the two quarks are not always on opposite sides.   
We will thus include in the OS category all
algorithms that try to deduce the initial flavour of the B meson under
study by identifying the flavour of the other $b$-hadron in the event.
In the SS category we include
all algorithms that look directly at the particles accompanying the
B meson which has decayed in the channel under investigation (also
called {\em signal} B in the following).

It can be shown that the statistical error of an asymmetry measurement
is inversely proportional to the quantity 
$(1-2\omega)\sqrt{\epsilon N}$, where $N$ is the total (untagged) number
of events, $\epsilon$ is the tagging efficiency and $\omega$ is the
wrong-tag fraction. For this reason, tagger-cuts are chosen in order to 
maximize the {\em quality factor} $Q=\epsilon D^2$, where $D=1-2\omega$ is
called tagger-{\em dilution}. Approximate numbers for efficiencies and 
dilutions for the algorithms described below are listed in 
Tab.~\ref{tab:tags}. Further developments and cut optimization might be 
needed to improve the performance of the tagging algorithms already
studied and to bring all of them at the same level of understanding.
The majority of the presented results refers to the 
$B^0_d \rightarrow J/\psi K^0_s$ sample (see Sec.~\ref{subsec:BdpsiKS}).
Variations from sample to sample have been observed. 
Because of this and because of differences in the simulation details,
trigger selections and analysis cuts, a direct comparison between 
tagger potentials (and experiment performance) is not straightforward.

\subsubsection{Opposite Side Tagging}

The OS techniques which have been studied up to now by the LHC
experiments are: lepton (muon or electron) tagging, kaon tagging 
(LHCb only) and jet-charge tagging.

In the lepton-tagging method, one looks for a lepton in the 
event coming from the semileptonic decay of the other $b$ quark in the
event: $b \rightarrow l$. This method has a low efficiency (due to the 
relatively low $b$ semileptonic branching ratio of about 10\%), but good
purity. Furthermore a significant enhancement arises through
the trigger, where for all the experiments leptons are used. 
The main contributions to the mistag rate are due to flavour mixing 
of the neutral B mesons and to cascade decays 
$b \rightarrow c \rightarrow l$. Wrong tags from cascade decays can be 
reduced by increasing the lepton $p_T$ threshold. It has also been 
shown~\cite{ATtag} that the mistag rate increases with increasing
$p_T$ of the signal B for a fixed lepton-tag transverse-momentum threshold.
For the studies presented
in this Chapter, the threshold has been set to 5 GeV for both electrons and 
muons in the ATLAS analysis, to 2 (2.5) GeV for muons (electrons) in CMS and 
to 1.5 GeV for both muons and electrons in LHCb.

Kaon tagging exploits the decay chain $b \rightarrow c \rightarrow s$ to
identify the flavour of the $b$ quark from the charge of the kaon produced
in the cascade decay.
This method can be only used by LHCb as it requires the particle identification
capability of the RICH detector. 
Candidate kaons are searched for down to a $p_T$ of
0.4 GeV and are required to have impact parameter significance incompatible 
with the reconstructed primary vertex at the 3$\sigma$ level. For
kaon tagging (as well as for lepton tagging), if more than one candidate
survives all cuts, the one with the highest $p_T$ is chosen. 

Jet-charge tagging deduces the flavour of the other $b$ quark in the event
by looking at the total charge of the tracks which belong to the $b$ 
fragmentation. At LEP, where this algorithm was first developed, 
the identification of the opposite-side jet in 
$Z \rightarrow b \overline{b}$ events was almost straightforward.
At the LHC, the other $b$-jet may escape the detector-acceptance 
and can be identified only by dedicated jet-clustering algorithms.
These algorithms
are usually based on track clustering possibly seeded by displaced tracks.  
Once the jet has been found, the jet total charge,
$Q_{jet}$, is defined by an average of the track's charge in the cluster, 
weighted by a function of their momenta. The right (wrong) sign events are 
then defined as those with $Q_{jet} > +c$ ($Q_{jet} < -c$), where $c$
is a tunable cut. 
Although investigated in the past, OS jet charge is not used in the ATLAS
analyses presented in this report. The LHCb numbers for this tagging method, 
which are calculated for events where no other type of tag has been found, 
are preliminary and are not used for the results presented in this Chapter.

\subsubsection{Same Side Tagging}

The SS techniques presented in this section exploits production and 
fragmentation properties of the B meson to deduce its flavour. These
techniques are not affected by mistags due to mixing. Moreover, as they
apply to the same B meson whose decay is under investigation, there is
no loss of efficiency due to the identification of the other $b$ jet in
the event.

During the process of a $\overline{b}$ quark fragmentation to produce 
a $B^0_d$
meson, pions which are charge-correlated to the flavour of the B meson, can 
be produced by two mechanisms~\cite{GroNiRo}. 
The $\overline{b}$ quark can pick up a $d$ quark from the
quark sea to form a $B^0_d$, thus making available a $\overline{d}$ quark to
form a $\pi^+$. Another mechanism proceeds through production of orbitally
excited states of B mesons, called $B^{**}$, which then decay to $B^0_d$:
$B^{**} \rightarrow B^{(*)0} \pi^+$. If a $B^{*0}$ is produced, it 
decays radiatively as $B^{*0} \rightarrow B^0 \gamma$.  

The $B$--$\pi$ {\em correlation} method, studied by ATLAS, exploits these
correlations by searching for low-$p_T$ pions, compatible with coming from 
the primary vertex, in proximity of the decayed B meson. Tracks belonging 
to the B decay products are excluded and what it is called {\em pion} is 
actually a generic charged track, as no $\pi / K$ separation is used. In
this method, both production mechanisms described above contribute correlated 
pions and no attempt is made to separate these two contributions.

The CMS experiment prefers to concentrate on the explicit reconstruction of
the $B^{**}$ resonance ($B^{**}$ {\em method}). In the
Monte Carlo, these resonance have been modelled  
according to~\cite{Bstarstar}.  In
this method, pions with $p_T > 1$ GeV are combined with a $B^0_d$ to give 
a $B^{**}$ meson with a mass between 5.6 and 5.9 GeV. As above, the charge
sign of the associated pion gives the tag. No attempt is made to reconstruct
the low-$p_T$ photon which is present when a $B^{*0}$ is produced in the 
cascade and to resolve the different peaks which superimpose in the 
$B^{**}$ mass spectrum. It would also be possible to study the mistag rate 
from the data itself by looking at the {\em side-bands} of the mass 
resonance, so that one need not rely only on the 
Monte Carlo modelling of the process.

Similar to the $B$--$\pi$ correlation method, the $B^0_s$ tagging method, 
which is under investigation by LHCb, consists in looking for a primary 
kaon in the vicinity of the $B^0_s$ meson. Efficiency and dilution for this 
tagger, which is not used for the results presented in this Chapter, are
preliminary.

In a different approach, it is possible to use jet-charge tagging also on 
the {\em same side}. In this case, similarly to the OS jet-charge tagging, 
the jet charge is a weighted average of the charge of the tracks in the jet,
but the tracks belonging to the B meson decay products are excluded from the
sum. The weights are functions of the momentum of the track and are often
written in the form $w(p)^k$, where $w(p)$ can be chosen as the transverse
momentum, the projection of the momentum along the $B$ direction or a more 
complicated function of them. The parameter $k$ controls the relative 
influence of soft and hard tracks in the total charge.

\begin{table}[t]
\begin{center}
\begin{tabular}{|c|l|c||c|c|c|c|c|c|}
\hline
 \multicolumn{3}{|c||}{Tagging} & \multicolumn{2}{|c|}{ATLAS} & 
 \multicolumn{2}{|c|}{CMS}     & \multicolumn{2}{|c|}{LHCb} \\
\cline{4-9}
  \multicolumn{3}{|c||}{Method} & $\epsilon$ & $D$ & $\epsilon$ & $D$ & 
                                 $\epsilon$ & $D$ \\
\hline
\hline
     & Lepton Tag   & e & 0.016 & 0.46 & 0.027 & 0.44  & &  \\
\cline{3-7}
 OS    &     & $\mu$ & 0.025 & 0.52 & 0.034 & 0.44  & 0.40 & 0.40  \\
\cline{2-7}
    & \multicolumn{2}{|l||}{Kaon Tag} & \multicolumn{2}{|c|}{n/a} & 
                      \multicolumn{2}{|c|}{n/a} & &      \\
\cline{2-9}
     & \multicolumn{2}{|l||}{Jet Charge}  & \multicolumn{2}{|c|}{n/a} &  
     0.70 & 0.18 & 0.60 & 0.16 \\
\hline
\hline
     & \multicolumn{2}{|l||}{$B$--$\pi$}  &  0.82 & 0.16 & 
     \multicolumn{2}{|c|}{n/a} &  \multicolumn{2}{|c|}{n/a}   \\
\cline{2-9}
 SS & \multicolumn{2}{|l||}{$B^{**}$}  & \multicolumn{2}{|c|}{n/a} & 
   0.22  & 0.32 & \multicolumn{2}{|c|}{n/a} \\
\cline{2-9}
     & \multicolumn{2}{|l||}{Jet Charge}  & 0.62 & 0.23 & 0.50 & 0.23 & 
                    \multicolumn{2}{|c|}{n/a} \\
\cline{2-9}
     & \multicolumn{2}{|l||}{$B^0_s$ tag}  & \multicolumn{2}{|c|}{n/a}  &
                      \multicolumn{2}{|c|}{n/a} & 0.11 & 0.34 \\
\hline
\end{tabular}
\end{center}
\vspace*{-0.5cm}
\caption{Efficiencies ($\epsilon$) and dilutions ($D$) for the 
flavour-tagging algorithms described in the text. The shorthand ``n/a'' ({\em
not available}) means that one tagger either cannot be used or has not yet
been fully studied by a particular experiment. The LHCb numbers for
lepton and kaon tagging refer to the combined algorithm described 
in the text.} \label{tab:tags}
\begin{center}
\begin{tabular}{|c|c|c|c|c|c|}
\hline
$A$ & $B$ & $\epsilon(A)$ & $\epsilon(B)$ & 
  $\epsilon(A)+\epsilon(B)$ & $\epsilon(A \cup B)$ \\
\hline
\hline
 Lepton Tag  & $B^{**}$ & 0.06\phantom{0}  & 0.215 & 0.275 & 0.26\phantom{0}\\
\hline
 Lepton Tag  & SS Jet Charge    & 0.06\phantom{0}  & 0.5\phantom{00}
 & 0.56\phantom{0}  & 0.53\phantom{0}\\\hline
 Lepton Tag  & OS Jet Charge    & 0.06\phantom{0}  & 0.7\phantom{00}   
& 0.76\phantom{0}  & 0.72\phantom{0}\\\hline
  $B^{**}$   & SS Jet Charge    & 0.215 & 0.5\phantom{00}   & 0.715 
& 0.56\phantom{0}\\\hline
  $B^{**}$   & OS Jet Charge    & 0.215 & 0.7\phantom{00}   & 0.915 
& 0.76\phantom{0}\\\hline
  SS Jet Charge & OS Jet Charge & 0.5\phantom{00}   & 0.7\phantom{00}  
& 1.2\phantom{00}   & 0.845\\
\hline
\end{tabular}
\end{center}
\vspace*{-0.5cm}
\caption{Combined tagging efficiencies from CMS Monte Carlo. The last
column shows the combined efficiency of algorithms A and B when the
overlap has been subtracted: 
$\epsilon(A \cup B) = \epsilon(A)+\epsilon(B)-\epsilon(A \cap B)$.}
\label{tab:corr}
\end{table}

\subsubsection{Combined Tagging}\label{comb_tags}

The {\em best} tagging strategy would combine all taggers, weighted by their 
dilutions, simultaneously
on both sides on an event-by-event basis. This requires, however, a full
understanding of tagger correlations. The CMS experiment has perfomed a
preliminary study of these correlations for four of the tagger algorithms 
described
above (lepton tag, $B^{**}$, SS jet charge and OS jet charge). The results 
are summarized in Tab.~\ref{tab:corr}, where, for pairs of algorithms, the 
combined efficiency is shown, taking into account overlaps. 
Correlations are sizeable (e.g.\ between $B^{**}$ and SS jet charge tags,
as expected) and need to be properly taken into account in combining taggers.

In a simplified 
approach, overlaps can be avoided by applying taggers one after the other,
by applying the second tagger on the sample not tagged by the first one (and 
so on). 
The LHCb experiment combines lepton and kaon tagging: if more than one
tag is present in one event, the {\em best} tag is chosen in the following
order: muon, electron and kaon. The combined efficiency and dilution of
this algorithm is reported in Tab.~\ref{tab:tags}.
In ATLAS, only lepton tagging and $B$--$\pi$ tagging have been statistically
combined so far. Lepton tagging (which has the highest purity) is applied 
first and then, on the 
remaining events, tagging pions are searched for.  
Similarly, CMS combines four algorithms in the following order (of
decreasing dilution): lepton tagging, $B^{**}$, SS jet 
charge and OS jet charge. Each tagger is applied, with its own dilution,
on the sample not tagged by the previous one; in the end, a total number
of events four times the initial lepton tagged samples is selected.

\setcounter{equation}{0}
\section[BENCHMARK CP MODES]{BENCHMARK CP 
MODES\protect\footnote{Section coordinators: 
R. Fleischer and  G. Wilkinson.}}\label{sec:benchmark}
This section considers the use of benchmark $B$ decays to 
explore CP violation and to extract the angles of the unitarity triangles. 
By `benchmark' we mean modes that are well established in
the literature.  Some, but by no means all, of these channels will be
first probed at experiments that run before the LHC starts to operate.
To be specific, we will discuss the extraction of $\beta$ from mixing-induced
CP violation in the ``gold-plated'' decay $B_d\to J/\psi\,K_{\rm S}$, the 
prospects to probe $\alpha$ with $B_d\to\pi^+\pi^-$ and $B\to\rho\pi$ modes, 
as well as extractions of $\gamma$ from $B_d\to D^{\ast\pm}\pi^\mp$ and 
$B_s\to D_s^\pm K^\mp$ decays. Finally, we will also give a discussion
of the determination of $\gamma$ from $B\to DK$ decays. Since 
$B_s\to J/\psi\,\phi$ -- another benchmark CP mode -- is of
particular interest for the LHC, we have devoted a separate section 
to the discussion of the physics potential of this ``gold-plated'' mode 
for the LHC experiments: Sec.~\ref{sec:Bspsiphi}.

\subsection[Extracting $\beta$ from $B_d\to J/\psi\,K_{\rm S}$]{Extracting
  \protect\boldmath $\beta$ from \protect\boldmath 
$B_d\to J/\psi\,K_{\rm S}$\protect\footnote{With help from P. Colrain, 
Y. Lemoigne and G.F.\ Tartarelli.}}\label{subsec:BdpsiKS}
Probably the most important application of the formalism discussed in
Sec.~\ref{intro-neutCP} is the decay $B_d\to J/\psi\,K_{\rm S}$
\cite{bisa}, which is a transition into a CP eigenstate with eigenvalue 
$-1$ and originates from $\bar b\to\bar c\,c\,\bar s$ quark-level decays. 

\begin{figure}
\begin{center}
\leavevmode
\epsfysize=4.5truecm 
\epsffile{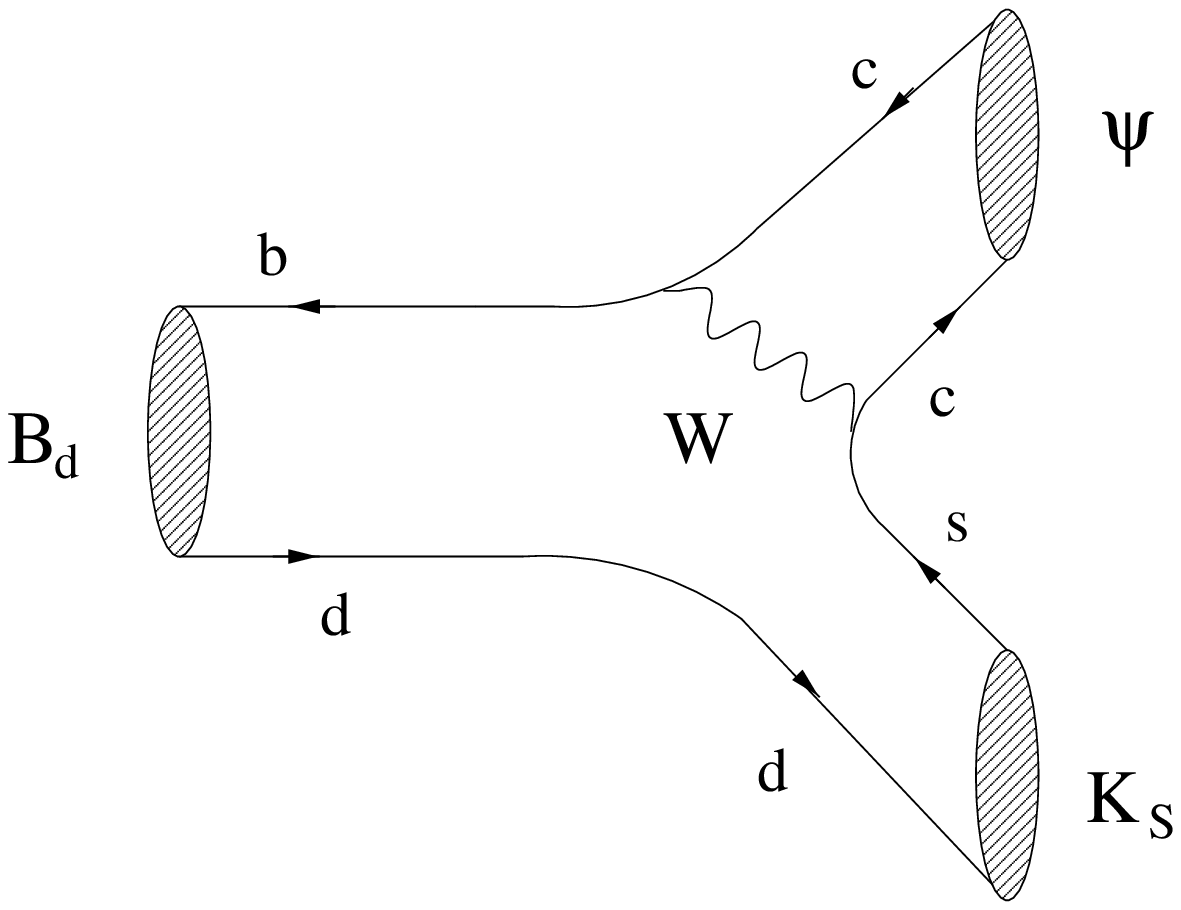} \hspace*{1truecm}
\epsfysize=4.5truecm 
\epsffile{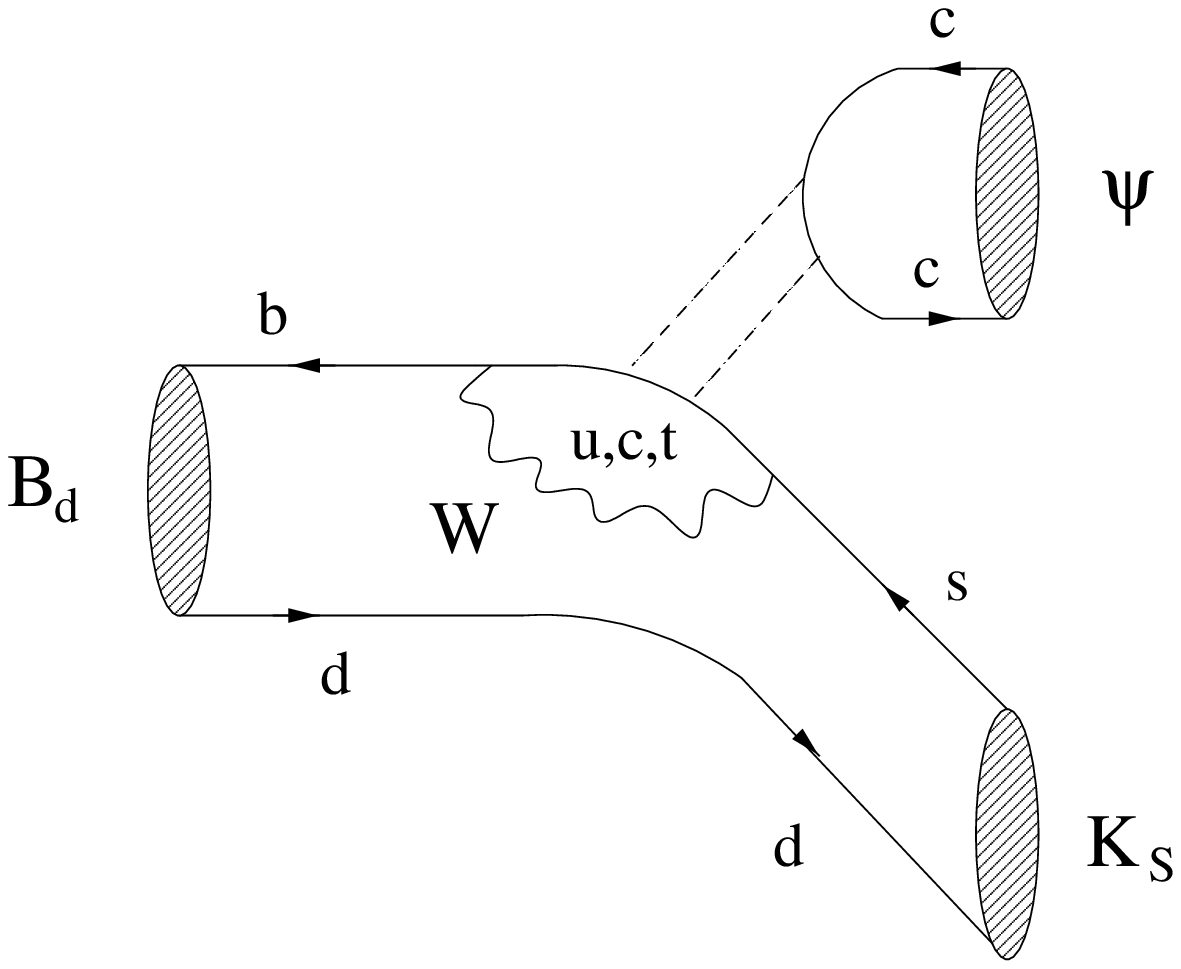}
\end{center}
\vspace*{-0.5truecm}
\caption{Feynman diagrams contributing to $B_d\to J/\psi\,K_{\rm S}$, 
consisting of colour-suppressed tree-diagram-like and penguin topologies.
The dashed lines in the penguin topology represent a colour-singlet 
exchange.}\label{fig:BdPsiKS}
\end{figure}

\subsubsection{Theoretical Aspects}
In the case of $B_d\to J/\psi\,K_{\rm S}$, we have to deal both with 
current--current, i.e.\ tree-diagram-like, and with penguin contributions, 
as can be seen in Fig.~\ref{fig:BdPsiKS}. The corresponding transition 
amplitude can be written as follows \cite{RF-BdsPsiK}:
\begin{equation}\label{Bd-ampl1}
A(B_d^0\to J/\psi\, K_{\rm S})=\lambda_c^{(s)}\left(A_{\rm cc}^{c'}+
A_{\rm pen}^{c'}\right)+\lambda_u^{(s)}A_{\rm pen}^{u'}
+\lambda_t^{(s)}A_{\rm pen}^{t'}\,,
\end{equation}
where $A_{\rm cc}^{c'}$ denotes the current--current contributions,
i.e.\ the ``tree'' processes in Fig.\ \ref{fig:BdPsiKS}, and the amplitudes 
$A_{\rm pen}^{q'}$ describe the contributions from penguin topologies with 
internal $q$ quarks ($q\in\{u,c,t\})$. These penguin amplitudes take into 
account both QCD and electroweak penguin contributions. The primes in 
(\ref{Bd-ampl1}) remind us that we are dealing with a $\bar b\to\bar s$
transition, and the
$
\lambda_q^{(s)}\equiv V_{qs}V_{qb}^\ast
$
are CKM factors. 
If we make use of the unitarity of the CKM matrix and apply the
Wolfenstein parametrization \cite{wolf}, generalized to include
non-leading terms in $\lambda$ \cite{BLO}, we obtain
\begin{equation}\label{Bd-ampl2}
A(B_d^0\to J/\psi\, K_{\rm S})=\left(1-\frac{\lambda^2}{2}\right){\cal A'}
\left[1+\left(\frac{\lambda^2}{1-\lambda^2}\right)a'e^{i\theta'}e^{i\gamma}
\right],
\end{equation}
where
\begin{equation}\label{Aap-def}
{\cal A'}\equiv\lambda^2A\left(A_{\rm cc}^{c'}+A_{\rm
    pen}^{ct'}\right)~~\mbox{and}~~
a'e^{i\theta'}\equiv R_b\left(\frac{A_{\rm pen}^{ut'}}{A_{\rm cc}^{c'}+
A_{\rm pen}^{ct'}}\right)
\end{equation}
with $A_{\rm pen}^{ct'}\equiv A_{\rm pen}^{c'}-A_{\rm pen}^{t'}$.
The quantity $A_{\rm pen}^{ut'}$ is defined in analogy to $A_{\rm pen}^{ct'}$,
and the CKM factor $A$ is given as follows:
\begin{equation}\label{CKM-exp}
A\equiv\frac{1}{\lambda^2}\left|V_{cb}\right|=0.81\pm0.06\,;
\end{equation}
the definition of $R_b=0.41\pm0.07$ can be found in (\ref{Rb-intro}).

It is very difficult to calculate the ``penguin'' parameter $a'e^{i\theta'}$,
which introduces the CP-violating phase factor $e^{i\gamma}$ into
the $B_d^0\to J/\psi\, K_{\rm S}$ decay amplitude and represents -- sloppily 
speaking -- the ratio of the penguin to tree contributions. However, this 
parameter, and therefore also $e^{i\gamma}$, enters in (\ref{Bd-ampl2}) 
in a doubly Cabibbo-suppressed way. Consequently, to a very good
approximation, 
$B_d^0\to J/\psi\, K_{\rm S}$ is dominated by only one CKM amplitude,
so that, from (\ref{ee7}) and (\ref{ee10}):
\begin{equation}\label{e12}
{\cal A}^{\mbox{{\scriptsize mix}}}_{\mbox{{\scriptsize
CP}}}(B_d\to J/\psi\, K_{\mbox{{\scriptsize S}}})=+\sin[-(\phi_d-0)]=
-\sin(2\beta)\,.
\end{equation}
Since (\ref{ee10}) applies with excellent accuracy to $B_d\to J/\psi\, 
K_{\mbox{{\scriptsize S}}}$, as penguins
enter essentially with the same weak phase as the leading tree
contribution, it is referred to as the 
``gold-plated'' mode to determine the $B^0_d$--$\overline{B^0_d}$ mixing
phase \cite{bisa}. Strictly speaking, mixing-induced CP violation in 
$B_d\to J/\psi\, K_{\rm S}$ probes $\sin(\phi_d+\phi_K)$, where $\phi_K$ 
is related to the CP-violating weak $K^0$--$\overline{K^0}$ mixing phase. 
Similar modifications must also be performed for other final-state 
configurations containing $K_{\rm S}$- or $K_{\rm L}$-mesons. However, 
$\phi_K$ is negligible in the SM, and -- owing to the 
small value of the CP-violating parameter $\varepsilon_K$ of the neutral 
kaon system -- can only be affected by very contrived models of new 
physics \cite{nir-sil}.

First attempts to measure $\sin(2\beta)$ through the CP asymmetry 
(\ref{e12}) have recently been performed by the OPAL, CDF and ALEPH 
collaborations \cite{sin2b-exp}:
\begin{equation}\renewcommand{\arraystretch}{1.1}
\sin(2\beta)=\left\{\begin{array}{l@{\quad}l}
3.2^{+1.8}_{-2.0}\pm0.5&\mbox{(OPAL Collaboration)}\\
0.79^{+0.41}_{-0.44}&\mbox{(CDF Collaboration)}\\
0.93^{+0.64+0.36}_{-0.88-0.24}&\mbox{(ALEPH Collaboration).}
\end{array}\right.
\renewcommand{\arraystretch}{1}\end{equation}
Although the experimental uncertainties are very large, it is interesting 
to note that these results favour the SM expectation of a 
{\it positive} value of $\sin(2\beta)$. In the $B$-factory era, an 
experimental uncertainty of $\left.\Delta\sin(2\beta)\right|_{\rm exp}=0.05$ 
appears to be achievable, whereas the experimental uncertainty at the LHC
is expected to be one order of magnitude smaller, as discussed on
page~\pageref{page:sin2b}.

In addition to (\ref{e12}), one more important implication of the SM
is 
\begin{equation}\label{e13}
{\cal A}^{\mbox{{\scriptsize dir}}}_{\mbox{{\scriptsize
CP}}}(B_d\to J/\psi\, K_{\mbox{{\scriptsize S}}})\approx0\approx
{\cal A}_{\mbox{{\scriptsize CP}}}(B^+\to J/\psi\, K^+),
\end{equation}
which is interesting for the search of new physics. An observation of
these direct CP asymmetries at the level of 10\% would be a strong 
indication for physics beyond the SM. 

In view of the 
tremendous experimental accuracy that can be achieved in the LHC era, 
it is an important issue to investigate the theoretical accuracy of 
(\ref{e12}) and (\ref{e13}), which is a very challenging theoretical
task. An interesting channel in this respect is $B_s\to J/\psi\,K_{\rm S}$ 
\cite{RF-BdsPsiK}, allowing one to control the (presumably very 
small) penguin uncertainties in the determination of $\beta$ from 
CP-violating effects in $B_d\to J/\psi\,K_{\rm S}$, and to extract the
angle $\gamma$. We shall come back to this strategy in 
Sec.~\ref{subsec:BsdPsiKS}.

\def\Bdpsiks{$B^0_d\rightarrow J/\psi \, K^0_S\,$}
\def\Bdpsiksvis{$B^0_d\rightarrow J/\psi[\to
\mu^+\mu^- \,\mbox{or} \, e^+e^-] \, 
K^0_S[\rightarrow \pi^+\pi^-]$}
\def\Jtomumu{$J/\psi \rightarrow \mu^+\mu^-$}
\def\Jtoee{$J/\psi \rightarrow e^+e^-$}

\subsubsection{Experimental Studies}
\label{sec_bdpsiKS_exp}

As well as being theoretically "gold-plated",  
the decay \Bdpsiks, with $J/\psi \rightarrow \mu^+\mu^-$ or $J/\psi 
\rightarrow e^+e^-$ is experimentally 
clean, and can be reconstructed with relatively low background.
The $B^0_d\rightarrow J/\psi K^0$ branching ratio is 
measured to be $\rm (8.9 \pm 1.2) \times 10^{-4}$~\cite{PDG}, 
yielding a visible branching ratio $B$(\Bdpsiksvis) of $\rm 1.8 \times 
10^{-5}$. For a complete account of each of the analyses described below, see 
Refs.~\cite{ATtag,CMtag,LHCbTP}.

\subsection*{Selection}

\begin{table}
\begin{center}
\begin{tabular}{|l||c|r|r|r|r|r|} \hline
Selection stage & \multicolumn{2}{c|}{ATLAS} &
\multicolumn{2}{c|}{CMS} & 
\multicolumn{2}{c|}{LHCb} \\ \cline{2-7} \cline{2-7}
                          & \multicolumn{1}{c|}{$ \mu^+
                  \mu^-$} & \multicolumn{1}{c|}{$e^+ e^-$}  &
                \multicolumn{1}{c|}{$\mu^+ \mu^-$} & 
\multicolumn{1}{c|}{$e^+ e^-$} & \multicolumn{1}{c|}{$\mu^+ \mu^-$} & 
\multicolumn{1}{c|}{$e^+ e^-$} \\ \hline
First trigger level       &  733k & 48.9k & 3485k & 893k  & 818k  & 425k  \\
Second trigger level    &  536k & 16.8k & 1394k & 353k  & 116k  & 60k   \\
$B^0_d$ reconstruction &  160k & 4.8k  &  384k & 49k   & 73k   & 15k
\\ \hline \hline
Signal/Background & \phantom{0}31 & 16 & 8 & 2 & 7 & 2 \\ \hline
\end{tabular}
\end{center}
\vspace*{-0.5cm}
\caption[]{\Bdpsiks\ event yields at different stages of the selection 
procedure and S/B ratio for one year's data. The events are untagged 
apart from the ATLAS \Jtoee\  
sample which is automatically tagged by the Level-1 trigger
muon. The ATLAS 
Level-2 trigger numbers also include the $J/\psi$ 
reconstruction offline cuts.}\label{tab_bdpsiks_yield}
\end{table}

In each experiment the event samples were generated using
PYTHIA and the full detector response was simulated using the GEANT 
program.
For the ATLAS analyis, electron and muon identification efficiencies 
are parametrized as a function of $p_T$ and $\eta$ using separate samples 
of fully simulated calorimeter and muon chamber data and then applied to
the \Bdpsiks\ sample.

Trigger strategies for the three experiments are summarized 
in Sec.~\ref{exp:trig}: here triggers relevant for the \Bdpsiks\ analysis are
briefly recalled. In the ATLAS analysis, a single muon with $p_T >6$ GeV
and $|\eta|<2.4$ is required at Level-1. To increase statistics, a 
dimuon trigger (with $\eta$--dependent thresholds) is under study. 
At Level-2, the trigger requires either a second muon with $p_T>3$ GeV, an
electron with $p_T>5$ GeV or a $e^+ e^-$ pair, with electron $p_T$
thresholds at 0.5 GeV. In CMS, the following Level-1 triggers are 
available: 1 $\mu$ with $p_T > 7$ GeV, 2 $\mu$'s with 
$p_T > 2\,  \textrm{or} \, 4$ GeV (depending of $\eta$), 1 $e$ with
$p_T > 12$ GeV, 2 $e$ with $p_T > 5$ GeV and  an $e-\mu$ pair with
$p_T(e) > 4.5$ GeV and $p_T(\mu) > 2\,  \textrm{or} \, 4$ GeV.

The first step in reconstructing \Bdpsiks\ decays is the 
selection of oppositely charged lepton pairs originating from a common vertex
and with a mass close to the $J/\psi$ mass. Next,  
$K^0_S$ candidates are selected and combined with those from $J/\psi$
to form $B^0_d$ candidates. 
In ATLAS, the same lepton trigger $p_T$-cuts are applied in the offline 
selection. In CMS and LHCb, no offline cuts are applied to the lepton-$p_T$ 
after pattern recognition.

For the \Jtoee\ selection, both ATLAS and CMS use an asymmetric window
for the reconstructed $J/\psi$ mass in order to account for bremsstrahlung 
energy-losses which produce a long tail at small invariant masses. Cuts on 
the $J/\psi$ decay-length remove the prompt $J/\psi$ background. 
In LHCb, 
to guarantee a good vertex resolution, the tracks are required to have
hits in the vertex detector.

In ATLAS and CMS, the $K^0_S$ candidates are reconstructed from all 
oppositely 
charged track pairs originating from a common vertex and with a mass
close to that of the kaon. In LHCb, the charged tracks are required to
be identified as pions in the RICH system.
To reduce combinatorial background, the $K^0_S$-candidate vertices are
required to 
be well separated from the primary vertex. 
The leptons and pions from the surviving $J/\psi$ and $K^0_S$ candidates
are then used
to reconstruct candidate \Bdpsiks\ decays using a three-dimensional kinematic 
fit to the four tracks and applying vertex and mass constraints on both the 
lepton-lepton and $\pi^+\pi^-$ system. 
Finally, the fully reconstructed 
$B^0_d$ is required to point to the reconstructed primary vertex.
The event yields (untagged) at various stages in the selection of the three 
experiments are shown in Tab.~\ref{tab_bdpsiks_yield}. The final $B^0_d$ mass 
and proper time resolutions are shown in Tab.~\ref{tab_bdpsiks_attr}.
\begin{table}
\begin{center}
\begin{tabular}{|l||c|c|r|r|r|r|} \hline
 & \multicolumn{2}{c|}{ATLAS} & \multicolumn{2}{c|}{CMS} & 
\multicolumn{2}{c|}{LHCb} \\ \hline \hline
                & $ \mu^+ \mu^-$ & $e^+ e^-$  &
\multicolumn{1}{c|}{$\mu^+ \mu^-$} & \multicolumn{1}{c|}{$e^+ e^-$} & 
\multicolumn{1}{c|}{$\mu^+ \mu^-$} & \multicolumn{1}{c|}{$e^+ e^-$} \\ 
\cline{2-7}
Mass resolution [MeV/c$\rm ^2$]   & 18  & 24  & 16 & 22  & 7 & 20 \\
Proper time resolution [ps]  & 73  & 73  & 61 & 61  & 36  & 44 \\
\hline
\end{tabular}
\end{center}
\vspace*{-0.5cm}
\caption[]{Mass and proper time resolution of the reconstructed 
$B^0_d$ meson after all offline selection 
cuts for each experiment.}\label{tab_bdpsiks_attr}
\end{table}
\begin{figure}
~\\[-1cm]
$$
\subfigure[ATLAS]
{\epsfig{file=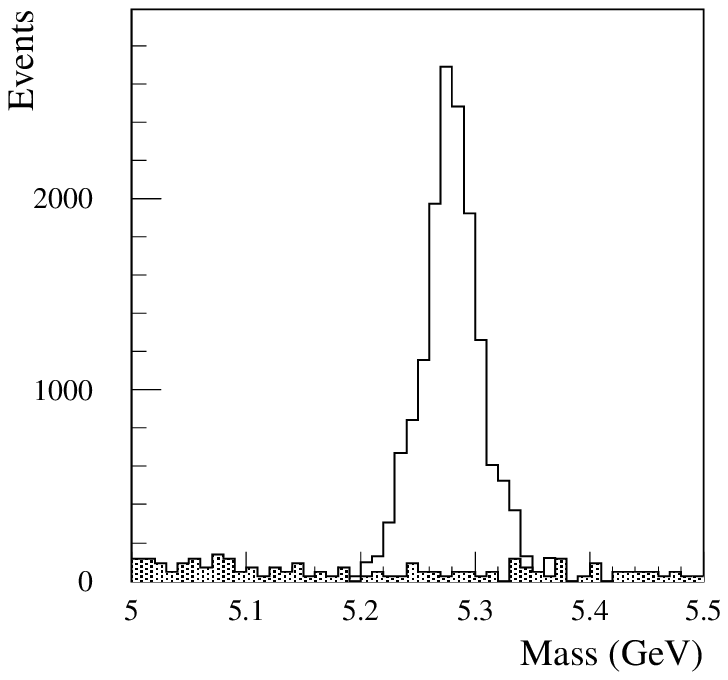,width=0.31\textwidth}}\quad
\subfigure[CMS]
{\epsfig{file=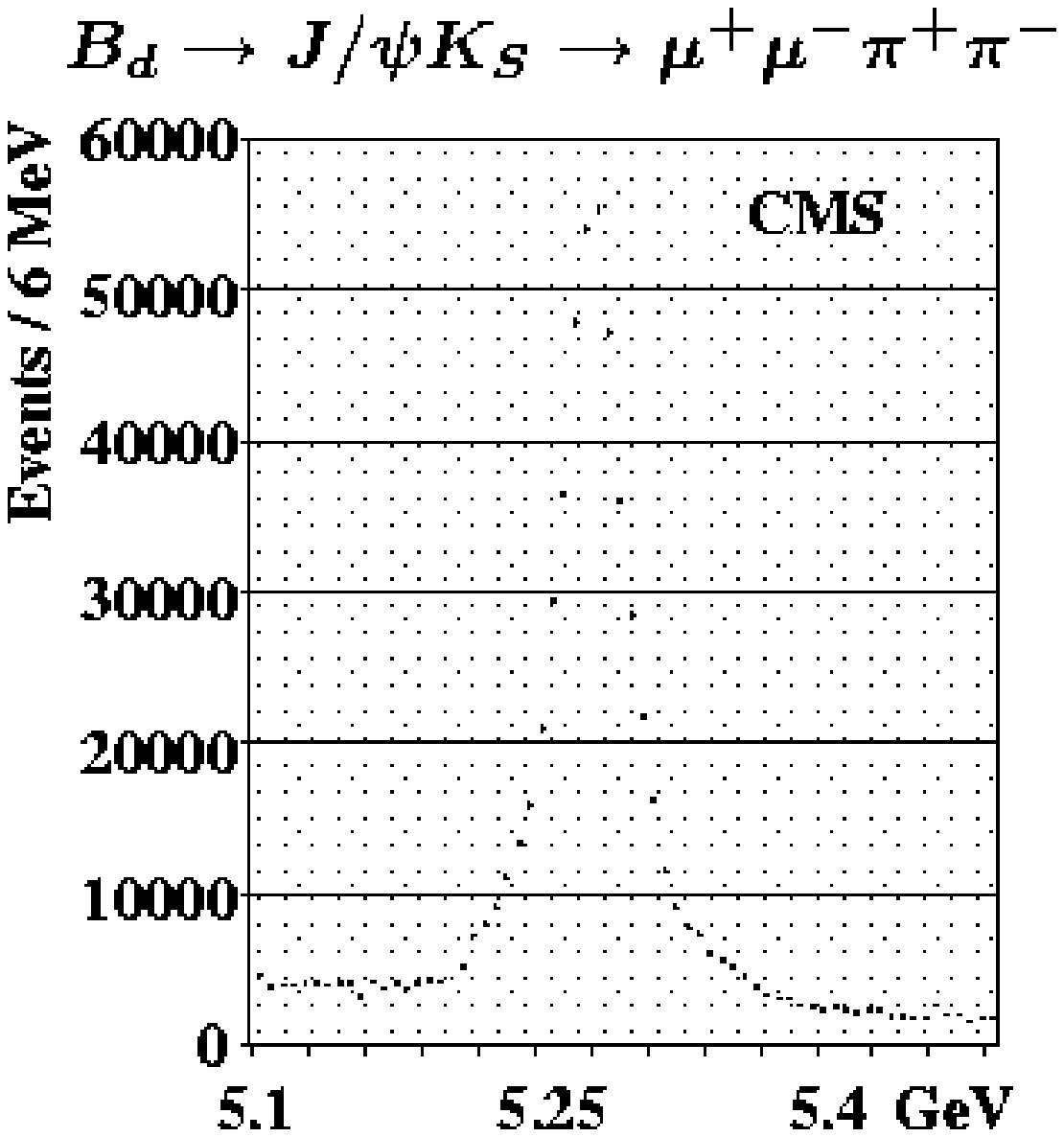,width=0.28\textwidth}}\quad
\subfigure[LHCb]
{\epsfig{file=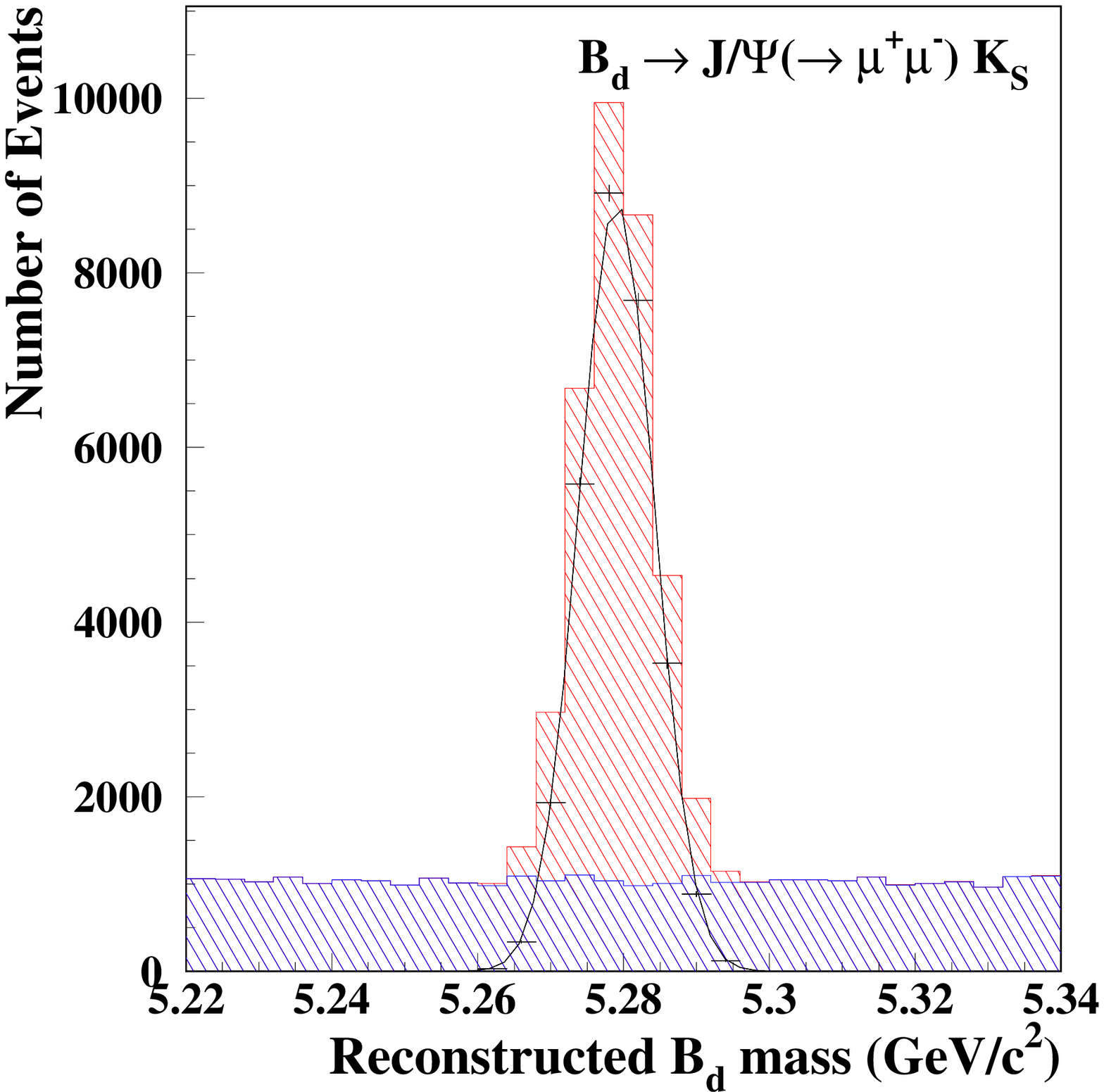,width=0.31\textwidth}}
$$
\vspace*{-1cm}
\caption[]{Example reconstructed \Bdpsiks\ mass peaks (signal and background)
for the three LHC experiments: (a) ATLAS $J/\psi 
\rightarrow e^+e^-$ sample after 3 years of data; (b) CMS $J/\psi 
\rightarrow \mu^+\mu^-$ sample after 1 year of data; 
(c) LHCb  $J/\psi \rightarrow \mu^+\mu^-$ sample after 1 year of data.}
\label{bspsiks_masspk}
\end{figure}

In each experiment, the 
dominant source of background arises from the 
combination of a 
true $J/\psi$ from $B$ decay and any other $K^0_S$ within the event, 
which can originate  from fragmentation,
from other $B$ decays or be a fake $K^0_S$. 
In LHCb, thanks to the $\pi/K$ separation available in the RICH, 
the fake $K^0_s$
contribution is reduced and the only significant
background is due to real $J/\psi$ from $B$ combined with
a real $K^0_S$.
However, this background is rather large due to the large number of $K^0_S$
mesons from fragmentation produced in the forward direction within the 
LHCb acceptance.
ATLAS used fast simulation programs (after careful comparison with full
simulation results) to generate large samples
of all backgrounds, whereas LHCb used smaller samples of fully
simulated events
 and extrapolated to higher statistics. CMS used a combination of the
two approaches.
The signal/background ratios obtained in the three
 experiments after all offline selection-cuts and before any 
flavour tagging (except for the ATLAS \Jtoee\ sample, where the
flavour is tagged automatically by the Level-1 trigger muon) are
summarized in Table~\ref{tab_bdpsiks_yield}.
Figure~\ref{bspsiks_masspk}
shows example $B^0_d$ mass peaks (signal and background) after all offline 
cuts: the background levels are low in all cases. 

\subsection*{Tagging}

Some of the flavour tagging strategies introduced in Sec.~\ref{exp_ft}
have been studied in particular detail for the \Bdpsiks\ channel.
All three experiments use the lepton from the semileptonic decay of the other
$b$ hadron (the opposite side $b$) 
in the event to tag the flavour of $B^0_d$ at production. In the ATLAS
\Jtoee\ sample, the Level-1 trigger muon provides a 100\% efficient tag.
Using the $\pi$--K separation provided by its RICH detector, 
LHCb can also use kaons to tag the flavour of the opposite side $b$ quark.
In addition to the lepton tag, ATLAS and CMS studied jet-charge tagging 
(both on the opposite and same side) and $B$--$\pi$ correlation tagging. 
The same-side 
jet-charge tags and the $B$--$\pi$ tags are highly correlated. For this 
reason, ATLAS chose to use only the higher purity $B$--$\pi$ tag. 
It has not yet been demonstrated that the same side $B$--$\pi$ tag 
method will work in LHCb since the track densities encountered there
are large.  
All three experiments plan to 
combine all tagging information in each event in order to obtain optimal
statistical precision. 
The efficiencies and mistag rates of all tagging methods are shown 
in Tab.~\ref{tab_bdpsiks_tags}. 
For the LHCb  study, the overall tagging efficiency 
and dilution of the combined lepton and kaon tagging method 
(see Sec.~\ref{comb_tags}) have been used.
\begin{table}
\begin{center}
\begin{tabular}{|l||c|c|c|c|c|c|} \hline
Tagging method & \multicolumn{2}{c|}{ATLAS} & \multicolumn{2}{c|}{CMS}
& 
\multicolumn{2}{c|}{LHCb} \\ \hline \hline
                & efficiency & dilution & efficiency & dilution &
                efficiency 
& dilution \\ \cline{2-7}
electron        & 0.012/--\phantom{.00} & 0.46/--\phantom{.00}  & 
0.024/0.035 & 0.44  & n/a  & n/a \\
muon            & 0.025/1.\phantom{00} & 0.52/0.57 & 0.033/0.035        
& 0.44  & n/a  & n/a \\
$B-\pi$ (or $B^{**}$) & 
  \phantom{0}0.82/0.80 & 0.16/0.14 & 0.21 & 0.32  & n/a  & n/a \\
jet charge (SS) & \phantom{0}0.64/0.71 & 0.17/0.12 & 0.5\phantom{0}
&0.23& 
n/a & n/a \\ 
jet charge (OS) & n/a & n/a & 0.70 & 0.18 & n/a & n/a \\ 
lepton and kaon & n/a & n/a & n/a  & n/a  & 0.40  & 0.40 \\ \hline
\end{tabular}
\end{center}
\vspace*{-0.5cm}
\caption[]{Tagging efficiencies and dilution factors for each of the
                tagging methods used by the three collaborations in the
\Bdpsiks\ analysis. 
Numbers before and after the slash (/) are for the \Jtomumu\ 
and \Jtoee\ samples, respectively. ATLAS uses $B$--$\pi$ and CMS
uses $B^{**}$ tagging (see Sec.~\ref{exp_ft}). The shorthand ``n/a''
means {\em not available} or {\em not applied} in this analysis by
a particular experiment.}\label{tab_bdpsiks_tags}
\end{table}

\subsection*{Sensitivity to \protect\boldmath $\beta$}

\begin{figure}[bth]
\begin{center}
\begin{tabular}{ccc}
   \epsfysize=6cm
   \epsffile{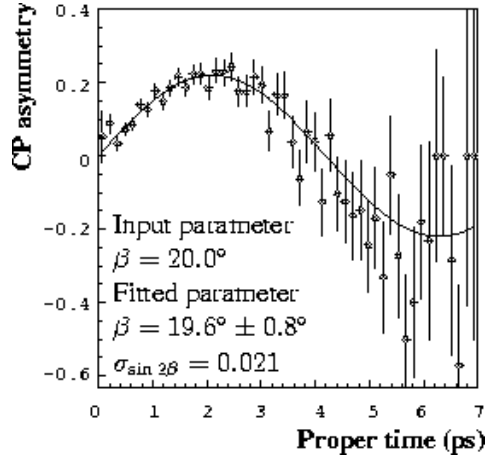}
\end{tabular}
\end{center}
\vspace*{-0.8cm}
\caption[]{Example time-dependent fit for $\beta$ to the asymmetry of
   Eq.~\protect{(\ref{eq:shit})} for LHCb with one year's data.}
\label{lhcb_beta_fit}
\end{figure}

The CKM parameter $\beta$ is extracted from a fit to the measured 
time-dependent asymmetry with a function of the form: 
\begin{equation}\label{eq:shit}
{\cal A}_{CP}(\rm B^0_d \rightarrow J/\psi K^0_s) = D \sin (2\beta)
\sin \Delta m t, 
\end{equation}
where $D$ is the overall dilution factor due to both tagging and background. 
Here any direct CP violation is neglected,  and the only 
free parameter in the fit is $\sin 2 \beta$. The background is assumed to
have no asymmetry. Figure~\ref{lhcb_beta_fit} shows an example fit to the LHCb
time-dependent CP asymmetry distribution after one year of data taking.

Table~\ref{tab_bdpsuks_beta} summarizes the sensitivity of the three
experiments to $\sin{2\beta}$ using the different tagging
methods studied by each experiment. The ATLAS lepton-tagged events have 
been removed from the $B$--$\pi$ tagged sample to yield two 
statistically independent samples. The four separate CMS results are 
statistically correlated. However, to obtain the final precision on
$\sin{2\beta}$, the 
analysis was performed using only those events not tagged by another 
method, as explained in Sec.~\ref{comb_tags}.

All experiments estimate a statistical error on $\sin 2\beta$ which is
independent of the input value for $\beta$.
\begin{table}
\begin{center}
\begin{tabular}{|l||c|c|c|c|c|c|} \hline
Tagging method & \multicolumn{2}{c|}{ATLAS} & \multicolumn{2}{c|}{CMS}
& 
\multicolumn{2}{c|}{LHCb} \\ \hline \hline
                & $ \mu^+ \mu^-$ & $e^+ e^-$  &
                $\mu^+ \mu^-$ & 
$e^+ e^-$ & $\mu^+ \mu^-$ & $e^+ e^-$ \\ \cline{2-7}
Lepton          & 0.039  & 0.031  & \multicolumn{2}{c|}{0.031}     &
n/a  & n/a \\
$B$--$\pi$      & 0.026  &  n/a      & \multicolumn{2}{c|}{0.023}  
& n/a  & n/a \\
SS Jet charge   &  n/a   &  n/a      & \multicolumn{2}{c|}{0.021}  
& n/a  & n/a \\
OS Jet charge   &  n/a   &  n/a      & \multicolumn{2}{c|}{0.023}  
& n/a  & n/a \\
Lepton and kaon &  n/a   &  n/a      & \multicolumn{2}{c|}{ n/a   }  &
0.023  & 0.051 
\\ \hline \hline
Total           & \multicolumn{2}{c|}{0.017} &
\multicolumn{2}{c|}{0.015}  & 
\multicolumn{2}{c|}{0.021} \\ \hline
\end{tabular}
\end{center}
\vspace*{-0.5cm}
\caption[]{Sensitivity to $\sin{2\beta}$ after one year of data taking at
the LHC. For the ATLAS  $J/\psi \rightarrow \mu^+ \mu^-$ sample, lepton
tags have been removed from the $B$--$\pi$ tagged sample. The four partial
CMS results are correlated, but the total sensitivity has been obtained
subtracting overlaps. The shorthand ``n/a'' means {\em not available}
or {\em not applied} in this analysis by a particular 
experiment.}\label{tab_bdpsuks_beta}
\end{table}
Combining the statistical precision achievable after 3 years of running 
of ATLAS and CMS with 5 years of running at LHCb, a total statistical 
precision on $\sin 2\beta$ of 0.005\label{page:sin2b} can be obtained. 
This precision is  one
order of magnitude better than the expected statistical precision at
the $e^+ e^-$ B factories.    With this sensitivity, the experiments
can also probe for a direct CP violating contribution, 
${\cal A}^{\rm dir}_{CP}(\rm B^0_d \rightarrow J/\psi K^0_s)$, to
the asymmetry.   Fitting
an additional term  to account for such
a contribution degrades the precision on 
$\sin 2 \beta$ by $\sim 30\%$ and gives a similarly small uncertainty on
${\cal A}^{\rm dir}_{CP}(\rm B^0_d \rightarrow J/\psi K^0_s)$.

\subsection*{Systematic Uncertainties}

In order not to compromise the excellent statistical precision obtainable
on the determination of $\sin{2\beta}$ at the LHC, a similar or better
control of the systematic uncertainties must be achieved. 

A detailed discussion on systematic errors on CP-violation measurements
and strategies to control them are presented in Sec.~\ref{sec:syst}.
As theoretical
uncertainties are expected to be very small, the main contribution to the
systematic error comes from the initial-state production asymmetry and
from experimental factors. The latter ones include tagging uncertainties
and uncertainties from background.

ATLAS have performed a preliminary estimate of such uncertainties
using $B^+ \rightarrow J/\psi(\mu \mu) K^+$ and 
$B^0_d \rightarrow J/\psi(\mu \mu) K^{\star 0}$ control 
samples~\cite{ATtag}. It is estimated that for a statistical error of
$\sin{2\beta}=0.010\,$(stat.), achievable after 3 years running, a
corresponding systematic error of $\sin{2\beta}=0.005\,$(sys.), coming
from the limited size of the control channels, can be obtained.

\subsection[Probing $\alpha$ with 
$B_d\to\pi^+\pi^-$]{Probing \protect\boldmath $\alpha$ with
  \protect\boldmath $B_d\to\pi^+\pi^-$\protect\footnote{With 
help from J. Charles, D. Rousseau and A. Starodumov.}}\label{subsec:Bpipi}
Another benchmark CP mode is $B_d\to\pi^+\pi^-$, which allows one to
probe the CKM angle $\alpha$. Unfortunately, penguin topologies make
the interpretation of the CP-violating $B_d\to\pi^+\pi^-$ observables
in terms of $\alpha$ difficult. 

\begin{figure}[htb]
\begin{center}
\leavevmode
\epsfysize=4.0truecm 
\epsffile{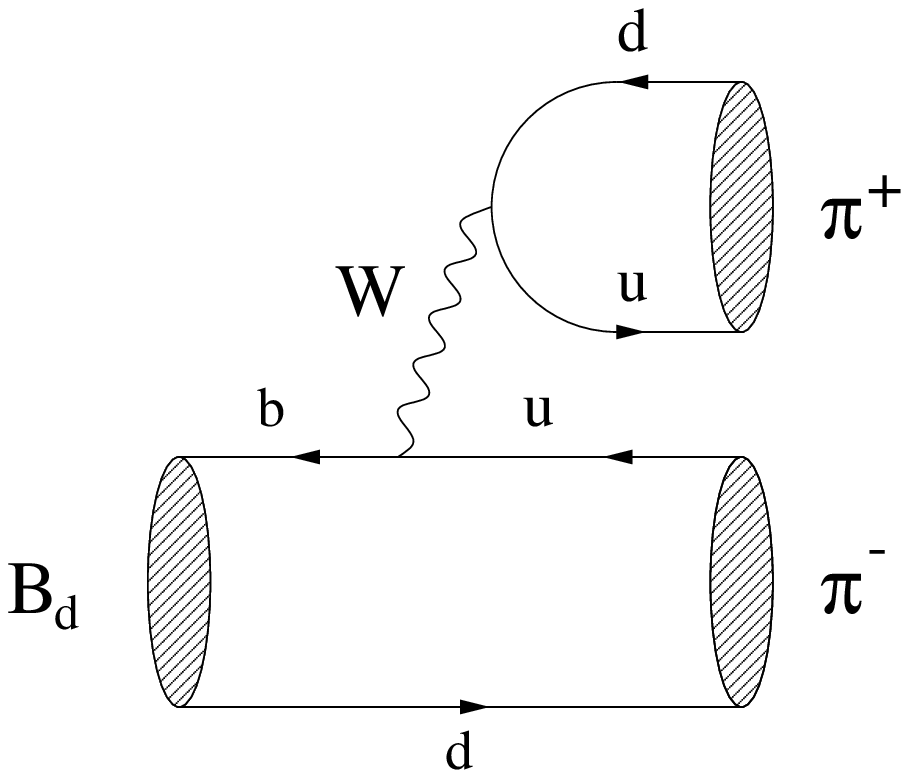} \hspace*{1truecm}
\epsfysize=4.5truecm 
\epsffile{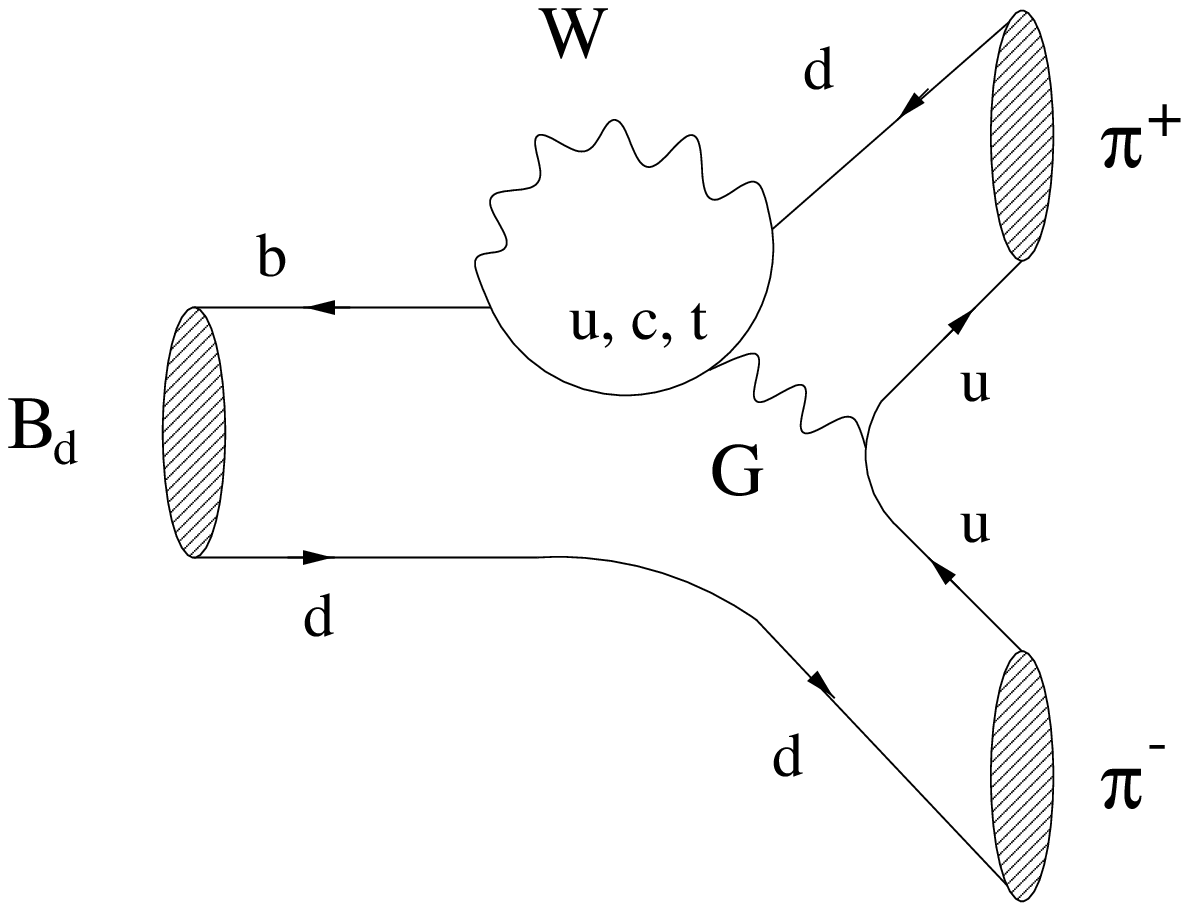}
\end{center}
\vspace*{-0.5truecm}
\caption{Feynman diagrams contributing to 
$B_d^0\to\pi^+\pi^-$.}\label{fig:bpipi}
\end{figure}

\subsubsection{Theoretical Aspects}
In the case of $B_d^0\to\pi^+\pi^-$, we have to deal with the Feynman diagrams
shown in Fig.~\ref{fig:bpipi}, and in analogy to (\ref{Bd-ampl1}),
the corresponding decay amplitude can be expressed as
\begin{equation}\label{Bdpipi-ampl1}
A(B_d^0\to\pi^+\pi^-)=\lambda_u^{(d)}\left(A_{\rm cc}^{u}+A_{\rm pen}^{u}
\right)+\lambda_c^{(d)}A_{\rm pen}^{c}+\lambda_t^{(d)}A_{\rm pen}^{t}\,.
\end{equation}
If this mode did not receive penguin contributions, its mixing-induced
CP asymmetry would allow a measurement of $\sin 2\alpha$, in complete
analogy to $B_d\to J/\psi K_S$:
\begin{equation}
{\cal A}^{\mbox{{\scriptsize mix}}}_{\mbox{{\scriptsize
CP}}}(B_d\to\pi^+\pi^-)=-\sin[-(2\beta+2\gamma)]=-\sin 2\alpha.
\end{equation}
However, this relation is strongly affected by penguin effects, 
which were analysed by many authors \cite{alpha-uncert,charles}. There 
are various methods on the market to control the corresponding hadronic
uncertainties; unfortunately, these strategies are usually rather 
challenging from an experimental point of view. 

The best-known approach was proposed by Gronau and London \cite{gl}. 
It makes use of the SU(2) isospin relation
\begin{equation}
\sqrt{2}\,A(B^+\to\pi^+\pi^0)=A(B^0_d\to\pi^+\pi^-)+
\sqrt{2}\,A(B^0_d\to\pi^0\pi^0),
\end{equation}
and of its CP-conjugate, which form two triangles in the complex
plane. The sides of these triangles can be determined through the 
corresponding branching ratios, while their relative orientation can be 
fixed by measuring the CP-violating observable ${\cal A}^{\mbox{{\scriptsize 
mix}}}_{\mbox{{\scriptsize CP}}}(B_d\to\pi^+\pi^-)$. 
Following these lines, it is in principle possible to take into account the 
QCD penguin effects in the extraction of $\alpha$. It should be noted that 
electroweak penguins cannot be controlled with the help of this isospin 
strategy. However, their effect is expected to be rather small, and -- as was 
pointed out recently \cite{BF,GPY} -- can be included through additional 
theory input. Unfortunately, the Gronau--London approach suffers from an 
experimental problem, since the measurement of $B(B_d\to\pi^0\pi^0)$,
which is expected to be of ${\cal O}(10^{-6})$ or smaller, is very difficult. 
However, upper bounds on the CP-averaged $B_d\to\pi^0\pi^0$ branching 
ratio may already be useful to put upper bounds on the QCD penguin 
uncertainty 
that affects the determination of $\alpha$ \cite{charles,gq-alpha}.

Alternative methods to control penguin uncertainties are very desirable. One 
of them is provided by $B\to\rho\,\pi$ modes \cite{Brhopi,SQ},
and will be discussed in more detail in the following subsection. As 
we shall see in Sec.~\ref{subsec:BsKK}, another interesting strategy
is to use the CP-violating observables of
$B_s\to K^+ K^-$ {\em together} with those of $B_d\to\pi^+\pi^-$,
which allows a
simultaneous determination of $\beta$ and $\gamma$ {\em without} any 
assumptions about penguin topologies. 

The observation of $B_d\to\pi^+\pi^-$ was announced by the CLEO 
collaboration in the summer of 1999 \cite{cleo-Bpipi}, with a branching 
ratio of
\begin{equation}
B(B_d\to\pi^+\pi^-)=\left(0.47^{+0.18}_{-0.15}\pm0.13\right)\times
10^{-5}. 
\end{equation}
Other CLEO results on $B\to\pi K$ modes indicate that QCD penguins play 
in fact an important r\^{o}le, and that we definitely have to worry about 
them in the extraction of $\alpha$ from $B_d\to\pi^+\pi^-$ \cite{RF-Bpipi}. 
In order to discuss penguin effects in a quantitative way, 
we use once more the unitarity of the CKM matrix, and rewrite 
(\ref{Bdpipi-ampl1}) as follows:
\begin{equation}\label{j_tp}
A(B_d^0\to\pi^+\pi^-)=e^{i\gamma}\,T+e^{-i\beta}\,P\,,
\end{equation}
where the complex quantities
\begin{equation}
T\equiv-|\lambda_u^{d}|\left[A_\mathrm{cc}^u+A_\mathrm{pen}^u-
A_\mathrm{pen}^c\right], \qquad
P\equiv-|\lambda_t^{d}|\left[A_\mathrm{pen}^t-A_\mathrm{pen}^c\right],
\end{equation}
denote the $B_d^0\to\pi^+\pi^-$ ``tree'' and ``penguin'' amplitudes,
respectively. The CP-conjugate amplitude can be obtained 
straightforwardly from (\ref{j_tp}) by replacing
$\beta$ by $-\beta$ and $\gamma$ by $-\gamma$. 
For the following considerations, 
also the CP-conserving strong phase
$
\delta\equiv\mathrm{Arg}(PT^\ast)
$
plays an important r\^{o}le. Since the $B^0_d$--$\overline{B^0_d}$ mixing
phase is given by $2\beta$ in the SM, the unitarity relation
$\alpha+\beta+\gamma=180^\circ$ allows one to express the CP-violating
observables $\mathcal{A}^\mathrm{dir}_\mathrm{CP}(B_d^0\to\pi^+\pi^-)$ and
$\mathcal{A}^\mathrm{mix}_\mathrm{CP}(B_d^0\to\pi^+\pi^-)$ as functions
of the CKM angle $\alpha$, and the hadronic parameters $|P|/|T|$ and
$\delta$. Consequently, we have at our disposal  two observables that depend
on three ``unknowns''. Eliminating the CP-conserving strong phase 
$\delta$, one obtains \cite{charles}:
\begin{equation}\label{j_aeff}
\mathcal{A}^\mathrm{mix}_\mathrm{CP}(B_d^0\to\pi^+\pi^-)
=-\sqrt{1-{\mathcal{A}^\mathrm{dir}_\mathrm{CP}}^2}\,\,\sin
2\alpha_\mathrm{\mbox{\scriptsize eff}}\,,
\end{equation}
where
\begin{equation}\label{j_alpha_p/t}
\cos(2\alpha-2\alpha_\mathrm{\mbox{\scriptsize eff}})=\frac{1}
{\sqrt{1-{\mathcal{A}^\mathrm{dir}_\mathrm{CP}}^2}}
\left[1-\left(1-\sqrt{1-{\mathcal{A}^\mathrm{dir}_\mathrm{CP}}^2}\right)
\left|\frac{P}{T}\right|^2\right]
\label{eq:ptallorder}
\end{equation}
with 
$2\alpha_\mathrm{\mbox{\scriptsize eff}}\equiv\mathrm{Arg}
\left[-\xi_{\pi^+\pi^-}^{(d)}\right]$.
$\xi_f^{(q)}$ was defined in Eq.~(\ref{xi-expr}).
The quantity $2\alpha_\mathrm{\mbox{\scriptsize eff}}$ reduces to
$2\alpha$ if penguin topologies
are neglected. Once the time-dependent CP-asymmetry (\ref{ee6}) has been 
measured, Eqs.~(\ref{j_aeff}) and (\ref{j_alpha_p/t}) allow one to fix 
contours in the $(|P|/|T|,2\alpha)$ plane. This plot constitutes a
model-independent representation of the experimental data in terms of the 
SM parameters. In order to simplify the experimental discussion
of the following subsection, we keep only leading order terms in
$|P|/|T|$, which yields~\cite{FM-Bpipi}
\begin{eqnarray}
{\cal A}_{\rm CP}^{\rm dir}(B_d\to\pi^+\pi^-)&=&2\left|\frac{P}{T}\right|\,
\sin\delta\sin\alpha+{\cal O}((|P|/|T|)^2),\nonumber\\
{\cal A}_{\rm CP}^{\rm mix}(B_d\to\pi^+\pi^-)&=&
-\sin(2\alpha)-2\left|\frac{P}{T}\right|\cos\delta\,\cos(2\alpha)\sin\alpha
+{\cal O}((|P|/|T|)^2),
\label{eq:ptorder1}
\end{eqnarray}
and leave the analysis of the exact results given in \cite{charles}
for further studies. Unfortunately, a theoretically reliable 
prediction for the ``penguin'' to ``tree'' ratio $|P|/|T|$, which would allow
the extraction of $\alpha$, is very challenging. An interesting new 
approach in this context was recently proposed in Ref.~\cite{BBNS}.
We shall come back to it in Sec.~\ref{sec:nonlept}. 
Let us finally note that any 
QCD-based approach to calculate $|P|/|T|$ requires also knowledge of  
$|V_{td}/V_{ub}|$. This input can be avoided, if all CP-violating weak 
phases are expressed in terms of the Wolfenstein parameters $\overline{\rho}$ 
and $\overline{\eta}$, allowing one to fix contours in the 
$\overline{\rho}$--$\overline{\eta}$ plane \cite{charles}.

\def\Bpipi{$\rm B^0_d\rightarrow \pi^+\pi^-$}
\def\BKpi{$\rm B^0_d\rightarrow K^\pm\pi^\mp$}
\def\BsKK{$\rm B^0_s\rightarrow K^+K^-$}
\def\BsKpi{$\rm B^0_s\rightarrow K^\pm \pi^\mp$}

\subsubsection{Experimental Studies}
\label{sec_bpipi_exp}

Low branching ratio and lack of any sub-mass constraint makes the 
reconstruction of \Bpipi\ a very demanding task.  Additional problems
are posed by isolating the signal from other two-body topologies,  such
as \BKpi, \BsKK, \BsKpi, $\rm \Lambda_b \rightarrow \, p \pi^-$ and
$\rm \Lambda_b  \rightarrow \, pK^-$ decays.
Despite these challenges,
extensive simulation studies have demonstrated the substantial potential
of the LHC experiments in this mode.   
Following recent measurements \cite{cleo-Bpipi},
these studies have assumed branching ratios
of $0.5\times10^{-5}$ for \Bpipi\ and \BsKpi,  
$1.9 \times 10^{-5}$ for \BKpi\ and \BsKK\ and
$8 \times 10^{-5}$ for $\rm \Lambda_b \rightarrow \, p \pi^-$ and
$\rm \Lambda_b  \rightarrow \, pK^-$.   
Note that much of the following discussion is
also relevant for the topics considered in Secs.~\ref{subsec:BpiK}
and \ref{subsec:BsKK} .

\subsection*{Selection}

The expected event-yields passing the early trigger levels 
are shown in Tab.~\ref{tab_bpipi_yield}.  In this mode 
LHCb in particular benefits from the high efficiency of its hadron trigger.
For ATLAS and CMS, the triggering muon will be used to flavour-tag the
events,  whereas for LHCb lepton and kaon tags will be used.

The higher level trigger and reconstruction cuts are optimised to fight 
combinatoric background from other $b\overline{b}$ events
and select genuine two-body B decays.   In these, the
requirements on the secondary vertex are the most powerful,  but isolation
and kinematic cuts also play a r\^{o}le.    The details of the selection are
discussed in Refs.~\cite{ATLASPTDR,CMSPTDR,LHCbTP}.  The event yields after
two-body selection are shown in Tab.~\ref{tab_bpipi_yield}.

\begin{table}
\begin{center}
\begin{tabular}{|l||c|c|c|} \hline
Selection stage         &  ATLAS &  CMS   & LHCb     \\ \hline \hline
First trigger level     & \phantom{.}46k  & \phantom{.}52k  & 149.9k \\
Second trigger level    & 4.2k & 4.3k  & \phantom{0}67.5k \\
Two-body selection      & 2.3k & 1.6k & \phantom{0}14.5k \\
$\pi^+ \pi^-$ selection & 2.3k & 0.9k  & \phantom{00}4.9k  \\
                        &      & (2.6k) &                   \\ \hline
\end{tabular}
\end{center}
\vspace*{-0.5cm}
\caption[]{Event yields in \Bpipi\ at various stages of the selection
procedure for one year's operation.   The final yields are for 
flavour-tagged events (an alternative yield is given for CMS, 
in brackets,  for a selection assuming
no dE/dx information).
}\label{tab_bpipi_yield}
\begin{center}
\begin{tabular}{|l||c|c|c|} \hline
                                   &  ATLAS   &  CMS            
& LHCb      \\ \hline \hline
Mass resolution  [MeV/c$^2$]   &  70    &  27        &   17    \\
Proper time resolution [ps]        &  0.065 &  0.060     &  0.04   \\
Signal / two-body background       &  0.19  &  1.6       &  15     \\
                                   &        &  (0.33)    &         \\
Signal / other background          &  1.6   &  5         &  $>\,$1 \\ 
Tagging dilution                   &  0.56  &  0.56      &  0.40
\\\hline
\end{tabular}
\end{center}
\vspace*{-0.5cm}
\caption[]{Attributes of the \Bpipi\ samples for the three experiments
(an alternative signal/two-body background number is given for CMS, 
in brackets, assuming no dE/dx information).
Note that ATLAS performs a fit to all events passing its two-body 
selection;  the background levels shown here are for illustration,
imposing a 1$\,\sigma$ mass window.}\label{tab_bpipi_props}
\end{table}

In order to reject non-$\pi^+\pi^-$
two-body background, LHCb exploits its powerful
RICH system, demanding that both tracks be identified as a pion or lighter
particle.   This and a window of $\pm 30 \, \rm MeV/c^2$ around the
$\rm B^0_d$ mass reduces the contamination by such decays to 7\%.
As explained in Sec.~\ref{sec:expover}, CMS will achieve 
a certain level
of $\pi$--K separation from the dE/dx information available from the tracker,
and the numbers and fit results presented here rely on this assumption,
although Tabs.~\ref{tab_bpipi_yield} and~\ref{tab_bpipi_props} contain
alternative numbers for a hadron-blind selection.
The requirement that both particles have an ionization within 
$^{+\, \infty}_{- \, 0.75 \sigma}$ of the expected pion energy loss,
and an invariant mass within $^{+ \, 54}_{- \, 40}~\, \rm MeV/c^2$ of the
nominal $\rm B^0_d$ mass, is expected 
to result in a final contamination of 40\%.   ATLAS chooses to make
no further cuts,  but rather to exploit the remaining discriminant 
information in a multi-parameter fit, in particular the limited
dE/dx information discussed in Sec.~\ref{sec:expover}.
At present, only ATLAS and the CMS hadron-blind analysis
have considered the background
contribution from $\rm \Lambda_b$ decays.

In addition to two-body contamination,  there will be some
residual combinatoric background passing the final cuts.
This is expected to be dominated by events with 
a false vertex being faked by two unrelated high impact parameter tracks.
The low branching ratio of the signal process renders estimates 
of the level of the combinatoric background very difficult.  ATLAS and 
CMS have used a combination of fast and full 
simulation techniques, whereas LHCb has 
extrapolated from a large sample of fully GEANT simulated events.
All experiments conclude that this background should be at lower
level than the signal.

\begin{figure}
\begin{center}
\subfigure[ATLAS]
{\epsfig{file=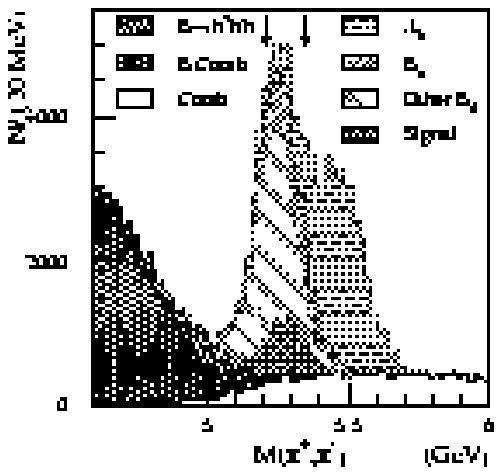,width=0.32\textwidth}}
\subfigure[LHCb before RICH]
{\epsfig{file=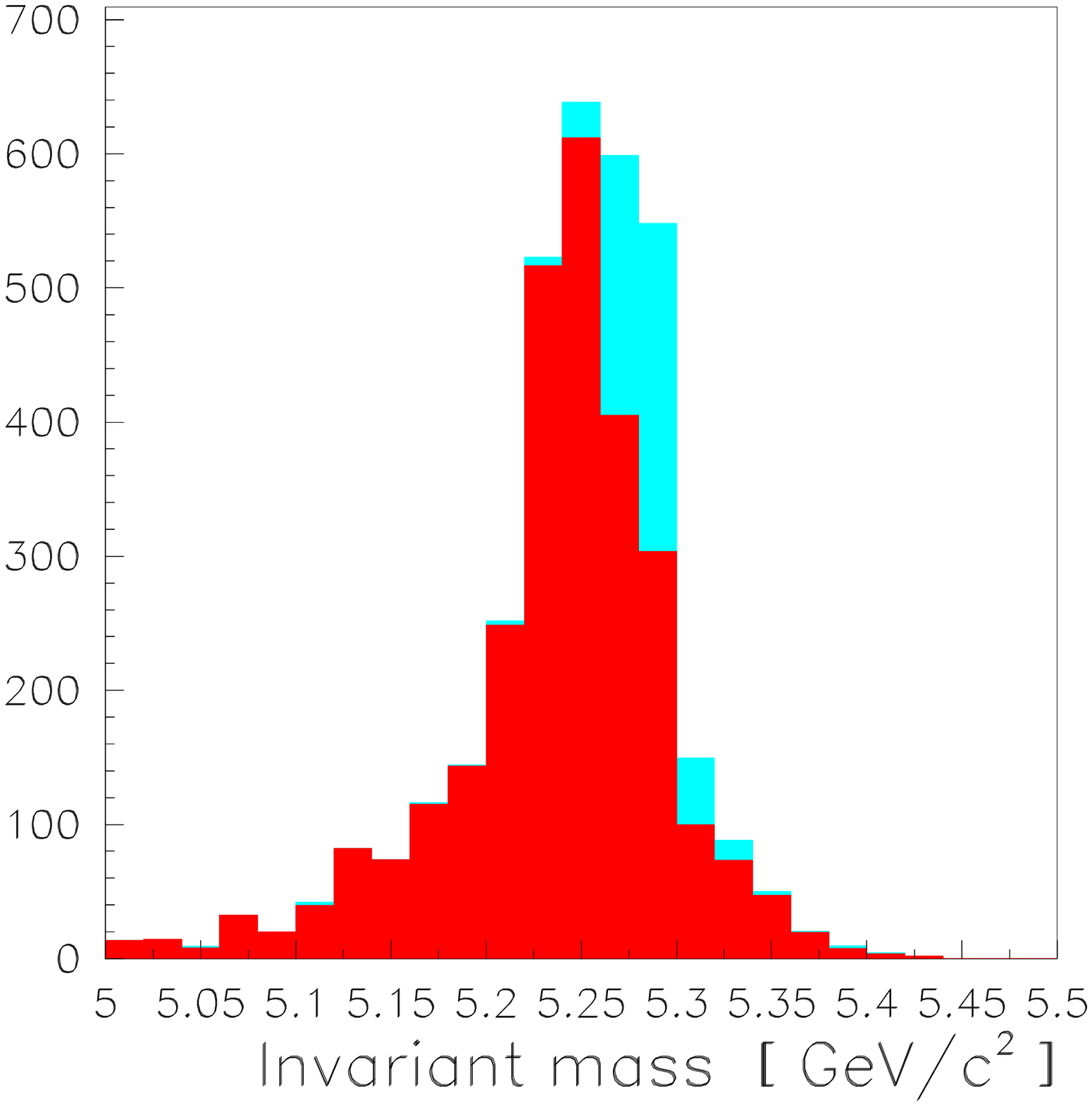,width=0.32\textwidth}}
\subfigure[LHCb after RICH]
{\epsfig{file=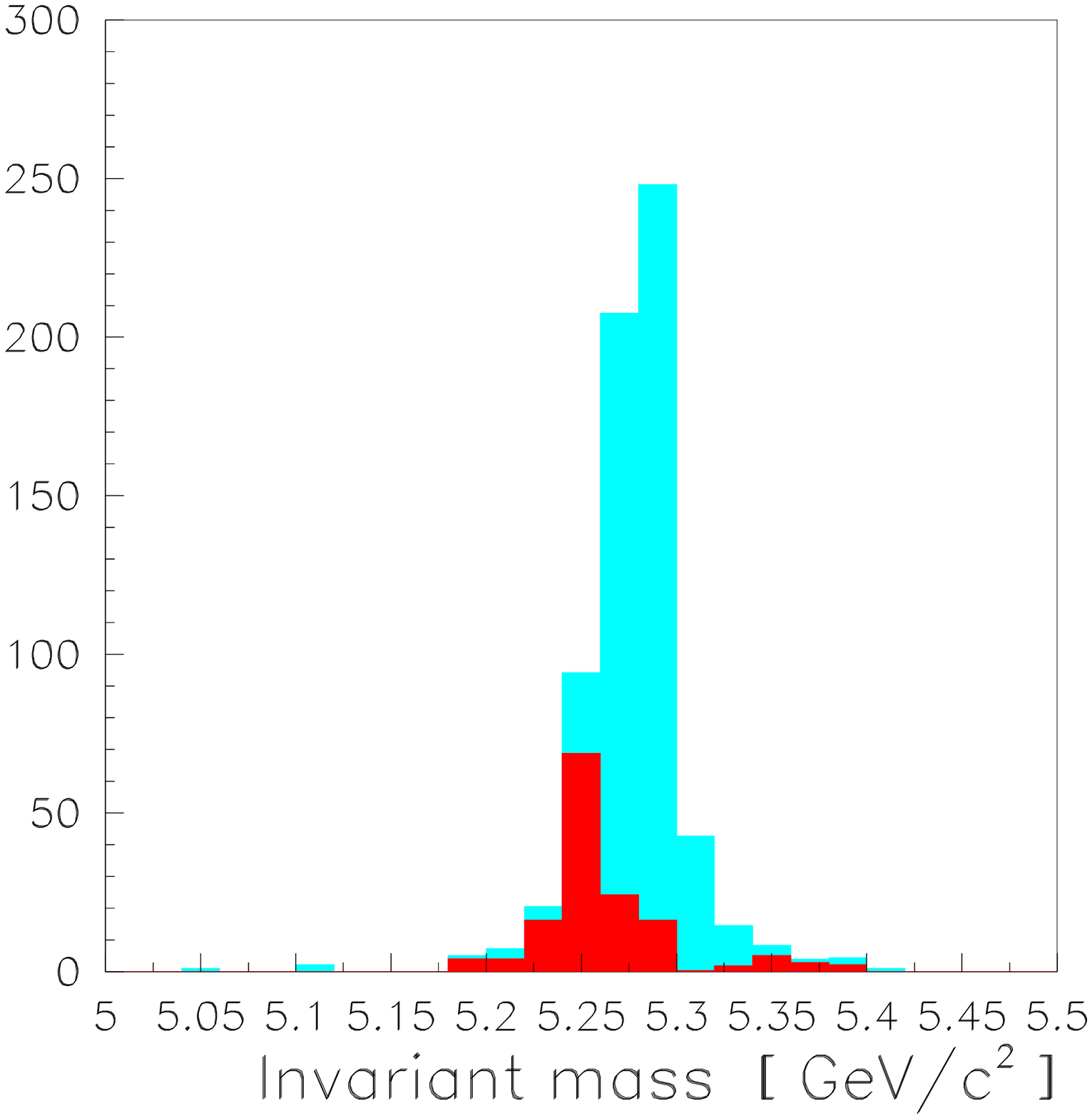,width=0.32\textwidth}}
\end{center}
\vspace*{-1cm}
\caption[]{$\pi^+\pi^-$ invariant mass peaks as simulated by ATLAS and
  LHCb. The LHCb
plots show the spectra before and after the application of RICH information,
with signal indicated by the light shading.
The ATLAS plot also contains $\Lambda_b$ decays and combinatoric background.}
\label{fig_bpipi_masspks}
\end{figure}

Some attributes of the final selected samples are given in 
Tab.~\ref{tab_bpipi_props},
and example mass peaks are shown in 
Fig.~\ref{fig_bpipi_masspks}.

\subsection*{Fitting the CP Asymmetry}

Assuming the performance figures presented above,  the experiments
have used Monte Carlo techniques to estimate their expected
sensitivity to the CP aymmetries  $\cal{A}^{\rm mix}_{\rm \pi^+\pi^-}$ 
and $\cal{A}^{\rm dir}_{\rm \pi^+\pi^-}$ from time-dependent fits, where these
are defined in the usual manner:
\begin{eqnarray}
{\cal A}_{\rm CP} (\rm B^0_d\rightarrow \pi^+\pi^-) (t) &=& 
{\cal A}^{\rm dir}_{\rm \pi^+\pi^-} \cos\Delta m t 
\, + \, 
{\cal A}^{\rm mix}_{\rm \pi^+\pi^-} \sin \Delta m t.
\end{eqnarray}
For the present study, LHCb has considered two-parameter fits
of ${\cal A}^{\rm mix}_{\rm \pi^+\pi^-}$  and 
${\cal A}^{\rm dir}_{\rm \pi^+\pi^-}$ 
to \Bpipi\ candidates passing  tight cuts.  
Any CP asymmetry in the
background has been neglected, assuming that these effects
can be controlled with sufficient precision through a study of separate
samples isolated by the RICH system.   
The CMS sensitivity with the dE/dx selection 
has been evaluated, also assuming any background asymmetry to be known.
The uncertainties obtainable with one year's statistics 
are shown in Tab.~\ref{tab_bpipi_prec}: they are found to be
independent of the values of the true CP asymmetries, symmetric
and Gaussian.  
The low frequency of the oscillations means that there is significant
correlation between  ${\cal A}^{\rm mix}_{\rm \pi^+\pi^-}$  and 
${\cal A}^{\rm dir}_{\rm \pi^+\pi^-}$.

ATLAS has developed a sophisticated method to extract the \Bpipi\
asymmetries, whereby they are determined in an unbinned
maximum-likelihood fit,  simultaneously with the asymmetries
of the other two-body classes.   Considering the allowed
$\rm \pi \pi$, $\rm \pi K$ and $\rm K K$  modes, 
$\rm \Lambda_b \rightarrow \, p \pi^-, \, pK^-$ decays
and the combinatoric background give nine coefficients.
The likelihood of  a given
decay hypothesis is computed using the event fraction,
the proper time, the invariant mass
of the two tracks under the hypothesis, the measured specific
ionization and the flavour at production and decay time.
It is assumed that the branching ratios will be known with
fractional errors of 5\%, but there is no assumption on the 
value of any possible asymmetries in the background. 
The uncertainties on the \Bpipi\
coefficients with one year's statistics  are shown
in Tab.~\ref{tab_bpipi_prec}.  
Without the $0.8\,\sigma$ $\pi/K$ separation provided by the 
ionization information, the sensitivity is about 20\% worse. 

\begin{table}
\begin{center}
\begin{tabular}{|l||c|c|c|} \hline
    &  ATLAS   &  CMS   & LHCb      \\ \hline \hline
 ${\cal A}^{\rm dir}_{\rm \pi^+\pi^-}$    &  \phantom{--}0.16  & 
\phantom{--}0.11 & \phantom{--}0.09    \\
 ${\cal A}^{\rm mix}_{\rm \pi^+\pi^-}$    &  \phantom{--}0.21  & 
\phantom{--}0.14 & \phantom{--}0.07    \\
 Correlation coefficient           &  --0.25 & --0.51 & --0.49  \\ \hline
\end{tabular}
\end{center}
\vspace*{-0.5cm}
\caption[]{Expected sensitivities for the \Bpipi\ CP asymmetry coefficients
${\cal A}^{\rm dir}_{\rm \pi^+\pi^-}$ and 
${\cal A}^{\rm mix}_{\rm \pi^+\pi^-}$
with one year's data taking, and correlation between the fitted 
parameters (the CMS numbers assume a selection exploiting dE/dx 
information).}\label{tab_bpipi_prec}
\end{table}

\subsection*{Sensitivity to \protect\boldmath $\alpha$}

\def\ApAt{$|P/T|$}

Present studies  to estimate the combined LHC precision for $\alpha$
rely on the sensitivities given in Tab.~\ref{tab_bpipi_prec}
and Eqs.~(\ref{eq:ptorder1});
they are being extended to include the full expression
(\ref{eq:ptallorder}).   The simpler expression gives rise
to `singularities' in the precision for $\alpha$ for certain parameter values
\cite{FM-Bpipi} which are not likely to occur with the full treatment.

Simulated measurements have shown that the sensitivities to the
 CP asymmetry coefficients quoted in
Tab.~\ref{tab_bpipi_prec},  estimated from the $\chi^2$ parabolic 
approximation, describe correctly the spread of experimental results.
Also, the sensitivities do not depend on the actual values of the
asymmetries, so that the numbers in Tab.~\ref{tab_bpipi_prec}
{\it  with correlations}
are sufficient to summarize 
the experimental precision of the measurements.

In contrast, the sensitivity to the parameters 
$\alpha$ and $\delta$ depends on the chosen set of parameters $\alpha$, 
$\delta$, \ApAt\ and on the theoretical uncertainty of \ApAt, so that 
the sensitivity to $\alpha$ can only be given for specific scenarios.
Also, Eqs.~(\ref{eq:ptorder1}) entail a four-fold discrete ambiguity in
$\alpha$.  Here sensitivites are given under the assumption
that this ambiguity can be correctly resolved.

\begin{figure}
$$
\begin{array}{c@{\qquad}c}
&\\[-1.7cm]
\epsfig{file=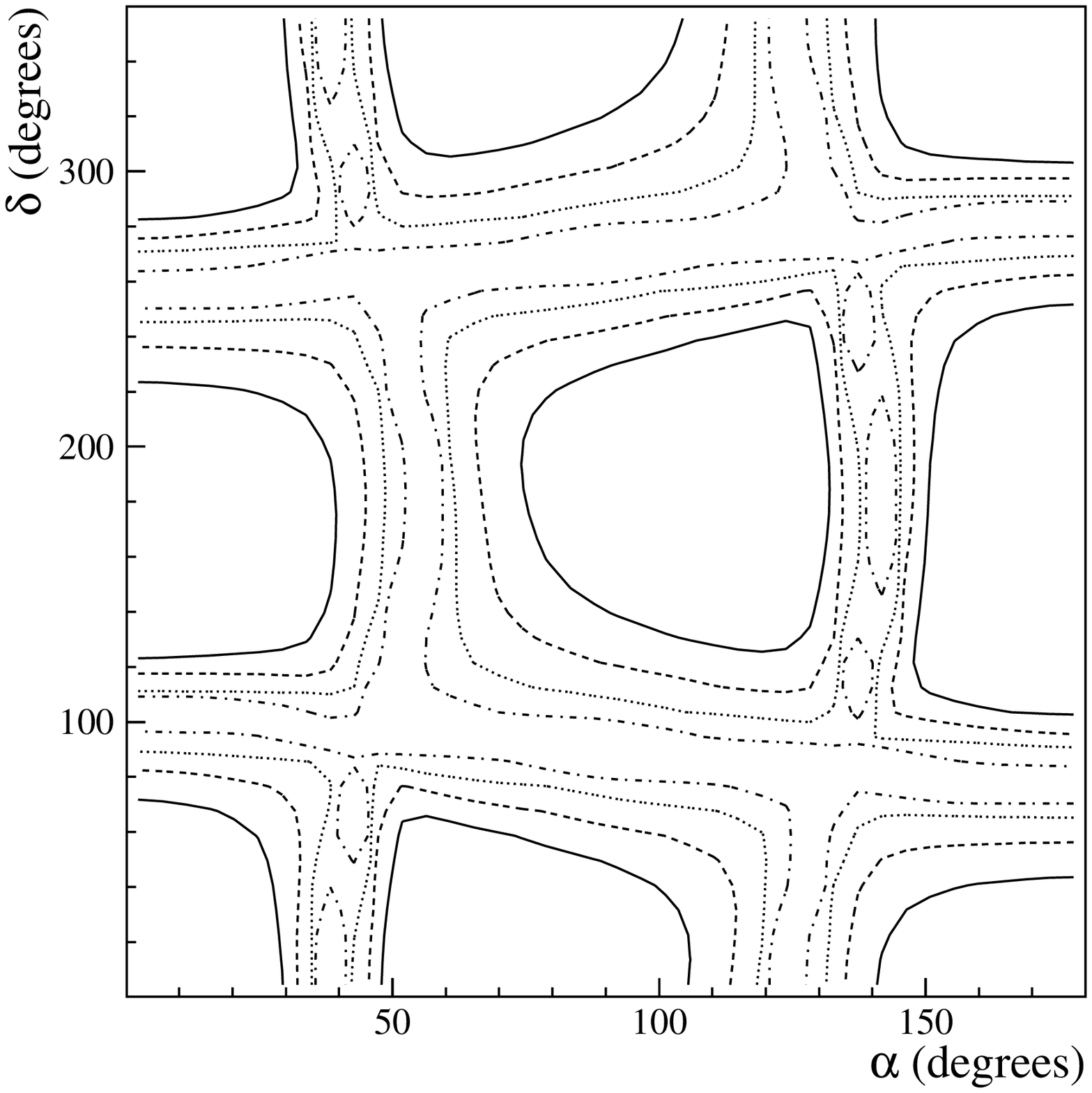,height=0.47\textwidth} &
\epsfig{file=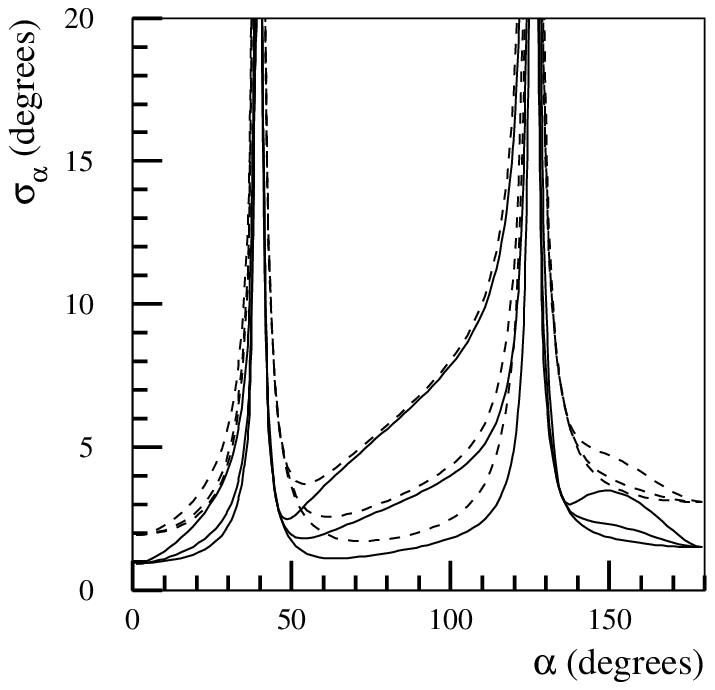,width=0.48\textwidth}
\end{array}
$$
\vspace*{-0.8cm}
\caption[]{Combined LHC sensitivity to $\alpha$: (a)  Sensitivity to
  $\alpha$ 
as a function of $\alpha$ and
 $\delta$ for a given \ApAt$\,=0.2\pm0.02$, for extended running at the LHC.
The contour lines correspond to a sensitivity of $2^\circ$ (solid), 
$3^\circ$ (dashed), $5^\circ$ (dotted) and $10^\circ$ (dashed-dotted). 
(b) Sensitivity to $\alpha$ as a function of $\alpha$, 
for $\delta=30^\circ$, \ApAt$\,$=$\,$0.2 after one year (dashed
lines) and five years (solid lines). In both cases, the curves are
given from bottom to top for an uncertainty of \ApAt\ of 0.02, 0.05 
and 0.1.}\label{fig_bpipi_salpha}
\end{figure}

Figure~\ref{fig_bpipi_salpha}(a) shows the expected sensitivity to 
$\alpha$ as a function of $\alpha$ and $\delta$ for a given 
\ApAt$=0.2\pm0.02$, after extended LHC running (3 years of low luminosity
running of ATLAS and CMS combined with 5 years of LHCb). The sensitivity 
is around $2^\circ$  in the larger part of the plane, except around 
lines corresponding to $\delta=90^\circ$ and $270^\circ$, 
and $\alpha=45^\circ$ and $135^\circ$. For these values of $\delta$
and $\alpha$, the leading-order term
in $|P/T|$ of the mixing-induced CP asymmetry
${\cal A}_{\rm CP}^{\rm mix}(B_d\to\pi^+\pi^-)$ vanishes, as can be
seen in (\ref{eq:ptorder1}).

Figure~\ref{fig_bpipi_salpha}(b) shows the expected sensitivity 
to $\alpha$ as a function of $\alpha$ for a given value $\delta=30^\circ$, 
\ApAt$\,$=$\,$0.2 and different values of the uncertainty on \ApAt\  
and for different integrated luminosities. It appears that for 
values of $\alpha$ around $90^\circ$, the sensitivity to $\alpha$ is
already limited after one year if the uncertainty on \ApAt\ is
not better than 10\%. The effect of the uncertainty on \ApAt\ is less 
dramatic for values of $\alpha$ around $0^\circ$ or $180^\circ$,
which are  disfavoured by current SM fits.

\subsubsection{Conclusions}

At the LHC it should be possible to measure the \Bpipi\ CP-violating
observables with high precision.  Interpreting these observables in terms
of the angle $\alpha$, however,  requires external information
on the strength of the penguin contributions.  This information has to
be rather precise if one is to fully exploit LHC's powerful reach.
Although exact conclusions depend on the particular parameter set,
it appears more promising to analyse the observables of \Bpipi\
and other two-body decays in the context of the approach discussed in
Sec.~\ref{subsec:BsKK}.

\newcommand{\Bo}   {\ensuremath{\mathrm{B}^{0}_{\mathrm{d}}}}
\newcommand{\Bobar}   {\ensuremath{\overline{\mathrm{B}^{0}_{\mathrm{d}}}}}
\newcommand{\rhoo}   {\ensuremath{\rho^{0}}}
\newcommand{\rhop}   {\ensuremath{\rho^{+}}}
\newcommand{\rhom}   {\ensuremath{\rho^{-}}}
\newcommand{\pio}   {\ensuremath{\pi^{0}}}
\newcommand{\pip}   {\ensuremath{\pi^{+}}}
\newcommand{\pim}   {\ensuremath{\pi^{-}}}
\newcommand{\pipm}   {\ensuremath{\pi^{\pm}}}
\subsection[Extracting $\alpha$ from 
 $B\to\rho\pi$ Modes]{Extracting \protect\boldmath
  $\alpha$ from \protect\boldmath $B\to\rho\pi$ Modes\protect\footnote{With 
help from J. Charles, A. Jacholkowska and J. Libby.}}\label{subsec:Brhopi}

\subsubsection{Theoretical Introduction}
The analysis of the decays $B_d\to
\rho^\pm\pi^\mp$ allows, in principle, the extraction of 
$\alpha$~\cite{j_alek}.
However, the simplest approach, where the $\rho$ is considered as
stable particle, is plagued by both high order discrete ambiguities
and penguin pollution, like in 
$B_d^0\to\pip\pim$. To solve either problem, Snyder and Quinn 
\cite{SQ} proposed a full three-body analysis of the decay 
$B_d^0\to\pi^+\pi^-\pi^0$ in the $\rho$ resonance region, taking into
account interference effects between vector mesons of different
charges. The knowledge of the strong decay $\rho\to\pi\pi$,
parametrized as a Breit-Wigner amplitude, allows the extraction of all
 parameters that describe both the tree and penguin contributions
to  $B_d\to\rho\pi$, including $\alpha$,  from a multi-dimensional
likelihood fit.

The two-body $B_d\to\rho\pi$ amplitudes can be written as:
\begin{equation}\label{j_tprhopi}
A^{\pm\mp}(B_d^0\to\rho^\pm\pi^\mp)=e^{-i\alpha}\,T^{\pm\mp}+\,
P^{\pm\mp}\,,\qquad
A^{00}(B_d^0\to\rho^0\pi^0)=e^{-i\alpha}\,T^{00}+\,P^{00}.
\end{equation}
The CP-conjugate amplitudes $\overline{A}^{ij}\equiv
A(\overline{B_d^0}\to \rho^i\pi^j)$ are obtained by
changing the sign of the weak phases. The full three-body
$B_d\to\pi^+\pi^-\pi^0$ amplitude takes the form:
\begin{equation}\label{j_3bamp}
A(B_d\to\pi^+\pi^-\pi^0)=A^{+-}f_++A^{-+}f_-+A^{00}f_0\, ,
\end{equation}
when $\rho$-dominance is assumed.
Here $f_i$ stands for the Breit-Wigner amplitude for the decay of the
$\rho^i$, and is a function of the two independent variables of the
three-pion Dalitz plot, which are chosen as the invariant masses
$s^\pm=(p_{\pi^\pm}+p_{\pi^0})^2$. The Breit-Wigner
parametrization is not unique; in the following we take:
\begin{equation}\label{j_bw}
f_+\propto\frac{\cos\theta^\ast}{s^+-m_\rho^2+im_\rho\Gamma_\rho}\,,
\end{equation}
where $\theta^\ast$ is the helicity angle of the $\rho$ decay which is 
given in terms of $(s^+,s^-)$ by the standard formulae. This dependence
has the property of enhancing the number of events in the
corners of the Dalitz plot, where interferences are maximal.

The time-dependent analysis of the event distribution in the
 Dalitz plot allows one to extract $|{A}(B_d^0\to\pi^+\pi^-\pi^0)|$,
$|\overline{A}(\overline{B_d^0}\to\pi^+\pi^-\pi^0)|$ and Im$\,[
\frac{q}{p}\overline{A}{A}^\ast]$
as functions of $(s^+,s^-)$. Using 
(\ref{j_3bamp}) and (\ref{j_bw}), it is straightforward to show
that the magnitudes and the relative phases of the two-body amplitudes 
$A^{ij}$ and $\overline{A}^{ij}$ can be obtained~\cite{SQ}; this amounts
to determining 11 independent parameters, taking into account that one overall
phase is irrelevant, and including the overall normalization. In
addition, assuming isospin symmetry and neglecting electroweak
penguins, the relation~\cite{Brhopi} 
\begin{equation}
P^{00}=-\frac{1}{2}(P^{+-}+P^{-+})
\end{equation}
allows a further reduction in the number of independent parameters
that describe $A^{ij}$ and $\overline{A}^{ij}$. These 
parameters can be chosen as $\alpha$ and the complex 
amplitudes $T^{-+}$, $T^{00}$, $P^{+-}$ and $P^{-+}$. 
It is important to note that $A^{ij}$ and
$\overline{A}^{ij}$ are determined without
discrete ambiguity in the general case, such that both $\cos2\alpha$
and $\sin2\alpha$ (and thus $\alpha$ in $[0,\pi]$) are
accessible~\cite{SQ}.  This resolves in particular 
the ambiguity between $\alpha$ and $\pi/2-\alpha$.

\subsubsection{Experimental Studies}

\subsection*{Selection}

\begin{figure}
\begin{minipage}[t]{0.64\textwidth}
\begin{center}
\epsfig{figure=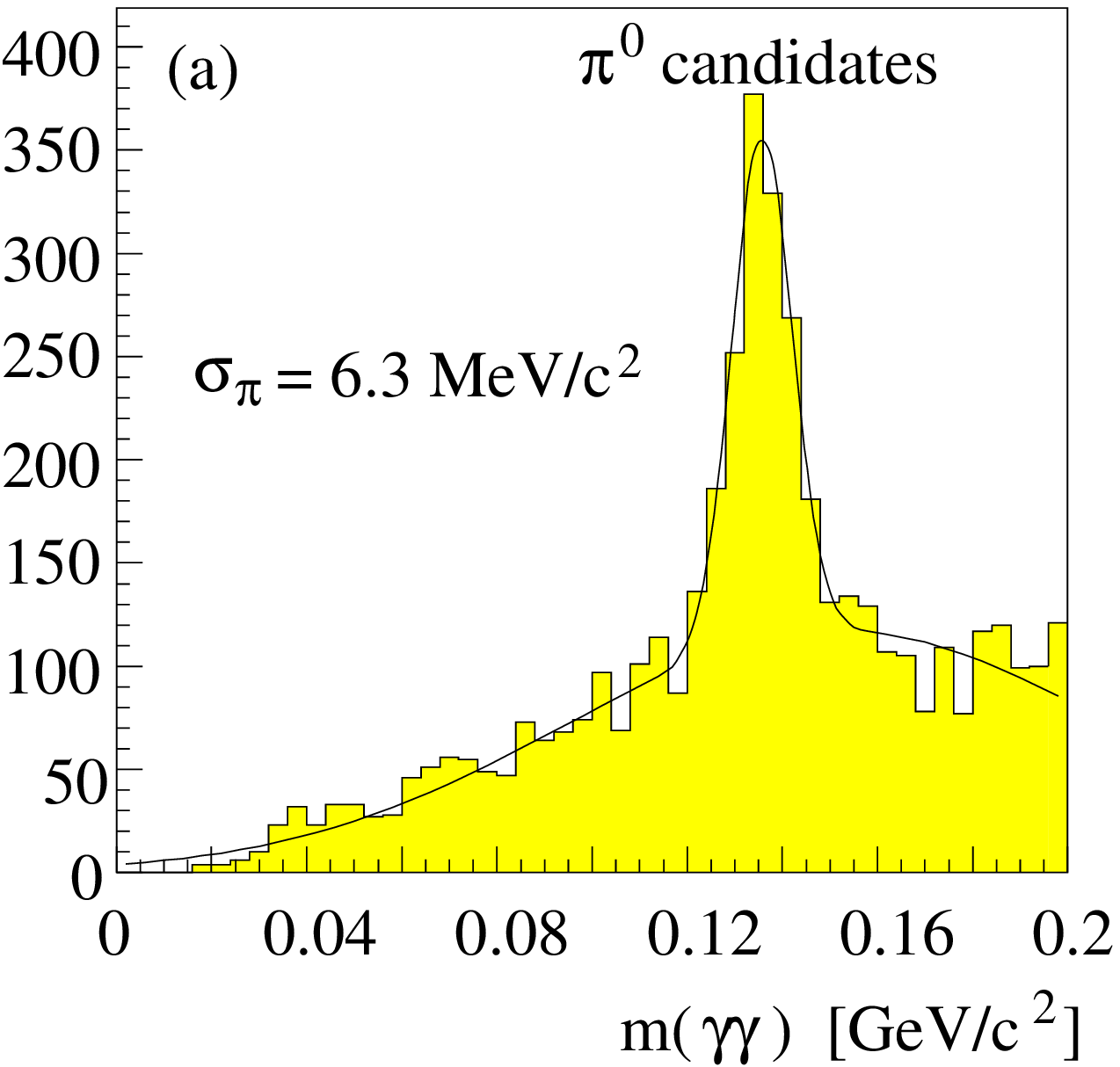,width=0.46\textwidth,height=0.46\textwidth}
\hspace*{5pt} 
\epsfig{figure=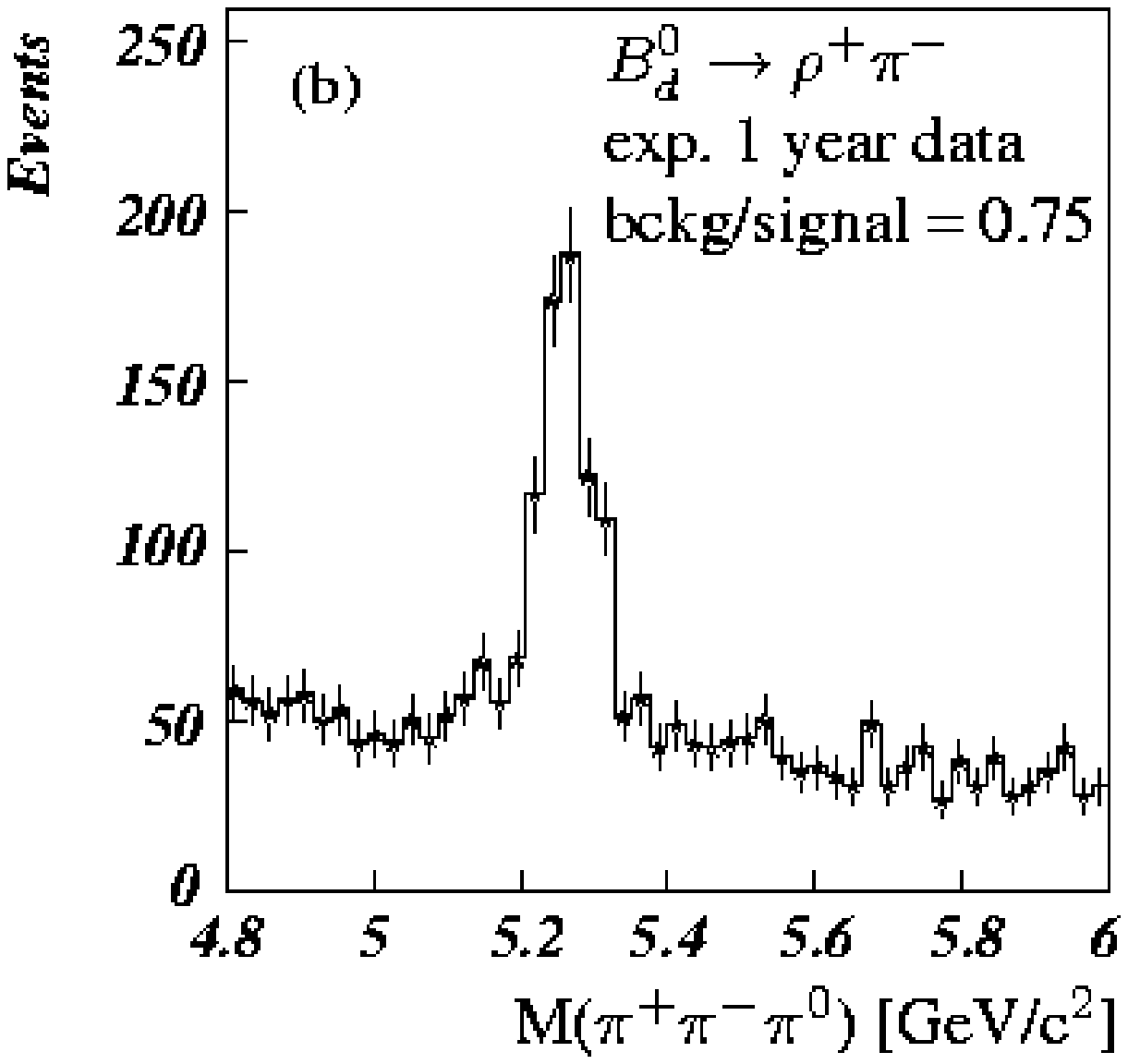,width=0.46\textwidth,height=0.46\textwidth}
\end{center}
\vspace*{-0.8cm}
\caption[]{
The invariant mass for (a) $\pi^0$ candidates in 
B$^{0}_{d} \rightarrow \pi^{+}\pi^{-}\pi^{0}$ events;
(b)  B$^{0}_{d} \rightarrow \rho\pi$ candidates reconstructed in LHCb.
The combinatorial background comes mainly
 from inclusive $b\bar{b}$ events.} \label{fig:brhopi1}
\end{minipage}
\hspace*{10pt}
\begin{minipage}[t]{0.32\textwidth}
\begin{center}
\epsfig{figure=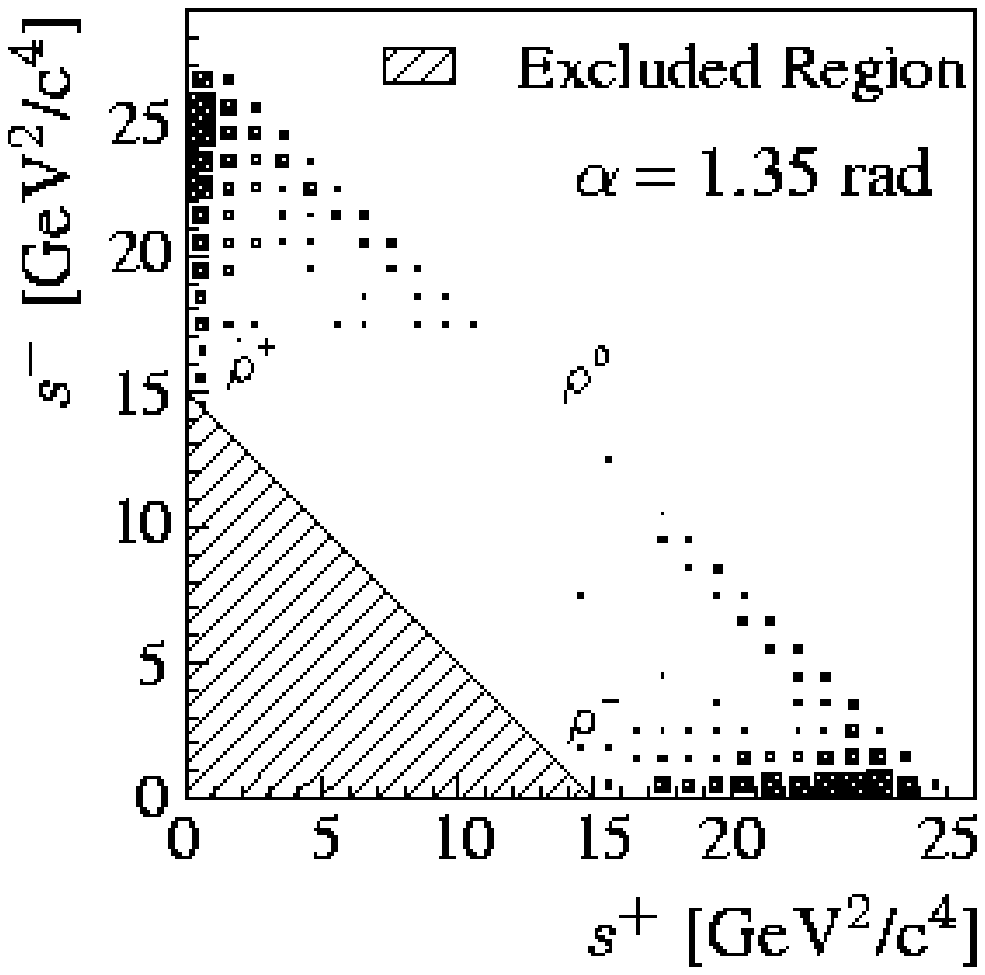,width=0.92\textwidth,height=0.92\textwidth}
\end{center}
\vspace*{-0.8cm}
\caption[]{
The Dalitz plot for B$^{0}_{d} \rightarrow \pi^{+}\pi^{-}\pi^{0}$
 decays after ac\-cep\-tance-cuts for events generated using the 
stand-alone simulation.}\label{fig:brhopi2}
\end{minipage}
\end{figure}

The LHCb collaboration has
performed full simulation studies on the selection of the 
B$^{0}_{d} \rightarrow \pi^{+}\pi^{-}\pi^{0}$ channel.
The charged pions are reconstructed in the tracking devices
and are identified in the RICH detectors.
At present, only $\pi^{0}$s built from two resolved photons 
are used in the analysis.
Figure \ref{fig:brhopi1}(a) shows the two photon invariant mass in 
B$^{0}_{d} \rightarrow \pi^{+}\pi^{-}\pi^{0}$ events,
for photons with energy above 2~GeV.
The resolution of the $\pi^{0}$
mass varies between 5 and 7 MeV, depending 
on the $\pi^{0}$ production angle.
The overall efficiency for $\pi^{0}$ reconstruction is $25\%$, 
with a signal to combinatorial background ratio of approximately 1.
The measured $\pi^{0}$ mass is used in further B$^{0}_{d}$ mass
 reconstruction. 

The background 
comes from combinatorics and from inclusive $b\bar{b}$ events.
For its  suppression, the following 
qualitative selection cuts have been applied:
\begin{itemize}
\item 
a pre-selection for charged pions and photons which required the
momentum or energy to exceed a value 
depending on the polar angle of the candidate. 
For  charged pions, the momentum cut varied between 1 and 2$\,$GeV 
and for photons the energy cut varied 
between 2 and 6$\,$GeV;
\item
selection of signal-like events based 
on a discriminant variable built
from kinematic variables of $\pi$, $\rho$ and B$^{0}_{d}$; 
\item 
selection based on the reconstructed secondary 
vertex for a $\pi^{+}\pi^{-}$
combination;
\item
Dalitz plot cuts to eliminate low energy $\pi^{0}$ combinatorial
 background due to  particles from the primary vertex.
\end{itemize}
These selection criteria result in a
combinatorial background suppression factor
of the order of $10^{7}$ and give an acceptance for  triggered
and tagged events of 1\%. 
Figure~\ref{fig:brhopi1}(b) shows the expected  
$\pi^{+}\pi^{-}\pi^{0}$ invariant
mass distribution after one year of data
 taking.  The  measured B$^{0}_{d}$ width is 50 MeV/c$^2$. 
The  annual event-yields for triggered,
fully reconstructed and tagged events are given in Tab.~\ref{tab:new}.
\begin{table}[htb]
\begin{center}
\begin{tabular}{|l||c|c|c|}
\hline
Channel & $B^0\to\rho^+\pi^-$ & $B^0\to\rho^-\pi^+$ &
$B^0\to\rho^0\pi^0$\\ \hline
$BR$ & 44$\,\times\,$10$^{-6}$  & 10$\,\times\,$10$^{-6}$&
1$\,\times\,$10$^{-6}$\\
Event Yield & 1000 & 200 & 100\\ \hline
\end{tabular}
\end{center}
\vspace*{-0.5cm}
\caption[]{Annual event-yields for $B\to\rho\pi$ decays. The branching
  fractions are crude estimates used in {\sc BaBar}'s study of these
  decays \protect\cite{BaBar}.}\label{tab:new}
\end{table}

Figure~\ref{fig:brhopi2} shows the Dalitz plot for the
B$^{0}_{d} \rightarrow \pi^{+}\pi^{-}\pi^{0}$ channel after 
acceptance cuts. Helicity effects enhance the population 
in the interference
 regions, in particular in the 
most critical $\rho^{\pm}$--$\rho^{0}$ regions,
where the sensitivity to the $\alpha$ parameter is highest. 
The $\rhop$--$\rhom$ interference region is 
not accessible due to the dominance of combinatorial background in the
corresponding area of the Dalitz space.  

\subsection*{Sensitivity to \protect\boldmath $\alpha$}

 A stand-alone simulation which introduces the weak phase $\alpha$ as
 well as the relative tree and penguin amplitudes 
was used to generate events for the fitting studies. 
Cuts in the Dalitz space have been made 
to eliminate the \rhom--\rhop\ interference region. 
Furthermore, cuts are applied to the invariant mass of a $\rho$ 
candidate to select only 
resonant decays. However, the full LHCb acceptance has not yet been 
simulated and backgrounds have not been considered. 

\begin{table}[t]
\begin{minipage}[b]{0.4\textwidth}
\begin{tabular}{|c||c|} \hline
   Parameter & Value \\ \hline
   $\alpha$  & 0.9, 1.35 or 1.95 radians \\ \hline
   $T^{+-}$  & 1.00                       \\
   $T^{-+}$  & 0.47                       \\
   $T^{00}$  & 0.14                       \\
   $P^{+-}$  & --0.20$\,e^{\rm -0.5i}$           \\
   $P^{-+}$  & 0.15$\,e^{\rm 2.0i}$             \\ \hline
\end{tabular} 
\caption[]{The three values of $\alpha$ and the amplitudes used 
in the generation of the studied samples.\\~}\label{tab:brhopi1} 
\end{minipage}
\hspace*{0.4cm}
\begin{minipage}[b]{0.55\textwidth}
\begin{center}
  \begin{tabular}{|c||c|c||c|c|} \hline
    &  \multicolumn{2}{c||}{1 year} & \multicolumn{2}{c|}{5
     years} \\ 
\cline{2-5}\ 
 $\alpha$ &  $\langle\alpha\rangle$  & $\langle\sigma_{\alpha}\rangle$
     &  $\langle\alpha\rangle$  & 
$\langle\sigma_{\alpha}\rangle$ \\ 
    ($^{\circ}$)  &   ($^{\circ}$)  & ($^{\circ}$)        & ($^{\circ}$)
     &  
($^{\circ}$)       \\ \hline\hline
   \phantom{1}51.6 & \phantom{1}51.6     &     4.9 &
     \phantom{1}51.0 &2.1     \\
   \phantom{1}77.3     &     \phantom{1}76.2     &     2.5
     &  
  \phantom{1}76.2      &
     1.1     
          \\    
   111.7    &    102.6     &     4.3             &    102.0     &
     1.8     
          \\ \hline    
  \end{tabular}
\end{center}\vspace*{-0.2cm}
  \caption[]{The mean fitted values of $\alpha$,
     $\langle\alpha\rangle$, 
and the mean error on $\alpha$, $\langle\sigma_{\alpha}\rangle$, for samples 
approximating 1 or 5 years data taking for LHCb at $\alpha$ = 0.90,
1.35 and 1.95 radians $(51.6^{\circ}, 77.3^{\circ}$ and 
$111.7 ^{\circ})$.}
\label{tab:brhopi2} 
 \end{minipage}
\end{table}

 The amplitudes used for these studies contain
 a large penguin contribution 
and are identical to those studied by Babar~\cite{BaBar}. 
Their values are given Tab.~\ref{tab:brhopi1}.  
Samples of 10$^5$ events were generated for each value of $\alpha$.
 An unbinned maximum likelihood fit was used to extract the
 parameters. The form of the 
used likelihood is: 
\begin{eqnarray*}
 -2\ln{\mathcal L} & = & -2\sum^{N_{\Bo}}_{i=1} 
\ln{(\frac{|{A}(s^{+}_i,s^{-}_i,t_i;\alpha)|^{2}}
{{\mathcal N(\alpha)}})} -2\sum^{N_{\Bobar}}_{j=1}
\ln{(\frac{|\bar{{ A}}(s^{+}_j,s^{-}_j,t_j;\alpha)|^{2}}
{{\mathcal N(\alpha)}})} , 
\end{eqnarray*}
where $N_{\Bo}$ and $N_{\Bobar}$ are the number of \Bo\ and \Bobar\
 events, respectively, and ${\mathcal N}$ is the normalization.
It is given by $(|{ A}|^2+ |\bar{{ A}}|^2)$, integrated over the 
Dalitz plot acceptance and was 
calculated numerically using a sub-sample of 20000 simulated events. 
The fit was performed on 75 sub-samples of 1000 events, to simulate 
approximately 1 year data taking, and 15 samples of 5000 events 
to simulate 5 years data taking. The mean fitted value of $\alpha$ and
 the mean error are given in 
Tab.~{\ref{tab:brhopi2}}. The error 
varies with the true value of $\alpha$ as expected \cite{SQ},
and the fitted values are unbiased for $\alpha=$ 0.9 and 1.35 radians. 
The bias of $\sim\,$0.15 radians for  $\alpha=\,$1.95 radians was 
not observed when fits were made to samples where no Dalitz plot
 selection was made. 
Therefore, this bias appears to be related to the exclusion of the 
$\rhop$--$\rhom$ interference region and needs further investigation.
Correction for this bias will be required to extract $\alpha$  from 
the final data sample and will introduce systematic uncertainties which 
may be of a magnitude similar to the statistical precision. 

\begin{figure}[t]
\centerline{\epsfig{figure=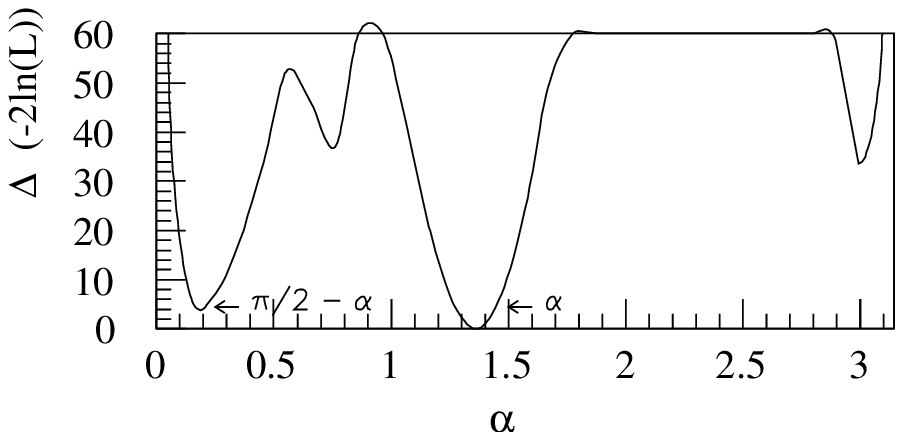,width=0.45\textwidth}}
\vspace*{-0.3cm}
\caption[]{An example likelihood scan-curve for 1000 fitted 
LHCb events, generated with $\alpha=\,$1.35. $\alpha$ was fixed to 
 40 values between 0 and $\pi$ radians and then 
 the negative likelihood was minimised
with respect to the other 8 parameters.}
\label{fig:brhopi4}
\vspace*{-1cm}
\begin{center}
\subfigure[LHCb 1 year data samples]
{\epsfig{figure=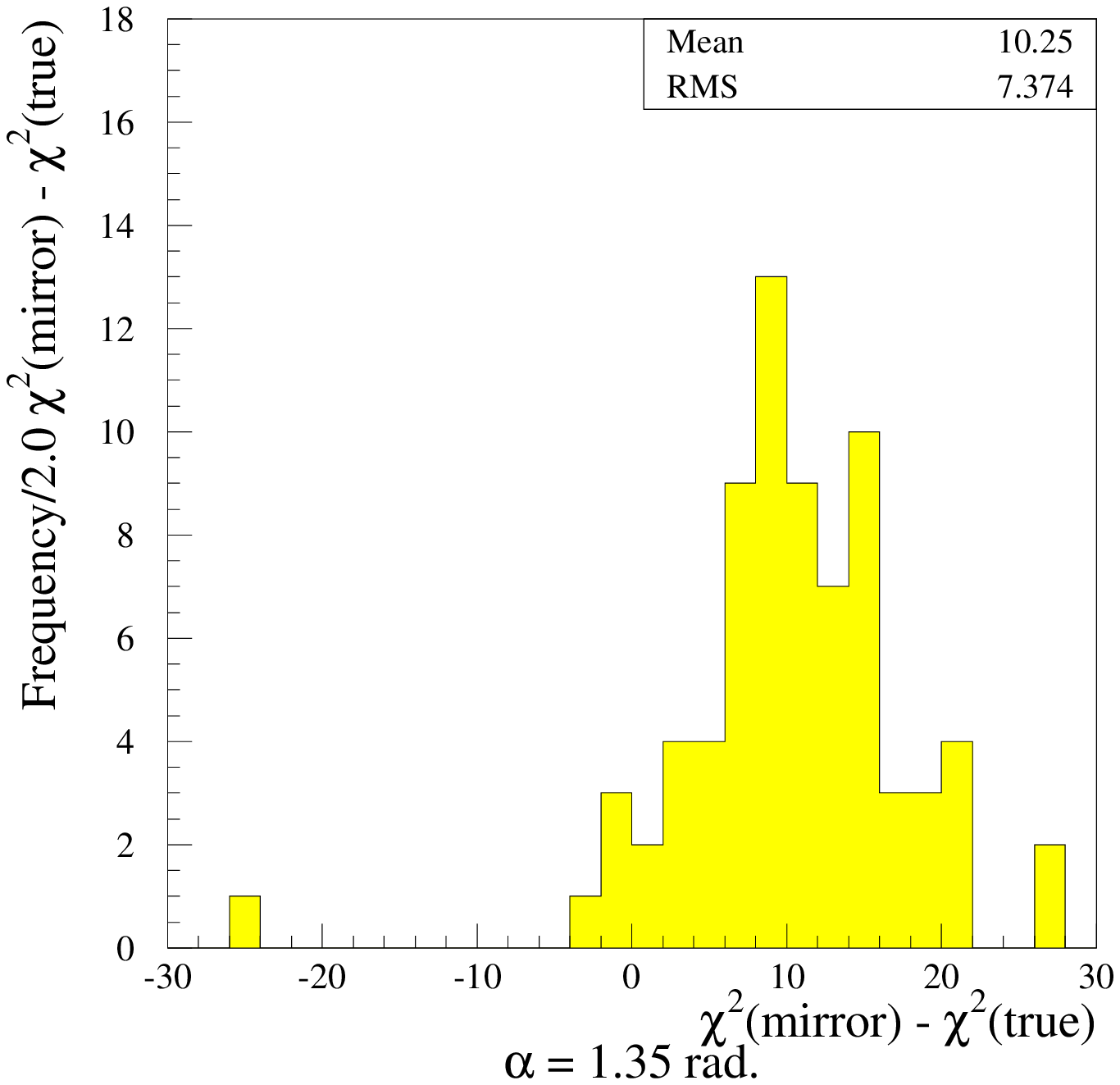,width=0.48\textwidth}}
\subfigure[LHCb 5 year data samples]
{\epsfig{figure=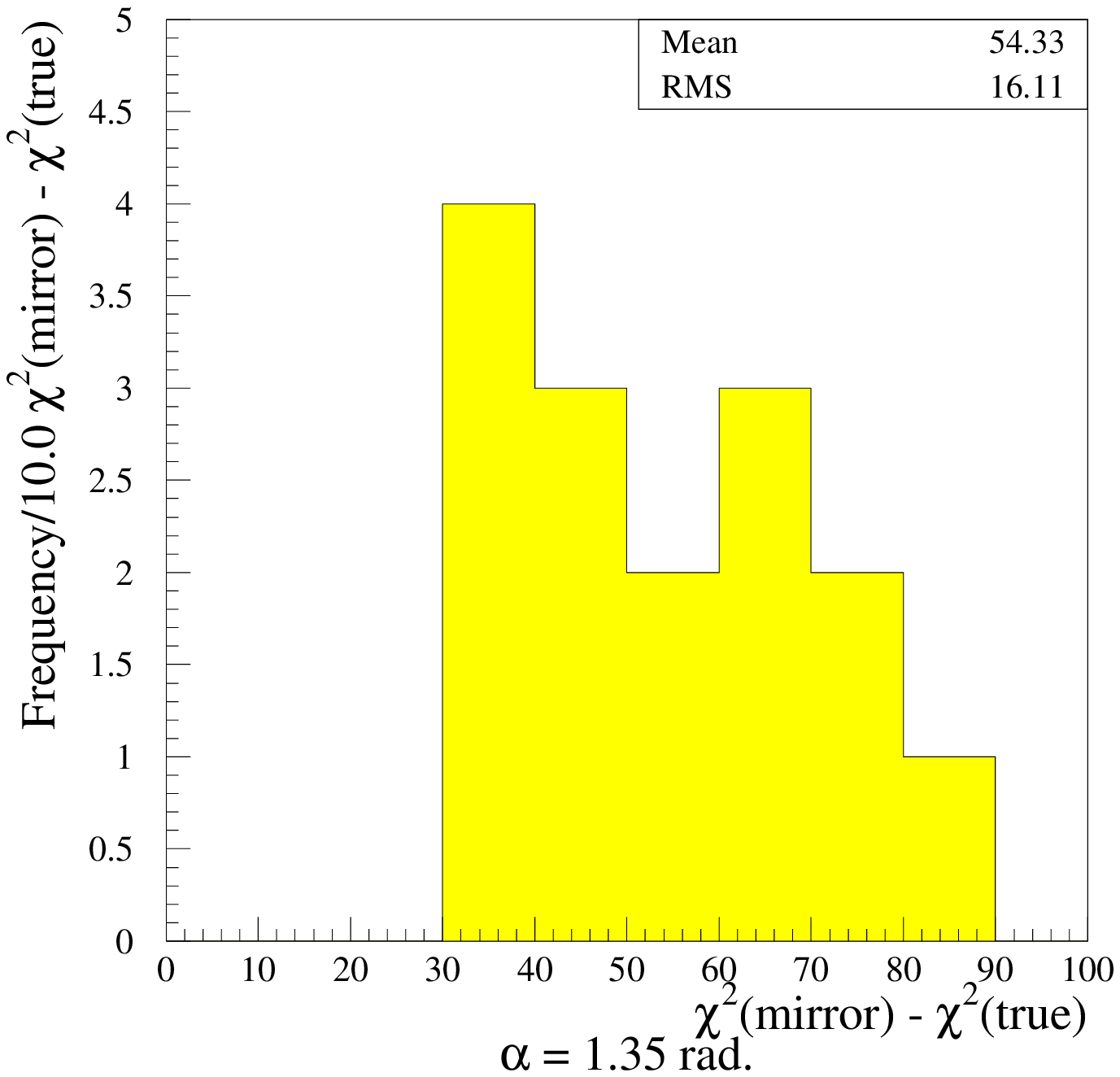,width=0.48\textwidth}}
\end{center}
\vspace*{-0.8cm}
\caption[]{The difference in $-2\ln{\mathcal L}$ between the true and 
mirror solution minima.}\label{fig:brhopi5}
\end{figure}

In Fig.~\ref{fig:brhopi4} an example likelihood scan curve is given
for 1000 fitted events generated with $\alpha=\,$1.35 
radians. The fake mirror solution at $\frac{\pi}{2} - \alpha$ gives
a local minimum in the likelihood curve. 
The difference in the likelihood, expressed as 
$\chi^2$ $(= -2\ln{\mathcal L})$, between the true and the mirror
solution for the 75 one year data samples are displayed in 
Fig.~\ref{fig:brhopi5}(a). In approximately 10\% of all cases the
mirror solution is the global 
minimum or is separated by less than 1$\,\sigma$ from the true
solution. The same quantity for the 15 five year data samples is shown in 
Fig.~\ref{fig:brhopi5}(b). The mirror and true solution minima are now
well 
separated.    
   
\subsubsection{Conclusions}

{}From the theoretical point of view, the main advantage of the isospin
analysis of the decay $B_d\to\pi^+\pi^-\pi^0$ in the $\rho$-dominance
assumption, with respect to its analogue in the two-pion channel, is
the determination of the penguin amplitudes and the resolution of 
discrete ambiguities. {}From the experimental side, it benefits from
larger branching ratios~\cite{cleo-Bpipi} and from the interference,
which entails that the sensitivity of the analysis is directly proportional to 
the colour-suppressed channel $B\to\rho^0\pi^0$. This can be compared to the 
Gronau-London branching ratio construction~\cite{gl} in 
$B_d\to\pi\pi$ which has a sensitivity 
proportional to the amplitude squared of $B_d\to\pi^0\pi^0$.

Preliminary studies for LHCb have shown that $\Bo\to\pip\pip\pio$ events can 
be reconstructed and selected in sufficient numbers, so that an unambiguous 
value for $\alpha$ can be extracted without the problems that afflict
the $\Bo\to\pip\pim$ channel.
 It should be stressed that the fitting studies are preliminary and are 
optimistic in the fact that the exact LHCb acceptance has not been used and 
backgrounds have not been included. 
Also, the biases observed are likely to introduce significant
systematic uncertainties. Furthermore, several important issues remain
to be considered, 
which already have been studied in the specific context of $e^+e^-$ 
B factories \cite{j_sv,jc_phd,j_4}. One may cite, among
others, various points: the influence of higher resonances ($\rho^\prime$,
$\rho^3$\dots), the influence of the exact parametrization of the Breit-Wigner
amplitude, the existence of bounds on the penguin-induced error on $\alpha$,
when the $\rho^0\pi^0$ channel is too scarce to achieve the full analysis, and 
the r\^ole of electroweak penguins. All these issues will be further
investigated  in the future. 

 There are also some topics, yet to be investigated, which should enhance the 
precision on $\alpha$: the determination of 
the branching fractions from $e^+ e^-$ experiments provide additional 
constraints on the fit and the untagged sample can be used to 
determine parameters other than $\alpha$. 
It is to be expected that after several years of data
taking at $e^+e^-$ experiments and/or at the LHC era, the above issues will be
much better understood.

\subsection[Extracting $2\beta+\gamma$ from 
 $B_d\to D^{(\ast)\pm}\pi^{\mp}$
Decays]{Extracting \protect\boldmath $2\beta+\gamma$ from 
\protect\boldmath $B_d\to D^{(\ast)\pm}\pi^{\mp}$ 
decays\protect\footnote{With help from J. Rademacker.}}\label{subsec:BDpi}

So far, we have put a strong emphasis on neutral $B$ decays into final 
CP eigenstates. However, in order to extract CKM phases, there are also 
interesting decays of $B_{d,s}$ mesons into final states that are {\it not} 
eigenstates of the CP operator. An important example are the decays
$B_d\to D^{(\ast)\pm}\pi^{\mp}$, which receive only contributions from
tree-diagram-like topologies, and are the topic of this subsection.
 
\begin{figure}
\begin{center}
\leavevmode
\epsfysize=3.8truecm 
\epsffile{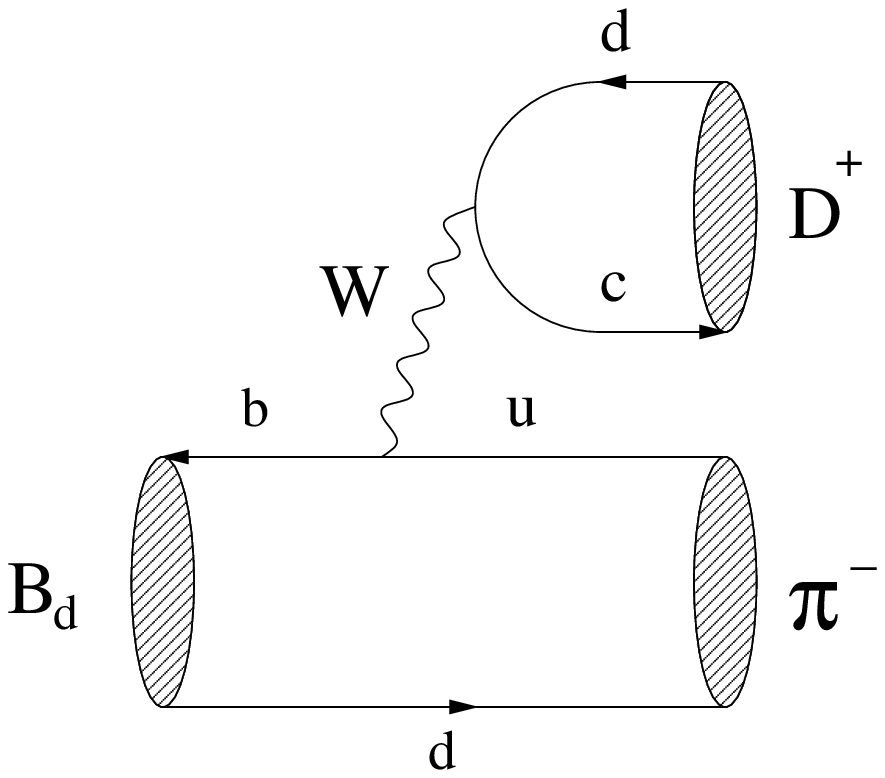} \hspace*{1truecm}
\epsfysize=3.8truecm 
\epsffile{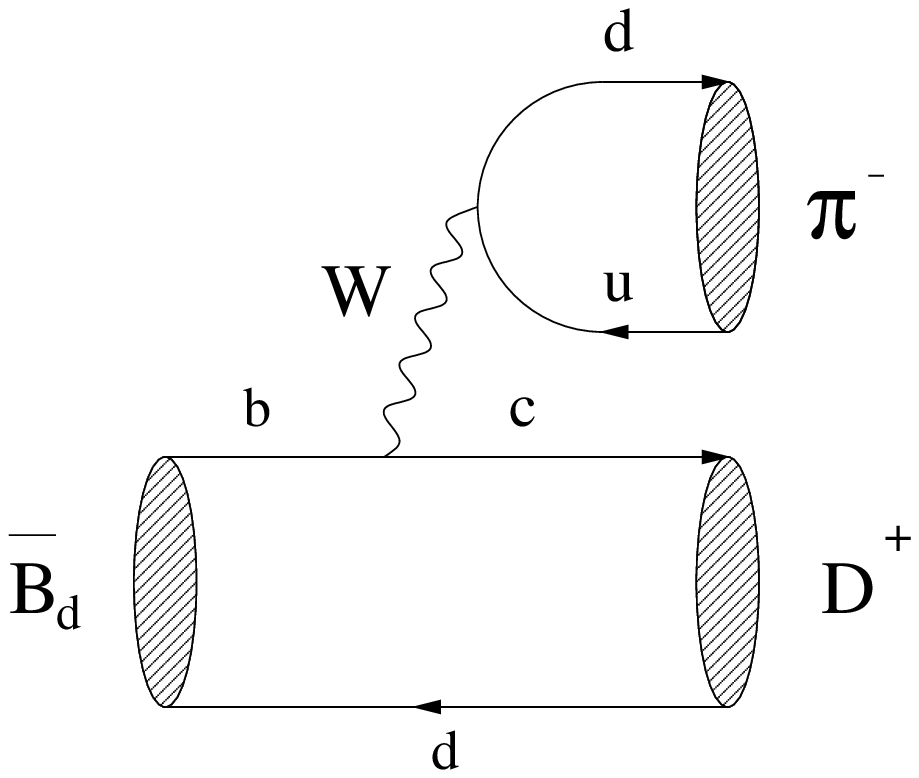}
\end{center}
\vspace*{-0.4truecm}
\caption[]{Feynman diagrams contributing to $B^0_d, \overline{B^0_d}\to 
D^{(\ast)+}\pi^-$.}\label{fig:BDpi}
\end{figure}

\subsubsection{Theoretical Aspects}

As can be seen in Fig.~\ref{fig:BDpi}, $B^0_d$- and $\overline{B^0_d}$-mesons 
may both decay into $D^{(\ast)+}\pi^-$, thereby leading to interference 
effects between $B^0_d$--$\overline{B^0_d}$ mixing and decay processes. 
Consequently, the time-dependent decay rates for initially, i.e.\ at time 
$t=0$, present $B^0_d$- or $\overline{B^0_d}$-mesons decaying into the 
final state $f\equiv D^{(\ast)+}\pi^-$ allow one to determine the 
observable \cite{RF-rev}
\begin{equation}
\xi_f^{(d)}=-\,e^{-i\phi_d}
\frac{A(\overline{B^0_d}\to f)}{A(B^0_d\to f)}
=-\,e^{-i(\phi_d+\gamma)}\left(\frac{1-\lambda^2}{\lambda^2R_b}\right)
\frac{\overline{M}_f}{M_{\overline f}},
\end{equation}
whereas those corresponding to $\bar f\equiv D^{(\ast)-}\pi^+$ allow one 
to extract
\begin{equation}
\xi_{\bar f}^{(d)}=-\,e^{-i\phi_d}
\frac{A(\overline{B^0_d}\to \bar f)}{A(B^0_d\to \bar f)}=
-\,e^{-i(\phi_d+\gamma)}\left(\frac{\lambda^2R_b}{1-\lambda^2}\right)
\frac{M_{\overline f}}{\overline{M}_f}.
\end{equation}
Here, $R_b$ is the usual CKM factor (see (\ref{CKM-exp})), and
\begin{equation}
\overline{M}_f\equiv\Bigl\langle f\Bigl|\overline{O}_1(\mu){\cal C}_1(\mu)+
\overline{O}_2(\mu){\cal C}_2(\mu)\Bigr|\overline{B^0_d}\Bigr\rangle,\quad
M_{\overline{f}}\equiv\Bigl\langle\overline{f}\Bigl|O_1(\mu){\cal C}_1(\mu)+
O_2(\mu){\cal C}_2(\mu)\Bigr|\overline{B^0_d}\Bigr\rangle
\end{equation}
are hadronic matrix elements of the following current--current operators:
\begin{equation}
\begin{array}{rclrcl}
\overline{O}_1&=&(\bar d_\alpha u_\beta)_{\mbox{{\scriptsize 
V--A}}}\left(\bar c_\beta b_\alpha\right)_{\mbox{{\scriptsize V--A}}},&
\overline{O}_2&=&(\bar d_\alpha u_\alpha)_{\mbox{{\scriptsize 
V--A}}}\left(\bar c_\beta b_\beta\right)_{\mbox{{\scriptsize V--A}}},\\
O_1&=&(\bar d_\alpha c_\beta)_{\mbox{{\scriptsize V--A}}}
\left(\bar u_\beta b_\alpha\right)_{\mbox{{\scriptsize V--A}}},&
O_2&=&(\bar d_\alpha c_\alpha)_{\mbox{{\scriptsize V--A}}}
\left(\bar u_\beta b_\beta\right)_{\mbox{{\scriptsize 
V--A}}},
\end{array}
\end{equation}
where $\alpha$ and $\beta$ denote colour indices, and 
V--A refers to the Lorentz structures $\gamma_\mu(1-\gamma_5)$.
The observables $\xi_f^{(d)}$ and $\xi_{\bar f}^{(d)}$ allow 
a {\it theoretically clean} extraction of the weak phase $\phi_d+\gamma$ 
\cite{BDpi}, as the hadronic matrix elements $\overline{M}_f$ and 
$M_{\overline{f}}$ cancel in the following combination:
\begin{equation}\label{Prod}
\xi_f^{(d)}\times\xi_{\bar f}^{(d)}=e^{-2i(\phi_d+\gamma)}.
\end{equation}
Since the $B^0_d$--$\overline{B^0_d}$ mixing phase $\phi_d$, i.e.\ $2\beta$,
can be determined rather straightforwardly with the help of the 
``gold-plated'' mode $B_d\to J/\psi\, K_{\rm S}$ (see 
Sec.~\ref{subsec:BdpsiKS}), we may extract the CKM angle $\gamma$ from 
(\ref{Prod}). As the $\bar b\to\bar u$ quark-level transition in 
Fig.~\ref{fig:BDpi} is doubly Cabibbo-suppressed by 
$\lambda^2R_b\approx0.02$ with respect to the $b\to c$ transition, 
the interference effects are tiny. However, the branching ratios 
are large, i.e.\ of order $10^{-3}$, and the $D^{(\ast)\pm}\pi^\mp$ states 
can be reconstructed with a good efficiency and modest background. 
Consequently, $B_d\to D^{(\ast)\pm}\pi^\mp$ decays offer an interesting 
strategy to determine $\gamma$, as we will discuss in the following.

\subsubsection{Experimental Studies}
\label{sec:bdstarpi}

\newcommand{\BDpi}{\ensuremath{B_d^0  \to D^{*\mp}\pi^{\pm}}}
\newcommand{\BDa}{\ensuremath{B_d^0  \to D^{*\mp} a_1^{\pm}}}
\newcommand{\BDpipipi}{\ensuremath{B_d^0  \to
    D^{*\mp}\pi^{\pm}\pi^{\pm}\pi^{\mp}\pi{\pm}}}
\newcommand{\BDthreepi}{\ensuremath{B_d^0  \to
    D^{*}3\pi}}
\newcommand{\BDpia}{\ensuremath{B_d^0  \to D^{*\mp}\pi(a_1)^{\pm}}}
\newcommand{\pion}{\ensuremath{\pi}}
\newcommand{\Dz}{\ensuremath{D^0}}
\newcommand{\Dzb}{\ensuremath{\overline{D^0}}}
\newcommand{\Dst}{\ensuremath{D^*}}
\newcommand{\Dstpm}{\ensuremath{D^{*\pm}}}
\newcommand{\Dstp}{\ensuremath{D^{*+}}}
\newcommand{\Dstm}{\ensuremath{D^{*-}}}
\newcommand{\Bz}{\ensuremath{B^0}}
\newcommand{\Bzb}{\ensuremath{\overline{B^0}}}

\newcommand{\un}[2]{\ensuremath{#1\,\mbox{#2}}}

LHCb have investigated the potential of measuring $\gamma$ through
\BDpi\, with the $D^*$ decaying strongly to a $D^0$ meson.
As interference effects are tiny, a very
large data sample is necessary to extract $\gamma$ with an interesting
precision.   Two methods have been studied: first 
a conventional exclusive reconstruction  with $\overline{D^0}\to K^+\pi^-$
and second a partial 
reconstruction approach in order to boost statistics.
The reconstruction study has also been extended to \BDa\ decays,
but such events have not yet been considered for the
extraction of CKM phases.

\subsection*{Exclusive Reconstruction}

Loose RICH criteria were used to select the candidate \Dzb\ decay
products. To identify \Dstm, the difference between the
reconstruced \Dstm\ and \Dzb\ mass was required to lie within
a \un{3}{MeV} wide window around its nominal value of \un{144}{MeV}, just
above the pion mass. Figure~\ref{dst_massdiff}(a) shows the signal
peak ($\sigma=\un{1}{MeV}$) with the background superimposed in
arbitray units.
The usual \Bz\ cuts (high $p_T$ and detached vertex) were applied to
the pion coming from the \Bz. The final \Bz\ mass peak has a width of
\un{13}{MeV}. Selecting events within a window of $\pm \un{30}{MeV}$
results in 84k selected events (triggered \& tagged) per year with a
S/B of $\sim 12$.

\subsection*{Partial Reconstruction}

Instead of reconstructing the full
decay chain, one can obtain all necessary information from 
the pion coming directly from the \Bz\ 
( the `fast pion', $\pi_f$ ) and the pion coming from the \Dstm\ 
(the `slow pion', $\pi_s$). 
As shown below, one
can reconstruct the full \Bz\ momentum from the momenta of $\pi_f$ and
$\pi_s$ and the direction of the \Bz. This direction can be inferred
from the position of the primary vertex and the decay vertex of the
\Bz,  the latter being defined by the crossing point
of fast and slow pion. 

\begin{figure}
\begin{center}
\epsfig{file=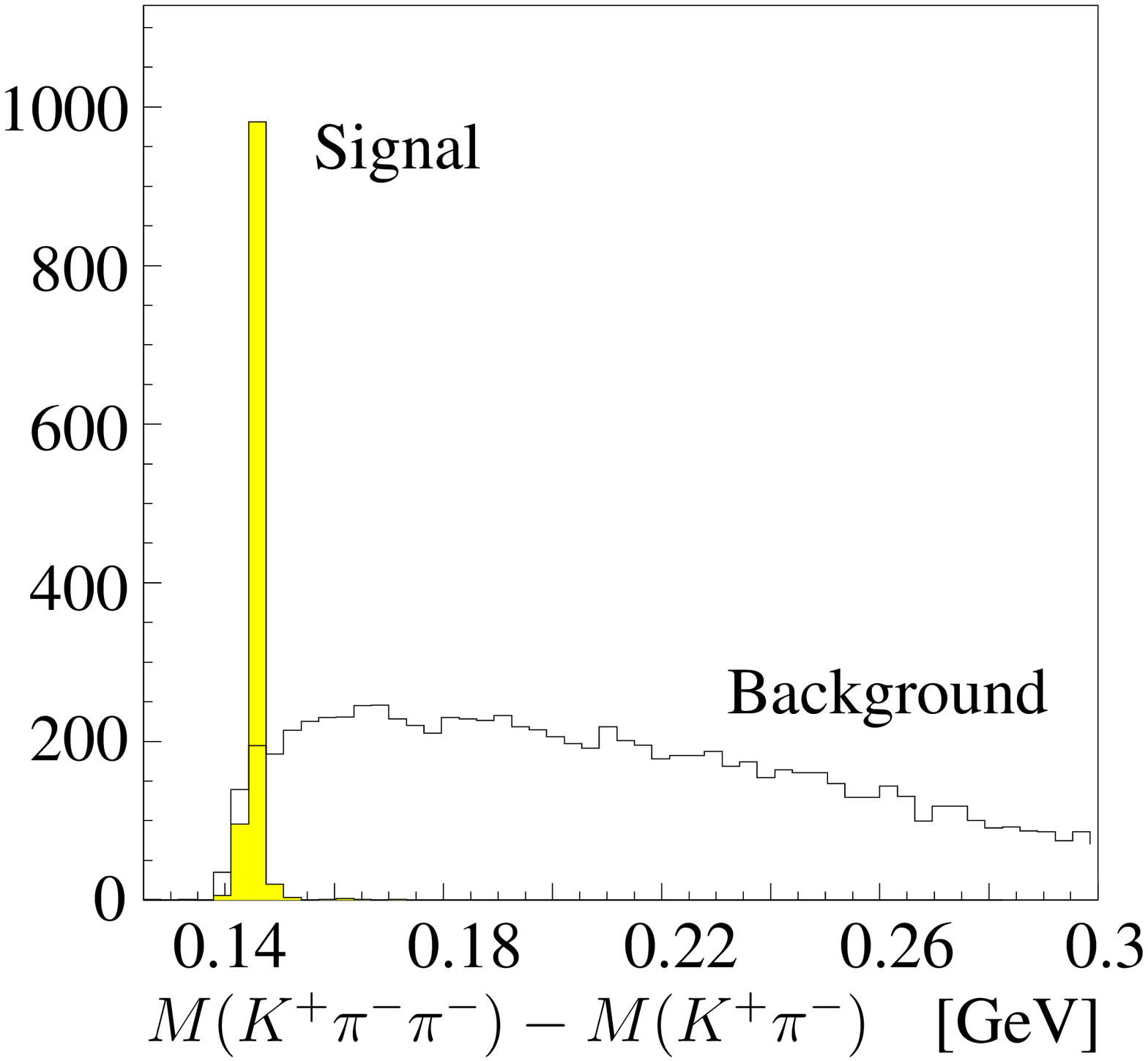, width=5cm}$\quad$
\epsfig{file=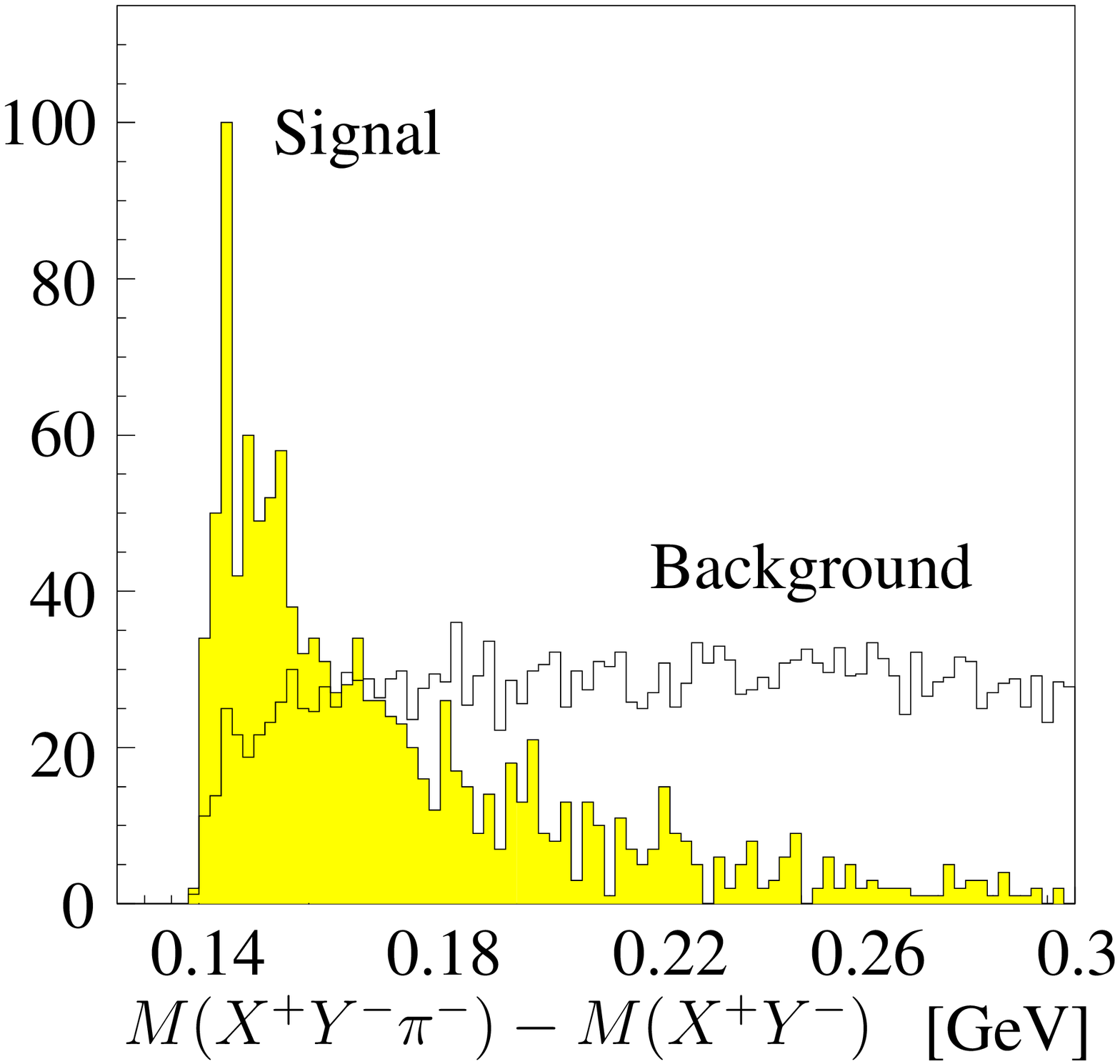, width=5cm}
\end{center}
\vspace*{-0.6cm}
\caption[]{The difference between reconstructed \Dstm\ and \Dzb\ mass for
  the exclusive and inclusive reconstruction. The background is
  superimposed with arbitrary normalization. For the exclusive
  reconstruction $\Delta m \in [\un{143.5}{MeV},\un{146.5}{MeV}]$, for
  the inclusive reconstruction $\Delta m \in
  [\un{144}{MeV},\un{160}{MeV}]$.}\label{dst_massdiff}
~\\
\begin{tabular}{cc}
    \parbox{10.4cm}{
    \epsfig{file=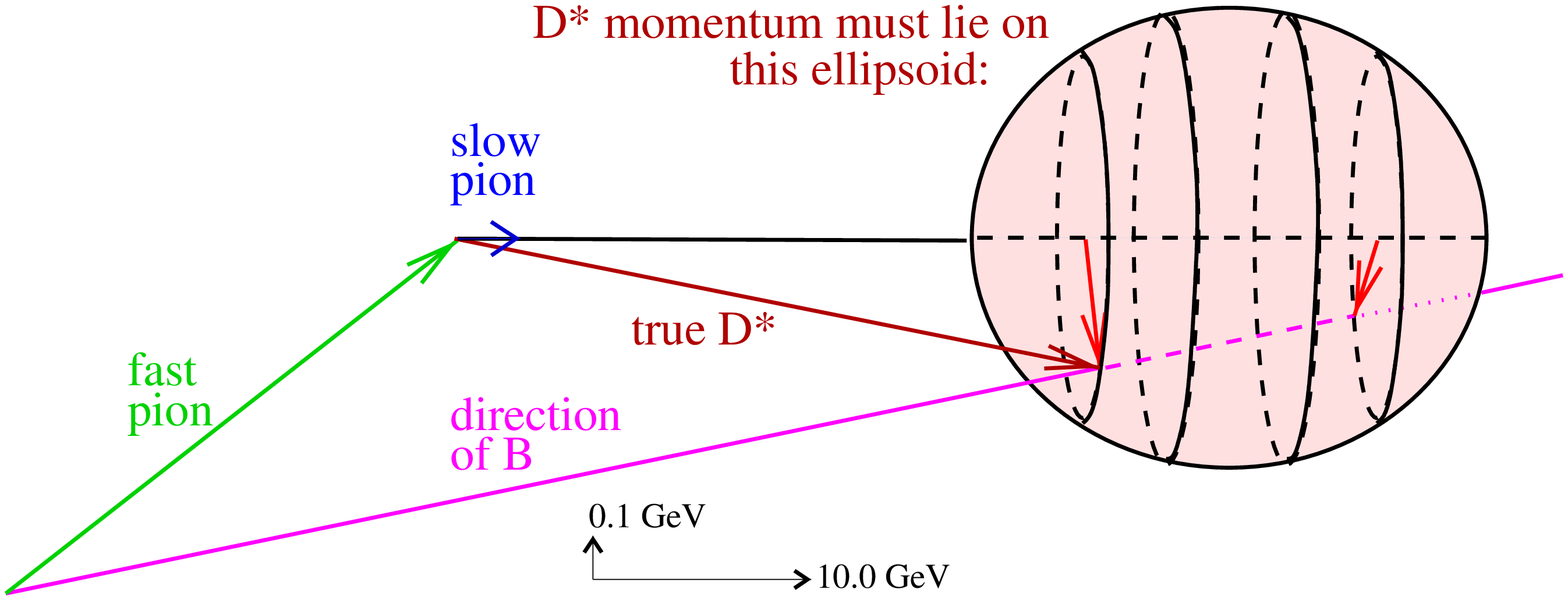, width=10.4cm}
    \hspace{-2em}(a)}
  & \parbox{4.6cm}{
    \epsfig{file=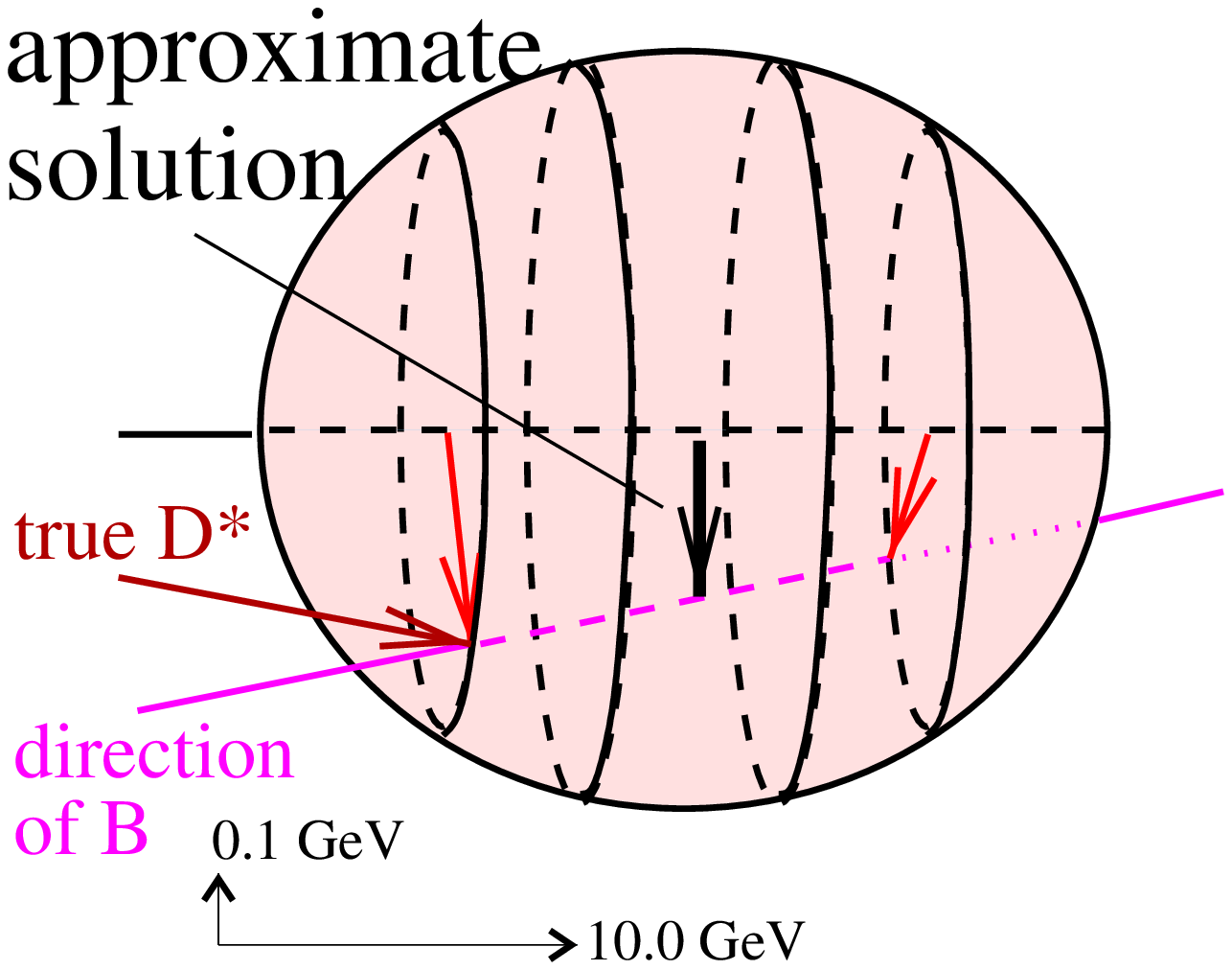, width=4.6cm}
    \hspace{-2em}(b)}
\end{tabular}
\vspace*{-0.3cm}
\caption[]{Schematic of the kinematics in the 
\BDpi\ decay,  showing in (a) the two possible solutions and in (b)
the approximation that is actually used in the analysis.}\label{rugby_ball}
\end{figure}

To reconstruct the \Dst\ (and then \Bz) momentum from this limited
information, we use  the fact that knowing the $\pi_s$ momentum restricts
the possible \Dst\ momenta to a two-dimensional
 surface.    This surface is shown
schematically in Fig.~\ref{rugby_ball}(a).   Kinematics defines
two possible solutions,  but in practice the solutions lie very close,
and it suffices to approximate with the distance of closest approach
between the slow pion and $B^0$ vectors as shown in 
Fig.~\ref{rugby_ball}(b).   In order to suppress background,
a probability distribution is cut on,  which exploits the allowed ranges
and expected correlations between the parameters in this reconstruction.

To further reduce background, one can use a  cut similar to that
on the mass difference 
between the \Dstm\ and the \Dzb\ as applied in the
exclusive case.
Instead of fully reconstructing the \Dzb, one tries to
identify two charged decay products of the \Dzb\ ($X^+,Y^-$) and cuts
on the difference $\Delta m$ between the pseudo masses:
\begin{equation}
\label{def_delta_m}
\Delta m= M(X^+Y^-\pi^-) - M(X^+Y^-).
\end{equation}
$\Delta m$ would be the mass difference between the \Dst\ and the \Dz\ 
if $X^+$ and $Y^-$ were the only decay products of the \Dz. In general,
though, there will be some missing momentum. Fortunately the missing
momentum cancels to some extent in Eq.~(\ref{def_delta_m}), so
that even for the partially reconstructed \Dzb\ this remains a
powerful cut as shown in Fig.~\ref{dst_massdiff}(b).

After all cuts,
260k reconstructed, triggered and tagged
events per year are expected inside the mass window of $\pm
\un{200}{MeV}$ with a S/B $\sim 3$.  The reconstruction returns a mass
peak of width \un{200}{MeV}.

\subsection*{\protect\boldmath $B_d^0\to D^{*\mp} a_1^\pm$}

The same inclusive analysis was performed for the
channel \BDa, with $a_1^\pm\to \rho^0 \pi^\pm$, which has $\sim 3$
times as high a branching ratio as \BDpi. As expected, the efficiency
for this channel is lower, as there are more particles to reconstruct,
while the mass resolution is slightly improved ($\sigma\approx
\un{180}{MeV}$), due to better reconstruction of the \Bz\ decay
vertex from 4 instead of only 2 particles. 360k reconstructed,
triggered and tagged events are expected within a $\pm \un{200}{MeV}$
mass window per year, with a S/B of $\sim 4$.

\vspace*{0.5cm}

\noindent 
The yield in all analyses is summarized in Tab.~\ref{tab:bdpi},  with a
   total that assumes negligible correlation between the selections.

\subsection*{Sensitivity to \protect\boldmath $\gamma$}

For \BDpi\ decays the parameters $\xi_f^{(d)}$ and $\xi_{\bar f}^{(d)}$
can in principle be completely determined by fitting the two time-dependent
asymmetries 
\begin{eqnarray}
A_{D^{*-}}(\tau) 
&=& 
\frac{     \Gamma_{\tau}\left(          B_d^0  \to D^{*-}\pi^{+}\right)
         - \Gamma_{\tau}\left(\overline{B_d^0} \to D^{*-}\pi^{+}\right)
      }{
           \Gamma_{\tau}\left(          B_d^0  \to D^{*-}\pi^{+}\right)
         + \Gamma_{\tau}\left(\overline{B_d^0} \to D^{*-}\pi^{+}\right)
      }\nonumber\\
&=&
\frac{  \left( 1- \left|\xi_{\bar f}^{(d)}\right|^2 \right) \cos(\Delta m \tau)
       -   2\left|\xi_{\bar f}^{(d)}\right|
           \sin\left(-\left(\phi_d + \gamma \right)
                     +\Delta_{\mbox{\scriptsize S}}\right) 
           \sin(\Delta m \tau)
     }{
       1+ \left|\xi_{\bar f}^{(d)}\right|^2
     }\,,\\
A_{D^{*+}}(\tau) 
&=& 
\frac{     \Gamma_{\tau}\left(\overline{B_d^0} \to D^{*+}\pi^{-}\right)
         - \Gamma_{\tau}\left(          B_d^0  \to D^{*+}\pi^{-}\right)
      }{
           \Gamma_{\tau}\left(\overline{B_d^0} \to D^{*+}\pi^{-}\right)
         + \Gamma_{\tau}\left(          B_d^0  \to D^{*+}\pi^{-}\right)
      }\nonumber\\
&=&
\frac{  \left( 1- \left|\xi_{f}^{(d)}\right|^2 \right) \cos(\Delta m \tau)
       -   2\left|\xi_{f}^{(d)}\right|
           \sin\left(+\left(\phi_d + \gamma \right)
                     +\Delta_{\mbox{\scriptsize S}}\right) 
          \sin(\Delta m \tau)
     }{
       1+ \left|\xi_{f}^{(d)}\right|^2
     }\,,
\end{eqnarray}
\noindent where $\Delta_{\mbox{\scriptsize S}}$ is 
a possible strong phase shift
entering $\xi_f^{(d)}$ via $\overline{M}_f/M_{\overline f}$.

Acceptance effects cancel in each of the two asymmetries.  In
practice, as the interference effect is so tiny, $|\xi_{\bar
    f}^{(d)}| = 1/|\xi_{f}^{(d)}|$ needs to be
constrained.  Fits therefore have been performed assuming this
parameter is known with a relative precision of $\epsilon_{|\xi|}$.
This uncertainty translates directly into a relative uncertainty on $
\sin( \Delta_{\mbox{\scriptsize S}}\pm \{\phi_d + \gamma\})$.
Throughout, a plausible true value of $|\xi_{\bar f}^{(d)}|
= 0.016$ has been assumed; the final resolution on $\gamma$ turns out
to be directly proportional to this value (if $\epsilon_{|\xi|}=
0$), i.e.\ $\sigma_{\gamma}\propto 1/|\xi_{\bar f}^{(d)}|$.

\begin{figure}
\begin{minipage}[t]{0.32\textwidth}
\begin{center}\epsfig{file=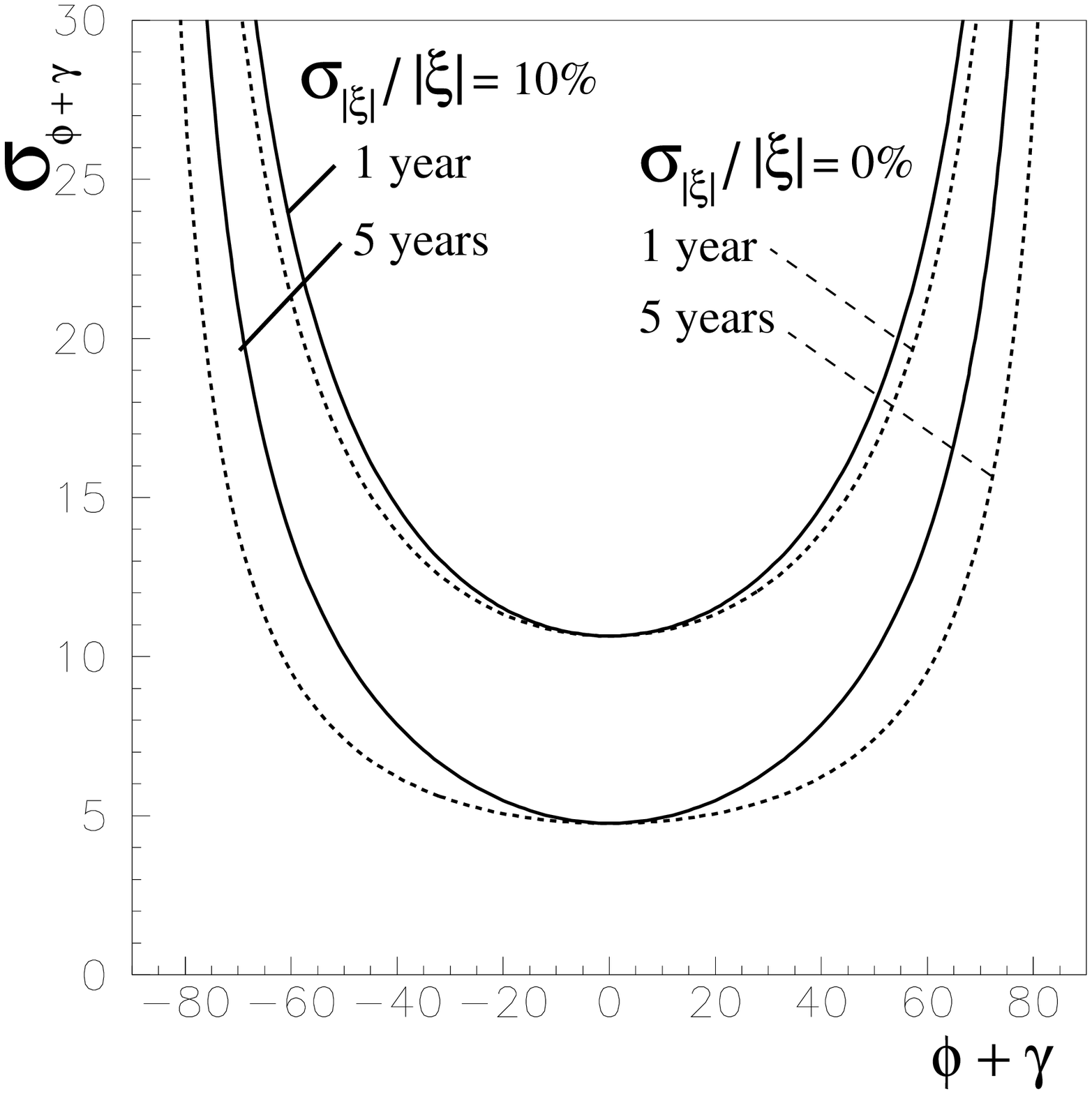,width=5cm,height=5cm}\end{center}
\vspace*{-0.5cm}
\caption[]{
  Error on $\phi_d+\gamma$ as a function of $\phi_d+\gamma$ for
  $\Delta_{\mbox{\scriptsize S}}=0$, after 1 and 5 years of data
  taking, assuming  
  that $|\xi_{\bar f}^{(d)}|$ is known perfectly or up to 10\%.}\label{results}
\end{minipage}
\hspace*{5pt}
\begin{minipage}[t]{0.64\textwidth}
~\\[-6.8cm]
\begin{center}
\subfigure[No RICH]
{\epsfig{file=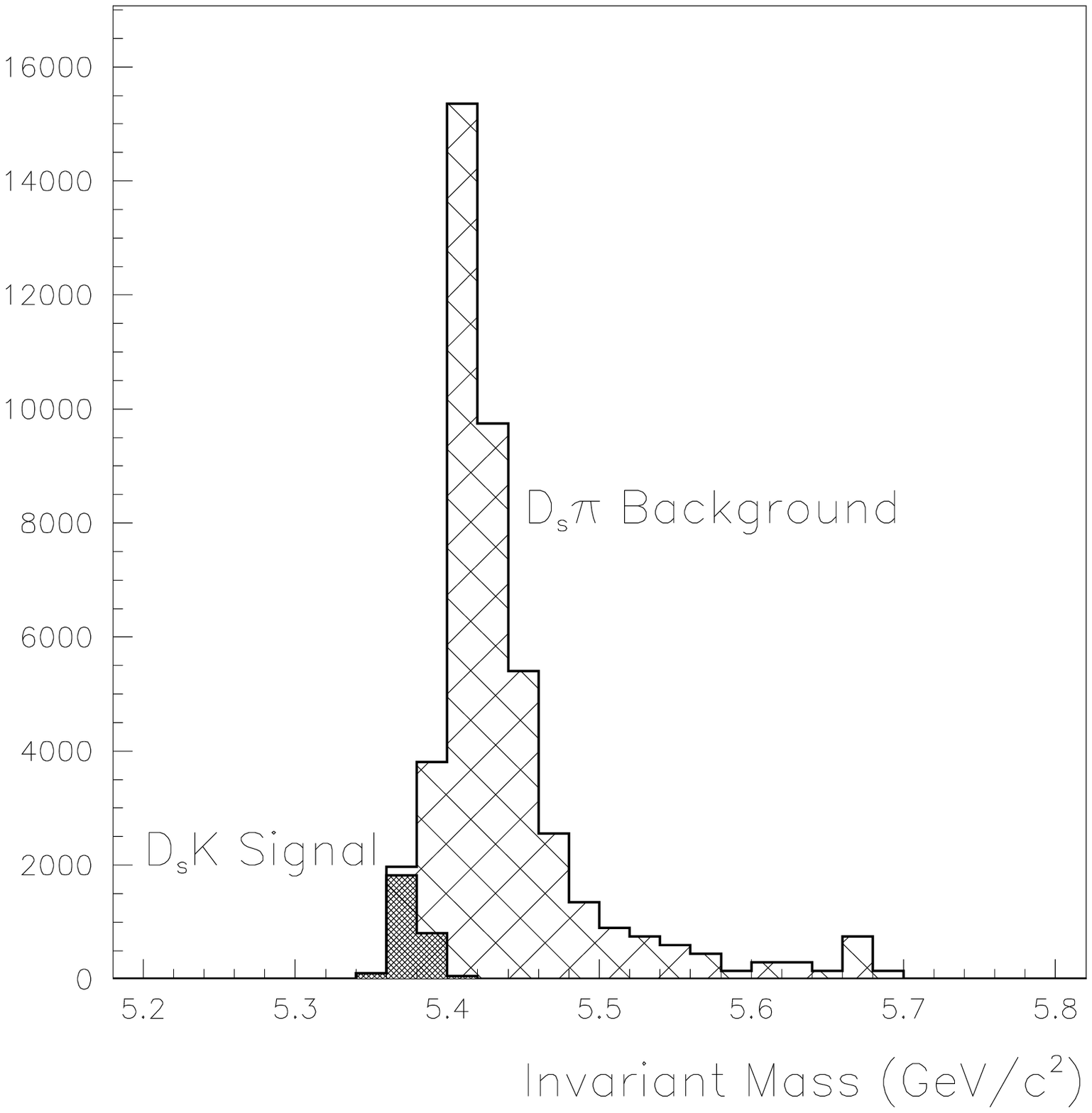,width=5cm,height=5.5cm}}
\subfigure[With RICH]
{\epsfig{file=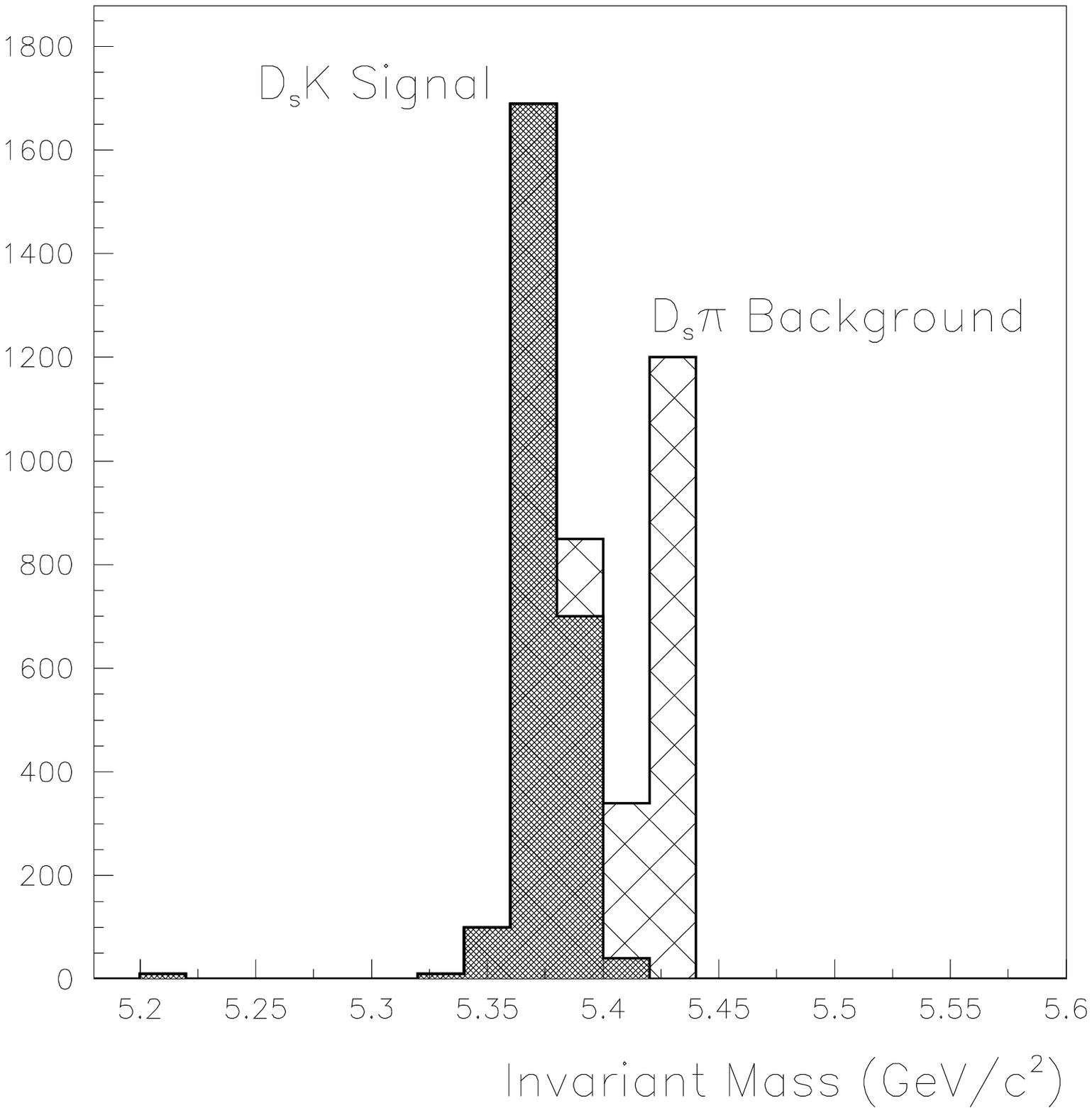,width=5cm,height=5.5cm}}
\end{center}
\vspace*{-0.5cm}
\caption{LHCb reconstruction of $B_s\to D_s^{\pm} K^{\mp}$,  showing
the contribution of  $B_s\to D_s^{\pm} \pi^{\mp}$ background before (a)
and after (b) the application of RICH information.}
\label{fig:bsdskrich}
\end{minipage}
\end{figure}

Using a stand-alone MC simulation and feeding it with the
parameters, event yields (340k) and S/B ratios ($\sim 3$) for \BDpi\ 
as discussed above, the statistical error on $
\sin(\Delta_{\mbox{\scriptsize S}}\pm\{\phi_d + \gamma \})
$ is found to be (for $\epsilon_{|\xi|}=0$):
\begin{equation}
\sigma_{\mbox{sin}} = \frac{0.26}{\sqrt{\mbox{no. of years}}}\,,
\end{equation}
independent of the input values for $(\phi_d + \gamma)$ and
$\Delta_{\mbox{\scriptsize S}}$.  Translating this into 
$\gamma$--$\Delta_{\mbox{\scriptsize S}}$
space, the resolution now does depend on the input values; an
uncertainty in $|\xi_{\bar f}^{(d)}|$ also introduces a
dependence on \( \sin(\Delta_{\mbox{\scriptsize S}}\pm\{\phi_d + \gamma
  \}) \). Figure~\ref{results} shows the error on
$(\phi_d+\gamma)$ as a function of $(\phi_d+\gamma)$ for
$\Delta_{\mbox{\scriptsize S}}=0$, after 1 and after 5 years of LHCb
data taking,
for the cases that $|\xi_{\bar f}^{(d)}|$ is known exactly
(broken lines) and that the uncertainty in 
$|\xi_{\bar f}^{(d)}|$ is 10\% (solid lines).
Assuming that $\phi_d$ can be fixed with negligible uncertainty from
$\rm B^0_d \, \rightarrow J/\psi K^0_s$ decays,  this error will
apply to $\gamma$ itself.

\begin{table}
\begin{center}
  \begin{tabular}{ |c |r |r | }
    \hline
    Channel        & S/B & Yield \\
    \hline \hline
    \BDpi\ (excl)   & 12  &   83k\\
    \BDpi\ (incl)   &  3  &  260k\\
    \BDa\  (incl)   &  4  &  360k\\
  \hline
    Total  & \multicolumn{2}{r|}{703k}\\
  \hline
  \end{tabular}
\end{center}
\vspace*{-0.5cm}
\caption[]{Expected S/B and yields in reconstructed, triggered and
tagged events in a single year of LHCb data taking.}
\label{tab:bdpi}
\end{table}

Presumably the large yield in \BDa\ events can also be exploited to
obtain additional sensitivity to $\gamma$.  However the presence of
two spin-1 particles in the decay complicates the extraction,
requiring that an angular analysis be performed to disentangle the
final-state configurations  (see~\cite{FD2} for the discussion of
an analogous problem).   This study has not yet been performed.

\subsubsection{Conclusions}

It can be seen that the large statistics at the LHC offers the possibility
of measuring $\gamma$ with very interesting precision from
\BDpia\ decays,  despite the expected low value of interference effects.


\subsection[Extracting $\gamma-2\delta\gamma$
from $B_s\to D_s^{\pm} K^{\mp}$
Decays]{Extracting \protect\boldmath $\gamma-2\delta\gamma$ from 
\protect\boldmath $B_s\to D_s^{\pm} K^{\mp}$
Decays}\label{subsec:BsDsK}
 
\subsubsection{Theoretical Aspects}

The decays $B_s\to D_s^{\pm} K^{\mp}$, which receive only contributions from 
tree-diagram-like topologies, are the $B_s$ counterparts of the 
$B_d\to D^{(\ast)\pm}\pi^\mp$ modes discussed in 
Sec.~\ref{subsec:BDpi}, and probe the CKM combination 
$\gamma-2\delta\gamma$ instead of $\gamma+2\beta$ in a {\it theoretically 
clean} way \cite{adk}. As we will see in the following section, the
CP-violating weak $B^0_s$--$\overline{B^0_s}$ mixing phase 
$\phi_s=-2\delta\gamma$ can be extracted with the help of the decay 
$B_s\to J/\psi\,\phi$. Since one decay path in $B^0_s$, 
$\overline{B^0_s}\to D^+_sK^-$ is only suppressed 
by $R_b\approx0.41$, and not doubly Cabibbo-suppressed by 
$\lambda^2 R_b$, as in the case of $B_d\to D^{(\ast)\pm}\pi^\mp$, 
the interference effects in $B_s\to D_s^\pm K^\mp$ are much larger. A
similar strategy to determine $\gamma-2\delta\gamma$ is also provided
by the colour-suppressed decays $B_s\to D\phi$ \cite{GroLoBs}. In 
Ref.~\cite{dun}, untagged data samples of these decays were considered 
to extract CKM phases, and angular distributions of untagged decays of
the kind $B_s\to D^{\ast\pm}K^{\ast\mp}$, $B_s\to D^{\ast}\phi$ were 
considered in \cite{FD2}.

\subsubsection{Experimental Studies}

LHCb have investigated the expected event yields in $B_s\to D_s^{\pm} K^{\mp}$
and resulting sensitivity to $\gamma$--$2\delta\gamma$~\cite{LHCbTP}.
An experimental challenge in selecting this mode is the need to
reject the about 10 times more  abundant $B_s\to D_s^{\pm} \pi^{\mp}$ events.
Figure~\ref{fig:bsdskrich} shows the event sample before and after
the application of information from the RICH detector.  It can be seen
that with such $\pi$--$K$ discrimination the  $B_s\to D_s^{\pm} \pi^{\mp}$
contamination can be adequately suppressed.
2.4k reconstructed and tagged events are expected in one year,
with a low background. 

The CKM phase $\gamma$--$2\delta\gamma$ can be determined from a fit to such
a sample,  in a manner directly analogous to that described in
Sec.~\ref{subsec:BDpi}.  In this case however, the intrinsic sensitivity
is  higher due to the larger interference effects.   As always, the 
precision on the CKM phase depends on the value of the parameters, which
here include $\Delta \Gamma_s / \Gamma_s$ and $\Delta m_s$.        
In one year's operation it 
is typically $8^\circ \,(\mbox{mean}) \, \pm 2^\circ (\mbox{rms})$ 
for scenarios with $\Delta m_s = 15 \, \mbox{ps}^{-1}$,  and degrades to
$\sim 12^\circ$ at $\Delta m_s = 45 \, \mbox{ps}^{-1}$.  Full tables can
be found in~\cite{LHCbTP}.   Assuming that $2\delta\gamma$ can be 
constrained from measurements in $\rm B_s \to J/\psi \phi$ decays,  
then this channel will provide a very clean and competitive measurement
of the angle $\gamma$.

\subsection[Extracting $\gamma$ from 
 $B\to D K$ Decays]{Extracting \protect\boldmath
  $\gamma$ from \protect\boldmath $B\to D K$ 
Decays}\label{subsec:BDK}
During the recent years, relations among amplitudes of nonleptonic $B$ decays
have been very popular to develop strategies for extracting the angles of
the unitarity triangles, in particular for $\gamma$. The prototype of
this approach involves charged $B^\pm\to D K^\pm$ decays \cite{gw}.
 
\begin{figure}[htb]
\begin{center}
\leavevmode
\epsfysize=3.8truecm 
\epsffile{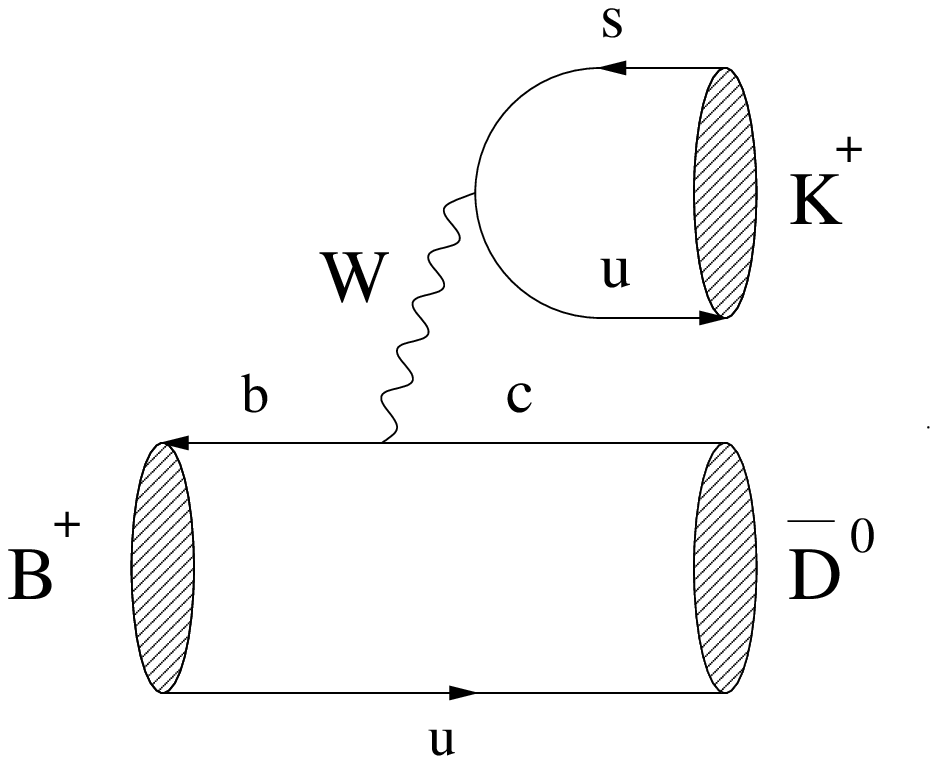} \hspace*{1truecm}
\epsfysize=4.5truecm 
\epsffile{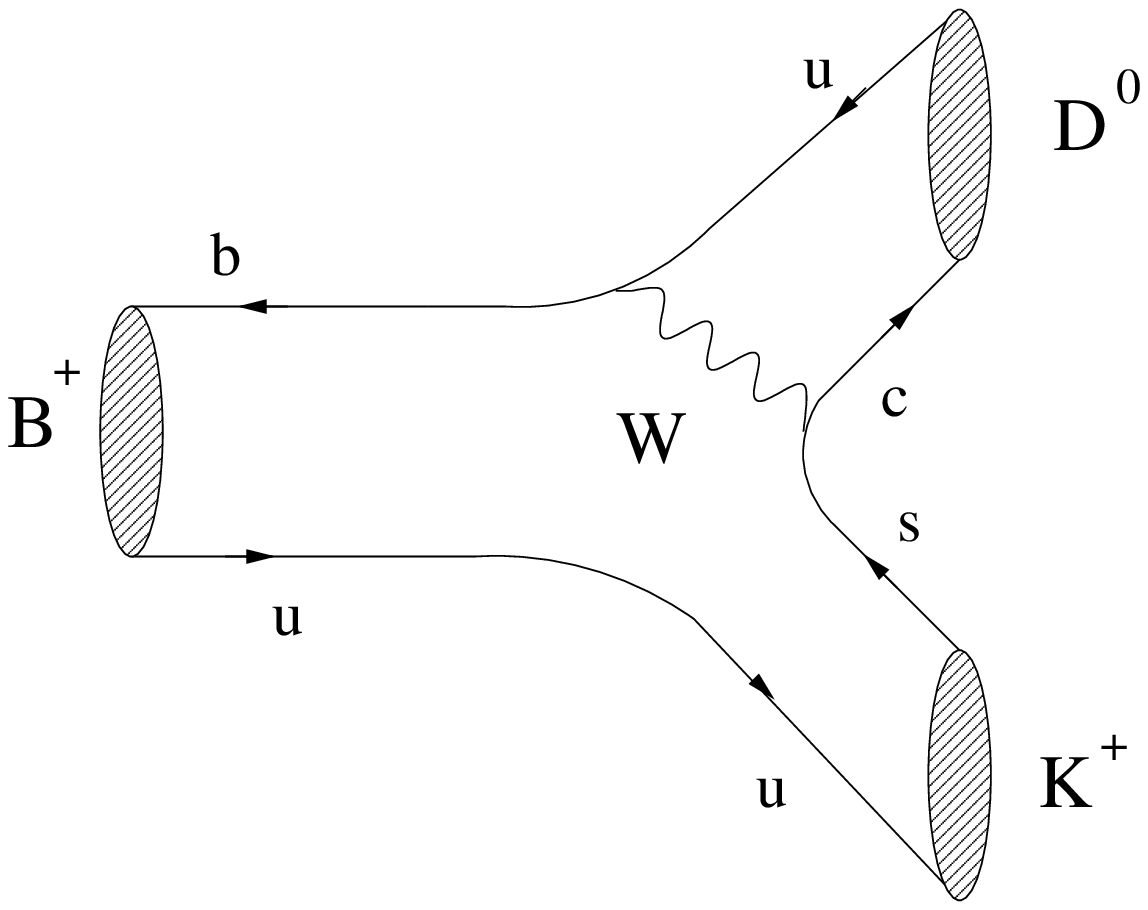}
\end{center}
\vspace*{-0.8truecm}
\caption{Feynman diagrams contributing to the decays 
$B^+\to \overline{D^0}K^+$ and $B^+\to D^0K^+$.}\label{fig:BDK}
\end{figure}

\subsubsection{Theoretical Aspects}
\begin{figure}
\centerline{
\epsfysize=2.4truecm
{\epsffile{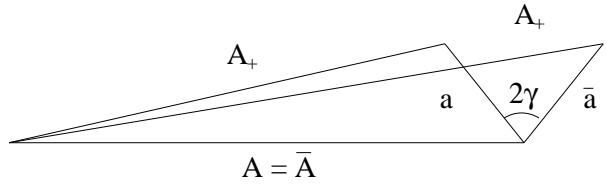}}}
\vspace*{-0.3cm}
\caption{Triangle relations between charged $B^\pm\to DK^\pm$ decay 
amplitudes.}\label{fig:BDK-triangle}
\end{figure}

The decays $B^+\to \overline{D^0}K^+$ and $B^+\to D^0K^+$, which are
pure ``tree'' decays, as can be seen in Fig.~\ref{fig:BDK}, provide an 
interesting strategy to extract $\gamma$, if we make in addition
use of the transition $B^+\to D^0_+K^+$. Here, $D^0_+$ denotes the 
CP eigenstate of the neutral $D$-meson system with CP eigenvalue $+1$,
which is given by
\begin{equation}\label{ED85}
\left|D^0_+\right\rangle=\frac{1}{\sqrt{2}}\left(\left|D^0\right
\rangle+\left|\overline{D^0}\right\rangle\right)
\end{equation}
and leads to the following amplitude relations:
\begin{eqnarray}
\sqrt{2}A(B^+\to D^0_+K^+)&=&A(B^+\to D^0 K^+)+A(B^+\to\overline{D^0}K^+),
\label{ED86}\\
\sqrt{2}A(B^-\to D^0_+K^-)&=&A(B^-\to\overline{D^0} K^-)+A(B^-\to D^0K^-).
\label{ED87}
\end{eqnarray}
Since we are dealing with pure ``tree'' decays that are caused by 
$\bar b\to\bar c\, u\,\bar s,\, \bar u\,c\,\bar s$ quark-level transitions, 
we have
\begin{eqnarray}
a\,\,\,\equiv\,\,\, A(B^+\to D^0K^+)&=&A(B^-\to \overline{D^0}K^-)\times
e^{2i\gamma},\\
A\,\,\,\equiv\,\,\, A(B^+\to \overline{D^0}K^+)&=&A(B^-\to D^0K^-),
\end{eqnarray}
allowing a {\it theoretically clean} determination of $\gamma$ with the
help of the triangle construction shown in Fig.~\ref{fig:BDK-triangle}.
Unfortunately, we have to deal with rather squashed triangles, since 
$a\equiv A(B^+\to D^0K^+)$ is colour-suppressed with respect to
$A\equiv A(B^+\to \overline{D^0}K^+)$:
\begin{equation}
\frac{|a|}{|A|}=\frac{|\overline{a}|}{|\overline{A}|}\approx
\frac{1}{\lambda}\frac{|V_{ub}|}{|V_{cb}|}\times\frac{a_2}{a_1}
\approx0.41\times\frac{a_2}{a_1}\approx0.1\,,
\end{equation}
where $a_1$ and $a_2$ are the usual phenomenological colour factors.

In 1998, the CLEO collaboration has reported the observation of 
$B^+\to \overline{D^0}K^+$ \cite{cleo-BDK}:
\begin{equation}
B(B^+\to \overline{D^0}K^+)=
(0.257\pm0.065\pm0.032)\times10^{-3}.
\end{equation}
Using arguments based on ``colour suppression'', we expect
\begin{equation}
B(B^+\to D^0K^+)\approx 10^{-2}\times
B(B^+\to \overline{D^0}K^+).
\end{equation}
While the branching ratio $B(B^+\to \overline{D^0}K^+)$ can be 
measured using conventional methods, the measurement of 
$B(B^+\to D^0K^+)$ suffers from considerable experimental 
problems \cite{ads}:
\begin{itemize}
\item If the branching ratio of $B^+\to D^0K^+$ is measured through 
hadronic decays of the $D^0$-meson, e.g.\ through 
$B^+\to D^0[\to K^-\pi^+]\,K^+$, we have 
large interference effects of ${\cal O}(1)$ with the decay chain 
$B^+\to\overline{D^0}[\to K^-\pi^+]\,K^+$ (note that the $\overline{D^0}$
decay is doubly Cabibbo-suppressed). 
\item All possible hadronic tags of the $D^0$ in $B^+\to D^0K^+$ will 
be affected by such interference effects.
\item Such problems can in principle be avoided by using semi-leptonic tags 
$D^0\to l^+\nu_l X_s$. However, here there will be large backgrounds due 
to $B^+\to l^+\nu_l X_c$, which may be difficult to control.
\end{itemize}
Moreover, decays of neutral $D$-mesons into CP eigenstates, such as  
$D^0_+\to\pi^+\pi^-,K^+K^-$, are experimentally challenging. Consequently, 
the original method proposed by Gronau and Wyler \cite{gw}
will unfortunately be very difficult in practice. A variant of this 
approach was proposed by Atwood, Dunietz and Soni in \cite{ads}. In order 
to overcome the problems discussed above, the following decay chains can 
be considered:
\begin{equation}
B^+\to\overline{D^0}\,[\to f_i]\,K^+,\quad B^+\to D^0 \,[\to f_i]\,K^+,
\end{equation}
where $f_i$ denotes doubly Cabibbo-suppressed (Cabibbo-favoured) non-CP 
modes of the $\overline{D^0}$ ($D^0$), for instance, $f_i=K^-\pi^+$, 
$K^-\pi^+\pi^0$. In order to extract $\gamma$, at least two different 
final states $f_i$ have to be considered. In this method, one makes 
use of the large interference effects, which spoil the hadronic tag 
of the $D^0$ in the original Gronau--Wyler method. In contrast to the
case of $B^+\to D^0_+K^+$ discussed above, here both contributing decay 
amplitudes should be of comparable size, thereby leading to potentially 
large CP-violating effects. Furthermore, the  
branching ratio $B(B^+\to D^0K^+)$, which is difficult to measure, 
is not required, but can 
rather be determined as a by-product. Unfortunately, this approach is also
challenging, since many channels are involved, with total branching ratios 
of ${\cal O}(10^{-7})$ or even smaller. An accurate determination of 
the relevant $D$ branching ratios $B(D^0\to f_i)$ and 
$B(\overline{D^0}\to f_i)$ is also essential for this method.

\begin{figure}[htb]
\begin{center}
\leavevmode
\epsfysize=4.5truecm 
\epsffile{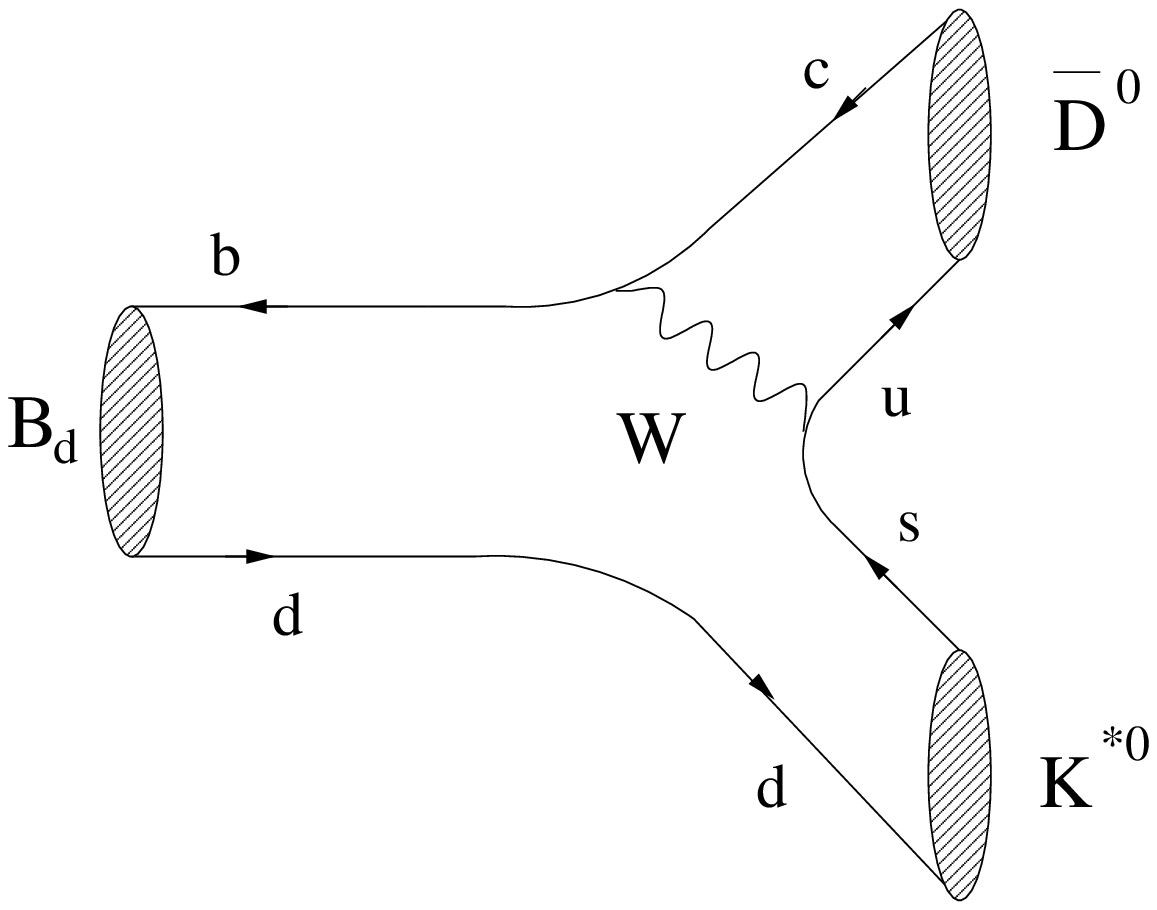} \hspace*{1truecm}
\epsfysize=4.5truecm 
\epsffile{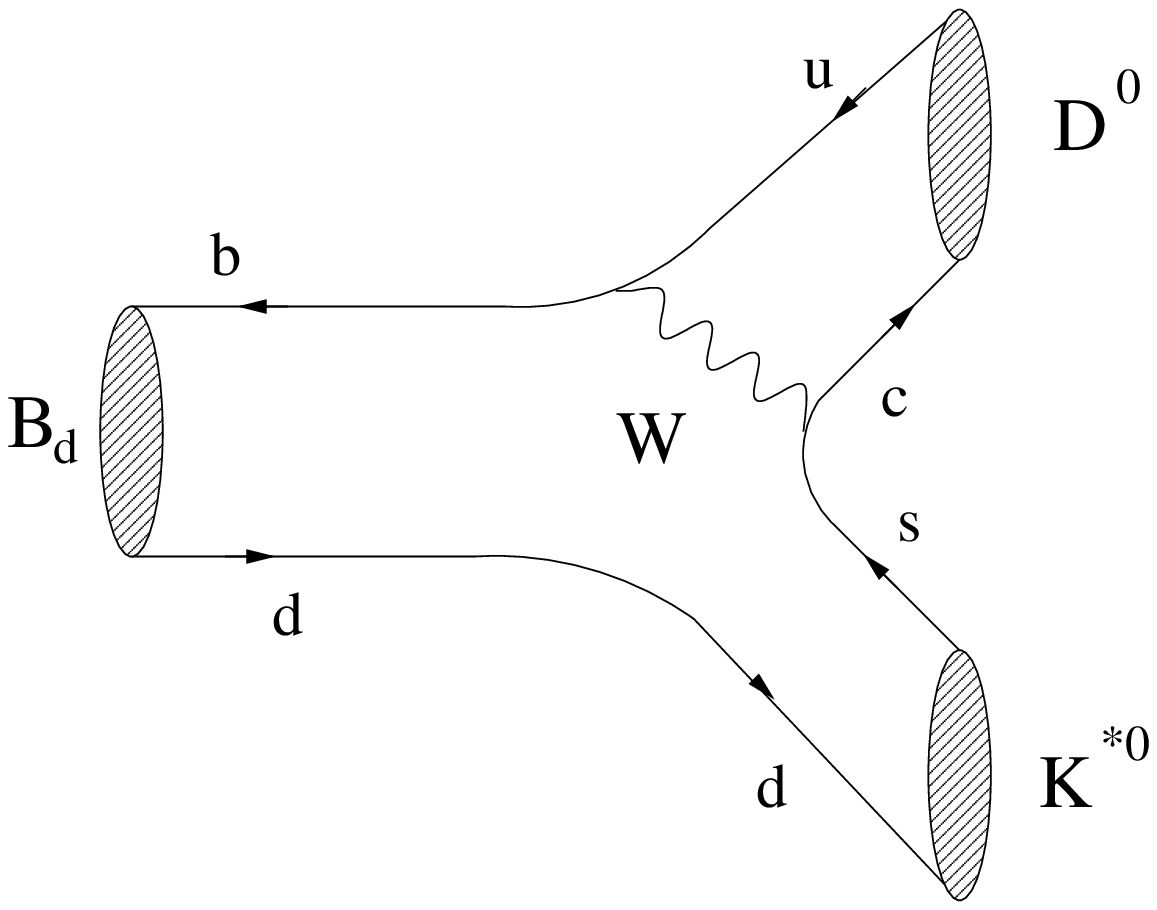}
\end{center}
\vspace*{-0.5truecm}
\caption{Feynman diagrams contributing to the decays 
$B^0_d\to\overline{D^0}K^{\ast0}$ and 
$B^0_d\to D^0K^{\ast0}$.}\label{fig:BDK-neut}
\end{figure}

\begin{figure}[htb]
\centerline{
\epsfysize=3.0truecm
{\epsffile{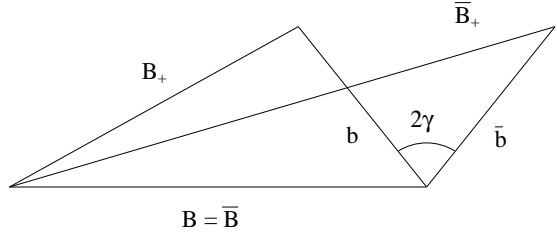}}}
\vspace*{-0.3cm}
\caption{Triangle relations between neutral $B_d\to DK^\ast$ decay 
amplitudes.}\label{fig:BDKneut-triangle}
\end{figure}

So far, we have only considered charged $B^\pm\to DK^\pm$ decays. However, 
also neutral decays of the kind $B^0_d\to D K^{\ast0}$, which are shown
in Fig.~\ref{fig:BDK-neut}, allow one to extract $\gamma$ \cite{dun-BDK}. 
As these modes are ``self-tagging'' through $K^{\ast0}\to K^+\pi^-$, no 
time-dependent measurements are required in this case. If we make again 
use of the CP eigenstate $D^0_+$ of the neutral $D$-meson system, we 
obtain the following amplitude relations:
\begin{eqnarray}
\sqrt{2}A(B^0_d\to D^0_+K^{\ast0})&=&A(B^0_d\to\overline{D^0}K^{\ast0})+
A(B^0_d\to D^0K^{\ast0}),\\
\sqrt{2}A(\overline{B^0_d}\to D^0_+\overline{K^{\ast0}})&=&
A(\overline{B^0_d}\to D^0\overline{K^{\ast0}})+
A(\overline{B^0_d}\to\overline{D^0}\,\overline{K^{\ast0}}).
\end{eqnarray}
Moreover, we have
\begin{eqnarray}
b\equiv A(B^0_d\to D^0K^{\ast0})&=&A(\overline{B^0_d}\to \overline{D^0}\,
\overline{K^{\ast0}})\times e^{2i\gamma},\\
B\equiv A(B^0_d\to \overline{D^0}K^{\ast0})&=&A(\overline{B^0_d}\to 
D^0\overline{K^{\ast0}}),
\end{eqnarray}
allowing one to extract $\gamma$ from the triangle construction shown in
Fig.~\ref{fig:BDKneut-triangle}, which is completely analogous to the
$B^\pm\to DK^\pm$ case. However, there is an important difference, 
which is due to the fact that both decays $B^0_d\to \overline{D^0}K^{\ast0}$
and $B^0_d\to D^0K^{\ast0}$ are ``colour-suppressed'', as can be seen in
Fig.~\ref{fig:BDK-neut}:
\begin{equation}
\frac{|A(B^0_d\to D^0K^{\ast0})|}{|A(B^0_d\to \overline{D^0}K^{\ast0})|}
\approx\frac{1}{\lambda}\frac{|V_{ub}|}{|V_{cb}|}\,\frac{a_2}{a_2}
\approx0.41\,.
\end{equation}
Consequently, the triangles are expected to be not as squashed as in the 
$B^\pm\to DK^\pm$ case. The corresponding branching ratios are expected
to be of ${\cal O}(10^{-5})$. However, we have also to deal with the 
difficulties of detecting the neutral $D$-meson CP eigenstate $D_+^0$.

\subsubsection{Experimental Studies}

Both ATLAS~\cite{ATLASPTDR} and LHCb~\cite{LHCbTP} have investigated
the possibility of determining $\gamma$ through amplitude relations in
the family of $B^0_d\to D K^{\ast0}$ decays.
Both experiments have demonstrated that it will be possible to
reconstruct samples of such events, with LHCb in particular benefiting
from its hadron trigger.
However,  with the branching ratios that
have been assumed, the yields are still low,  with only a few 10's
of events expected in the $D_1 K^{\ast0}$ and 
$D_1 \overline{K^{\ast0}}$ modes.   At this level
several years are required to integrate sufficient statistics for a 
meaningful measurement.   The experiments will continue to investigate 
this,  and associated $\rm B \rightarrow DK$  measurements,  and 
search for possible improvement.

\setcounter{equation}{0}
\newcommand{\Bstl} {{\mbox{$\mathrm {B_{s}\to J/\psi( l^+l^-)\ \phi( 
K^{+}K^{-})\ }$}}}
\newcommand{\DG}{\mbox{$\mathrm \Delta\Gamma_{s}\ $}}
\newcommand{\Bs}{\mbox{$\mathrm B^{0}_{s}\ $}}
\newcommand{\aBs}{\mbox{$\mathrm \overline B^{0}_{s}\ $}}
\newcommand{\Bh}{\mbox{$\mathrm B^{0}_{H}\ $}}
\newcommand{\Bl}{\mbox{$\mathrm B^{0}_{L}\ $}}
\newcommand{\Gh}{\mbox{$ \Gamma_{\mathrm{H}}\ $}}
\newcommand{\Gl}{\mbox{$ \Gamma_{\mathrm{L}}\ $}}
\newcommand{\DGdef}{\mbox{$ \Delta\Gamma_{\mathrm{s }}=\Gamma_{\mathrm
      {H}}-
\Gamma_{\mathrm {L}}$}}
\newcommand{\Gdef}{\mbox{$ \Gamma_{s}=(\Gamma_{\mathrm{ H}}+
\Gamma_{\mathrm {L}})/2$}}
\newcommand{\Ght}{\mbox{$ \Gamma_{\mathrm{H}}t\ $}}
\newcommand{\Glt}{\mbox{$ \Gamma_{\mathrm{L}}t\ $}}
\newcommand{\xis}{\mbox{$ x_{\mathrm s}\ $}}
\newcommand{\dMss}{\mbox{$  {\Delta}M_{s}  \ $}}
\newcommand{\dMsst}{\mbox{$ {\Delta}mt$}}
\newcommand{\Bsto}{\mbox{$\mathrm \Bs \rightarrow \JpsiPhi \ \rightarrow
 \mu^{+} \mu^{-} K^{+}K^{-} \  $}}
\newcommand{\Bstoo}{\mbox{$\mathrm \Bs \rightarrow \rm J/\psi (
 \mu^{+} \mu^{-})  \phi ( \rm K^+K^-)$}}
\newcommand{\Bst}{\mbox{$\mathrm \Bs \rightarrow \JpsiPhi$}}
\newcommand{\aBst}{\mbox{$\mathrm \aBs \rightarrow \JpsiPhi \   $}}
\newcommand{\JpsiPhi}{\mbox{$\mathrm J/\psi \phi$}}
\newcommand{\BstoDs}{\mbox{$\mathrm \Bs \rightarrow D_{s} \pi \ $}}
\newcommand{\BstoDsa}{\mbox{$\mathrm \Bs \rightarrow D_{s} a_{1} \ $}}
\newcommand{\BtoKs}{\mbox  {$\mathrm {B^{0}_{d} \rightarrow \mathrm 
J/\psi K^{*}}$}}
\newcommand{\BtoK} {\mbox{$\mathrm B_{d} \rightarrow J/\psi K^{0}  \ $ }}
\newcommand{\BtoKpi} {\mbox{$\mathrm B_{d} \rightarrow J/\psi
 K^{+}\pi^{-}  \ $ }}
\newcommand{\Lamzero}{\mbox{$\mathrm \Lambda\ $}}
\newcommand{\Lamzerob}{\mbox{$\mathrm \Lambda^{0}_{b}$}}
\newcommand{\Lamzeroto}{\mbox{$\mathrm \Lamzero \rightarrow p \pi^{-}\ $}}
\newcommand{\Lamzerobtoo}{\mbox{$\mathrm \Lamzerob \rightarrow \LamzeroJpsi$}}
\newcommand{\Lamzerobtooo}{\mbox{$\mathrm \Lamzerob \rightarrow
 \LamzeroJpsi\ \rightarrow p\pi^{-}\mu^{+}\mu^{-}\ $}}
\newcommand{\aLamzero}{\mbox{$\mathrm \overline{\Lambda^{0}}\ $}}
\newcommand{\LamzeroJpsi}{\mbox{$\mathrm \Lamzero \rm J/\psi$}}
\newcommand{\LamzeroJpsitol}{\mbox{$\mathrm \LamzeroJpsi \rightarrow
 p\pi^{-}\mu^{+}\mu^{-}\ $}}
\newcommand{\LamzeroJpsitomu}{\mbox{$\mathrm \LamzeroJpsi \rightarrow
 p\pi^{-}\mu^{+}\mu^{-}\ $}}
\newcommand{\LamzeroJpsitoe}{\mbox{$\mathrm \LamzeroJpsi \rightarrow p\pi^{-}
 \mathrm{e}^{+}\mathrm{e}^{-}\ $}}
\newcommand{\aLamzeroJpsi}{\mbox{$\mathrm \aLamzero J/\psi$}}
\newcommand{\Xizerob}{\mbox{$\mathrm \Xi^{0}_{b}\ $}}
\newcommand{\Ximinusb}{\mbox{$\mathrm \Xi^{-}_{b}\ $}}
\newcommand{\aXizerob}{\mbox{$\mathrm \overline{\Xi^{0}_{b}}\ $}}
\newcommand{\aXiplus}{\mbox{$\mathrm \overline{\Xi^{+}_{b}}\ $}}
\newcommand{\aSigmaminus}{\mbox{$\mathrm \overline{\Sigma^{-}_{b}}\ $}}
\newcommand{\Xizerobto}{\mbox{$\mathrm \Xizerob \rightarrow \LamzeroJpsi\ $}}
\newcommand{\Xinulabto}{\mbox{$\mathrm \Xizerob \rightarrow \Xi^{0} J/\psi\ $}}
\newcommand{\Xizeroto}{\mbox{$\mathrm \Xi^{0} \rightarrow \Lamzero \pi^{0}\ $}}
\newcommand{\Ximinusbto}{\mbox{\mathrm $\Xi^{-}_{b} \rightarrow
 \Xi^{-} 
J/\psi\ $}}
\newcommand{\Ximinusto}{\mbox{$\mathrm \Xi^{-} \rightarrow \Lamzero
 \pi^{-}\ 
$}}
\newcommand{\aXizerobto}{\mbox{\mathrm $\aXizerob \rightarrow
 \aLamzeroJpsi\ 
$}}
\newcommand{\Bto}{\mbox{$\mathrm B_{b} \rightarrow \LamzeroJpsi\ $}}
\newcommand{\Jpsitomm}{\mbox{$\mathrm J/\psi \rightarrow \mu^{+} \mu^{-}\ $}}
\newcommand{\Jpsitoee}{\mbox{$\mathrm J/\psi \rightarrow
 \mathrm{e}^{+} \mathrm{e}^{-}\ $}}
\newcommand{\Jpsitol}{\mbox{$\mathrm   J/\psi \rightarrow \mu^{+} \mu^{-}\ $}}
\newcommand{\phitoKK}{\mbox{$\mathrm   \phi  \rightarrow K^{+} K^{-}\ $}}
\newcommand{\mumuppi}{\mbox{$\mathrm \mu^{+}\mu^{-}p\pi^{-}\ $}}
\newcommand{\mueeppi}{\mbox{$\mathrm \mu eep\pi^{-}\ $}}
\newcommand{\bquark}{\mbox{$\mathrm b\ $}}
\newcommand{\abquark}{\mbox{$\mathrm \overline{b}\ $}}
\newcommand{\bbarb}{\mbox{$\mathrm pp \rightarrow \bquark \abquark X\ $}}

\section[THE ``GOLD-PLATED'' DECAY 
\protect\boldmath $B_s\to J/\psi\,\phi$]{THE ``GOLD-PLATED'' DECAY 
\protect\boldmath $B_s\to J/\psi\,\phi$\protect\footnote{Section coordinators: 
R. Fleischer and M. Smizanska, with help from A. Dighe, P. Galumian and
N. Zaitsev.}}\label{sec:Bspsiphi}

\begin{figure}[bth]
\begin{center}
\leavevmode
\epsfysize=4.5truecm 
\epsffile{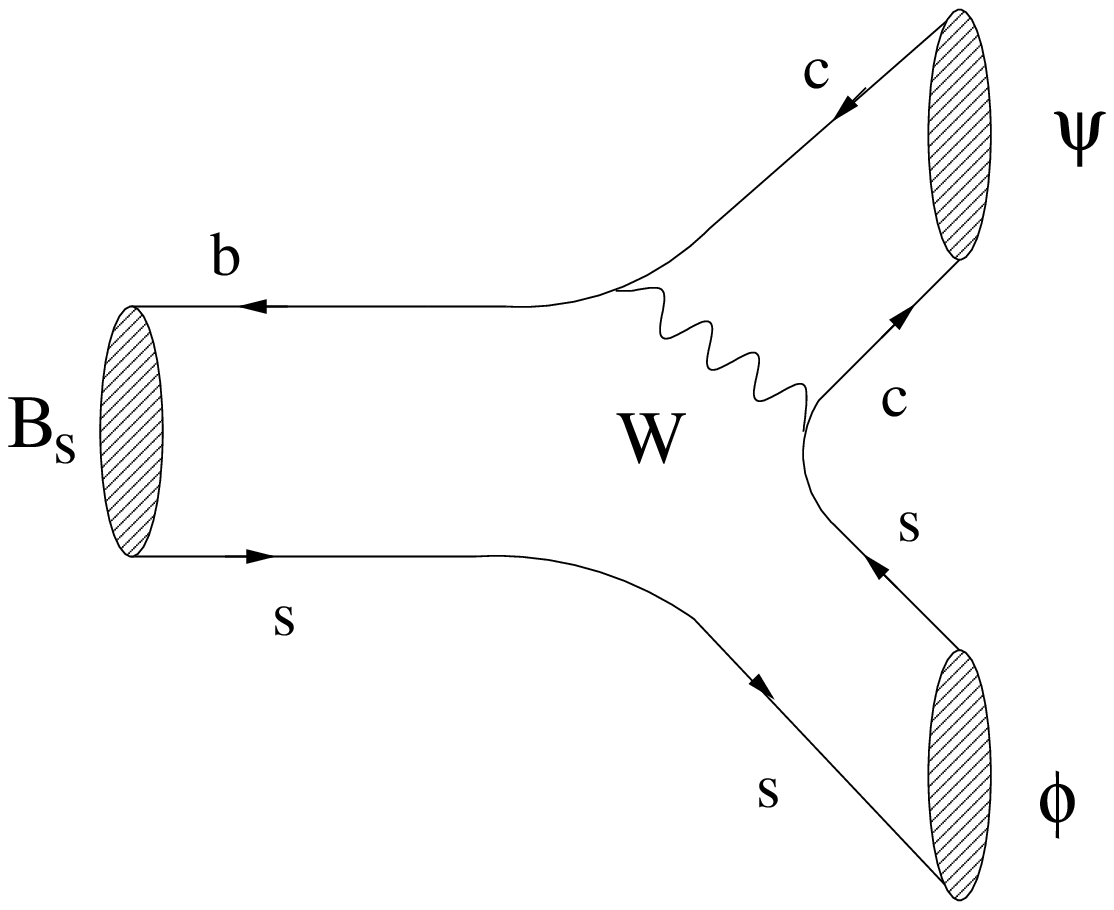} \hspace*{1truecm}
\epsfysize=4.5truecm 
\epsffile{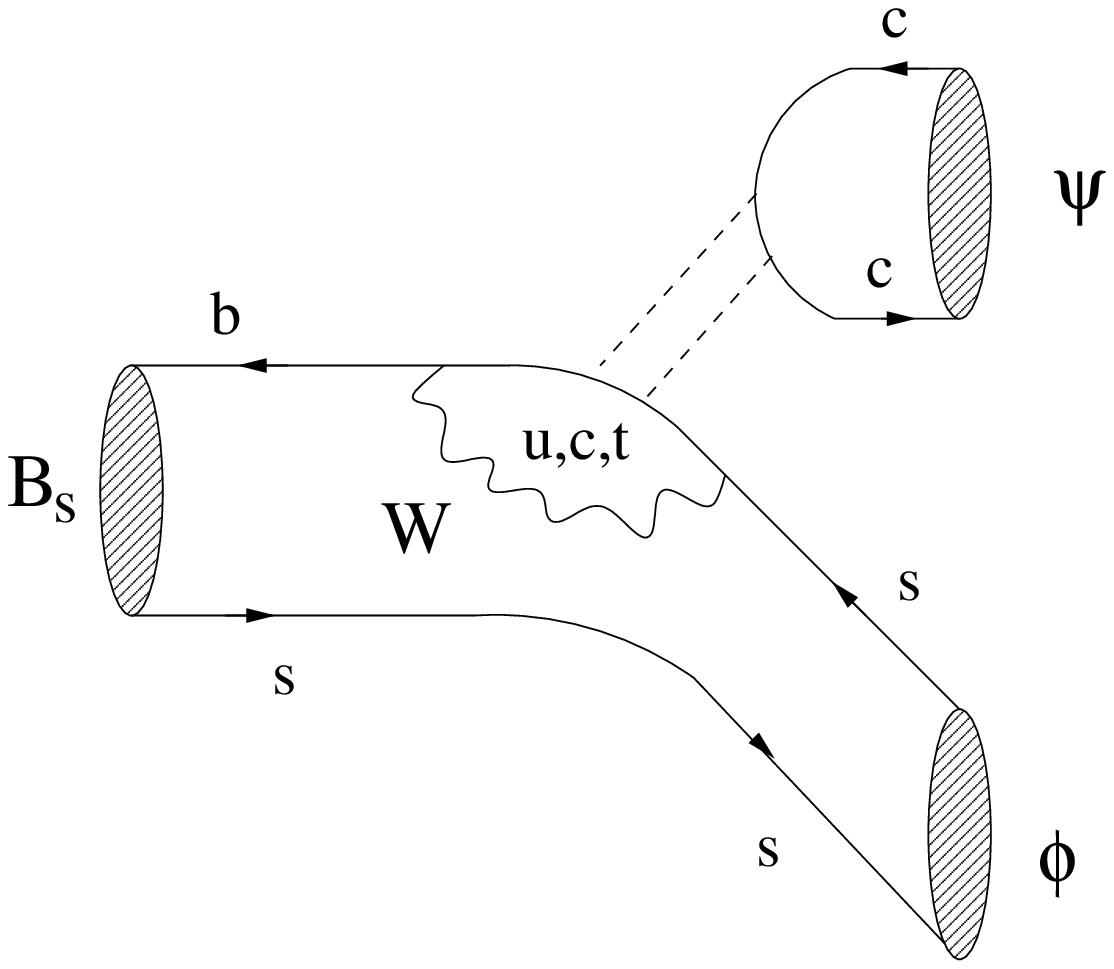}
\end{center}
\vspace*{-0.5truecm}
\caption[]{Feynman diagrams contributing to $\mathrm B_s^0\to
  J/\psi\,\phi$. The 
dashed lines represent a colour-singlet exchange.}\label{fig:Bspsiphi}
\end{figure}

The decay \Bst\ shown in Fig.~\ref{fig:Bspsiphi} is the 
$\mathrm B_s$ counterpart to the ``gold-plated'' mode 
$\mathrm B_d\to J/\psi\,K_{\rm S}$ 
and is particularly interesting  because of 
its rich physics potential.  A complete analysis of this decay
appears feasible at the LHC,  because of the large statistics and
good proper time resolution of the experiments.

\subsection{Theoretical Aspects}
In the case of $\mathrm B_s^0\to J/\psi\, \phi$, the final state is an 
admixture 
of different CP eigenstates. In order to disentangle them, an angular 
analysis of the decay products of \Bstl\ has to be performed
\cite{DDLR,ddf1}. In addition 
to interesting strategies to extract the $B^0_s$--$\overline{B^0_s}$ 
mixing parameters $\Delta\Gamma_s$ and $\Delta M_s$, we may also probe 
the weak mixing phase $\phi_s=-2\delta\gamma=-2\lambda^2\eta$, thereby 
allowing one to measure the Wolfenstein parameter $\eta$ \cite{FD1,ddf1}. 
A particularly interesting feature of \Bst\ decays is that 
they exhibit tiny CP-violating effects within the SM. 
Consequently, they represent a sensitive probe for CP-violating 
contributions from physics beyond the SM \cite{nir-sil,DFN}. 
Since new-physics contributions have to compete
with SM tree-diagram-like topologies, the natural place for any 
manifestation of new physics is in CP asymmetries induced by $B_s$
mixing.
Illustrations of the new-physics effects in \Bst\ for 
specific scenarios of new physics can be found 
in \cite{sil,bfNP} and are discussed in more detail below.


\subsubsection{General Structure of the Decay Probability Functions}
For an initially, i.e.\ at time $t=0$, present $\mathrm B_s^0$-meson, the 
time-dependent angular distribution of the decay chain 
\Bstl\ can be written 
generically as follows:
\begin{equation}\label{ang}
f( \Theta^{'},\Theta^{''},\chi; t)=\sum_k{\cal O}^{(k)}(t)\,
g^{(k)}(\Theta^{'},\Theta^{''},\chi),
\end{equation}
where we have denoted the angles describing the kinematics of the decay
products of $J/\psi\to l^+l^-$ and $\phi\to K^+K^-$ by $\Theta^{'}$, 
$\Theta^{''}$ 
and $\chi$. The functions ${\cal O}^{(k)}(t)$ describe the time evolution 
of the angular distribution (\ref{ang}), and can be expressed in terms of 
real or imaginary parts of bilinear combinations of decay amplitudes. 
In the case of decays into two vector mesons, such as
$\mathrm B_s^0\to J/\psi\, \phi$, it is convenient to introduce linear 
polarization 
amplitudes $A_0(t)$, $A_\parallel(t)$ and $A_\perp(t)$ \cite{pol}. Whereas 
$A_\perp(t)$ describes a CP-odd final-state configuration, both $A_0(t)$ 
and $A_\parallel(t)$ correspond to CP-even final-state configurations, 
i.e.\ to the CP eigenvalues $-1$ and $+1$, respectively. The 
${\cal O}^{(k)}(t)$ of the corresponding angular distribution are given by
\begin{equation}\label{obs1}
\left|A_f(t)\right|^2\quad\mbox{with}\quad f\in\{0,\parallel,\perp\},
\end{equation}
as well as by the interference terms
\begin{equation}\label{obs2}
\mbox{Re}\{A_0^\ast(t)A_\parallel(t)\}\quad\mbox{and}\quad
\mbox{Im}\{A_f^\ast(t)A_\perp(t)\} \quad\mbox{with}\quad f\in\{0,\parallel\}.
\end{equation}
These quantities are governed by  
\begin{equation}
\xi^{(s)}_{\psi\phi}\,\propto\, e^{-i\phi_s}\left[\frac{\lambda_u^{(s)*}
A_{\rm pen}^{ut'}+\lambda_c^{(s)*}\left(A_{\rm cc}^{c'}+A_{\rm pen}^{ct'}
\right)}{\lambda_u^{(s)}A_{\rm pen}^{ut'}+\lambda_c^{(s)}\left(A_{\rm cc}^{c'}
+A_{\rm pen}^{ct'}\right)}\right],
\end{equation}
where we have used the unitarity of the CKM matrix, the $\lambda_q^{(s)}$
are given by $V_{qs} V_{qb}^\ast$, and $A_{\rm pen}^{ut'}$ and 
$A_{\rm pen}^{ct'}$ denote the differences of penguin topologies with 
internal up- and top-quark and charm- and top-quark exchanges, respectively. 
The $A_{\rm pen}^{ut'}$ pieces are strongly CKM-suppressed by 
$|\lambda_u^{(s)}/\lambda_c^{(s)}|\approx0.02$; the penguin amplitudes 
are suppressed  even further because of their loop and colour structure. 
Yet, the ``current--current'' amplitudes are ``colour-suppressed'', 
and we may well have
\begin{equation}
\frac{\left|\lambda_u^{(s)}A_{\rm pen}^{ut}\right|}{\left|\lambda_c^{(s)}
\left(A_{\rm cc}^c+A_{\rm pen}^{ct}\right)\right|}={\cal O}(10^{-3}),
\end{equation}
yielding
\begin{equation}\label{xispsiphi}
\xi^{(s)}_{\psi\phi}\,\propto\, e^{-i\phi_s}
\left[1-2\,i\,\sin\gamma\times{\cal O}(10^{-3})\right].
\end{equation}
Since $\phi_s$ is of ${\cal O}(0.03)$ in the SM, there may well 
be hadronic uncertainties as large 
as ${\cal O}(10\%)$ in the extraction of $\phi_s$.
These hadronic uncertainties, which are an important issue for the LHC, 
can be controlled with the help of the decay $\mathrm B_d\to J/\psi\, \rho^0$ 
\cite{RF-ang}. Moreover, the angular distribution of this decay allows 
one to determine both $\sin\phi_d$ and $\cos\phi_d$, i.e.\ to fix $\phi_d$
{\it unambiguously}, and to extract $\gamma$, if penguin effects in 
$\mathrm B_d\to J/\psi\, \rho^0$ are sizeable. An unambiguous determination of
the $\mathrm B^0_d$--$\overline{B^0_d}$ mixing phase $\phi_d$ is also 
possible by combining the \Bst\ observables with those of the decay
$\mathrm B_d\to J/\psi(l^+l^-)\,K^{\ast0}( \pi^0K_{\rm S})$ \cite{ddf2};
other alternatives can be found in \cite{ambig}. For simplicity, we 
assume $\xi^{(s)}_{\psi\phi}\,\propto\, e^{-i\phi_s}$ in the following 
discussion, i.e.\ that the $\mathrm B_s^0\to J/\psi\, \phi$ decay amplitudes 
do not involve a CP-violating weak phase, which implies vanishing direct
CP violation; the question of the hadronic uncertainties 
affecting (\ref{xispsiphi}) is left for further studies. 


\subsubsection{Time Evolution of the Decay Probability 
Functions}\label{subsec:time-evol}
For our considerations, the time evolution of the decay probability 
functions specified in (\ref{obs1}) and (\ref{obs2}) plays a central 
r\^{o}le. In the case of (\ref{obs1}), we obtain (see also \cite{DFN})
\addtolength{\arraycolsep}{-3pt}
\begin{eqnarray}
|A_0(t)|^2&=&\frac{|A_0(0)|^2}{2} \left[\left(1+\cos\phi_s\right)
e^{-\Gamma_{\rm L}^{(s)} t} + \left(1-\cos\phi_s\right)
e^{-\Gamma_{\rm H}^{(s)} t}+2e^{-\Gamma_s t}\sin(\Delta M_s t)
\sin\phi_s\right]~~~~\label{EQB}\\
|A_{\|}(t)|^2&=&\frac{|A_{\|}(0)|^2}{2} \left[\left(1+\cos\phi_s\right)
e^{-\Gamma_{\rm L}^{(s)} t} + \left(1-\cos\phi_s\right)
e^{-\Gamma_{\rm H}^{(s)} t}+
2e^{-\Gamma_s t}\sin(\Delta M_s t)\sin\phi_s\right]~~~~\label{EQB1}\\
|A_{\perp}(t)|^2&=&\frac{|A_{\perp}(0)|^2}{2} \left[\left(1-
\cos\phi_s\right)e^{-\Gamma_{\rm L}^{(s)} t} + \left(1+\cos\phi_s\right)
e^{-\Gamma_{\rm H}^{(s)} t}-
2e^{-\Gamma_s t}\sin(\Delta M_s
t)\sin\phi_s\right]\hspace*{-0.1truecm},
~~~~\label{EQB2}
\end{eqnarray}
whereas we have in the case of the interference terms (\ref{obs2}):
\begin{eqnarray}
\mbox{Re}\{A_0^*(t) A_{\|}(t)\}&=&\frac{1}{2}|A_0(0)||A_{\|}(0)|\cos(\delta_2-
\delta_1)\nonumber\\
&&\times\left[\left(1+\cos\phi_s\right)e^{-\Gamma_{\rm L}^{(s)} t} +
\left(1-\cos\phi_s\right)e^{-\Gamma_{\rm H}^{(s)} t}
 + 2e^{-\Gamma_s t}\sin(\Delta M_s t)\sin\phi_s\right]~~~~\label{EQB3}\\
\mbox{Im}\{A_{\|}^*(t)A_{\perp}(t)\}&=&|A_{\|}(0)||A_{\perp}(0)|\Bigl[
e^{-\Gamma_s t}\left\{\sin\delta_1\cos(\Delta M_s t)-\cos\delta_1
\sin(\Delta M_s t)\cos\phi_s\right\}\nonumber\\
&&-\frac{1}{2}\left(e^{-\Gamma_{\rm H}^{(s)} t}-
e^{-\Gamma_{\rm L}^{(s)} t}\right)\cos\delta_1\sin\phi_s\Bigr],\label{EQB4}\\
\mbox{Im}\{A_{0}^*(t)A_{\perp}(t)\}&=&|A_{0}(0)||A_{\perp}(0)|\Bigl[
e^{-\Gamma_s t}\left\{\sin\delta_2\cos(\Delta M_s t)-
\cos\delta_2\sin(\Delta M_s t)\cos\phi_s\right\}\nonumber\\
&&-\frac{1}{2}\left(e^{-\Gamma_{\rm H}^{(s)} t}-e^{-\Gamma_{\rm L}^{(s)} t}
\right)\cos\delta_2\sin\phi_s\Bigr].\label{EQE}
\end{eqnarray}
Here the CP-conserving strong phases $\delta_1$ and $\delta_2$ are 
defined as follows \cite{ddf1}:
\begin{equation}
\delta_1\equiv\mbox{arg}\Bigl\{A_{\|}(0)^\ast A_{\perp}(0)\Bigr\},\quad
\delta_2\equiv\mbox{arg}\Bigl\{A_{0}(0)^\ast A_{\perp}(0)\Bigr\}.
\end{equation}
The time evolutions (\ref{EQB})--(\ref{EQE}) generalize those given in 
\cite{ddf1} to the case of a sizeable 
$\mathrm B_s^0$--$\overline{\mathrm B_s^0}$ mixing 
phase $\phi_s$, thereby allowing one to include also new-physics effects
\cite{DFN}; an even more generalized formalism, taking into account 
also penguin contributions, can be found in \cite{RF-ang}. It should be 
noted that new physics is expected to manifest itself only in the 
decay probability functions ${\cal O}^{(k)}(t)$ and that the form of the 
$g^{(k)}(\Theta^{'},\Theta^{''},\chi)$ is not affected.

Since the meson content of the $J/\psi\, \phi$ final states is
independent of the flavour of the initial meson, $\mathrm B_s^0$ or 
$\overline{\mathrm B_s^0}$, we may use the same angles $\Theta^{'}$, 
$\Theta^{''}$ and $\chi$ to describe the
kinematics of the decay products of the CP-conjugate transition 
$\overline{\mathrm B_s^0}\to J/\psi\, \phi$. Consequently, we have 
\begin{equation}\label{ang-CP}
\overline{f}(\Theta^{'},\Theta^{''},\chi;t)=\sum_k
\overline{{\cal O}}^{(k)}(t)\,
g^{(k)}(\Theta^{'},\Theta^{''},\chi).
\end{equation}
Within this formalism, CP transformations relating 
$\mathrm B_s^0\to[J/\psi\, \phi]_f$ to $\overline{\mathrm
  B_s^0}\to[J/\psi\, 
\phi]_f$ 
($f\in\{0,\parallel,\perp\}$) are taken into account in the expressions 
for the ${\cal O}^{(k)}(t)$ and $\overline{{\cal O}}^{(k)}(t)$, 
and do not affect the form of the
$g^{(k)}(\Theta^{'},\Theta^{''},\chi)$. 
Therefore 
the same functions $g^{(k)}(\Theta^{'},\Theta^{''},\chi)$ are present
in 
(\ref{ang}) 
and (\ref{ang-CP}) (see also \cite{FD1}). The CP-conjugate functions
$\overline{{\cal O}}^{(k)}(t)$ take the following form:
\begin{eqnarray}
|\overline{A}_0(t)|^2&=&\frac{|A_0(0)|^2}{2} \left[\left(1+\cos\phi_s\right)
e^{-\Gamma_{\rm L}^{(s)} t} + \left(1-\cos\phi_s\right)
e^{-\Gamma_{\rm H}^{(s)} t}-2e^{-\Gamma_s t}\sin(\Delta M_s t)
\sin\phi_s\right]~~~~\label{EQBB1}\\
|\overline{A}_{\|}(t)|^2&=&\frac{|A_{\|}(0)|^2}{2} 
\left[\left(1+\cos\phi_s\right)
e^{-\Gamma_{\rm L}^{(s)} t} + \left(1-\cos\phi_s\right)
e^{-\Gamma_{\rm H}^{(s)} t}-
2e^{-\Gamma_s t}\sin(\Delta M_s t)\sin\phi_s\right]~~~~\label{EQBB2}\\
|\overline{A}_{\perp}(t)|^2&=&\frac{|A_{\perp}(0)|^2}{2} \left[\left(1-
\cos\phi_s\right)e^{-\Gamma_{\rm L}^{(s)} t} + \left(1+\cos\phi_s\right)
e^{-\Gamma_{\rm H}^{(s)} t}+
2e^{-\Gamma_s t}\sin(\Delta M_s
t)\sin\phi_s\right]~~~~\label{EQBB3}\\[-20pt] \nonumber
\end{eqnarray}
\begin{eqnarray}
\mbox{Re}\{\overline{A}_0^*(t) \overline{A}_{\|}(t)\}&=&\frac{1}{2}
|A_0(0)||A_{\|}(0)|\cos(\delta_2-\delta_1)\nonumber\\
&&\times\left[\left(1+\cos\phi_s\right)e^{-\Gamma_{\rm L}^{(s)} t} +
\left(1-\cos\phi_s\right)e^{-\Gamma_{\rm H}^{(s)} t}
 - 2e^{-\Gamma_s t}\sin(\Delta M_s t)\sin\phi_s\right]~~~~\label{EQBB4}\\
\mbox{Im}\{\overline{A}_{\|}^*(t)\overline{A}_{\perp}(t)\}&=&
-|A_{\|}(0)||A_{\perp}(0)|\Bigl[e^{-\Gamma_s t}
\left\{\sin\delta_1\cos(\Delta M_s t)-\cos\delta_1
\sin(\Delta M_s t)\cos\phi_s\right\}\nonumber\\
&&+\frac{1}{2}\left(e^{-\Gamma_{\rm H}^{(s)} t}-
e^{-\Gamma_{\rm L}^{(s)} t}\right)\cos\delta_1\sin\phi_s\Bigr]\label{EQBB5}\\
\mbox{Im}\{\overline{A}_{0}^*(t)\overline{A}_{\perp}(t)\}&=&
-|A_{0}(0)||A_{\perp}(0)|\Bigl[e^{-\Gamma_s t}
\left\{\sin\delta_2\cos(\Delta M_s t)-\cos\delta_2\sin(\Delta M_s t)
\cos\phi_s\right\}\nonumber\\
&&+\frac{1}{2}\left(e^{-\Gamma_{\rm H}^{(s)} t}-e^{-\Gamma_{\rm L}^{(s)} t}
\right)\cos\delta_2\sin\phi_s\Bigr].\label{EQBB6}
\end{eqnarray}
\addtolength{\arraycolsep}{3pt}

\subsubsection{Angular Distributions}

\def \beq{\begin{equation}}
\def \barr{\begin{eqnarray}}
\def \earr{\end{eqnarray}}
\def \eeq{\end{equation}}
\def \bph{{\bf \hat p}}
\def \bbbar{\stackrel{(-)}{B}}
\def \apar{A_{\|}(t)}
\def\aperp{A_{\perp}(t)}

\begin{figure}
\centerline{\includegraphics[width=8.5cm,height=4.6cm]{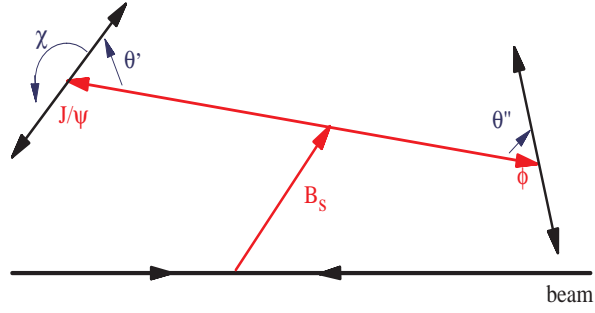}}
\vspace*{-0.5cm}
\caption{Angular conventions for the 
decay $\rm B^0_s \rightarrow J/\psi\, \phi$.}
\label{ang-conv}
\end{figure}

The full angular distribution of \Bstl\
involves three physical angles. The convention used is as follows (see 
Fig.~\ref{ang-conv}): the  $z'(z'')$-axis is defined to
be the direction of $p_{J/\psi} (p_\phi)$ in the rest frame of the 
$B_s^0$. Let the $x'(x'')$-axis be any arbitrarily fixed direction   
in the plane normal to the $z'(z'')$ axis. The $y'(y'')$-axis
is then fixed uniquely. Let $(\Theta^{'},\varphi')$ specify  the direction 
of the $\ell^+$ in the $J/\psi$ rest frame, and let $(\Theta^{''},\varphi'')$
be the direction of the $K^+$ in the $\phi$ rest frame. Since the
orientation of the $x'$ and $x''$ axes is a matter of convention,
only the combination $\chi \equiv \varphi'+\varphi''$ of the
two azimuthal angles is physical. 
The full angular distribution
in terms of the three physical angles $(\Theta^{'},\Theta^{''},\chi)$ 
 (normalized  such that
$\Gamma = |A_{0}(t)|^2 + |A_{\|}(t)|^2 + |A_{\perp}(t)|^2 $)
is given by
\addtolength{\arraycolsep}{-3pt}
\begin{eqnarray}
\lefteqn{W^{+}(\Omega,t)=\frac{d^3 \Gamma}{d\cos\Theta^{'}\,
d\cos\Theta^{''}\,d\chi} = \frac{9}{64 \pi} 
\left\{ 4 |A_{0}(t)|^2 \sin^2 \Theta^{'} \cos^2 \Theta^{''} \right.
} \hspace*{1cm}\nonumber \\
& + & 
        |\apar|^2 [ (1 + \cos^2 \Theta^{'}) \sin^2 \Theta^{''}
        - \sin^2 \Theta^{'} \sin^2 \Theta^{''} \cos 2\chi ] \nonumber\\
        &+& |\aperp|^2  [ (1 + \cos^2 \Theta^{'}) \sin^2 \Theta^{''}
        + \sin^2 \Theta^{'} \sin^2 \Theta^{''} \cos 2\chi ]  \nonumber \\
       & + &
       2 \mbox{Im} (\apar ^* \aperp)
        \sin^2 \Theta^{'} \sin^2 \Theta^{''} \sin 2\chi
        - \sqrt{2} \mbox{Re}(A_0^{*}(t) \apar) \sin 2\Theta^{'} \sin
        2\Theta^{''} \cos \chi \nonumber \\
        & + &\left.  
\sqrt{2} \mbox{Im}(A_0^{*}(t) \aperp) \sin 2\Theta^{'} \sin
2\Theta^{''} 
\sin \chi\right\},\label{w}
\end{eqnarray}

\noindent where the  bilinear combinations of the complex functions 
 $A_{0}(t)$, $A_{\|}(t)$ and  $A_{\perp}(t)$ are defined in 
(\ref{EQB}) to (\ref{EQE}). The angular distribution $W^{-}(\Omega,t)$ 
of the  CP-conjugate transition $\overline{\mathrm B_s^0}\to J/\psi\, \phi$ 
is analogous to (\ref{w}), using the bilinear combinations of 
$\overline{A}_{0}(t)$, $\overline{A}_{\|}(t$) and  
$\overline{A}_{\perp}(t)$ defined in  Eqs.~(\ref{EQBB1}) to (\ref{EQBB6}).


\subsubsection{An Illustration of New-Physics Effects}

As we have already noted, a very important feature of 
$B_s\to J/\psi\,\phi$ decays is that they represent a sensitive probe 
for CP-violating contributions to $B^0_s$--$\overline{B^0_s}$ mixing from 
physics beyond the SM. Let us illustrate these effects 
in this subsection, where we shall follow closely Ref.~\cite{bfNP}, for a 
particular scenario of new physics, the symmetrical 
SU$_{\rm L}$(2)$\,\times\,$SU$_{\rm R}$(2)$\,\times\,$U(1) model with 
spontaneous CP violation (SB--LR) \cite{LR-refs,JMF}. Needless to note
that there are also other scenarios for physics beyond the SM 
 which are interesting in this respect, for example models allowing
mixing to a new isosinglet down quark, as in $\mbox{E}_6$ \cite{sil}. 

In a recent paper \cite{my}, the SB--LR model has been investigated in 
the light of current experimental constraints from $K$- and $B$-decay 
observables. In a large region of parameter space, the model mainly affects 
neutral-meson mixing, but does not introduce sizeable ``direct'' CP 
violation. The sensitive observables constraining the model are thus the 
meson mass difference in the kaon sector 
$\Delta M_K$, those in the B sector $\Delta M_{d}$, $\Delta M_{s}$, 
the ``indirect'' CP-violating parameter $\epsilon_K$ of the neutral kaon 
system, and the mixing-induced CP asymmetry 
${\cal A}^{\rm mix}_{\rm CP}(B_d\to J/\psi\, K_{\rm S})$.
In particular, it was found that, for a set of fixed CKM parameters and 
quark masses, the model predicts a small value for 
$|{\cal A}^{\rm mix}_{\rm CP}(B_d\to J/\psi\, K_{\rm S})|$ below 10\%,
which is in agreement at the 2$\sigma$ level with the CDF measurement
$0.79^{+0.41}_{-0.44}$, but at variance with the SM 
expectation $0.73\pm0.21$ \cite{AL}. 

\begin{figure}
\subfigure[Allowed values for ${\cal A}^{\rm mix}_{\rm CP}
(B_d\to J/\psi K_{\rm S})$ and 
${\cal A}^{\rm mix}_{\rm CP}(B_s\to f)$, with 
$f=D_s^+D^-_s$, $J/\psi\, \eta^{(')}$.]
{\epsfig{file=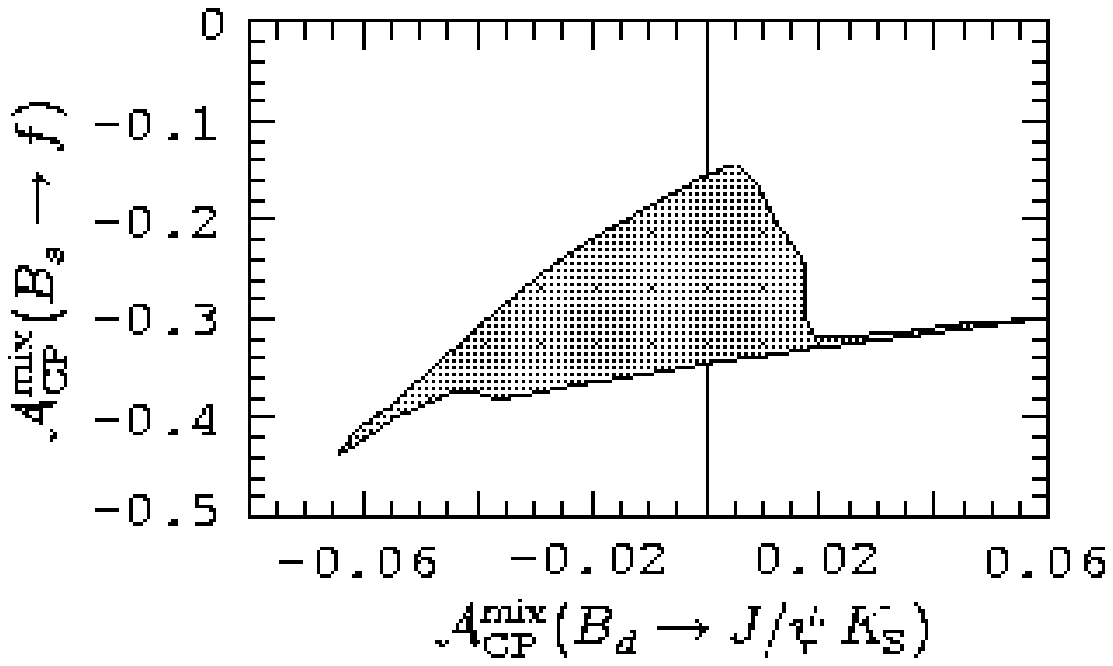,width=0.33\textwidth}}
\subfigure[$A_{\rm CP}(B_s\to J/\psi\phi)$ as a function
of the hadronic parameter $D$.]
{\epsfig{file=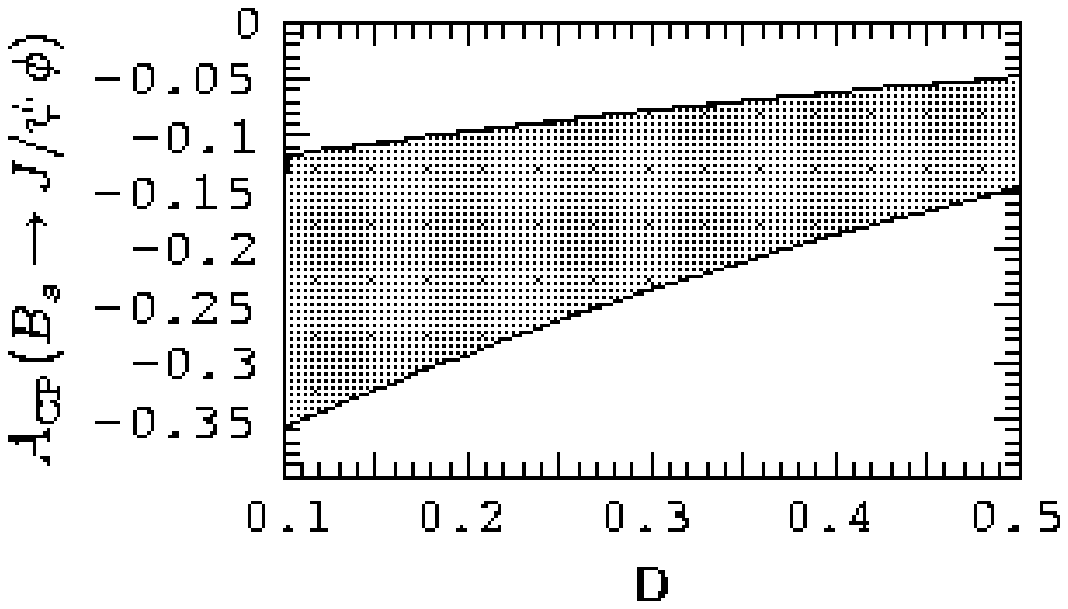,width=0.33\textwidth}}
\subfigure[Correlation between $\Delta M_s$ and $\Delta\Gamma_s$, 
normalized to their SM values.]
{\epsfig{file=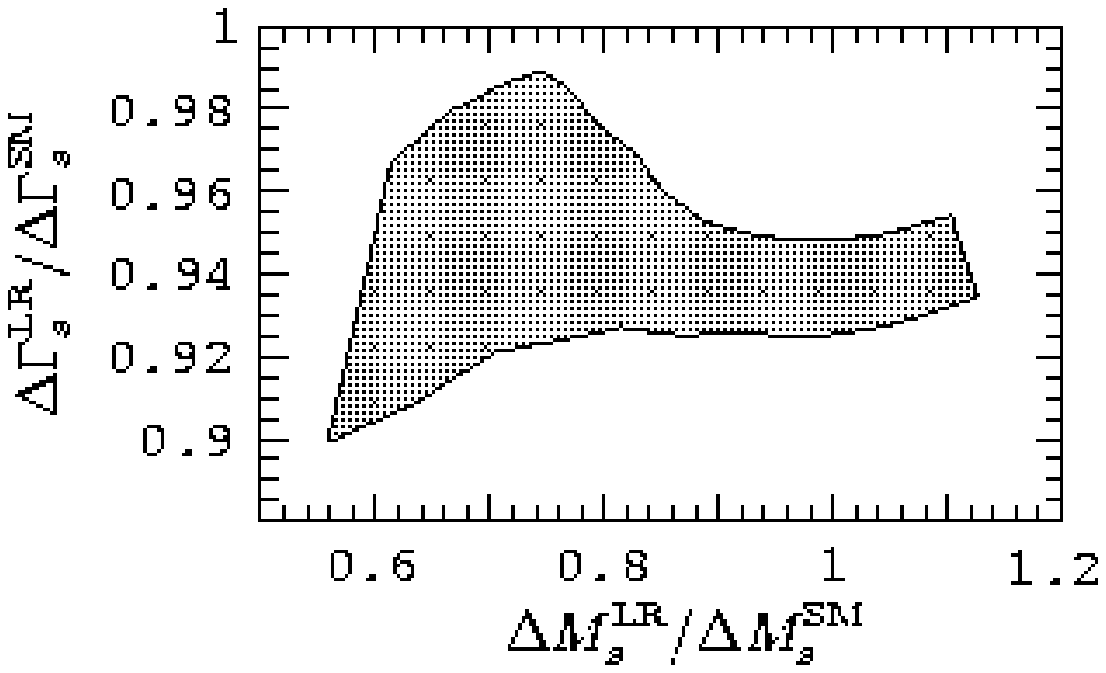,width=0.33\textwidth}}
\vspace*{-0.8cm}
\caption[]{Predictions of the left-right symmetric model for several
  CP observables.}\label{fig:LRSM}
\end{figure}

As was pointed out in \cite{bfNP}, the SB--LR model predicts also values 
for the mixing-induced CP asymmetries of $B_s\to J/\psi\, \phi$ -- and
similar modes, such as $B_s\to D_s^+D^-_s$ and $J/\psi\, \eta^{(')}$ --
that largely deviate from the SM expectation of very small 
CP-violating effects. In the case of the latter modes, which are decays
into CP eigenstates with CP eigenvalue $+1$, we simply have 
\begin{equation}
{\cal A}^{\rm mix}_{\rm CP}(B_s\to f)=\sin\phi_s,~\mbox{where}~
\phi_s\equiv\phi_s^{\rm SM}+\phi_s^{\rm NP}=-2\lambda^2\eta+\phi_s^{\rm NP},
\end{equation} 
with $\phi_s^{\rm NP}$ originating from new physics. In 
Fig.~\ref{fig:LRSM}(a), we show the allowed region for 
${\cal A}^{\rm mix}_{\rm CP}(B_s\to f)=\sin\phi_s$ and 
${\cal A}^{\rm mix}_{\rm CP}(B_d\to J/\psi\, K_{\rm S})$ in the SB--LR model;
the corresponding direct CP asymmetries remain very small, since new 
contributions to the decay amplitudes are strongly 
suppressed. The figure illustrates nicely that CP asymmetries as large as 
${\cal O}(40\%)$ may arise in the $B_s$ channels, whereas the SB--LR model
favours a small CP asymmetry in $B_d\to J/\psi\,K_{\rm S}$. 

In order to simplify the discussion of $B_s\to J/\psi\,\phi$, 
let us consider the CP asymmetry
\begin{equation}\label{CP-asym}
{\cal A}_{\rm CP}(B_s(t)\to J/\psi\,\phi)\equiv\frac{\Gamma(t)-
\overline{\Gamma}(t)}{\Gamma(t)+\overline{\Gamma}(t)}
=\left[\frac{1-D}{F_+(t)+D F_-(t)}\right]
\sin(\Delta M_s t)\,\sin\phi_s,
\end{equation}
where $\Gamma(t)$ and $\overline{\Gamma}(t)$
denote the time-dependent rates for decays of initially, i.e.\ at $t=0$, 
present $B^0_s$- and $\overline{B^0_s}$-mesons into $J/\psi\,\phi$ final 
states, respectively. The remaining quantities are defined as 
\begin{equation}
D\equiv\frac{|A_{\perp}(0)|^2}{|A_0(0)|^2 + |A_{\|}(0)|^2}~\mbox{and}~
F_{\pm}(t)\equiv\frac{1}{2}\left[\left(1\pm\cos\phi_s\right)
e^{+\Delta\Gamma_s t/2}+\left(1\mp\cos\phi_s\right)
e^{-\Delta\Gamma_s t/2}\right].
\end{equation}
Note that we have $F_+(t)=F_-(t)=1$ for a negligible width difference 
$\Delta\Gamma_s$. Obviously, the advantage of the ``integrated'' observable 
(\ref{CP-asym}) is that it can be measured {\it without} performing an
angular analysis. The disadvantage is of course that it also depends on 
the hadronic quantity $D$, which precludes a theoretically clean extraction 
of $\phi_s$ from (\ref{CP-asym}). However, this feature does not limit the 
power of this CP asymmetry to search for indications of new physics, which 
would be provided by a measured sizeable value of (\ref{CP-asym}). Model 
calculations of $D$, making use of the factorization hypothesis, typically 
give $D=0.1\ldots0.5$ \cite{ddf1}, which is also in agreement with a recent 
analysis of the $B_s\to J/\psi\, \phi$ polarization amplitudes performed
by the CDF collaboration \cite{CDF-schmidt}. In order to extract 
$\phi_s$ from CP-violating effects in the decay $B_s\to J/\psi\, \phi$ 
in a theoretically clean way, an angular analysis has to be performed, 
as is discussed in detail above. 

Although the $B^0_s$--$\overline{B^0_s}$ oscillations are very rapid, 
it should be possible to resolve them at the LHC (see Sec.~\ref{sec:mix}). 
The first extremal 
value of the time-dependent CP asymmetry (\ref{CP-asym}), corresponding 
to $\Delta M_st =\pi/2$, is given to a very good approximation by
\begin{equation}\label{ACP-def0}
A_{\rm CP}(B_s\to J/\psi\,\phi)=\left(\frac{1-D}{1+D}\right)\sin\phi_s,
\end{equation}
which would also fix the magnitude of (\ref{CP-asym}) in the case of a 
negligible width difference $\Delta\Gamma_s$. In Fig.~\ref{fig:LRSM}(b), we 
show the prediction of the SB--LR model for (\ref{ACP-def0}) as a function 
of the hadronic parameter $D$. For a value of $D=0.3$, this CP asymmetry 
may be as large as --25\%. The dilution through the hadronic parameter 
$D$ is not effective in the case of the CP-violating observables of the 
$B_s\to J/\psi[\to l^+l^-]\,\phi[\to K^+K^-]$ angular distribution, 
which allow one to probe $\sin\phi_s$ directly (see 
Sec.~\ref{subsec:time-evol}).
Predictions for other $B_s$ decays in the SB--LR model have been discussed
in \cite{quim}.

Let us finally note that new physics affects also the 
$B^0_s$--$\overline{B^0_s}$ mass and width differences. In the latter case,
we have \cite{grossman}
\begin{equation}\label{DGamNP}
\Delta\Gamma_s=\Delta\Gamma_s^{\rm SM}\cos\phi_s,
\end{equation}
where $\Delta\Gamma_s^{\rm SM}={\cal O}(-15\%)$ is the SM 
width difference \cite{DGamma-cal1,DGamma-cal2}. In
Fig.~\ref{fig:LRSM}(c), we show the 
correlation between $\Delta M_s$ and $\Delta\Gamma_s$ in the SB--LR model. 
The reduction of $\Delta\Gamma_s$ through new-physics effects, which is 
described by (\ref{DGamNP}), is fortunately not very effective in this case, 
whereas the mass difference $\Delta M_s$ may be reduced significantly.


\newcommand{\Vubd}{\mbox{$ V_{ub}^{\ast} V_{ud}\ $}}
\newcommand{\Vcbd}{\mbox{$ V_{cb}^{\ast} V_{cd}\ $}}
\newcommand{\AsAo}{\mbox{$ A_{s} A_{p 1/2}^{\ast}\ $}}
\newcommand{\AsAt}{\mbox{$ A_{s} A_{p 3/2}^{\ast}\ $}}
\newcommand{\AtAt}{\mbox{$ A_{p 3/2}^{2}\ $}}
\newcommand{\AoAt}{\mbox{$ A_{p 1/2} A_{p 3/2}^{\ast}\ $}}
\newcommand{\AlAj}{\mbox{$ A_{l} A_{j}^{\ast}\ $}}
\newcommand{\aAlAj}{\mbox{$ \overline A_{l} \overline A_{j}^{\ast}\ $}}
\newcommand{\AsAoAt}{\mbox{$ 2 A_{s}^{2} + 2 A_{p 1/2}^{2} + A_{p 3/2}^{2}\ $}}
\newcommand{\asaoat}{\mbox{$ 2 a_{s}^{2} + 2 a_{p 1/2}^{2} + a_{p 3/2}^{2}\ $}}
\newcommand{\sppd}{\mbox{$s, p_{1/2}, p_{3/2}, d_{3/2}$}}
\newcommand{\jnn}{\mbox{${1 \over N}$}}
\newcommand{\jnd}{\mbox{${1 \over 2}$}}
\newcommand{\jnt}{\mbox{${1 \over 3}$}}
\newcommand{\jnp}{\mbox{${1 \over 5}$}}
\newcommand{\jnsty}{\mbox{${1 \over 4}$}}
\newcommand{\jnde}{\mbox{${1 \over 9}$}}
\newcommand{\jndo}{\mbox{${1 \over 8}$}}
\newcommand{\jnpt}{\mbox{${1 \over 15}$}}
\newcommand{\jnsp}{\mbox{${1 \over 45}$}}
\newcommand{\snstp}{\mbox{${16 \over 135}$}}
\newcommand{\dnstp}{\mbox{${2 \over 135}$}}
\newcommand{\dnsp}{\mbox{${2 \over 45}$}}
\newcommand{\jns}{\mbox{${1 \over 4 \pi}$}}
\newcommand{\jnstt}{\mbox{${1 \over {(4 \pi)}^{2}}$}}
\newcommand{\jnst}{\mbox{${1 \over {(4 \pi)}^{3}}$}}
\newcommand{\tnd}{\mbox{${3 \over 2}$}}
\newcommand{\tnsd}{\mbox{${3 \over\sqrt{2}}$}}
\newcommand{\tnsN}{\mbox{${3 \over\sqrt{N}}$}}
\newcommand{\onde}{\mbox{${1 \over 9}$}}
\newcommand{\dendo}{\mbox{${9 \over 8}$}}
\newcommand{\psubt}{\mbox{$p_{ \mathrm{T}}\ $}}
\newcommand{\psubtt}{\mbox{$p_{ \mathrm{T}} $}}
\newcommand{\tabulka}[3]{
  \begin{table}[htb]
    \begin{center}
      #1
      \end{center}\vspace*{-0.5cm}
    \caption{#2}
    \label{#3}
  \end{table}}

\subsection{Experimental Studies}\label{sec:4.2}
 
The prospective performance of ATLAS, CMS and LHCb in analysing
\Bstoo\ has been studied in 
\cite{ATLASPTDR,ATLASJpsiphi,CMSJpsiphi,LHCbJpsiphi}.
   
\subsubsection{Expected Data Characteristics}

Despite different strategies, all three experiments expect a large number 
 of  events in this channel.
With present studies the highest yield is expected in CMS,
where a dimuon trigger is used.
 At higher trigger-level the  identification of two muons
 is  essential for \Jpsitomm\  
on-line selection in all three experiments. 
 The  reconstruction  is completed 
 in tracking  and vertex detectors  by fitting  muon candidate
 trajectories  
into a common vertex.
 For reconstructing $\phi$ mesons, pairs of oppositely charged particles are
 fitted 
into a common vertex
 and their invariant mass  calculated assuming the kaon hypotheses. In
 the case of 
LHCb, the RICHes are  used to separate charged K mesons from  $\pi$ mesons,
 allowing a reduction of the  backgrounds to \Bst. 
 As explained in Sec.~\ref{exp_id}, 
there is a limited possibility of charged hadron identification in
both ATLAS and CMS;  however this has not been exploited in the
 present studies. The  stronger solenoidal field in CMS  leads to
 better \Bs\ invariant mass resolution and lower 
 \Bst\  background then in ATLAS. 
The most significant difference between the experimental performance for this 
channel lies in the superior proper time resolution of LHCb.
 The  expected characteristics of  data  in the channel  \Bstoo\ and 
 of backgrounds are summarized in Tab.~\ref{table:abab}, under the assumptions
 presented in the workshop.  
 It is possible that the inclusion of low threshold dimuon triggers
 may also boost the final event yields in ATLAS and LHCb, as has been
 demonstrated to be the case in CMS.
  
 Flavour tagging is important to properly explore the physics of
 \Bst\ decays.
 This study considers only   lepton and charged  K mesons tags for LHCb  
 and  a jet charge method for ATLAS and CMS  (see Sec.~\ref{exp_ft}). 
 CMS are presently extending their study to include other tags.
 The efficiencies and the wrong tag  fractions in this channel are
 summarized in  Tab.~\ref{table:abab}. 
  
  The studies presented here  do not exhaust the whole potential of 
  the three experiments. Future studies can  be extended  in 
trigger and off-line selections
  as well as in  tagging methods.
      \tabulka{\begin{tabular}{|c||c|c|c|}
             \hline
                         &    ATLAS        &     CMS  & LHCb               
     \\   \hline\hline
Event yields             &  300,000        & 600,000  &  370,000          
     \\ \hline
Proper time resolution   &  0.063 ps       & 0.063 ps &  0.031 ps         
     \\ \hline
                         &  $\sim\,$ 15\%  & $\sim\,$ 10\%  & $\sim\,$
  3\%  
    \\ 
Background               &  dominated by   &  dominated  by &
  combinatorial  
  \\ 
                         &$\BtoKs$, $\rm J/\psi K^{+}\pi^{-}$ &
  $\BtoKs$, 
$\rm J/\psi K^{+}\pi^{-}$  &              \\ \hline
                         &  jet charge tag &   jet charge  tag &
  lepton tag +
 \\ 
                         &  $\epsilon \sim 63\%$   & $\epsilon \sim
  32\%$ & 
charged K tag \\   
Tagging                  &   wrong $38\%$    &  wrong $33\%$ &
  $\epsilon 
\sim 40\%$        \\ 
                         &           & lepton  tag      &  wrong $30\%$     \\ 
                         &           &  $\epsilon \sim 6.1\%$   &             
     \\
                         &           &  wrong $28\%$    &                    
   \\ \hline
             \end{tabular}}
           {Summary  of performance parameters for \Bstoo. The proper time
            resolutions have been determined by a single Gaussian fit.
            The event yields assume
            3 years operation for ATLAS \& CMS, and 5 years for LHCb.}
           {table:abab}

\subsubsection{Modelling \Bst\ Decays}

The  distribution (\ref{w}) of  the cascade decay \Bst \
contains eight unknown independent parameters. These  are
the  amplitudes $   |A_{||}(0)|$, $   |A_{\perp}(0)|$, the relative
strong phases $\delta_{1}$ and $\delta_{2}$, 
the decay rate difference, \DGdef,  and mean decay rate 
\Gdef \  of the mass eigenstates \Bh and \Bl,
their mass 
difference $\dMss = \xis/\Gamma_{s}$  and the  weak phase $\phi_{s}$.
These parameters  can be determined from
the measured three decay angles and lifetimes.
In the workshop two strategies were studied: the method of moments 
approach~\cite{CMSJpsiphi} and a maximum likelihood fit.

In the method of moments approach \cite{ddf1}, the terms bilinear in $A$ in
  (\ref{w})  are determined from the data using an appropriate set of 
weighting functions, which separate out the terms with different angular
dependences. 
The question of information-content loss in the angular moments analysis 
was investigated in \cite{DIGHESEN}.
 For the results presented in this report,  the likelihood approach is adopted.

We define a likelihood function by
\begin{equation}
L = \prod_{i=1}^{N} { \int_{0}^{\infty} 
\frac{{(\epsilon_{1} W^{+}(\it{t_{i}},\Omega_{i}) + \epsilon_{2} W^{-}
(\it{t_{i}},\Omega_{i}) + b \ \mathrm{e}^{-\Gamma_{0} 
\it{t_{i}}})\rho(\it{t}-\it{t_{i}}) }\,\mathrm{d}\it{t}}
{\int_{t_{\mathrm{min}}}^{\infty} {(\int_{0}^{\infty} {(\epsilon_{1} 
W^{+}(\it{t},\Omega) + \epsilon_{2} W^{-}(\it{t},\Omega) + 
b \ \mathrm{e}^{-\Gamma_{0} \it{t}}) 
\rho(\it{t}^{'}-\it{t})}\,\mathrm{d}\it{t}^{'})}\,\mathrm{d}\it{t}  }}\,,
\label{eq:likeli}
\end{equation}
where $\epsilon_{1}= \epsilon_{2}=0.5$ for untagged events,
 $   \epsilon_{1}= 1-w$, $     \epsilon_{2}=w $ for the case in which 
 the \Bs\ is tagged as a particle, 
 and  $   \epsilon_{1}= w$, $     \epsilon_{2}=1-w $ for the case in which 
 the \Bs\ is tagged as an antiparticle,
  $b$  is the  level  of  background and $   \Gamma_{\mathrm{0}}$ is  the 
   average decay rate of  background as 
    determined from simulation.  The time resolution function
   $   \rho(\it{t}-\it{t_{i}})$ was approximated by 
   a Gaussian of width $ \sigma = 0.063\,$ps and $\sigma = 0.031\,$ps 
for ATLAS/CMS and LHCb respectively.  The index $i$ runs over all $N$
 events. Finally, $t_{\mathrm{min}}$ is the minimum proper 
lifetime allowed in the event selection.
  
\begin{table}
\begin{center}
\begin{tabular}{|c||c|c|c|c|c|c|c|c|}
             \hline
    Parameter &
   $\displaystyle\left|{A_{||}(0) \over A_0(0)}\right|^{2}$ &
   $\displaystyle\left|{A_{\perp}(0) \over  A_0(0)}\right|^{2}$ & 
   $\delta_{1}$   & $\delta_{2}$ &
   \DG & $1/\Gamma_{s}$ &\xis &$\phi_{s}$\\
             \hline
Value\vphantom{$\displaystyle\left|{A_{||}(0) \over
    A_0(0)}\right|^{2}$} 
& 0.64 & 0.14 & 0 &  $\pi$ & 0.15$\, \times \Gamma_{s}$ & 
1.54$\,$ps  &
20--40 & 0.04--0.8\\ 
            \hline
             \end{tabular}
\end{center}\vspace*{-0.5cm}
\caption[]{Input values of theory parameters 
           used in simulating \Bst.}\label{table:input}     
\begin{center}
\begin{tabular}{|c||c|c|c|}
             \hline 
                         & ATLAS & CMS    &     LHCb \\ \hline\hline          
\DG                      &  12\% &   8\%  &    9\%   \\
 $\Gamma_{s}$            & 0.7\% & 0.5\%  &  0.6\%   \\
 $A_{||}$                & 0.8\% & 0.6\%  &  0.7\%   \\
 $A_{\perp}$             &   3\% &   2\%  &    2\%   \\ 
$\phi_{s}$ ($\xis = 20$) &  0.03 & 0.014  &   0.02   \\
$\phi_{s}$ ($\xis = 40$) &  0.05 & 0.03   &   0.03   \\
\hline 
 \end{tabular}
\end{center}\vspace*{-0.5cm}
\caption[]{Expected statistical uncertainties on 
\Bst\ parameters for each experiment under the assumptions given 
in the text. Apart from $\phi_{s}$,  the errors are 
relative.}\label{table:sumary}
\end{table}

\begin{figure}[tbp]
$$\epsfig{file=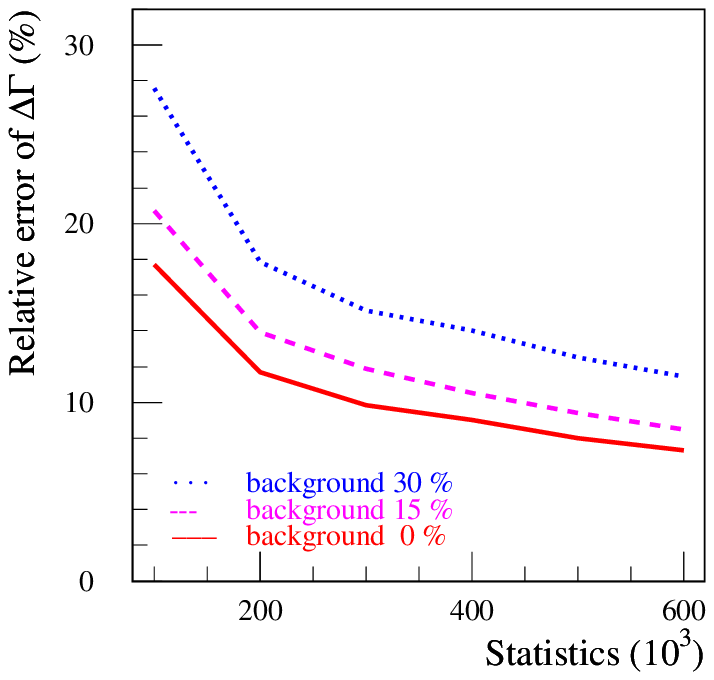,height=6cm,width=6cm} \qquad 
\epsfig{file=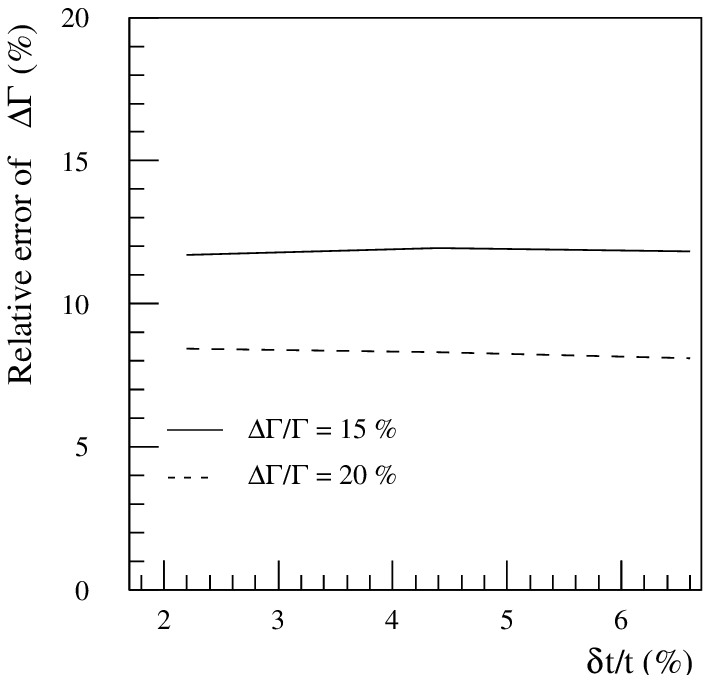,height=6cm,width=6cm}$$
\vspace*{-1cm}
\caption[]{Expected relative error on \DG\ from \Bst. (a) Estimate of
  the relative error of \DG\ as a function of signal 
statistics for several values of background. The expected background is 4
to 15\%. (b) Relative error of \DG\ as a function of relative 
precision of the lifetime measurement $\delta t/t$
 for two values of $\DG/\Gamma_{s}$. LHCb expects $\delta t/t=\,$2.2\%,
 ATLAS/CMS 4.4\%. 
A background of 15\% and statistics of 300,000 events is 
assumed.}\label{fig:nbg}
\vspace*{0.3cm}
\centerline{\epsfxsize=0.55\textwidth\epsffile{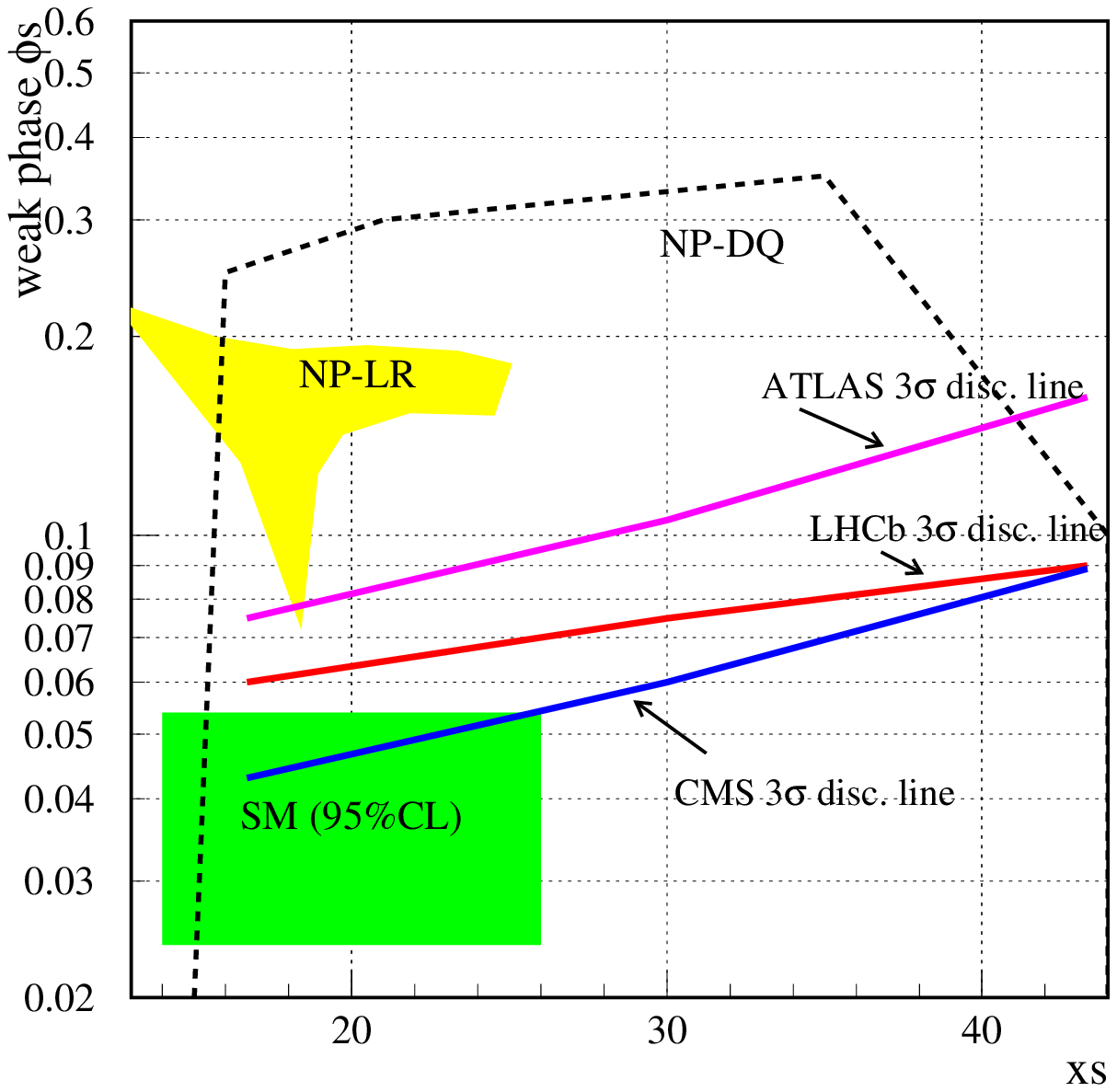}}
\vspace*{-0.8cm}
\caption[]{The $x_s$--$\phi_{s}$  region allowed in the SM, 
  the left-right symmetric model with spontaneous CP violation (NP-LR)
  and the
iso-singlet down quark mixing model (NP-DQ). Also shown is 
the region of  experimental sensitivity of  ATLAS and CMS, corresponding
  to 63 fs, and of LHCb with 31 fs. The NP--LR allowed region appears
  smaller than that of the SM, because it does not include all
  theory uncertainties.}
    \label{fig:conf2}
  \end{figure}
\subsubsection{Parameter Determination and Estimate of Precision}

 The expected experimental precision is not sufficient to allow
 simultaneous determination of eight unknown parameters. 
 Besides the limited statistics 
 there is a problem of the correlations between the parameters.
While in (\ref{w}) the eight parameters are 
independent, simulations with the maximum likelihood approach 
 showed that  in the experimental data 
 some of the parameters  have obvious correlations. 
 There is a strong correlation between 
 the two  relative phases $ \delta_{1}$ and $ \delta_{2}$ 
 which deteriorates a  simultaneous measurement
 of both of them with this method.  There is also a  correlation  
between another pair
 of parameters, $\dMss$   and  
the  weak phase $\phi_{s}$, that  depends on the 
values of $\dMss$ and   time resolution.
Consequently, the reduced set of parameters: \DG, $\Gamma_{s}$, $|A_{||}(0)|$, 
$|A_{\perp}(0)|$ and  $\phi_{s}$ 
 were determined in the fit and the other parameters  were fixed. For
 the strong phases the values
  $ \delta_{1}=0$ and $ \delta_{2}=\pi$ were used as suggested in 
Ref.~\cite{BSW}. For  $\dMss$  it is assumed that 
it can be determined from other channels, for instance \BstoDs,
although it should be stressed that \Bst\ is a very suitable channel for
such a measurement.

The choice of input values of the   unknown parameters, both fixed and free,  
based on the experimental results
 \cite{CLEO,CDF-schmidt,PDG} and theoretical considerations 
\cite{ddf1,BSW,BENEKE}
 is  summarized in   Tab.~\ref{table:input}.

The main results of the study are summarised in
Tab.~\ref{table:sumary} for each experiment.
With this method, the rate difference \DG could be determined 
 with a relative statistical 
 error which for LHCb, CMS and ATLAS  varies between 
 8 to 12\% for  $\DG/\Gamma_{s} = 0.15$, Fig.~\ref{fig:nbg}(a). 
 The   differences between the experiments 
 are small mainly because  the error 
 is not sensitive to the proper time precision differences between
 them, Fig.~\ref{fig:nbg}(b).
The statistical errors of $\Gamma_{s}$, $   |A_{||}(0)|$ and 
 $|A_{\perp}(0)|$ are typically a few percent.
The precision of the measurement of the weak phase $\phi_{s}$ strongly 
depends on the proper time resolution and $\xis$ (Fig.~\ref{fig:conf2}). 
There is sensitivity to the range of $\phi_{s}$ allowed in the SM,
and a clear potential for probing models containing new physics,
such as for instance the left-right 
symmetric model \cite{bfNP} or the isosinglet down quark  model \cite{sil}. 
If penguin contributions are non-negligible, the  number of 
parameters will increase. This will  necessitate   simultaneous analyses
of the \Bst\ and  the SU(3) related channels indicated earlier 
in the theoretical discussion.  
The combined LHC sensitivity to these parameters will be even better,
but this study has not yet been performed.

Studies with the method of moments approach gave results broadly in
agreement with the likelihood fits, but with certain differences 
which are yet to be resolved.  In particular, the moments analysis 
indicated that the
strong phases can be extracted simultaneously with the other parameters
through the separation of different angular terms~\cite{CMSJpsiphi}.
Future work will resolve these issues.

\subsection{Conclusions}

A rich variety of physics can be studied through the decay
\Bst\  and all the LHC experiments will be able to
perform powerful and interesting measurements.
More work is encouraged to extend still further the
potential of the experiments, in particular by
improving the sensitivity to the weak mixing phase  $ \phi_{s}$,
and to establish the optimum approach for analysing the data.

\setcounter{equation}{0}
\section[NEW STRATEGIES TO EXTRACT CKM PHASES]{NEW STRATEGIES TO 
EXTRACT CKM PHASES\protect\footnote{Section 
coordinators: R. Fleischer and G. Wilkinson.}}\label{sec:newstrat}
In addition to the refined studies of the usual benchmark CP modes
described above, an important goal of the workshop was to 
explore stategies for the extraction of CKM phases that had not been 
considered for ATLAS, CMS and LHCb before, and to search for new 
strategies. In this section, we will discuss extractions of $\gamma$ from
$B\to\pi K$ decays, which received a lot of attention in the literature
over the last couple of years \cite{BpiK-revs}, and new techniques 
\cite{new-over,RF-BdsPsiK,RF-ang,RF-BsKK}, which were developed during
this workshop and make use of certain U-spin related $B$ decays, where
all down and strange quarks are interchanged with each other 
\cite{dun-snowmass}. For the ``prehistory'' of the use of U-spin 
arguments to relate nonleptonic $B$ decays, the reader is referred 
to \cite{dun-snowmass}--\cite{pirjol}.

\subsection[Extracting $\gamma$ from 
 $B\to\pi K$ Decays]{Extracting \protect\boldmath
  $\gamma$ from \protect\boldmath $B\to\pi K$ 
Decays\protect\footnote{With help from 
C. Shepherd-Themistocleous.}}\label{subsec:BpiK}

\begin{figure}
\vskip 0.2truein
\begin{center}
\leavevmode
\epsfysize=4.5truecm 
\epsffile{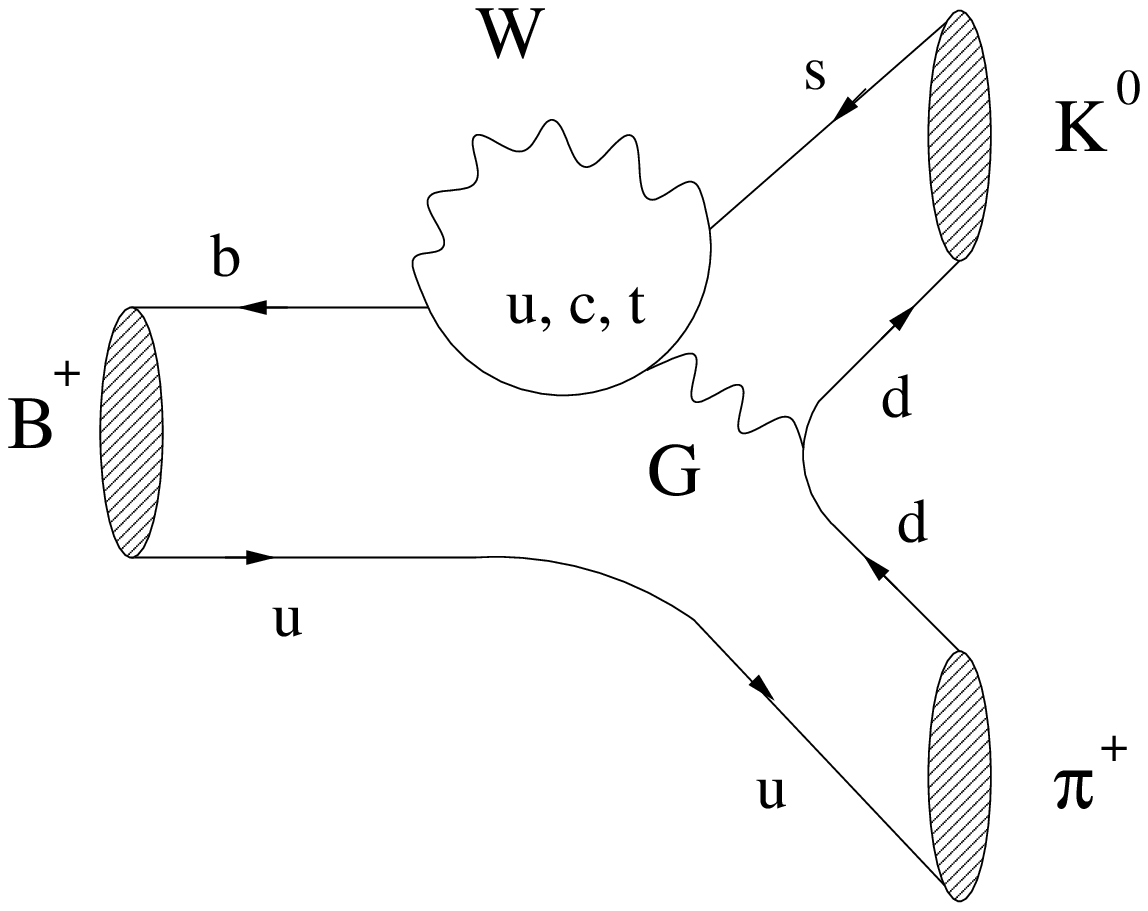} \hspace*{1truecm}
\epsfysize=4.5truecm 
\epsffile{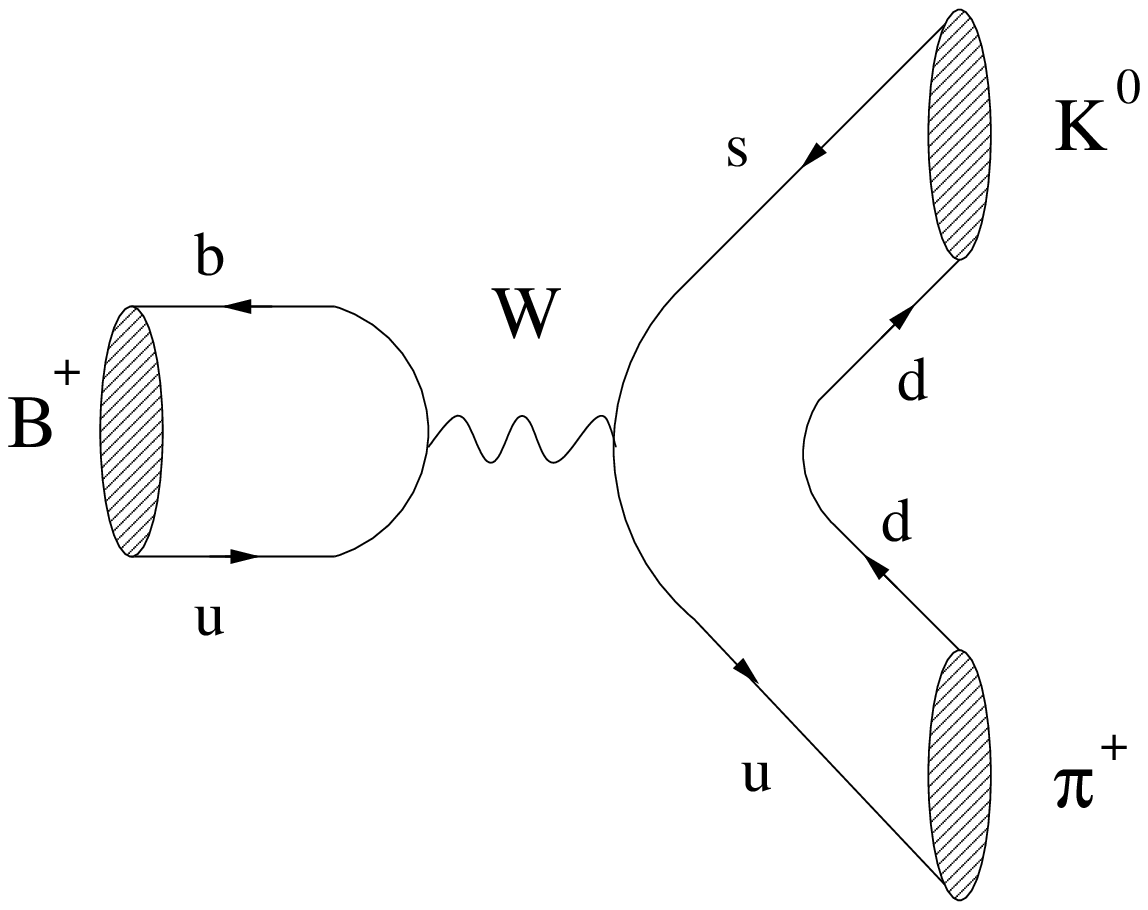}
\end{center}
\vspace*{-0.7truecm}
\caption{Feynman diagrams contributing to 
$B^+\to\pi^+K^0$.}\label{fig:BpiK-charged}
\end{figure}
\begin{figure}
\begin{center}
\leavevmode
\epsfysize=4.5truecm 
\epsffile{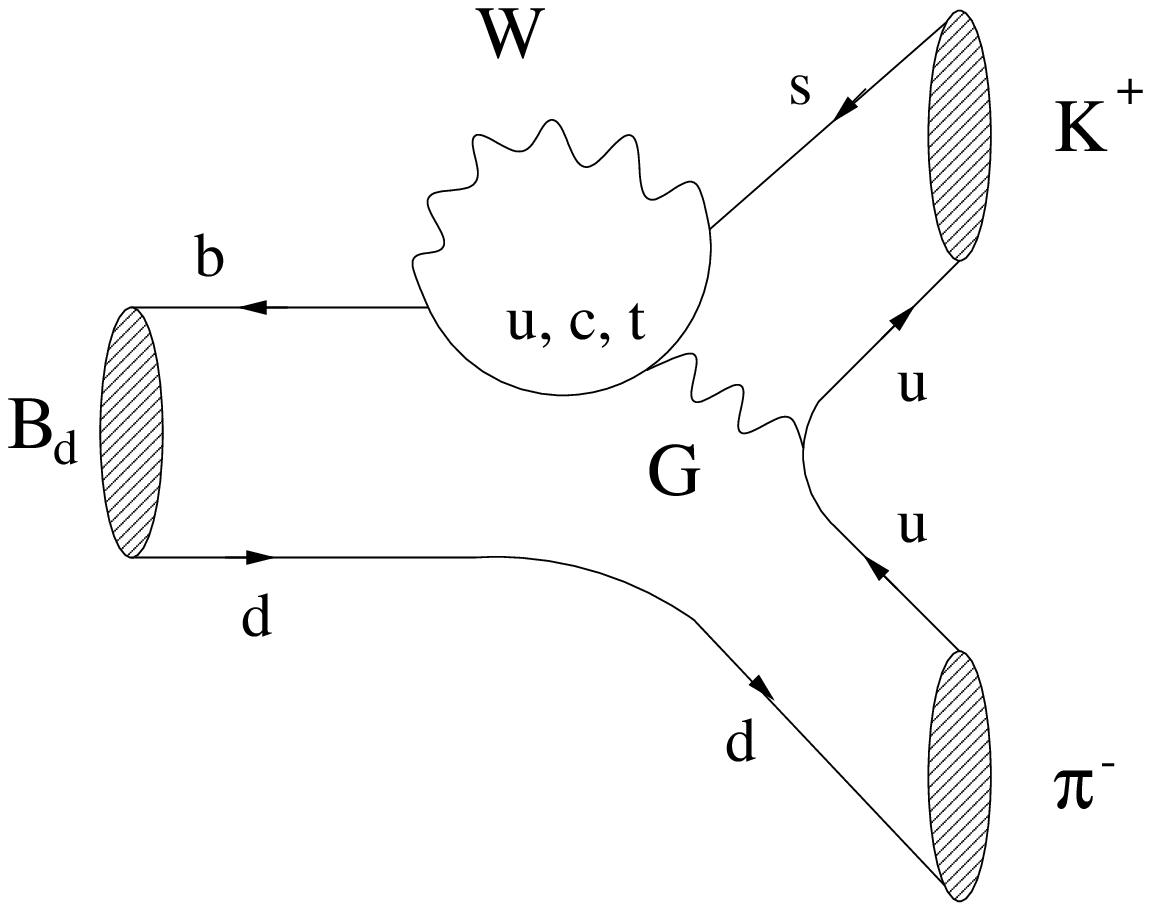} \hspace*{2truecm}
\epsfysize=4truecm 
\epsffile{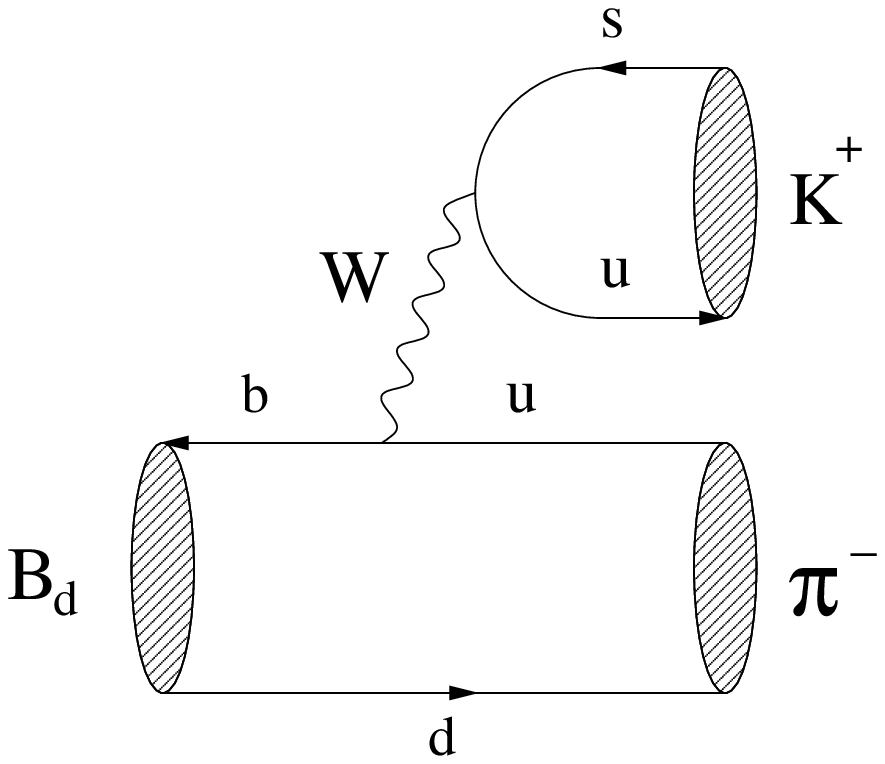}
\end{center}
\vspace*{-0.7truecm}
\caption{Feynman diagrams contributing to 
$B^0_d\to\pi^-K^+$.}\label{fig:BpiK-neutral}
\end{figure}

In order to obtain direct information on $\gamma$, $B\to\pi K$ decays 
are very interesting. These modes are not just an ``unwanted'' background 
for $B\to\pi\pi$, but have a very interesting physics potential. 
Fortunately, experimental data on these modes are now starting to become 
available. Since 1997, when the first results on the decays 
$B^\pm\to\pi^\pm K$ and $B_d\to\pi^\mp K^\pm$ were reported by the CLEO 
collaboration, there were several updated results for CP-averaged $B\to\pi K$
branching ratios at the $10^{-5}$ level  \cite{cleo-BpiK99}. Interestingly, 
 these CP-averaged branching ratios may lead  already to highly non-trivial 
constraints on $\gamma$ \cite{FMbound,NRbound}. Unfortunately, the present 
experimental uncertainties are too large to decide how effective these 
bounds actually are. The new results of the $e^+$--$e^-$ $B$-factories 
will certainly improve this situation, so that we should have a much better 
picture by the start of the LHC. In 1999, also the first preliminary results 
for CP-violating asymmetries in charmless hadronic $B$-meson decays were 
reported by the CLEO collaboration \cite{cleo-BpiK99}, which do not yet
indicate CP violation in such transitions. So far, to
probe $\gamma$, the following three 
combinations of $B\to\pi K$ decays were considered in the literature: 
$B^\pm\to\pi^\pm K$ and $B_d\to\pi^\mp K^\pm$ 
\cite{FMbound,PAPIII,GroRo}, $B^\pm\to\pi^\pm K$ and $B^\pm\to\pi^0 K^\pm$ 
\cite{GRL,NRbound,NR,BF}, as well as the combination of the neutral 
decays $B_d\to\pi^0 K_{\rm S}$ and $B_d\to\pi^\mp K^\pm$ \cite{BF}. 

Since the first combination does not involve a neutral pion, it is 
particularly promising for the LHC from an experimental point of view, 
although the other two combinations would have certain advantages from a 
theoretical point of view. In our experimental feasibility studies, we have 
therefore put a strong emphasis on that approach. Let us note, before having 
a closer look at this strategy, that $B\to\pi K$ decays play not only an 
important r\^{o}le to probe $\gamma$, but also to obtain insights into
the world 
of electroweak penguins. This interesting aspect is discussed in more detail
in \cite{RF-rev,PAPIII,BF,PAPI}.

\subsubsection{The $B^\pm\to\pi^\pm K$, $B_d\to\pi^\mp K^\pm$ Strategy}

Within the framework of the SM, the decays $B^+\to\pi^+K^0$
and $B^0_d\to\pi^-K^+$ receive contributions from Feynman diagrams of the 
type shown in Figs.~\ref{fig:BpiK-charged} and \ref{fig:BpiK-neutral},
respectively. Because of the tiny ratio 
$|V_{us}V_{ub}^\ast|/|V_{ts}V_{tb}^\ast|\approx0.02$, 
the QCD penguins play the dominant r\^{o}le in these decays, despite 
their loop suppression. If we make use of the SU(2) isospin 
symmetry of strong interactions to relate QCD penguin topologies, we 
may derive the following amplitude relations \cite{defan}:
\begin{equation}
A(B^+\to\pi^+K^0)\equiv P\label{rel0},\quad
A(B_d^0\to\pi^-K^+)=-\,\left[P+T+P_{\rm ew}^{\rm C}\right],\label{rel1}
\end{equation}
where 
\begin{equation}\label{T-Pew}
T\equiv|T|e^{i\delta_T}e^{i\gamma} \quad\mbox{and}\quad
P_{\rm ew}^{\rm C}\equiv-\,\left|P_{\rm ew}^{\rm C}\right|
e^{i\delta_{\rm ew}^{\rm C}}
\end{equation}
are due to tree-diagram-like topologies and EW penguins, respectively. 
The label ``C'' reminds us that only ``colour-suppressed''
EW penguin topologies contribute to $P_{\rm ew}^{\rm C}$. Making use of 
the unitarity of the CKM matrix and applying the Wolfenstein parametrization,
generalized to include non-leading terms in $\lambda$ \cite{BLO}, 
we obtain \cite{defan}
\begin{equation}
A(B^+\to\pi^+K^0)=-\left(1-\frac{\lambda^2}{2}\right)\lambda^2A\left[
1+\rho\,e^{i\theta}e^{i\gamma}\right]{\cal P}_{tc}\,,
\end{equation}
where
\begin{equation}
\rho\,e^{i\theta}=\left(\frac{\lambda^2 R_b}{1-\lambda^2}\right)
\left[1-\left(\frac{{\cal P}_{uc}+{\cal A}}{{\cal P}_{tc}}\right)\right].
\end{equation}
Here ${\cal P}_{tc}\equiv|{\cal P}_{tc}|e^{i\delta_{tc}}$ and ${\cal P}_{uc}$
describe differences of penguin topologies with internal top- and charm-quark
and up- and charm-quark exchanges, respectively, and ${\cal A}$ is due to 
the annihilation topology in Fig.~\ref{fig:BpiK-charged}. It is important 
to note that $\rho$ is strongly CKM-suppressed by $\lambda^2R_b
\approx0.02$. For the parametrization of $B^\pm\to \pi^\pm K$ and 
$B_d\to\pi^\mp K^\pm$ observables, it is convenient to introduce 
\begin{equation}
r\equiv\frac{|T|}{\sqrt{\langle|P|^2\rangle}}\,,\quad\epsilon_{\rm C}\equiv
\frac{|P_{\rm ew}^{\rm C}|}{\sqrt{\langle|P|^2\rangle}}
~~\mbox{with}~~ 
\langle|P|^2\rangle\equiv\frac{1}{2}\left(|P|^2+|\overline{P}|^2\right), 
\end{equation}
as well as the strong phase differences
\begin{equation}
\delta\equiv\delta_T-\delta_{tc}\,,\quad\Delta_{\rm C}\equiv
\delta_{\rm ew}^{\rm C}-\delta_{tc}\,.
\end{equation}
In addition to the ratio 
\begin{equation}\label{Def-R}
R\equiv\frac{B(B^0_d\to\pi^-K^+)+
B(\overline{B^0_d}\to\pi^+K^-)}{B(B^+\to\pi^+K^0)
+B(B^-\to\pi^-\overline{K^0})}
\label{eq:defofr}
\end{equation}
of CP-averaged branching ratios, also the ``pseudo-asymmetry'' 
\begin{equation}
A_0\equiv\frac{B(B^0_d\to\pi^-K^+)-B(\overline{B^0_d}\to
\pi^+K^-)}{B(B^+\to\pi^+K^0)+B(B^-\to\pi^-\overline{K^0})}
\label{eq:defofa0}
\end{equation}
plays an important r\^{o}le in probing $\gamma$. Here, we have neglected
tiny phase-space effects, which can be taken into account straightforwardly
(see \cite{FMbound}). Explicit expressions for $R$ and $A_0$ in terms of 
the parameters specified above are given in \cite{defan}. Using the 
presently available experimental results from the CLEO collaboration 
\cite{cleo-BpiK99}, we obtain
\begin{equation}\label{RFM-exp}
R=1.0\pm0.3,\quad A_0=0.04\pm0.18.
\end{equation}

\begin{figure}
\centerline{
\rotate[r]{
\epsfxsize=6.3truecm
{\epsffile{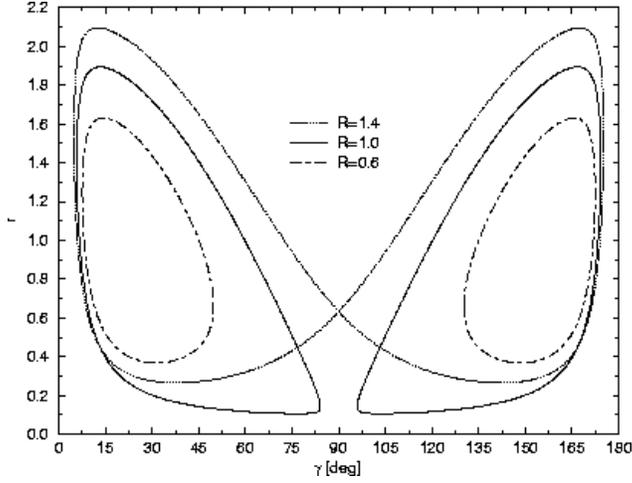}}}}
\vspace*{-0.5cm}
\caption[]{The contours in the $\gamma$--$r$ plane for $|A_0|=0.2$ 
($\rho=\epsilon_{\rm C}=0$).}\label{fig:g-r-cont}
\label{fig:a0contours}
\end{figure}

\begin{figure}
\centerline{
\rotate[r]{
\epsfxsize=3truecm
\epsffile{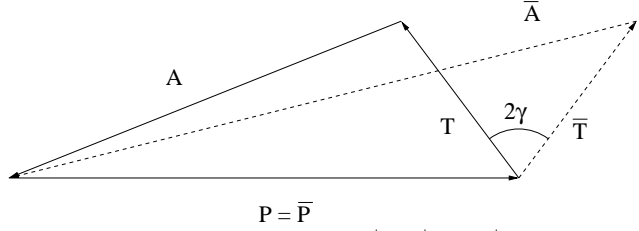}}}
\vspace*{-0.5cm}
\caption{Triangle construction to determine $\gamma$ from the
$B_d\to\pi^\mp K^\pm$, $B^\pm\to\pi^\pm K$ system in the case of 
$\rho=\epsilon_{\rm C}=0$. Here we have $A\equiv A(B^0_d\to\pi^-K^+)$
and $\overline{A}\equiv A(\overline{B^0_d}\to\pi^+K^-)$; note that 
$\rho=0$ implies $P=\overline{P}\equiv 
A(B^-\to\pi^-\overline{K^0})$.}\label{fig:BpiK-triang}
\end{figure}

The pseudo-asymmetry $A_0$ allows one to eliminate the strong phase $\delta$ 
in the expression for $R$, and to fix contours in the $\gamma\,$--$\,r$ 
plane \cite{defan}. These contours, which are illustrated in 
Fig.~\ref{fig:g-r-cont}, correspond to the mathematical implementation 
of a simple triangle construction \cite{PAPIII}, which is related to the
amplitude relation (\ref{rel1}), and is shown in Fig.~\ref{fig:BpiK-triang}. 
In order to determine $\gamma$, the quantity $r$, i.e.\ the magnitude of 
the ``tree'' amplitude $T$, has to be fixed. At this stage, a certain model 
dependence enters. An approximate way to fix this amplitude is to neglect 
``colour-suppressed'' current--current operator contributions to 
$B^+\to\pi^+\pi^0$, and to use  SU(3) flavour symmetry to relate the 
``colour-allowed'' current--current amplitude of that decay to $T$:
\begin{equation}
|T|\approx\lambda\frac{f_K}{f_\pi}\sqrt{2}|A(B^+\to\pi^+\pi^0)|\,.
\end{equation}
Another approach to obtain information on $|T|$ is to use 
``factorization'' \cite{facto}, leading to 
\begin{equation}\label{T-fact}
|T|_{{\rm fact}}=
\frac{G_{\rm F}}{\sqrt{2}}\,\lambda\,|V_{ub}|\,a_1\,\left(M_{B_d}^2-
M_\pi^2\right)\,f_{K}\,F_{B\pi}(M_K^2;0^+)\,,
\end{equation}
where $F_{B\pi}$ is a quark--current form factor and $a_1\approx1$ the usual
phenomenological colour factor. Using the form factor 
$F_{B\pi}(M_K^2;0^+)=0.3$, as obtained e.g.\ from QCD sum rules on the
light-cone \cite{bpiff,ball98}, yields
\begin{equation}
|T|_{{\rm fact}}=a_1\times\left[\frac{|V_{ub}|}{3.2\times
10^{-3}}\right]\times7.8\times10^{-9}\,\mbox{GeV}\,.
\end{equation}
As was pointed out in \cite{GroRo}, also semileptonic $B^0\to \pi^-l^+\nu_l$ 
decays may play an important r\^{o}le to fix $|T|$ with the help of arguments 
based on ``factorization''. Using (\ref{T-fact}), one finds \cite{FMbound}
\begin{equation}
r_{{\rm fact}}=0.18\times a_1\times
\left[\frac{|V_{ub}|}{3.2\times10^{-3}}\right]
\sqrt{\left[\frac{1.8\times10^{-5}}{B(B^\pm\to\pi^\pm K)}\right]
\times\left[\frac{\tau_{B_u}}{1.6\,\mbox{ps}}\right]}.
\end{equation}
Making use of such arguments based on ``factorization'', present data give
$r=0.18\pm0.05$. Although the factorization hypothesis \cite{facto}
may work reasonably well for ``colour-allowed'' tree-diagram-like topologies
\cite{bjorken}, $T$ may be shifted from its ``factorized'' value, as the 
properly defined amplitude $T$ does not only receive contributions from 
such ``tree'' topologies, but also from penguin and annihilation processes 
\cite{defan,bfm}, which are strongly related to rescattering processes 
\cite{bfm,FSI,neubert}. In an interesting recent paper by Beneke, Buchalla,
Neubert and Sachrajda \cite{BBNS}, it was pointed out that there is a 
heavy-quark expansion for nonleptonic $B$ decays into two light mesons, 
and that non-factorizable corrections, as well as rescattering processes, 
are suppressed by $\Lambda_{\rm QCD}/m_b$. This approach may turn out to 
be useful to fix the parameter $r$, which is required in order to determine 
$\gamma$ from $B_d\to\pi^\mp K^\pm$, $B^\pm\to\pi^\pm K$ decays.

Interestingly, it is possible to derive bounds on $\gamma$ that do {\it not}
depend on $r$ at all \cite{FMbound}. To this end, we eliminate again $\delta$ 
in $R$ through $A_0$. If we now treat $r$ as a ``free'' variable, we find 
that $R$ takes the minimal value \cite{defan} 
\begin{equation}\label{Rmin}
R_{\rm min}=\kappa\,\sin^2\gamma\,+\,
\frac{1}{\kappa}\left(\frac{A_0}{2\,\sin\gamma}\right)^2\geq
\kappa\,\sin^2\gamma,
\end{equation}
where
\begin{equation}\label{kappa-def}
\kappa=\frac{1}{w^2}\left[\,1+2\,(\epsilon_{\rm C}\,w)\cos\Delta_C+
(\epsilon_{\rm C}\,w)^2\,\right]~\mbox{with}~ 
w=\sqrt{1+2\,\rho\,\cos\theta\cos\gamma+\rho^2}. 
\end{equation}
The inequality in (\ref{Rmin}) arises if we keep both $r$ and $\delta$ as 
free parameters \cite{FMbound}. An allowed range for $\gamma$ is related 
to $R_{\rm min}$, since values of $\gamma$ implying $R_{\rm exp}<R_{\rm min}$ 
are excluded. In particular, $A_0\not=0$ would allow one to exclude a certain 
range of $\gamma$ around $0^\circ$ or $180^\circ$, whereas a measured value 
of $R<1$ would exclude a certain range around $90^\circ$, which would be of 
great phenomenological importance. The first results reported by 
CLEO in 1997 gave $R=0.65\pm0.40$ and led to great excitement, whereas the 
most recent update is the one given in (\ref{RFM-exp}). If the parameter $r$
is fixed, significantly stronger constraints on $\gamma$ can be obtained from 
the observable $R$ \cite{BF,GPY}. In particular, these constraints require 
only $R\not=1$ and are also effective for $R>1$.

\begin{figure}
\centerline{
\epsfysize=3.7truecm
{\epsffile{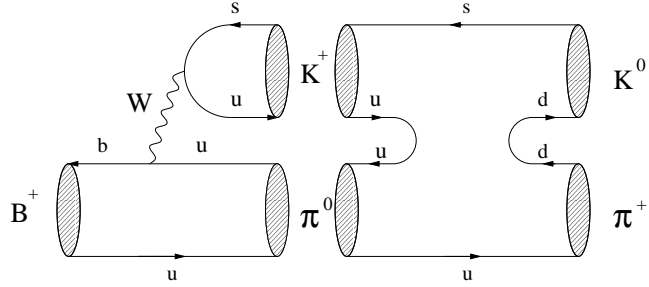}}}
\vspace*{-0.4cm}
\caption[]{Rescattering process contributing to 
$B^+\to\pi^+K^0$.}\label{fig:res}
\end{figure}

The theoretical accuracy of the strategies to probe $\gamma$ through the 
$B^\pm\to\pi^\pm K$, $B_d\to\pi^\mp K^\pm$ system is limited both 
by rescattering processes of the kind 
$B^+\to\{\pi^0K^+,\pi^0K^{\ast+},\ldots\}\to\pi^+K^0$ \cite{FSI,neubert}, 
which are illustrated in Fig.~\ref{fig:res}, and by the ``colour-suppressed'' 
EW penguin contributions described by the amplitude $P_{\rm ew}^{\rm C}$
\cite{GroRo,neubert}. In (\ref{Rmin}), these effects are described by the 
parameter $\kappa$. If they are neglected, we have $\kappa=1$. The 
rescattering effects -- it cannot be excluded that they may lead to values
of $\rho$ as large as ${\cal O}(0.1)$ -- can be controlled in the contours 
in the $\gamma$--$r$ plane and the constraints on $\gamma$ related to
(\ref{Rmin}) through experimental data on $B^\pm\to K^\pm K$ decays, 
which are the U-spin counterparts of $B^\pm\to\pi^\pm K$ \cite{defan,BKK}. 
Another important indicator for large rescattering effects are the
$B_d\to K^+K^-$ modes, for which there already exist stronger experimental 
constraints \cite{groro-FSI}.

An improved description of the EW penguins is possible if we use the 
general expressions for the corresponding four-quark operators
and perform appropriate Fierz transformations  \cite{defan,PAPIII,neubert}. 
Following these lines, we obtain
\begin{equation}\label{EWP-expr1}
q_{\rm C}\,e^{i\omega_{\rm C}}\equiv\frac{\epsilon_{\rm C}}{r}\,
e^{i(\Delta_{\rm C}-\delta)}=0.66\times \left[\frac{0.41}{R_b}\right]
\times a_{\rm C}\,e^{i\omega_{\rm C}},
\end{equation}
where 
$a_{\rm C}\,e^{i\omega_{\rm C}}=a_2^{\mbox{\scriptsize eff}}/
a_1^{\mbox{\scriptsize eff}}$ is the
ratio of certain generalized ``colour factors''. Experimental data on 
$B\to D^{(\ast)}\pi$ decays imply $a_2/a_1={\cal O}(0.25)$. A first step 
to fix the hadronic parameter $a_{\rm C}\,e^{i\omega_{\rm C}}$ experimentally 
is provided by the mode $B^+\to\pi^+\pi^0$ \cite{defan}; interesting 
constraints were derived in \cite{GPY}. For a detailed discussion of the 
impact of rescattering and EW penguin effects on the strategies to probe 
$\gamma$ with $B^\pm\to\pi^\pm K$ and $B_d\to\pi^\mp K^\pm$ decays, the 
reader is referred to \cite{defan,BKK,BF,Neubert-BpiK}. In order to 
control these hadronic uncertainties -- in addition to the full experimental 
picture of all $B\to\pi K$, $K\overline{K}$ decays -- also the theoretical 
approach to deal with nonleptonic $B$ decays into two light mesons 
developed recently in Ref.~\cite{BBNS} may play an important r\^{o}le.

\subsubsection{The Charged $B^\pm\to \pi^\pm K$, $B^\pm\to\pi^0K^\pm$
Strategy}

Several years ago, Gronau, Rosner and London proposed an SU(3) strategy 
to determine $\gamma$ from the charged decays 
$B^{\pm}\to\pi^{\pm} K$, $\pi^0K^{\pm}$, $\pi^0\pi^{\pm}$ \cite{GRL}. 
However, as was pointed out by Deshpande and He \cite{dh}, this elegant 
approach is unfortunately spoiled by EW penguins \cite{GHLR-EWP}, 
which play an important r\^{o}le in several nonleptonic $B$-meson decays 
because of the large top-quark mass \cite{RF-EWP1,RF-EWP2}. 
Recently, this approach 
was resurrected by Neubert and Rosner \cite{NRbound,NR}, who pointed out 
that the EW penguin contributions can be controlled in this case by using 
only the general expressions for the corresponding four-quark operators, 
appropriate Fierz transformations, and the SU(3) flavour symmetry 
of strong interactions (see also \cite{PAPIII}). 

In the case of $B^+\to\pi^+K^0$, $\pi^0K^+$, SU(2) isospin 
symmetry implies
\begin{equation}\label{charged-iso}
A(B^+\to\pi^+K^0)\,+\,\sqrt{2}\,A(B^+\to\pi^0K^+)=
-\left[(T+C)\,+\,P_{\rm ew}\right].
\end{equation}
The phase structure of this relation is completely analogous to 
$B^+\to\pi^+K^0$, $B^0_d\to\pi^-K^+$, as can be seen by
comparing with (\ref{rel0}) and (\ref{T-Pew}):
\begin{equation}
T+C=|T+C|\,e^{i\delta_{T+C}}\,e^{i\gamma},\quad
P_{\rm ew}=-\,|P_{\rm ew}|e^{i\delta_{\rm ew}}\,.
\end{equation}
In order to probe $\gamma$, it is useful to introduce the following
observables \cite{BF}:
\begin{eqnarray}
R_{\rm c}&\equiv&2\left[\frac{B(B^+\to\pi^0K^+)+
B(B^-\to\pi^0K^-)}{B(B^+\to\pi^+K^0)+
B(B^-\to\pi^-\overline{K^0})}\right],\label{Rc-def}\\
A_0^{\rm c}&\equiv&2\left[\frac{B(B^+\to\pi^0K^+)-
B(B^-\to\pi^0K^-)}{B(B^+\to\pi^+K^0)+
B(B^-\to\pi^-\overline{K^0})}\right],\label{A0c-def}
\end{eqnarray}
which correspond to $R$ and $A_0$; general expressions can be 
obtained from those for $R$ and $A_0$ with the following replacements:
\begin{equation}
r\to r_{\rm c}\equiv\frac{|T+C|}{\sqrt{\langle|P|^2\rangle}}\,, \quad
\delta\to \delta_{\rm c}\equiv\delta_{T+C}-\delta_{tc}\,,\quad
P_{\rm ew}^{\rm C}\to P_{\rm ew}.
\end{equation}
Using the presently available experimental results from the CLEO 
collaboration \cite{cleo-BpiK99}, one finds
\begin{equation}\label{RBF-exp}
R_{\rm c}=1.3\pm0.5,\quad A_0^{\rm c}=0.35\pm0.34.
\end{equation}
The observables $R_{\rm c}$ and $A_0^{\rm c}$ allow one to fix 
contours in the $\gamma$--$r_c$ plane, in complete analogy to the
$B^\pm\to\pi^\pm K$, $B_d\to\pi^\mp K^\pm$ strategy. However, the
charged $B\to\pi K$ approach has certain advantages from a theoretical 
point of view:
\begin{itemize}
\item  SU(3) flavour symmetry allows one to fix the parameter 
$r_c\propto|T+C|$ as follows \cite{GRL}:
\begin{equation}\label{SU3-rel1}
T+C\approx-\,\sqrt{2}\,\frac{V_{us}}{V_{ud}}\,
\frac{f_K}{f_{\pi}}\,A(B^+\to\pi^+\pi^0)\,,
\end{equation}
where $r_c$ thus determined is -- in contrast to $r$ -- not affected 
by rescattering effects; present data give $r_c=0.21\pm0.06$. The factor
$f_K/f_{\pi}$ takes into account factorizable SU(3) breaking. 
\item In the strict SU(3) limit, we have \cite{NRbound}
\begin{equation}\label{SU3-rel2}
q\,e^{i\omega}\equiv\left|\frac{P_{\rm ew}}{T+C}\right|\,
e^{i(\delta_{\rm ew}-\delta_{T+C})}=0.66\times
\left[\frac{0.41}{R_b}\right],
\end{equation}
which does -- in contrast to (\ref{EWP-expr1}) -- not involve a
hadronic parameter. Taking into account factorizable SU(3) breaking 
and using present data gives $q=0.63\pm0.15$.
\end{itemize}
The contours in the $\gamma$--$r_c$ plane may be affected -- in analogy 
to the $B^\pm\to\pi^\pm K$, $B_d\to\pi^\mp K^\pm$ case -- by rescattering 
effects \cite{BF}. They can be taken into account with the help of 
additional experimental data \cite{defan,BKK,FSI-recent}, and if we 
use the observable
\begin{equation}
B_0^{\rm c}\equiv A_0^{\rm c}-\left[\frac{B(B^+\to\pi^
+K^0)-B(B^-\to\pi^-\overline{K^0})}{B(B^+\to\pi^+K^0)+
B(B^-\to\pi^-\overline{K^0})}\right]
\end{equation}
instead of $A_0^{\rm c}$, the terms of ${\cal O}(\rho)$, which describe 
the rescattering effects, are suppressed by $r_{\rm c}$ \cite{Neubert-BpiK}.
The major theoretical advantage of the $B^+\to\pi^+K^0$, $\pi^0K^+$ strategy 
with respect to $B^\pm\to\pi^\pm K$, $B_d\to\pi^\mp K^\pm$ is that $r_c$ and 
$P_{\rm ew}/(T+C)$ can be fixed by using {\it only} SU(3) arguments, i.e.\
 no additional dynamical arguments have to be employed.  
Consequently, the theoretical accuracy is mainly limited by non-factorizable 
SU(3) breaking effects. The approach developed recently in \cite{BBNS}
may help to reduce these uncertainties. 

Let us finally note that the observable $R_{\rm c}$ may also imply 
interesting constraints on $\gamma$ \cite{NRbound}. These bounds, which 
are conceptually quite similar to \cite{FMbound}, are related to the 
extremal values of $R_{\rm c}$ that arise if we keep only the strong phase 
$\delta_{\rm c}$ as an ``unknown'' free parameter. As the resulting general 
expression is rather complicated \cite{BF,Neubert-BpiK}, let us expand it 
in $r_c$ \cite{NRbound}. If we keep only the leading-order terms and make 
use of the SU(3) relation (\ref{SU3-rel2}), we obtain
\begin{equation}\label{Rc-expansion}
\left.R_c^{\rm ext}\right|_{\delta_c}^{\rm LO}=
1\,\pm\,2\,r_c\,|\cos\gamma-q|.
\end{equation}
Interestingly, there are no terms of ${\cal O}(\rho)$ present in this
expression, i.e.\ rescattering effects do not enter at this level 
\cite{NRbound,NR}. However, final-state-interaction processes may still 
have a sizeable impact on the associated bounds on $\gamma$. Several 
strategies to control these uncertainties were considered in the recent 
literature \cite{BF,Neubert-BpiK,FSI-recent}, and also the approach of
Ref.~\cite{BBNS} may shed light on these issues. 

Unfortunately, the neutral pions appearing in $B^\pm\to\pi^0K^\pm$
make the charged approach challenging  experimentally. The
strategy using the neutral decays $B_d\to \pi^0 K_{\rm S}$ and 
$B_d\to\pi^\mp K^\pm$ to extract $\gamma$, which was proposed in \cite{BF},
is even worse in this respect, and we will not discuss it here in more 
detail, although it would have an interesting theoretical advantage 
concerning the impact of rescattering effects.

\begin{figure}
\begin{minipage}[t]{0.33\textwidth}
   \epsfxsize=\textwidth
   \centerline{\epsffile{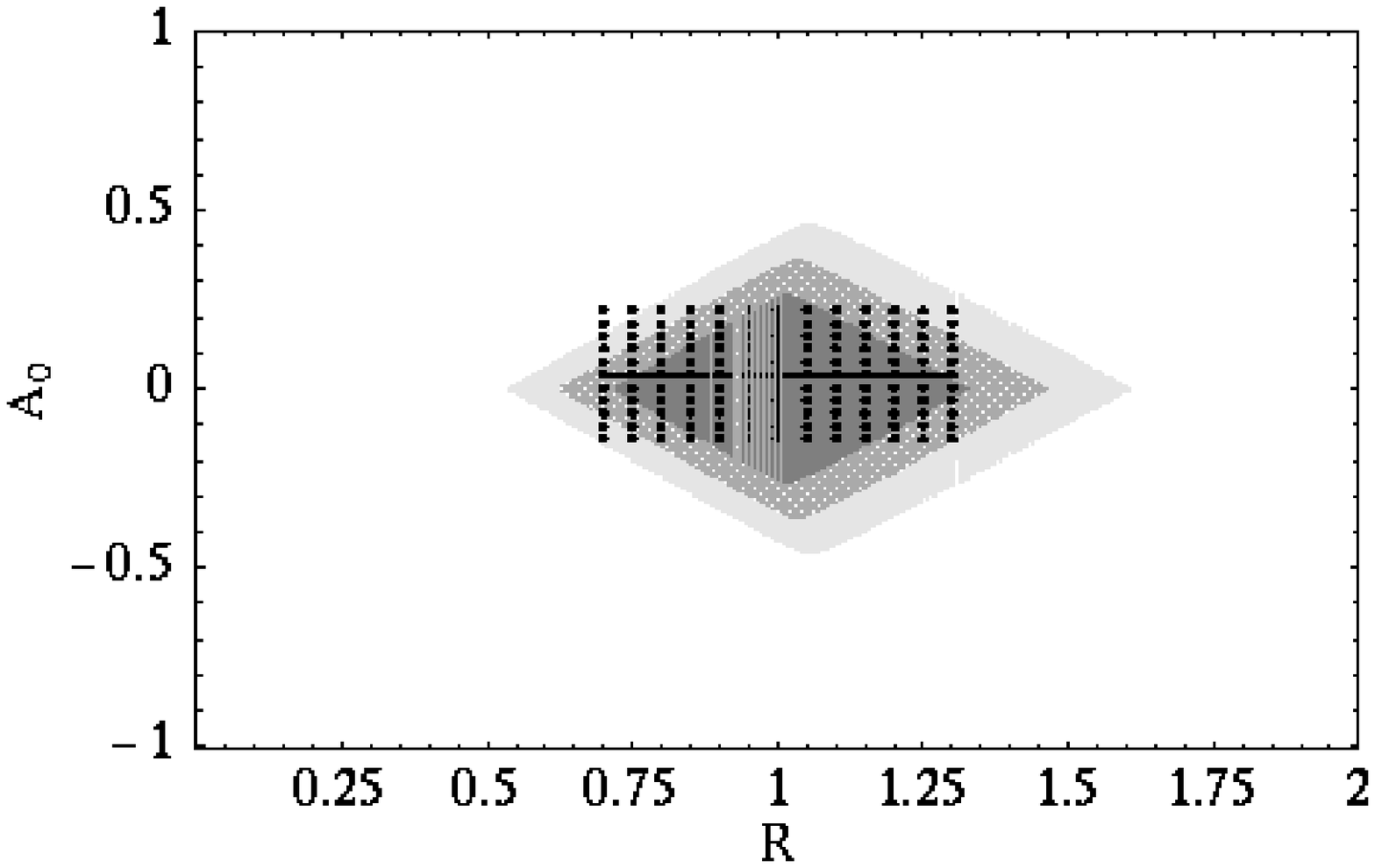}}
\vspace*{-0.5cm}
\caption[dummy]{Allowed region in the $R$--$A_0$ plane, 
characterizing $B^\pm\to\pi^\pm K$, $B_d\to\pi^\mp K^\pm$
in the SM. $0.13\leq r\leq0.23$, 
$q_{\rm C}=0.17$. FSI are neglected.
\label{fig:BpiK-mix-cont}}
\end{minipage}
\hspace*{0.3cm}
\begin{minipage}[t]{0.63\textwidth}
\centerline{
\subfigure[$0.15\leq r_{\rm c}\leq0.27$, $q=0.63$]
{\epsfxsize=0.5\textwidth\epsffile{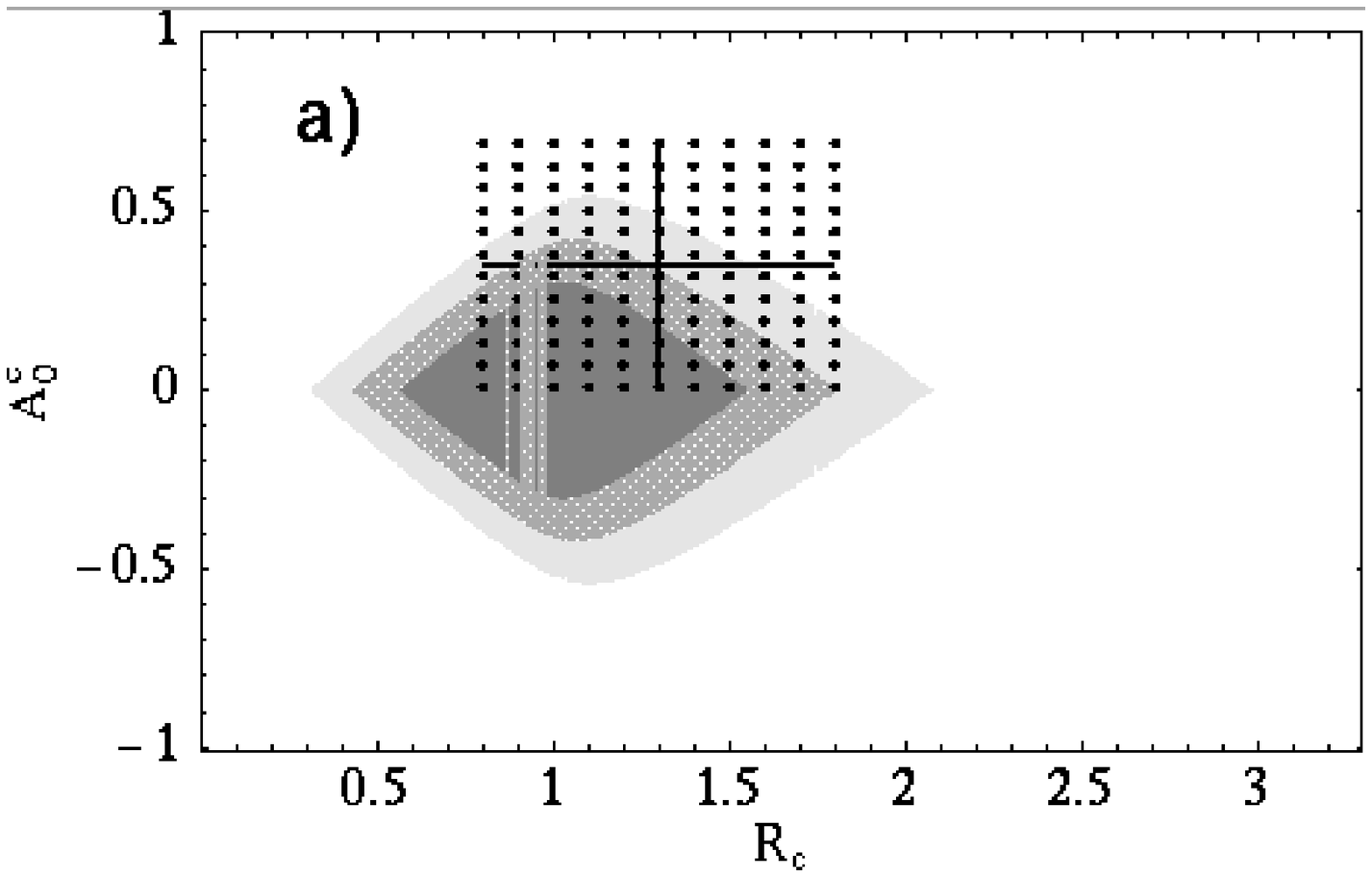}}
\subfigure[$r_{\rm c}=0.21$, $0.48\leq q\leq0.78$]
{\epsfxsize=0.5\textwidth\epsffile{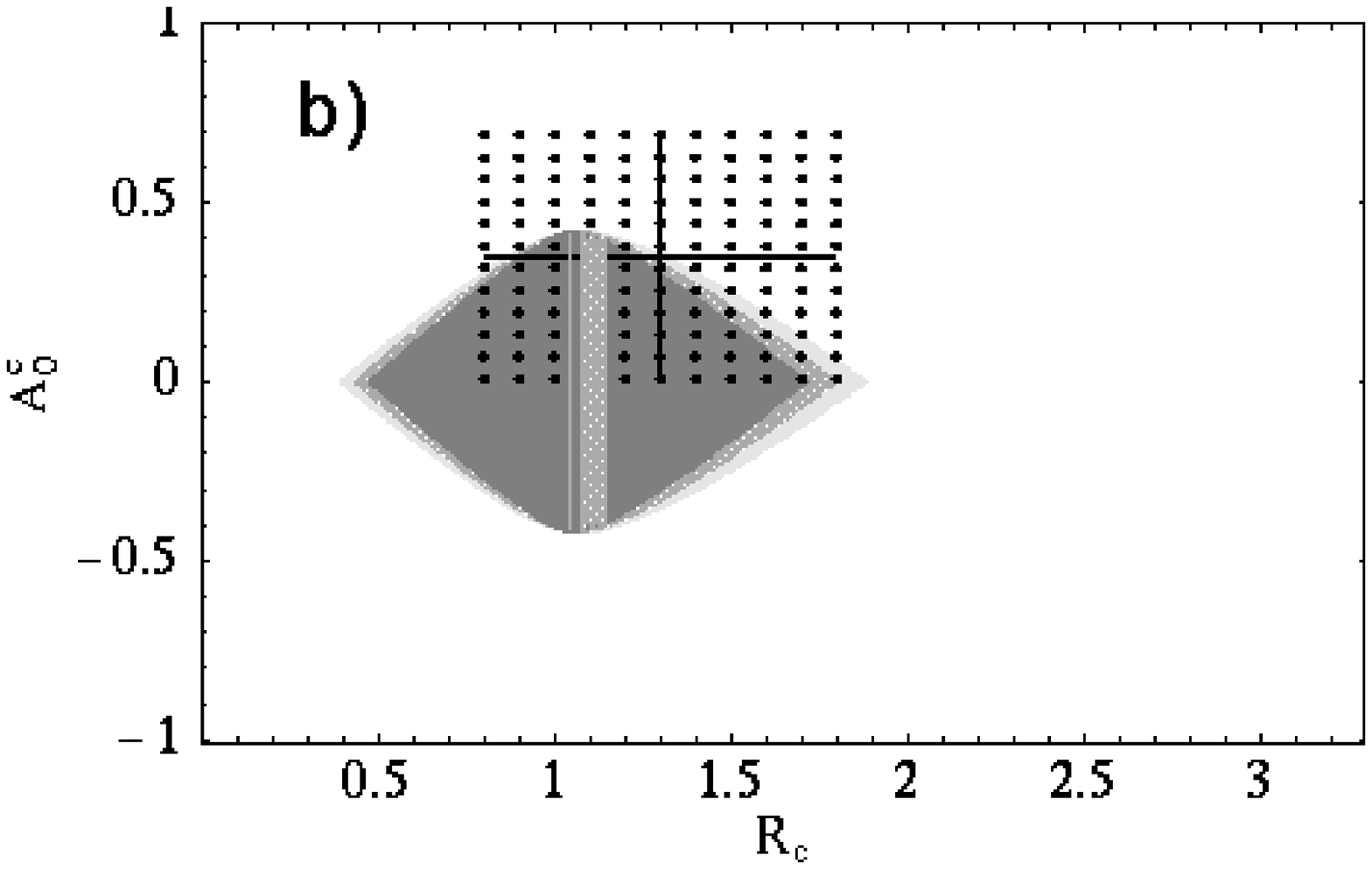}}
}
\vspace*{-0.5cm}
\caption[dummy]{Allowed region in the $R_{\rm c}$--$A_0^{\rm c}$ 
plane, characterizing $B^\pm\to\pi^\pm K$, $\pi^0K^\pm$ 
in the SM. FSI are neglected.
\label{fig:BpiK-char-cont}}
\end{minipage}
\end{figure}

\subsubsection{Some Remarks about New Physics}

Since $B^0_q$--$\overline{B^0_q}$ mixing ($q\in\{d,s\}$) is a ``rare'' 
flavour-changing neutral-current (FCNC) process, it is very likely that 
it is significantly affected by new physics, which may act upon the mixing 
parameters $\Delta M_q$ and $\Delta\Gamma_q$ as well as on the CP-violating
mixing phase $\phi_q$. Important examples for such scenarios of
new physics are non-minimal SUSY models, left--right-symmetric models,
models with extended Higgs sectors, four generations, or $Z$-mediated 
\mbox{FCNCs}~\cite{new-phys}. Since $B_d\to J/\psi\,K_{\rm S}$ and 
$B_s\to J/\psi\,\phi$ -- the benchmark modes to measure $\phi_d$ and 
$\phi_s$ -- are governed by current--current, i.e.\ ``tree'', processes, 
new physics is expected to affect their {\it decay amplitudes} in a 
minor way. Consequently, these modes still measure $\phi_d$ and $\phi_s$. 

In the clean strategies to measure $\gamma$ with the help of pure ``tree''
decays, such as $B\to DK$, $B_d\to D^{(\ast)\pm}\pi^\mp$ or 
$B_s\to D_s^\pm K^\mp$, new physics is also expected to play a very minor 
r\^{o}le. These strategies therefore provide a ``reference'' value for 
$\gamma$. 
Since, on the other hand, the $B\to\pi K$ strategies to determine $\gamma$ 
rely on the interference between tree and penguin contributions, 
discrepancies with the ``reference'' value for $\gamma$ may well show up 
in the presence of new physics \cite{BpiK-NP,GNK}. If we are lucky, we may 
even get immediate indications for new physics from $B\to\pi K$ decays 
\cite{FMat}, as the SM predicts interesting correlations 
between the corresponding observables that are shown in 
Figs.~\ref{fig:BpiK-mix-cont} and \ref{fig:BpiK-char-cont}. Here the 
dotted regions correspond to the CLEO results that were reported in 
1999 \cite{cleo-BpiK99}.

If future measurements should give results lying significantly 
outside the allowed regions shown in these figures, we would have an 
indication for new physics. On the other hand, if we should find values 
lying inside these regions, this would not automatically imply a 
confirmation of the SM. In this case, we would be in 
a position to extract a value for $\gamma$ by following the strategies 
described above, which may well lead to discrepancies with the ``reference'' 
values for $\gamma$ that are implied by the theoretically clean ``tree''
strategies, or with the usual fits of the unitarity triangle. In a 
recent paper \cite{GNK}, several specific models were employed to explore 
the impact of new physics on $B\to\pi K$ decays. For example, in models 
with an extra $Z'$ boson or in SUSY models with broken $R$-parity, the 
resulting electroweak penguin coefficients can be much larger than in the 
SM, since they arise already at tree level.

Interestingly, the present experimental range coincides perfectly 
with the SM region in Fig.~\ref{fig:BpiK-mix-cont}. 
This feature should be compared with the situation in 
Fig.~\ref{fig:BpiK-char-cont}. Unfortunately, the present experimental 
uncertainties are too large to speculate on new-physics effects. However, 
the experimental situation should improve considerably in the  
years before the start of the LHC. The strategies discussed in the 
following subsections are also well suited to search for new physics. 

\subsubsection{Experimental Studies}
\label{sec:bdpik}
\newcommand{\kpi}{\mbox{$\rm K^+\pi^-$}}
\newcommand{\kopi}{\mbox{$\rm K^0\pi^+$}}

Preliminary studies for the determination of $\gamma$ using the $\rm
K\pi$ decay modes of B mesons have been performed for the LHCb
experiment. As explained above, 
$\gamma$ may be determined using a number of strategies
that involve the final states \kpi, \kopi, $\rm K^+\pi^0$ and $\rm
K^0\pi^0$. Experimentally it is easiest to
reconstruct final states which contain charged particles and have
reconstructible decay vertices.  Clearly, therefore, the strategy
involving \kpi\ and \kopi\ final states provides the cleanest
experimental channel and this has been studied initially. 
Future work will involve a
study of the feasibility of reconstructing the $\rm K^+\pi^0$ mode. A
clean reconstruction of the $\rm K^0\pi^0$ mode is unlikely to be
possible at LHCb.

The experimental values that must be determined are the 
ratios $R$ and $A$ given in (\ref{eq:defofr}) and (\ref{eq:defofa0}).
which contain different final states in numerator and 
denominator. This means that the ratio 
of trigger and reconstruction efficiencies 
must be known for these final states. This is in contrast to 
most CP violation measurements where these quantities cancel 
and will be an additional source of systematic error which has yet to 
be investigated.

The principal features of the $\rm K\pi$ decays used for
reconstruction are well separated vertices and large impact
parameters. The particle identification provided by the RICH detectors
is vital for the \kpi\ mode and very helpful in the \kopi\ case. The
overall trigger efficiencies for the two channels are similar at
$\sim 0.3$,  where this value is defined relative 
to events decaying in the acceptance.
The net trigger and reconstruction efficiency is about 0.02 for the \kpi\
channel and 0.01 for \kopi.  The difference is mainly due to the
detector acceptance. Assuming
the latest CLEO branching ratio measurements of $(18.2 \pm 5) \times
10^{-6}$ for \kopi\ and $(18.8 \pm 3) 
\times 10^{-6}$ for \kpi~\cite{cleo-Bpipi},
results in about 90,000 events in the \kopi\
and 175,000 events in the \kpi\ channel per year. These numbers are 
rather preliminary since the background studies are still in an early stage,
and it may prove necessary to tighten the reconstruction cuts.

\begin{figure}[tbh]
\centerline{\epsfxsize=0.7\textwidth\epsffile{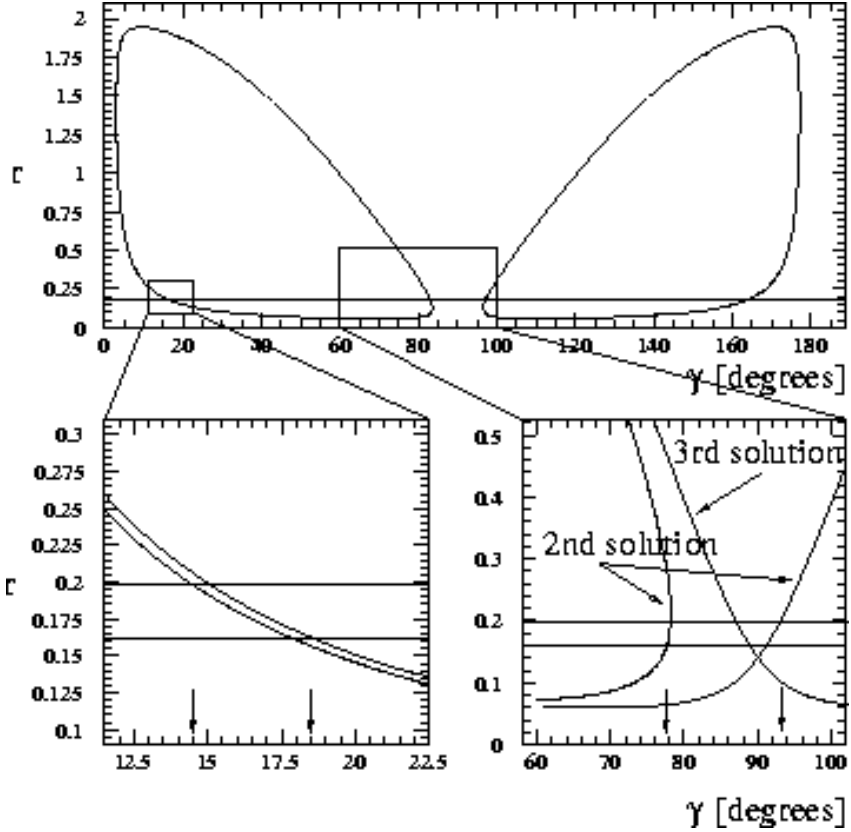}}
\caption[dummy]{\small An experimental study of the contours in the
$\gamma$--$r$ plane (see also Fig.\ \protect{\ref{fig:g-r-cont}}): 
the top plot corresponds to 
$R=1.0$, $A_0=0.1$ and $r=0.18$; in the blow-up plots, the bands indicate the
spread from correlated errors on $R$ and $A_0$ of 3\%,  and
$\pm\,$10\% on $r$; arrows
indicate the allowed range for $\gamma$. The
error on $\gamma$ from the first solution is $\pm 2^\circ$; the second
solution yields $77^\circ < \gamma < 93^\circ$. Note that for
our specific choice of input parameters the allowed band for the second
solution partially overlaps with that of the third one, starting at
$88^\circ$.
}\label{fig:bpkexp}
\end{figure}

Translating these numbers into final CP sensitivities is however not
trivial. The measured values of the ratios $R$ and $A$ define contours
in the $r$--$\gamma$ plane such as those in Fig.~\ref{fig:a0contours}.  
A value
for $\gamma$ can only be extracted once $r$ is known. This must be
determined theoretically. Experimental results indicating
large rescattering effects which would imply large errors in $r$ are, for
example, large CP violation in the $\rm B^+ \rightarrow \kopi$ channel
or larger than expected branching ratios for $\rm B^+ \rightarrow
K^+K^0$ and $\rm B^0 \rightarrow K^+K^-$.  The precise value for $r$
will have a large effect on the errors expected. There is also a four-fold
ambiguity for the value of $\gamma$. 
Figure~\ref{fig:bpkexp} illustrates the errors that might be expected 
assuming a value for $r$ of $0.18\pm 10\%$, for two of the possible 
solutions. For one of these solutions the error is $\sim\,$2$^\circ$,  
whereas for the other the error is $\sim\,$7$^\circ$.  
These uncertainties are mirrored in the remaining two  solutions.

In summary, from this preliminary study, it is expected that LHCb will
be able to provide determinations of the ratios $A$ and $R$ for the
strategy involving \kopi\ and \kpi\ final states with errors 
of the order of 3\%. As explained above this cannot simply be
translated into a CP sensitivity.  Work on these
promising decays is still under way  and will be extended to
include a study of the $\rm K^+\pi^0$ channel.

\subsection[Extracting  $\gamma$ from 
 $B_{s(d)}\to J/\psi\,K_{\rm S}$ Decays]{Extracting 
\protect\boldmath $\gamma$ from \protect\boldmath 
$B_{s(d)}\to J/\psi\,K_{\rm S}$ Decays\protect\footnote{With help from 
P. Colrain.}}\label{subsec:BsdPsiKS}

As we have already discussed in Sec.~\ref{subsec:BdpsiKS}, the 
``gold-plated'' mode $B_d\to J/\psi\, K_{\rm S}$ plays an outstanding 
r\^{o}le in the determination of the $B^0_d$--$\overline{B^0_d}$ mixing phase 
$\phi_d$, i.e.\ of the CKM angle $\beta$. In this subsection, we will have 
a closer look at the decay $B_s\to J/\psi\, K_{\rm S}$ \cite{RF-BdsPsiK}
(see also \cite{dun-snowmass}), which is related to 
$B_d\to J/\psi\, K_{\rm S}$ by interchanging all down 
and strange quarks (see Fig.~\ref{fig:BdPsiKS}), and may allow 
an interesting extraction of the CKM angle $\gamma$. 

\subsubsection{Theoretical Aspects}\label{subsec:BsdJpsiKs}

In analogy to (\ref{Bd-ampl2}), the $B_s\to J/\psi\, K_{\rm S}$ decay 
amplitude can be expressed as follows: 
\begin{equation}\label{Bs-ampl}
A(B_s^0\to J/\psi\, K_{\rm S})=-\lambda\,{\cal A}\left[1-a\, e^{i\theta}
e^{i\gamma}\right],
\end{equation}
where
\begin{equation}
{\cal A}\equiv\lambda^2A\left(A_{\rm cc}^{c}+A_{\rm pen}^{ct}\right)
~\mbox{and}~
a\, e^{i\theta}\equiv R_b\left(\frac{A_{\rm pen}^{ut}}{A_{\rm cc}^{c}+
A_{\rm pen}^{ct}}\right)\label{a-def}
\end{equation}
correspond to (\ref{Aap-def}). It should
be emphasized that (\ref{Bd-ampl2}) and (\ref{Bs-ampl}) rely only on 
the unitarity of the CKM matrix. In particular, these SM
parametrizations of the $B_{d(s)}^0\to J/\psi\, K_{\rm S}$ decay 
amplitudes also take into account final-state-interaction effects, which 
can be considered as long-distance penguin topologies with internal up- 
and charm-quark exchanges \cite{bfm}.

Comparing (\ref{Bd-ampl2}) with (\ref{Bs-ampl}), we
observe that the ``penguin parameter'' $a' e^{i\theta'}$ is doubly 
Cabibbo-suppressed in the $B_d^0\to J/\psi\, K_{\rm S}$ decay amplitude 
(\ref{Bd-ampl2}), whereas $a\, e^{i\theta}$ enters (\ref{Bs-ampl}) in a 
Cabibbo-allowed way. Consequently, there may be sizeable CP-violating 
effects in $B_s\to J/\psi\, K_{\rm S}$, which provide {\it two} independent 
observables, ${\cal A}_{\rm CP}^{\rm dir}(B_s\to J/\psi\, K_{\rm S})$ and 
${\cal A}_{\rm CP}^{\rm mix}(B_s\to J/\psi\, K_{\rm S})$, depending on the 
{\it three} ``unknowns'' $a$, $\theta$ and $\gamma$, as well as on the 
$B^0_s$--$\overline{B^0_s}$ mixing phase $\phi_s$. Consequently, in order 
to determine these ``unknowns'', we need an additional observable, which 
is provided by
\begin{equation}\label{H-def}
H\equiv\left(\frac{1-\lambda^2}{\lambda^2}\right)
\left(\frac{|{\cal A}'|}{|{\cal A}|}\right)^2
\frac{\langle\Gamma(B_s\to J/\psi\, K_{\rm S})\rangle}{\langle\Gamma
(B_d\to J/\psi\, K_{\rm S})\rangle},
\end{equation}
where the CP-averaged decay rates
$\langle\Gamma(B_s\to J/\psi\, K_{\rm S})\rangle$ and 
$\langle\Gamma(B_d\to J/\psi\, K_{\rm S})\rangle$ can be determined from 
the ``untagged'' rates introduced in (\ref{untag-def}) through
\begin{equation}\label{aver-rate}
\langle\Gamma(B_q\to f)\rangle\equiv\frac{\Gamma_q[f(0)]}{2}.
\end{equation}
In (\ref{H-def}), we have neglected tiny phase-space effects, which can
be included straightforwardly \cite{RF-BdsPsiK}.

Since the U-spin flavour symmetry of strong interactions implies
\begin{equation}\label{SU3-1}
|{\cal A}'|=|{\cal A}|
~\mbox{and}~
a'=a,\quad \theta'=\theta,
\end{equation}
we can determine $a$, $\theta$ and $\gamma$ as a function of the
$B_s^0$--$\overline{B_s^0}$ mixing phase $\phi_s$ by combining $H$ with 
${\cal A}_{\rm CP}^{\rm dir}(B_s\to J/\psi\, K_{\rm S})$ and
${\cal A}_{\rm CP}^{\rm mix}(B_s\to J/\psi\, K_{\rm S})$ or
${\cal A}_{\Delta\Gamma}(B_s\to J/\psi\, K_{\rm S})$. In contrast to 
certain isospin relations, electroweak penguins do not lead to any problems 
in these U-spin relations. As we have already noted, the 
$B^0_s$--$\overline{B^0_s}$ mixing phase $\phi_s=-2\delta\gamma$ is expected 
to be negligible in the SM. It can be probed with the help 
of $B_s\to J/\psi\,\phi$, Sec.~\ref{sec:Bspsiphi}. 
Strictly speaking, in the case of $B_s\to J/\psi\, K_{\rm S}$, we have 
$\phi_s\to-2\delta\gamma-\phi_K$, where $\phi_K$ is related to the 
$K^0$--$\overline{K^0}$ mixing phase and is negligible in the 
SM (see also the comment in Sec.~\ref{subsec:BdpsiKS}). 
Since the value of the CP-violating parameter $\varepsilon_K$ of the neutral 
kaon system is very small, $\phi_K$ can only be affected by very contrived 
models of new physics \cite{nir-sil}. 

Interestingly, the strategy to extract $\gamma$ from $B_{s(d)}\to J/\psi\, 
K_{\rm S}$ does not require a non-trivial CP-conserving strong phase 
$\theta$. However, its experimental feasibility depends strongly on the 
value of the quantity $a$ introduced in (\ref{a-def}). It is very difficult 
to estimate $a$ theoretically. In contrast to the ``usual'' QCD penguin 
topologies, the QCD penguins contributing to $B_{s(d)}\to J/\psi\, K_{\rm S}$
require a colour-singlet exchange, as indicated in Fig.~\ref{fig:BdPsiKS}
through the dashed lines, and are ``Zweig-suppressed''. Such a comment does 
not apply to the electroweak penguins, which contribute in ``colour-allowed'' 
form. The current--current amplitude $A_{\rm cc}^c$ is due to 
``colour-suppressed'' topologies, and the ratio 
$A_{\rm pen}^{ut}/(A_{\rm cc}^{c}+A_{\rm pen}^{ct})$, which governs $a$, 
may be sizeable. It is interesting to note that the measured branching 
ratio $B(B^0_d\to J/\psi\, K^0)=
2\,B(B_d^0\to J/\psi\, K_{\rm S})=(8.9\pm1.2)\times10^{-4}$ \cite{PDG} 
probes only the combination ${\cal A}'\propto\left(A_{\rm cc}^{c'}+
A_{\rm pen}^{ct'}\right)$ of current--current and penguin amplitudes, 
and obviously does not allow their separation. It would
be very important to have a better theoretical understanding of the
quantity $a\,e^{i\theta}$. However, such analyses are beyond the 
scope of this workshop, and are left for further studies. Let us note
that the measured $B_d^0\to J/\psi\, K_{\rm S}$ branching ratio implies,
if we use U-spin arguments, a $B_s\to J/\psi\, K_{\rm S}$ branching ratio 
at the level of $2\times10^{-5}$.

The general formalism to extract $\gamma$ from $B_{s(d)}\to J/\psi\,K_{\rm S}$ 
decays can be found in \cite{RF-BdsPsiK}. Although the corresponding formulae
are quite complicated, the basic idea is very simple: if $\phi_s$ is used
as an input, the CP-violating asymmetries 
${\cal A}_{\rm CP}^{\rm dir}(B_s\to J/\psi\, K_{\rm S})$ and
${\cal A}_{\rm CP}^{\rm mix} (B_s\to J/\psi\, K_{\rm S})$ allow one to
fix a contour in the $\gamma$--$a$ plane in a {\it theoretically clean}
way. Another contour can be fixed with the help of the U-spin relations
(\ref{SU3-1}) by combining the observable $H$ with
${\cal A}_{\rm CP}^{\rm mix}(B_s\to J/\psi\, K_{\rm S})$. Alternatively,
we may combine $H$ with ${\cal A}_{\Delta\Gamma}(B_s\to J/\psi\, K_{\rm S})$
to fix a third contour in the $\gamma$--$a$ plane. The intersection of these
contours then gives $\gamma$ and $a$. The general formulae simplify 
considerably, if we keep only terms linear in $a$. Within this approximation, 
we obtain 
\begin{equation}\label{gam-approx}
\tan\gamma\approx\frac{\sin\phi_s+
{\cal A}_{\rm CP}^{\rm mix}(B_s\to J/\psi\, K_{\rm S})}{(1-H)
\cos\phi_s}\,.
\end{equation}

\begin{figure}
~\\[-2cm]
\centerline{\rotate[r]{
\epsfysize=9truecm
{\epsffile{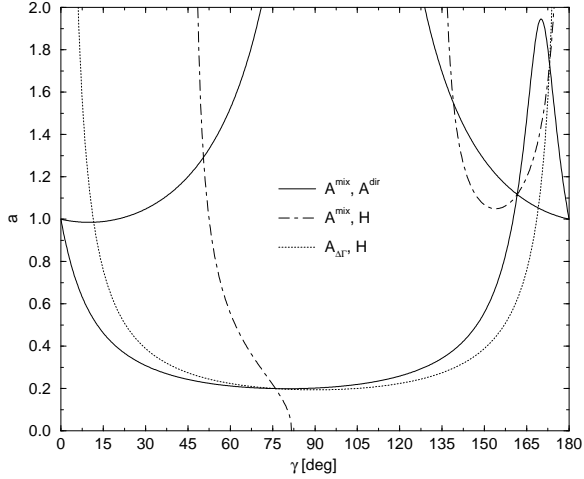}}}}
\vspace*{-0.8cm}
\caption{Contours in the $\gamma$--$a$ plane fixed through the 
$B_{s(d)}\to J/\psi\, K_{\rm S}$ observables for an example
discussed in the text.}\label{fig:BsdPsiKScont}
\end{figure}

Let us illustrate this approach by considering a simple example. Assuming
a negligible $B^0_s$--$\overline{B^0_s}$ mixing phase, i.e.\ $\phi_s=0$, 
and $\gamma=76^\circ$, which lies within the presently allowed ``indirect''
range for this angle, as well as $a=a'=0.2$ and $\theta=\theta'=30^\circ$, 
we obtain the following $B_{s(d)}\to J/\psi\, K_{\rm S}$ observables:
\begin{equation}\label{eq:paul_special}
\begin{array}{ll}
{\cal A}_{\rm CP}^{\rm dir}(B_s\to J/\psi\, K_{\rm S})=0.20,& 
{\cal A}_{\rm CP}^{\rm mix}(B_s\to J/\psi\, K_{\rm S})=0.33,\\ 
{\cal A}_{\Delta\Gamma}(B_s\to J/\psi\, K_{\rm S})=0.92, & H=0.95. 
\end{array}
\end{equation}
The corresponding contours in the $\gamma$--$a$ plane are shown 
in Fig.~\ref{fig:BsdPsiKScont}. Interestingly, in the case of these contours, 
we would not have to deal with ``physical'' discrete ambiguities for 
$\gamma$, since values of $a$ larger than 1 would simply appear unrealistic. 
If it should become possible to measure ${\cal A}_{\Delta\Gamma}$ with the 
help of the widths difference $\Delta\Gamma_s$, the dotted line could be 
fixed. In this example, the approximate expression (\ref{gam-approx}) 
yields $\gamma\approx82^\circ$, which deviates from the ``true'' value of 
$\gamma=76^\circ$ by only 8\%. It is also interesting to note that we have 
${\cal A}_{\rm CP}^{\rm dir}(B_d\to J/\psi\, K_{\rm S})=-0.98\%$ in our 
example.

An important by-product of the strategy described above is that 
the quantities $a'$ and $\theta'$ allow one to take into account the 
penguin contributions in the determination of $\phi_d$ from $B_d\to J/\psi\, 
K_{\rm S}$, which are presumably very small because of the strong
Cabibbo suppression in (\ref{Bd-ampl2}). However, as we have 
already noted in Sec.~\ref{subsec:BdpsiKS}, these uncertainties are
an important issue for the LHC because of the tremendously small experimental 
uncertainty for the CP-violating $B_d\to J/\psi\, K_{\rm S}$ observables.
Using (\ref{SU3-1}), we obtain an 
interesting relation between the direct CP asymmetries arising in the modes 
$B_d\to J/\psi\, K_{\rm S}$ and $B_s\to J/\psi\, K_{\rm S}$ and their 
CP-averaged rates:
\begin{equation}
\frac{{\cal A}_{\rm CP}^{\rm dir}(B_d\to J/\psi\, 
K_{\rm S})}{{\cal A}_{\rm CP}^{\rm dir}(B_s\to J/\psi\, K_{\rm S})}\approx
-\,\frac{B(B_s\to J/\psi\, K_{\rm S})}{B(B_d\to J/\psi\, 
K_{\rm S})}\,.
\end{equation}
Let us note that an analogous relation holds also between the CP-violating 
asymmetries in the decays $B^\pm\to\pi^\pm K$ and $B^\pm\to K^\pm K$ 
\cite{bfm,defan}.

Before turning to the experimental feasibility studies, let us say a few 
words on the SU(3) breaking corrections. Whereas the solid curves in 
Fig.\ \ref{fig:BsdPsiKScont} are {\it theoretically clean}, the
dot-dashed and dotted lines are affected by U-spin breaking corrections.
Because of the suppression of $a'e^{i\theta'}$ in (\ref{Bd-ampl2}) through 
$\lambda^2$, these contours are essentially unaffected by possible corrections 
to (\ref{SU3-1}), and rely predominantly on the U-spin relation 
$|{\cal A'}|=|{\cal A}|$. In the ``factorization'' approximation, we have
\begin{equation}\label{SU3-break}
\left.\frac{|{\cal A'}|}{|{\cal A}|}\right|_{\rm fact}=\,
\frac{F_{B_d^0K^0}(M_{J/\psi}^2;1^-)}{F_{B_s^0\overline{K^0}}
(M_{J/\psi}^2;1^-)}\,,
\end{equation}
where the form factors $F_{B_d^0K^0}(M_{J/\psi}^2;1^-)$ and 
$F_{B_s^0\overline{K^0}}(M_{J/\psi}^2;1^-)$ parametrize the 
quark--current matrix elements 
$\langle K^0|(\bar b s)_{\rm V-A}|B^0_d\rangle$ and 
$\langle\overline{K^0}|(\bar b d)_{\rm V-A}|B^0_s\rangle$, respectively 
\cite{BSW}. We are not aware of quantitative studies of (\ref{SU3-break}), 
which could be performed, for instance, with the help of sum rule or 
lattice techniques. In the light-cone sum-rule approach, sizeable 
SU(3) breaking effects were found for $B_{d,s}\to K^\ast$ 
form factors \cite{BB}. It should be emphasized that also non-factorizable 
corrections, which are not included in (\ref{SU3-break}), may play an 
important r\^{o}le. We are optimistic that 
SU(3) breaking will be under better control 
by the time the $B_s\to J/\psi\, K_{\rm S}$ measurements 
can be performed in practice. 

\def\Bdpsiks{$B_d\rightarrow J/\psi \, K_S \,\,$}
\def\Bdpsimmks{$B_d\rightarrow J/\psi(\rightarrow \mu^+ \mu^-) \, K_S \,\,$}
\def\Bspsiks{$B_s\rightarrow J/\psi \, K_S \,\,$}
\def\Bspsimmks{$B_s\rightarrow J/\psi(\rightarrow \mu^+ \mu^-) \, K_S \,\,$}

\subsubsection{Experimental Studies}

Both CMS and LHCb have performed preliminary studies of the 
feasibility of extracting the CKM angle $\gamma$ from a measurement of 
the time-dependent 
CP asymmetry in the decay \Bspsiks.   {}From these,  and the results
presented in Sec.~\ref{subsec:BdpsiKS},  the potential of ATLAS may
also be gauged.

The \Bspsiks\ branching ratio is expected to be at the level of 
$\rm 2.0 \times 10^{-5}$, see Sec.~\ref{subsec:BsdJpsiKs}, compared to 
$\rm (4.45 \pm 0.6) \times 10^{-4}$~\cite{PDG} for 
\Bdpsiks,  
and the $B_s$ production rate is 30\% of the $B^0_d$ rate.
Assuming the same  
selection procedure as used in the \Bdpsiks\ analysis, 
the \Bspsiks\ event yield will therefore be 1/74 that of the $B^0_d$ yield.
Experimentally the isolation of these events is challenging,  because of 
the large combinatoric background,  and the close
$B^0_d$ peak, only 90 $\rm MeV/c^2$ away.

\begin{figure}[tbh]
\begin{center}
\subfigure[CMS]
{\epsfig{file=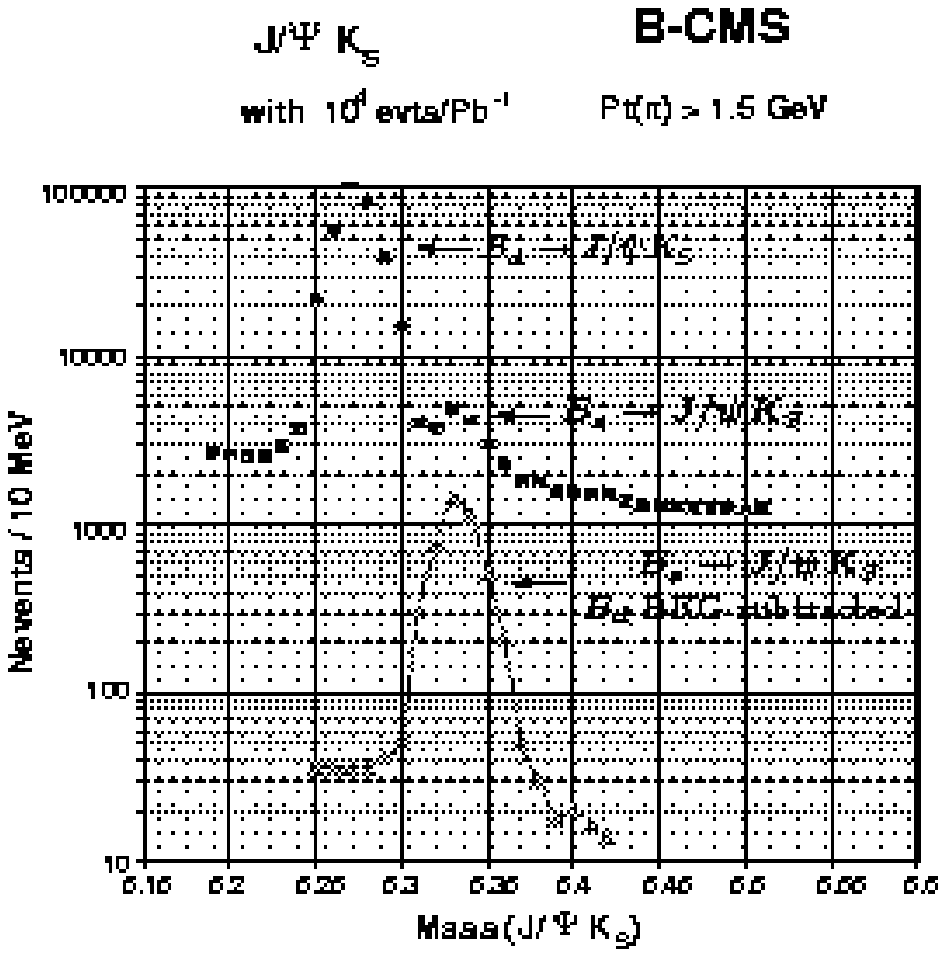,height=0.43\textwidth}}
\subfigure[LHCb]
{\epsfig{file=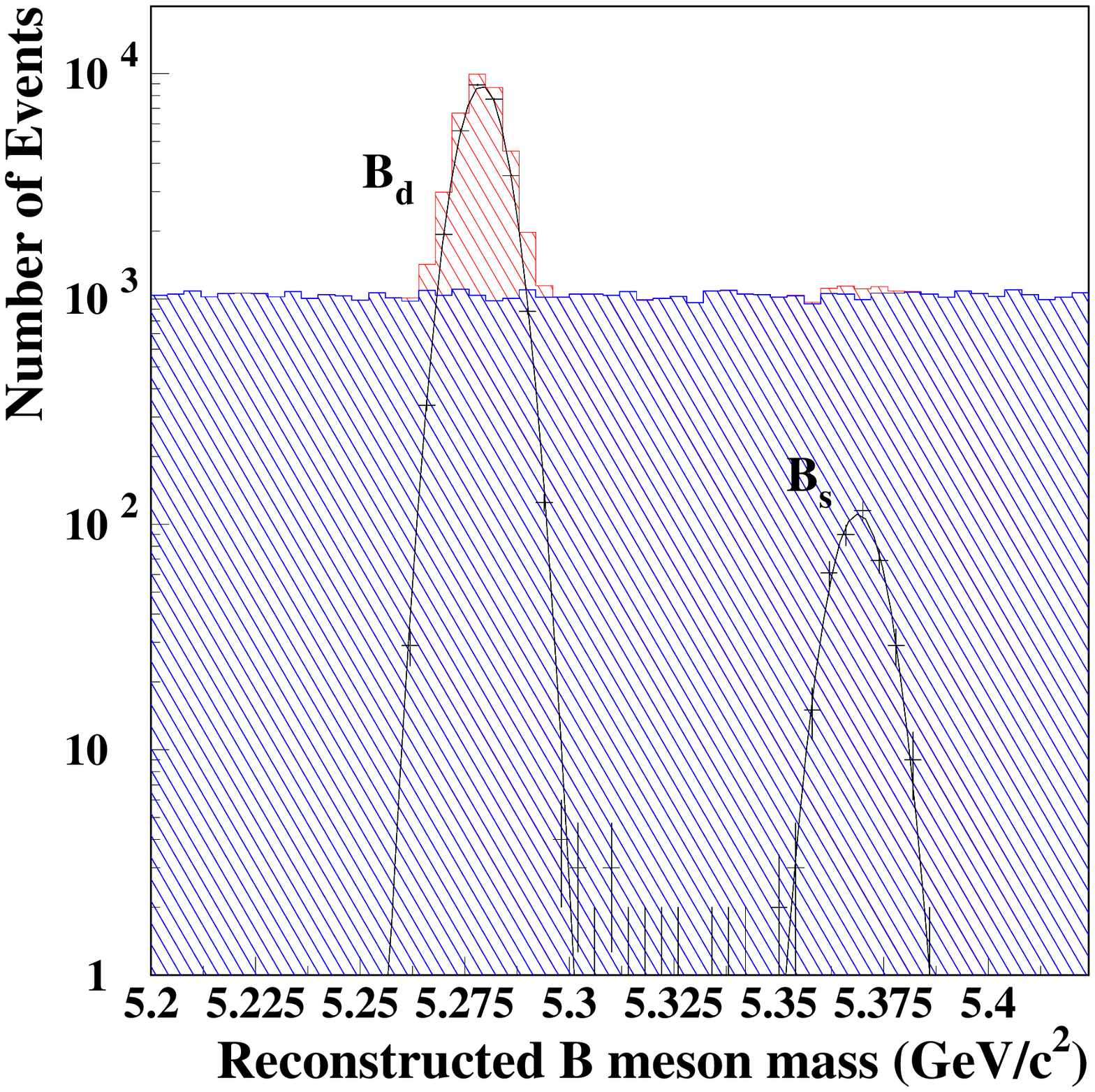,height=0.43\textwidth}}
\end{center}
\vspace*{-1cm}
\caption{\Bdpsimmks and \Bspsimmks mass peaks.}
\label{bspsi_masspk}
\end{figure}

CMS has developed a selection tailored to \Bspsiks\  decays.  The combinatoric
background can be heavily suppressed with a $p_T$ cut of $>1.5 \, {\rm GeV/c}$
on the pions from the $\rm K^0_s$ decays.   With such criteria a S/B of
$\approx \, 0.5$  can be achieved,  with an event yield of 4100 events per
year.  The mass resolution of $<20 \,{\rm MeV/c^2}$ is sufficient to separate
the events from those of the $B^0_d$ decay.  The reconstructed mass peaks
can be seen in Fig.~\ref{bspsi_masspk}(a).

LHCb has not yet investigated cuts specific to \Bspsiks.   As can be
seen from Fig.~\ref{bspsi_masspk}(b),  the standard \Bdpsiks\ selection
results in a combinatoric background which is an order of magnitude
above the \Bspsiks\ signal.  Further work will improve the selection
to suppress this contamination.   The $<10  \,{\rm MeV/c^2}$ resolution
on the invariant mass means that the $\rm B^0_d$ and $\rm B^0_s$ peaks 
are cleanly separated.

These studies indicate that a measurement of the CP asymmetry in
 \Bspsiks\  is feasible at the LHC,  so that $\gamma$ can be extracted
 from that decay.   For the parameter set considered in 
Sec.~\ref{subsec:BsdJpsiKs},  CMS estimate that a precision of 
$\sim 9^\circ$ is achievable in 3 years operation.

\subsection[Extracting  $\gamma$ from 
 $B_{d(s)}\to D_{d(s)}^+ D_{d(s)}^-$ 
Decays]{Extracting \protect\boldmath $\gamma$ from 
\protect\boldmath $B_{d(s)}\to D_{d(s)}^+ D_{d(s)}^-$ Decays
\protect\footnote{With help from 
V. Gibson.}}\label{subsec:BDD}

Usually, $B_d\to D_d^+D_d^-$ decays appear in the literature as a tool
to probe the $B^0_d$--$\overline{B^0_d}$ mixing phase $\phi_d$ 
\cite{CP-revs1,CP-revs2,BaBar}. 
In fact, if penguins played a negligible r\^{o}le in 
these modes, $\phi_d=2\beta$ could be determined from the corresponding 
mixing-induced CP-violating effects. However, penguin topologies, 
which contain also important contributions from final-state-interaction 
effects, may well be sizeable, although it is very difficult to calculate 
them in a reliable way. The strategy discussed in this subsection makes 
use of these penguin topologies \cite{RF-BdsPsiK}, allowing one to 
determine $\gamma$, if the overall $B_d\to D_d^+D_d^-$ normalization is 
fixed through the CP-averaged, i.e.\ the ``untagged'' $B_s\to D_s^+D_s^-$ 
rate, and if the $B^0_d$--$\overline{B^0_d}$ mixing phase $\phi_d$ is 
determined separately, for instance with the help of the ``gold-plated''
decay $B_d\to J/\psi\, K_{\rm S}$. It should be emphasized that no 
$\Delta M_st$ oscillations have to be resolved to measure the untagged 
$B_s\to D_s^+D_s^-$ rate.

\begin{figure} 
\begin{center}
\epsfig{file=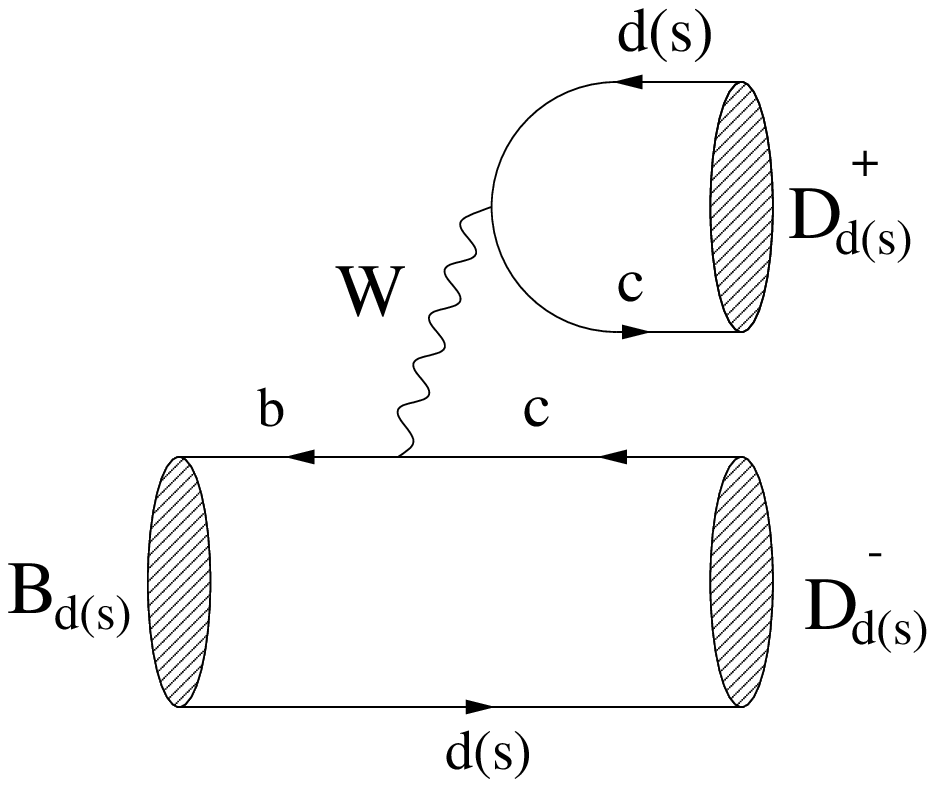,width=0.35\textwidth}
\epsfig{file=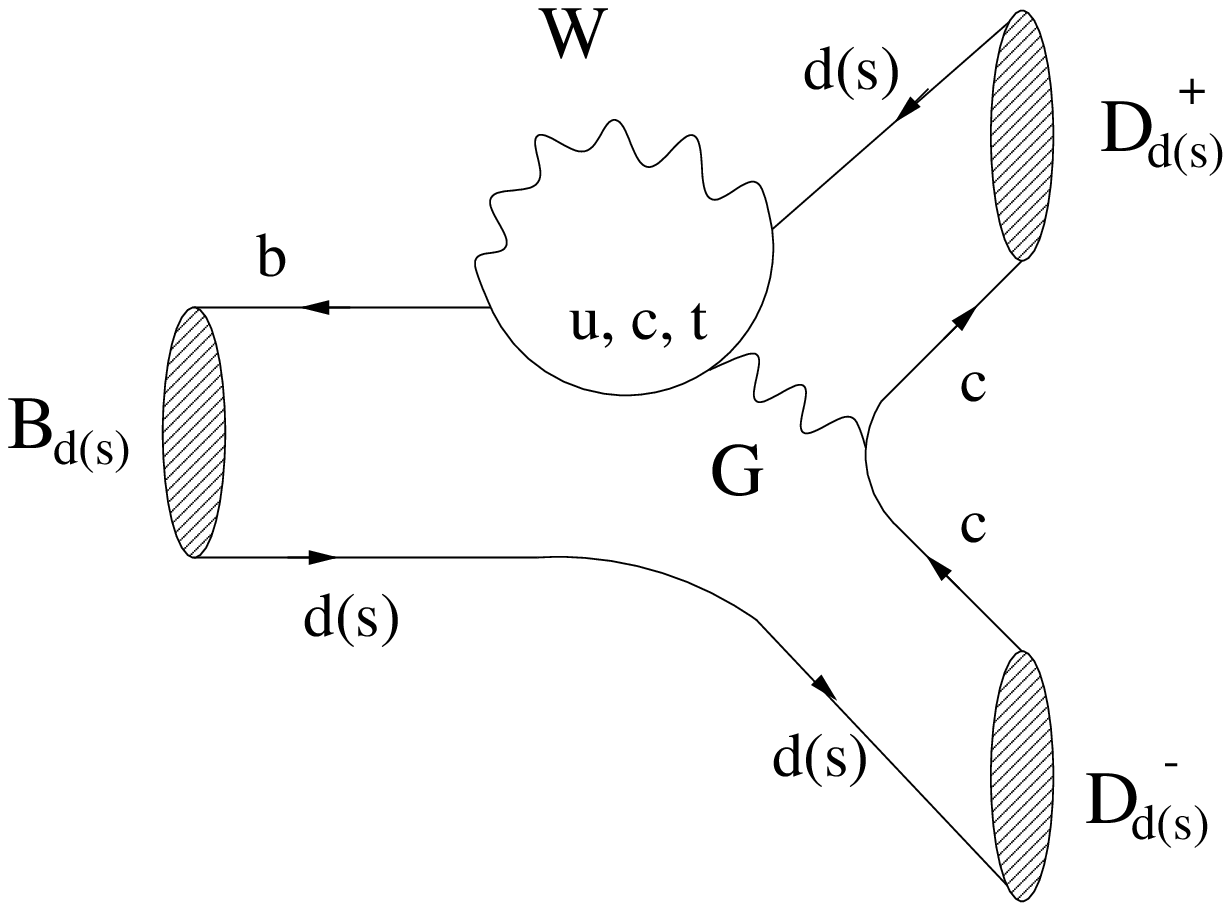,width=0.35\textwidth}
\end{center}
\vspace*{-0.5cm}
\caption{Feynman diagrams contributing to 
$B_{d (s)}^0\to D^{\,+}_{d(s)}\, D^{\,-}_{d(s)}$.}\label{fig:BDD}
\end{figure}

\subsubsection{Theoretical Aspects}

The decays $B_{d (s)}^0\to D^{\,+}_{d(s)}\, D^{\,-}_{d(s)}$ are 
transitions into a CP eigenstate with eigenvalue $+1$ and originate from 
$\bar b\to\bar c\,c\,\bar d\,(\bar s)$ quark-level decays. We have to 
deal both with current--current and with penguin contributions, as can be
seen in Fig.\ \ref{fig:BDD}. In analogy to (\ref{Bd-ampl2}) and 
(\ref{Bs-ampl}), the corresponding transition amplitudes can be written as
follows:
\begin{eqnarray}
A(B_s^0\to D_s^+D_s^-)&=&\left(1-\frac{\lambda^2}{2}\right)\tilde{\cal A}'
\left[1+\left(\frac{\lambda^2}{1-\lambda^2}\right)\tilde a'e^{i\tilde
\theta'}e^{i\gamma}\right]\label{BDDs-ampl}\\
A(B_d^0\to D_d^+D_d^-)&=&-\lambda\,\tilde{\cal A}\left[1-\tilde a\, 
e^{i\tilde\theta}e^{i\gamma}\right],\label{BDDd-ampl}
\end{eqnarray}
where the quantities $\tilde{\cal A}$, $\tilde{\cal A}'$ and 
$\tilde a\,e^{i\tilde\theta}$, $\tilde a'\,e^{i\tilde\theta'}$ take the
same form as for $B_{s (d)}\to J/\psi\, K_{\rm S}$. In contrast to
the decays $B_{s(d)}\to J/\psi\, K_{\rm S}$, there are ``colour-allowed''
current--current contributions to $B_{d (s)}\to D^{\,+}_{d(s)}\, 
D^{\,-}_{d(s)}$, as well as contributions from ``exchange'' topologies, 
and the QCD penguins do not require a colour-singlet exchange, i.e.\
they are not ``Zweig-suppressed''. 

Since the phase structures of the $B_d^0\to D_d^+D_d^-$ and 
$B_s^0\to D_s^+D_s^-$ decay amplitudes are completely analogous to those 
of $B_s^0\to J/\psi\, K_{\rm S}$ and $B_d^0\to J/\psi\, K_{\rm S}$, 
respectively, the approach discussed in the previous subsection can be 
applied after a straightforward replacements of variables. If we
neglect tiny phase-space effects, which can be taken into account 
straightforwardly (see \cite{RF-BdsPsiK}), we have
\begin{equation}
\tilde H=\left(\frac{1-\lambda^2}{\lambda^2}\right)
\left(\frac{|\tilde{\cal A'}|}{|\tilde{\cal A}|}\right)^2
\frac{\langle\Gamma(B_d\to D_d^+D_d^-)\rangle}{\langle
\Gamma(B_s\to D_s^+D_s^-)\rangle}\,,
\end{equation}
where the CP-averaged rates can be determined with the help of 
(\ref{aver-rate}). The $B_{d(s)}\to D_{d(s)}^+ D_{d(s)}^-$ counterpart 
to (\ref{gam-approx}) takes the following form:
\begin{equation}\label{gam-approx2}
\tan\gamma\approx\frac{\sin\phi_d-{\cal A}_{\rm CP}^{\rm mix}(B_d\to 
D_d^+D_d^-)}{(1-\tilde H)\cos\phi_d}\,,
\end{equation}
where the different sign of the mixing-induced CP asymmetry results from
the different CP eigenvalues of the $B_d\to D_d^+D_d^-$ and $B_s\to 
J/\psi\, K_{\rm S}$ final states. 

\begin{figure}
\centerline{\rotate[r]{
\epsfysize=9truecm
{\epsffile{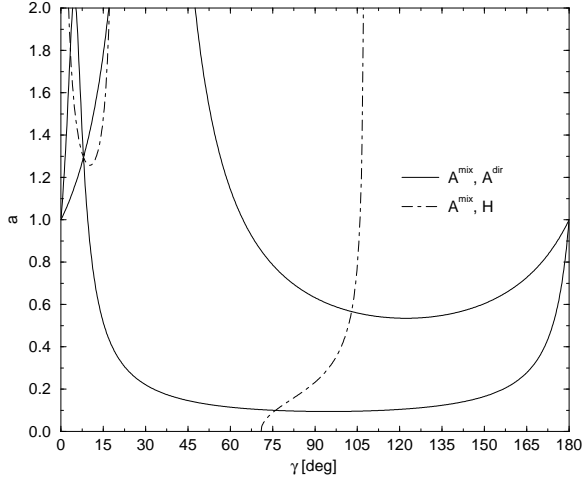}}}}
\vspace*{-0.5cm}
\caption[]{Contours in the $\gamma$--$\tilde a$ plane fixed through the 
$B_{d (s)}\to D^{\,+}_{d(s)}\, D^{\,-}_{d(s)}$ observables for an
example discussed in the text.}\label{fig:Bdcontours1}
\end{figure}

Let us illustrate the strategy to determine $\gamma$, again by considering 
a simple example. Assuming $\tilde a=\tilde a'=0.1$, $\tilde\theta=
\tilde\theta'=210^\circ$, $\gamma=76^\circ$ and a $B^0_d$--$\overline{B^0_d}$ 
mixing phase of $\phi_d=2\beta=53^\circ$, we obtain the following
observables:
\begin{equation}
{\cal A}_{\rm CP}^{\rm dir}(B_d\to D_d^+D_d^-)=-0.092,\quad
{\cal A}_{\rm CP}^{\rm mix}(B_d\to D_d^+D_d^-)=
0.88\quad\mbox{and}\quad \tilde H=1.05. 
\end{equation}
In this case, studies of CP violation in $B_d\to J/\psi\, K_{\rm S}$ would 
yield $\sin(2\beta)=0.8$, which is the central value of the most recent CDF 
analysis \cite{sin2b-exp}, implying $2\beta=53^\circ$ or 
$2\beta=180^\circ-53^\circ=127^\circ$. This twofold ambiguity can be
resolved experimentally, for example, by combining $B_s\to J/\psi\,\phi$ 
with $B_d\to J/\psi\,\rho^0$ \cite{RF-ang} (for alternatives, see 
\cite{ambig}), as noted in Sec.~\ref{sec:Bspsiphi}. In this example, 
we obtain the contours in the $\gamma$--$\tilde a$ plane shown in 
Fig.~\ref{fig:Bdcontours1}. Since values of $\tilde a={\cal O}(1)$ appear 
unrealistic, we would obtain a single ``physical'' solution of $76^\circ$ 
in this case. The approximate expression (\ref{gam-approx2}) gives
$\gamma\approx70^\circ$.

As in the $B_{s (d)}\to J/\psi\, K_{\rm S}$ case, only the contours 
involving the observable $\tilde H$, i.e.\ the dot-dashed lines in
Fig.~\ref{fig:Bdcontours1}, are affected by SU(3) breaking 
corrections, which are essentially due to the U-spin breaking corrections 
to $|\tilde{\cal A}'|=|\tilde{\cal A}|$. Within the ``factorization'' 
approximation, we have
\begin{equation}\label{SU3-breakBDD}
\left.\frac{|\tilde{\cal A'}|}{|\tilde{\cal A}|}\right|_{\rm fact}\approx\,
\frac{(M_{B_s}-M_{D_s})\,\sqrt{M_{B_s}M_{D_s}}\,(w_s+1)}{(M_{B_d}-M_{D_d})
\,\sqrt{M_{B_d}M_{D_d}}\,(w_d+1)}\frac{f_{D_s}\,\xi_s(w_s)}{f_{D_d}\,
\xi_d(w_d)}\,,
\end{equation}
where the restrictions form the heavy-quark effective theory for the 
$B_q\to D_q$ form factors have been taken into account by introducing 
appropriate Isgur--Wise functions $\xi_q(w_q)$ with $w_q=M_{B_q}/(2M_{D_q})$ 
\cite{neu-ste}. Studies of the light-quark dependence of the Isgur--Wise
function were performed within heavy-meson chiral perturbation theory, 
indicating an enhancement of $\xi_s/\xi_d$ at the level of $5\%$ 
\cite{HMChiPT1}. Applying the same formalism to $f_{D_s}/f_D$ gives values
at the 1.2 level \cite{HMChiPT2}, which is of the same order of magnitude as 
the results of recent lattice calculations \cite{lat1}. Further studies are 
needed to get a better picture of the SU(3) breaking corrections 
to the ratio $|\tilde{\cal A'}|/|\tilde{\cal A}|$. Since ``factorization'' 
may work reasonably well for $B_q\to D_q^+D_q^-$, the leading corrections 
are expected to be due to (\ref{SU3-breakBDD}).

The experimental feasibility of the strategy to extract $\gamma$ from
$B_{d (s)}\to D^{\,+}_{d(s)}\, D^{\,-}_{d(s)}$ decays depends strongly
on the size of the penguin parameter $\tilde a$, which is difficult to 
predict theoretically. The branching ratio for $B_d^0\to D_d^+D_d^-$ is 
expected at the $4\times10^{-4}$ level \cite{neu-ste}; the one for 
$B_s^0\to D_s^+D_s^-$ is enhanced by $1/\lambda^2\approx20$, and is 
correspondingly expected at the $8\times10^{-3}$ level.

\subsubsection{Experimental Studies}
\newcommand{\dd}{\mbox{$DD$}}
\newcommand{\dsds}{\mbox{$D_sD_s$}}

LHCb has conducted a preliminary feasibility study of this analysis,
considering the modes where 
the $D$ decays to $K\pi\pi$ and the
$D_s$ to $KK\pi$. For the $B_s \rightarrow \dsds$ decay only the total
rate is required, which is advantageous experimentally as it is neither
necessary to resolve the rapid  oscillations,  nor does flavour tagging
reduce the already suppressed yield in $B_s$ events. The observables
${\cal A}^{\rm mix}_{CP}$ and ${\cal A}^{\rm dir}_{CP}$ 
are extracted from a fit to the time
dependent CP asymmetry for $B\rightarrow DD$ decays. For this channel
 it is therefore necessary to obtain the decay time of the event and to
flavour tag the decays. These requirements entail that the analysis
exploits all the strengths of the LHCb detector, namely the specialized
trigger, the particle identification capability and the precise
vertexing.

The final states for both decays consist of six hadrons. The hadron
trigger is therefore vital and must be efficient for the low values of
$p_T$ which are a result of the high final state multiplicity. The
vertex trigger is particularly efficient for these channels as there
are two vertices containing three tracks ($D$ vertices) to trigger on
in each decay. The particle identification information from the RICH
detectors is important for background suppression and to eliminate
reflections from $KK\pi$ to $K\pi\pi$ and vice versa.

This analysis is at a preliminary level and is still underway, but
initial results look promising. The trigger efficiencies for both
channels are found to be about 25\% for events decaying within
the acceptance.  The reconstruction relies
principally on requiring well separated secondary vertices,
appropriate invariant masses and $p_T$ cuts. Reconstruction
efficiencies for the $B$ and $B_s$ of about 30\% have been found.
Using product branching ratios ($B(B\rightarrow X)\cdot B(X\rightarrow
Y))$ of $3.6\times10^{-5}$ for $B\rightarrow DD$ and
$3.2\times10^{-4}$ for $B_s\rightarrow \dsds$ gives about
$3\times10^5$ events per year for $B\rightarrow DD$,  after flavour tagging,
and $1.9\times10^5$ events per year for
$B_s\rightarrow \dsds$.  These estimates have been obtained by studying
signal Monte Carlo simulations only. A study of the effect of
backgrounds is currently underway. The errors achievable on
$\gamma$ depend on the specific values of $\gamma$ and $\beta$. For
$\gamma = 75^{\circ}$ and $\beta = 50^{\circ}$ an error of about
$1^{\circ}$ is expected. It should be emphasized that these numbers
are preliminary,  but it seems that the
potential of LHCb in this promising channel is good.

\subsection[A Simultaneous Determination of  $\beta$
and  $\gamma$ from  $B_d\to\pi^+\pi^-$ and  $B_s\to K^+K^-$]{A 
Simultaneous Determination of \protect\boldmath $\beta$ and 
\protect\boldmath $\gamma$ from 
\protect\boldmath $B_d\to\pi^+\pi^-$ and 
\protect\boldmath $B_s\to K^+K^-$\protect\footnote{With help from 
D. Rousseau and A. Starodumov.}}\label{subsec:BsKK}

In this subsection, we combine the CP-violating observables of the 
decay $B_d\to\pi^+\pi^-$ with those of the transition $B_s\to K^+K^-$,
which is the U-spin counterpart of $B_d\to\pi^+\pi^-$. Following 
these lines, a simultaneous determination of $\phi_d=2\beta$ and $\gamma$ 
becomes possible \cite{RF-BsKK}. This approach is not affected 
by any penguin topologies -- it rather makes use of them -- and does not 
rely on certain ``plausible'' dynamical or model-dependent assumptions. 
Moreover, final-state-interaction effects, which led to considerable 
attention in the recent literature in the context of the determination of 
$\gamma$ from $B\to\pi K$ decays (see Sec.~\ref{subsec:BpiK}), do 
not lead to any problems, and the theoretical accuracy is only limited by 
U-spin breaking effects. This strategy, which is furthermore very 
promising to search for indications of new physics \cite{FMat}, is 
conceptually quite similar to the extractions of $\gamma$ with the help of
the decays $B_{s(d)}\to J/\psi\, K_{\rm S}$ and $B_{d(s)}\to D_{d(s)}^+ 
D_{d(s)}^-$ discussed in Secs.~\ref{subsec:BsdPsiKS} and 
\ref{subsec:BDD}, respectively (see also \cite{dun-snowmass}).

\subsubsection{Theoretical Aspects}

As can be seen from Fig.~\ref{fig:bpipi}, $B_d\to\pi^+\pi^-$ 
and $B_s\to K^+K^-$ are related to each other by interchanging all down and
strange quarks, i.e.\ they are U-spin counterparts. If we make use of 
the unitarity of the CKM matrix and apply the Wolfenstein parametrization 
\cite{wolf}, generalized to include non-leading terms in $\lambda$ 
\cite{BLO}, the $B_d^0\to\pi^+\pi^-$ decay amplitude can be expressed as 
follows \cite{RF-BsKK}:
\begin{equation}\label{Bdpipi-ampl}
A(B_d^0\to\pi^+\pi^-)=e^{i\gamma}\,{\cal C}\left[1-d\,e^{i\theta}e^{-i\gamma}
\right],
\end{equation}
where
\begin{equation}\label{C-def}
{\cal C}\equiv\lambda^3A\, R_b\left(A_{\rm cc}^{u}+A_{\rm pen}^{ut}
\right),\quad d\,e^{i\theta}\equiv\frac{1}{R_b}
\left(\frac{A_{\rm pen}^{ct}}{A_{\rm cc}^{u}+A_{\rm pen}^{ut}}\right)
\end{equation}
with $A_{\rm pen}^{ut}\equiv A_{\rm pen}^{u}-A_{\rm pen}^{t}$.
In analogy to (\ref{Bdpipi-ampl}), we obtain for the $B_s^0\to K^+K^-$
decay amplitude 
\begin{equation}\label{BsKK-ampl}
A(B_s^0\to K^+K^-)=e^{i\gamma}\lambda\,{\cal C}'\left[1+\left(
\frac{1-\lambda^2}{\lambda^2}\right)d'e^{i\theta'}e^{-i\gamma}\right],
\end{equation}
where
\begin{equation}\label{dp-def}
{\cal C}'\equiv\left(\frac{\lambda^3A\,R_b}{1-\lambda^2/2}\right)
\left(A_{\rm cc}^{u'}+A_{\rm pen}^{ut'}\right)
~\mbox{and}~ 
d'e^{i\theta'}\equiv\frac{1}{R_b}
\left(\frac{A_{\rm pen}^{ct'}}{A_{\rm cc}^{u'}+A_{\rm pen}^{ut'}}\right)
\end{equation}
correspond to (\ref{C-def}). The general
expressions for the $B_d\to\pi^+\pi^-$ and $B_s\to K^+K^-$ observables
(\ref{ee7}) and (\ref{ADGam}) in terms of the parameters specified above 
can be found in \cite{RF-BsKK}. 

Since $B_d\to\pi^+\pi^-$ and $B_s\to K^+K^-$ are related to each other 
by interchanging all down and strange quarks, the U-spin flavour symmetry 
of strong interactions implies
\begin{equation}\label{U-spin-rel}
d'=d\quad\mbox{and}\quad\theta'=\theta.
\end{equation}
If we assume that the $B^0_s$--$\overline{B^0_s}$ mixing phase $\phi_s$ 
is negligible, or that it is fixed through $B_s\to J/\psi\,\phi$, 
the four CP-violating observables provided by $B_d\to\pi^+\pi^-$ and 
$B_s\to K^+K^-$ depend -- in the strict U-spin limit -- on the four 
``unknowns'' $d$, $\theta$, $\phi_d=2\beta$ and $\gamma$. We have
therefore sufficient observables at our disposal to extract these 
quantities simultaneously. In order to determine $\gamma$, it suffices 
to consider ${\cal A}_{\rm CP}^{\rm mix}(B_s\to K^+K^-)$ and the direct 
CP asymmetries ${\cal A}_{\rm CP}^{\rm dir}(B_s\to K^+K^-)$, 
${\cal A}_{\rm CP}^{\rm dir}(B_d\to\pi^+\pi^-)$. If we make use, in addition,
of ${\cal A}_{\rm CP}^{\rm mix}(B_d\to\pi^+\pi^-)$, $\phi_d$ can be determined 
as well. The formulae to implement this approach in a mathematical way
can be found in \cite{RF-BsKK}.

\begin{figure}
\centerline{\rotate[r]{
\epsfysize=9truecm
{\epsffile{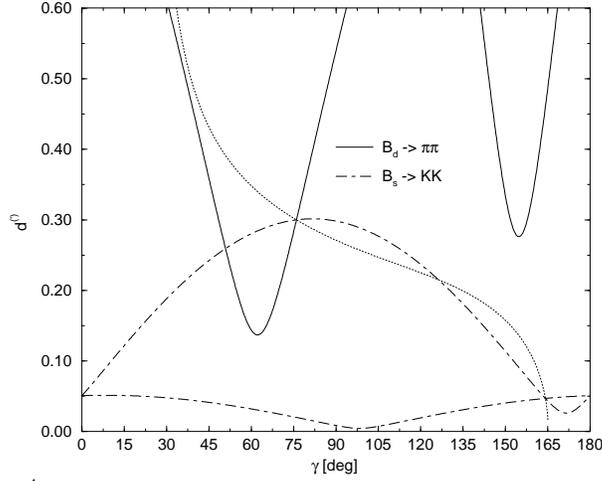}}}}
\vspace*{-0.5cm}
\caption[]{The contours in the $\gamma$--$d^{(')}$ planes fixed through the
CP-violating $B_d\to\pi^+\pi^-$ and $B_s\to K^+K^-$ observables for a 
specific example discussed in the text.}\label{fig:BsKKcont}
\end{figure}

The use of the U-spin flavour symmetry to extract $\gamma$ can 
be minimized, if we use not only $\phi_s$, but also the 
$B^0_d$--$\overline{B^0_d}$ mixing phase $\phi_d$ as an input.
Then, the CP-violating observables ${\cal A}_{\rm CP}^{\rm dir}
(B_d\to\pi^+\pi^-)$, ${\cal A}_{\rm CP}^{\rm mix}(B_d\to\pi^+\pi^-)$ and 
${\cal A}_{\rm CP}^{\rm dir}(B_s\to K^+K^-)$, 
${\cal A}_{\rm CP}^{\rm mix}(B_s\to K^+K^-)$ allow one to fix contours 
in the $\gamma$--$d$ and $\gamma$--$d'$ planes in a {\it theoretically 
clean} way. In order to extract $\gamma$ and the hadronic parameters 
$d$, $\theta$, $\theta'$ with the help of these contours, the U-spin 
relation $d'=d$ is sufficient. Let us illustrate this approach for a 
specific example: 
\begin{equation}\label{obs-examp}
\begin{array}{lcllcl}
{\cal A}_{\rm CP}^{\rm dir}(B_d\to\pi^+\pi^-)&=&+24\%,\,\,& 
{\cal A}_{\rm CP}^{\rm mix}(B_d\to\pi^+\pi^-)&=&+4.4\%,\\ 
{\cal A}_{\rm CP}^{\rm dir}(B_s\to K^+K^-)&=&-17\%,\,\,&
{\cal A}_{\rm CP}^{\rm mix}(B_s\to K^+K^-)&=&-28\%,
\end{array}
\end{equation}
corresponding to the input parameters $d=d'=0.3$, $\theta=\theta'=210^\circ$,
$\phi_s=0$, $\phi_d=53^\circ$ and $\gamma=76^\circ$. In 
Fig.~\ref{fig:BsKKcont}, the corresponding contours in the $\gamma$--$d$ 
and $\gamma$--$d'$ planes are represented by the solid and dot-dashed lines,
respectively. Their intersection yields a twofold solution for $\gamma$, 
given by $51^\circ$ and our input value of $76^\circ$. The dotted line
is related to 
\begin{equation}\label{K-intro}
K\equiv-\left(\frac{1-\lambda^2}{\lambda^2}\right)\left[
\frac{{\cal A}_{\rm CP}^{\rm dir}(B_d\to\pi^+\pi^-)}{{\cal A}_{\rm 
CP}^{\rm dir}(B_s\to K^+K^-)}\right],
\end{equation}
which can be combined with the mixing-induced CP asymmetry 
${\cal A}_{\rm CP}^{\rm mix}(B_s\to K^+K^-)$ through the U-spin relation 
(\ref{U-spin-rel}) to fix another contour in the $\gamma$--$d$
plane. Combining all contours in Fig.~\ref{fig:BsKKcont} with one another, 
we obtain a single solution for $\gamma$ in this example, which is given 
by the ``true'' value of $76^\circ$. 

It should be emphasized that the theoretical accuracy of $\gamma$ and
of the hadronic parameters $d$, $\theta$ and $\theta'$ is only limited
by U-spin breaking effects. In particular, it is not affected by
any final-state-interaction or penguin effects. A first consistency check
is provided by $\theta=\theta'$. Moreover, we may determine the normalization
factors ${\cal C}$ and ${\cal C}'$ of the $B^0_d\to\pi^+\pi^-$ and
$B^0_s\to K^+K^-$ decay amplitudes (see (\ref{Bdpipi-ampl}) and 
(\ref{BsKK-ampl})) with the help of the corresponding CP-averaged
branching ratios. Comparing them with the ``factorized'' result
\begin{equation}
\left|\frac{{\cal C}'}{{\cal C}}\right|_{\rm fact}=\,
\frac{f_K}{f_\pi}\frac{F_{B_sK}(M_K^2;0^+)}{F_{B_d\pi}(M_\pi^2;0^+)}
\left(\frac{M_{B_s}^2-M_K^2}{M_{B_d}^2-M_\pi^2}\right),
\end{equation}
we have another interesting probe for U-spin breaking effects. 
Interestingly, the relation
\begin{equation}
d'e^{i\theta'}=d\,e^{i\theta}
\end{equation}
is not affected by U-spin breaking corrections within a certain 
model-dependent approach (a modernized version \cite{RF-EWP1,pen-calc} of 
the ``Bander--Silverman--Soni mechanism'' \cite{bss}), making use -- 
among other things -- of the ``factorization'' hypothesis to estimate 
the relevant hadronic matrix elements \cite{RF-BsKK}. Although this 
approach seems to be rather simplified and may be affected by 
non-factorizable effects, it strengthens our confidence into the U-spin 
relations used for the extraction of $\beta$ and $\gamma$ from the decays 
$B_d\to\pi^+\pi^-$ and $B_s\to K^+K^-$. Further theoretical studies along 
the lines of Ref.~\cite{BBNS} to investigate the U-spin breaking effects 
in the $B_d\to\pi^+\pi^-$, $B_s\to K^+K^-$ system would be very interesting. 
In order to obtain further experimental insights, the $B_d\to\rho^+\rho^-$, 
$B_s\to K^{\ast+}\,K^{\ast-}$ system would be of particular interest, 
allowing one to determine $\gamma$ together with the mixing phases $\phi_d$ 
and $\phi_s$, and tests of several interesting U-spin relations 
\cite{RF-ang}. 

\begin{figure}
\vspace{-1cm}
   \epsfysize=9.3cm
   \centerline{\epsffile{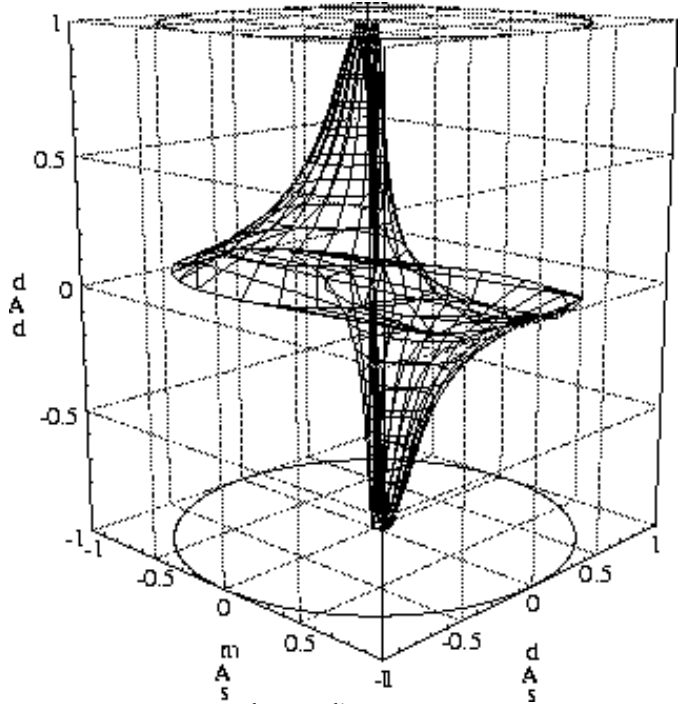}}
\vspace*{-0.5cm}
\caption[dummy]{\small The allowed region in the space of 
$A^{\rm d}_s\equiv{\cal A}_{\rm CP}^{\rm dir}(B_s\to K^+K^-)$, 
$A^{\rm m}_s\equiv{\cal A}_{\rm CP}^{\rm mix}(B_s\to K^+K^-)$ and 
$A^{\rm d}_d\equiv{\cal A}_{\rm CP}^{\rm dir}(B_d\to\pi^+\pi^-)$,  
characterizing $B_s\to K^+K^-$ and $B_d\to\pi^+\pi^-$ in the SM ($\phi_s=0$).
\label{fig:BsKK-SM}}
\end{figure}

Since penguin processes play an important r\^{o}le in the decays $B_s\to
K^+K^-$ and $B_d\to\pi^+\pi^-$, they -- and the strategy to determine
$\gamma$, where furthermore the unitarity of the CKM matrix is employed --
may well be affected by new physics. Interestingly, the SM 
implies a rather restricted region in the space of the CP-violating 
observables of the $B_s\to K^+K^-$, $B_d\to\pi^+\pi^-$ system \cite{FMat}, 
which is shown in Fig.~\ref{fig:BsKK-SM}. A future measurement of observables 
lying significantly outside of the allowed region shown in this figure would 
be an indication of new physics. Such a discrepancy could either be due to 
CP-violating new-physics contributions to $B^0_s$--$\overline{B^0_s}$ 
mixing, or to the $B_d\to\pi^+\pi^-$, $B_s\to K^+K^-$ decay amplitudes. 
The former case would also be indicated simultaneously by large 
CP-violating effects in the mode $B_s\to J/\psi\,\phi$, which would allow 
us to extract the $B^0_s$--$\overline{B^0_s}$ mixing phase $\phi_s$ (see
Sec.~\ref{sec:Bspsiphi}). A discrepancy between the measured 
$B_d\to\pi^+\pi^-$, $B_s\to K^+K^-$ observables and the region corresponding 
to the value of $\phi_s$ thus determined would then signal new-physics 
contributions to the $B_d\to\pi^+\pi^-$, $B_s\to K^+K^-$ decay amplitudes. 
On the other hand, if $B_s\to J/\psi\,\phi$ should exhibit negligible 
CP-violating effects, any discrepancy between the $B_d\to\pi^+\pi^-$, 
$B_s\to K^+K^-$ observables and the volume shown in Fig.~\ref{fig:BsKK-SM} 
would indicate new-physics contributions to the corresponding decay 
amplitudes. If, however, the observables should lie within the 
region predicted by the SM, we can extract a value for the 
CKM angle $\gamma$ by following the strategy discussed above, which may 
well be in disagreement with those implied by theoretically clean strategies 
making use of pure ``tree'' decays, thereby also indicating  the presence 
of new physics.

\def\Bpipi{$\rm B^0_d\rightarrow \pi^+\pi^-$}
\def\BKpi{$\rm B^0_d\rightarrow K^\pm\pi^\mp$}
\def\BsKK{$\rm B^0_s\rightarrow K^+ K^-$}
\def\BsKpi{$\rm B^0_s\rightarrow K^\pm \pi^\mp$}
\subsubsection{Experimental Studies}
\label{sec_bskk_exp}

It was demonstrated in Sec.~\ref{sec_bpipi_exp} that the LHC experiments
can expect large event yields in the two body decay \Bpipi.  With appropriately
modified selection similarly high statistics can be accummulated in \BsKK.
The excellent proper time resolution of the experiments then allows the
$\rm B^0_s$ oscillations to be distinguished,  and the CP asymmetry 
coefficients to be
measured.   By using the relationships presented above, the
\Bpipi\ and \BsKK\ observables can be used to cleanly extract CP phases, most 
interestingly the angle $\gamma$.   The potential of this approach has been
investigated by all three experiments.
  
\subsection*{Event Yields and Asymmetry Sensitivity}

Apart from the final requirements on the best particle hypothesis and
invariant mass of the two candidate tracks,  the CMS and LHCb 
isolation of \BsKK\ events is identical to the \Bpipi\ selection
described in Sec.~\ref{sec_bpipi_exp}.  After flavour tagging, LHCb
expects an annual yield of 4.6 events,  with a contamination from
other two body modes of 15\%.  The equivalent numbers for CMS are
960 and 540 respectively, assuming the dE/dx based selection.  
As explained previously,  ATLAS favours
an approach where the asymmetry of all selected two body events is fitted
simultaneously.  In this sample, 1.4k \BsKK\ events are expected within the one
sigma mass window.

The fit precision on the \BsKK\  CP parameters  
${\cal A}^{\rm mix}_{\rm K^+ K^- }$
and ${\cal A}^{\rm dir}_{\rm K^+ K^- }$ depends not only on the event yields,
but on the value of $\Delta m_s$, which governs the rapidity of 
the $\rm B^0_s \overline {B^0_s}$ oscillations.
Table~\ref{tab_bskk_prec} shows the precision expected for three different
values of  $\Delta m_s$ after an extended period of running.  The uncertainties
for one year's running scale in the expected statistical manner,  except that
ATLAS and CMS retain no sensitivity for $\Delta m_s =  30\,$ps$^{-1}$ 
with the smaller data set.

\begin{table}
\begin{minipage}[t]{0.46\textwidth}
\begin{center}
\begin{tabular}{|c||c|c|c|} \hline
   $\Delta m_s \, {\rm [ \mbox{ps}^{-1}]}$  
&  ATLAS   &  CMS & LHCb \\ \hline \hline
 15    &  0.09   & 0.10  & 0.034 \\
 20    &  0.13   & 0.13  & 0.047 \\
 30    &    /    & 0.33  & 0.068 \\ \hline
\end{tabular}
\end{center}
\vspace*{-0.5cm}
\caption[]{Expected sensitivities on the \BsKK\ CP asymmetry coefficients
${\cal A}^{\rm mix}_{\rm K^+ K^- }$ and ${\cal A}^{\rm dir}_{\rm K^+ K^- }$ for
3 (ATLAS/CMS) and 5 (LHCb) years' data taking, for different 
values of $\Delta m_s$ and $\Delta \Gamma_s = 0$.}\label{tab_bskk_prec}
\end{minipage}
\hspace*{9pt}
\begin{minipage}[t]{0.51\textwidth}
\begin{center}
\begin{tabular}{|c||c|c|c|} \hline
   $\Delta m_s \, {\rm [ \mbox{ps}^{-1}]}$  &  1 year & Extended  
running\\ \hline \hline
 15    &  3.7$^\circ$   & 1.9$^\circ$  \\
 20    &  4.8$^\circ$   & 2.4$^\circ$  \\
 30    &  7.4$^\circ$   & 3.4$^\circ$  \\ \hline
\end{tabular}
\end{center}
\vspace*{-0.5cm}
\caption[]{Expected sensitivities on the unitarity triangle 
angle $\gamma$ for the \Bpipi/\BsKK\ analysis for
LHC running after one year and 3 (ATLAS/CMS) / 5 (LHCb) years,
as a function of  $\Delta m_s$ and for
the parameter set specified in the text.}\label{tab_bskk_gam}
\end{minipage}
\end{table}

\subsection*{Sensitivity to the CP Violating Phases}

The sensitivity to which $\gamma$ can be determined  has been studied, assuming
the expected precisions on the \BsKK\  given in Tab.~\ref{tab_bskk_prec} and 
\Bpipi\ asymmetries, Tab.~\ref{tab_bpipi_prec}.
With the scenario given in previous subsection
($d=d'=0.3$, $\theta=\theta'=210^\circ$, $\phi_s=0$, $\phi_d=53^\circ$,
$\gamma=76^\circ$  and $\Delta m_s=15 \,$ps$^{-1}$, $\Delta
\Gamma_s=0$ and assuming an  uncertainty of 1\% on
$\sin(2\beta)=\sin(\phi_d)$), the sensitivity after one year at LHC is 
$\sigma_\gamma=3.7^\circ$, the constraints $d=d'$ and
$\theta=\theta'$ being applied. It improves to $\sigma_\gamma=1.9^\circ$
after 5 years.  Table~\ref{tab_bskk_gam} shows how these uncertainties
increase with $\Delta m_s$.  In the considered range of parameters,
the sensitivity is clearly impressive.

\begin{figure}
\begin{center}
\subfigure[Sensitivity on $\gamma$ for 5
years of LHC, with the constraints $\phi_s=0$, $d=d'$ and
$\theta=\theta'$ assuming an uncertainty of 1\% on
$\sin(2\beta)$, and with input values $d=d'=0.3$, 
and $\phi_d=53^\circ$, $\Delta m_s=15\, \mbox{ps}\rm{^{-1}}$ and
$\Delta\Gamma_s=0$. 
The contour lines correspond to sensitivities of $2^\circ$
(solid), $4^\circ$ (dashed), $6^\circ$ (dotted) and $8^\circ$ (dotted-dashed).]
{\epsfig{file=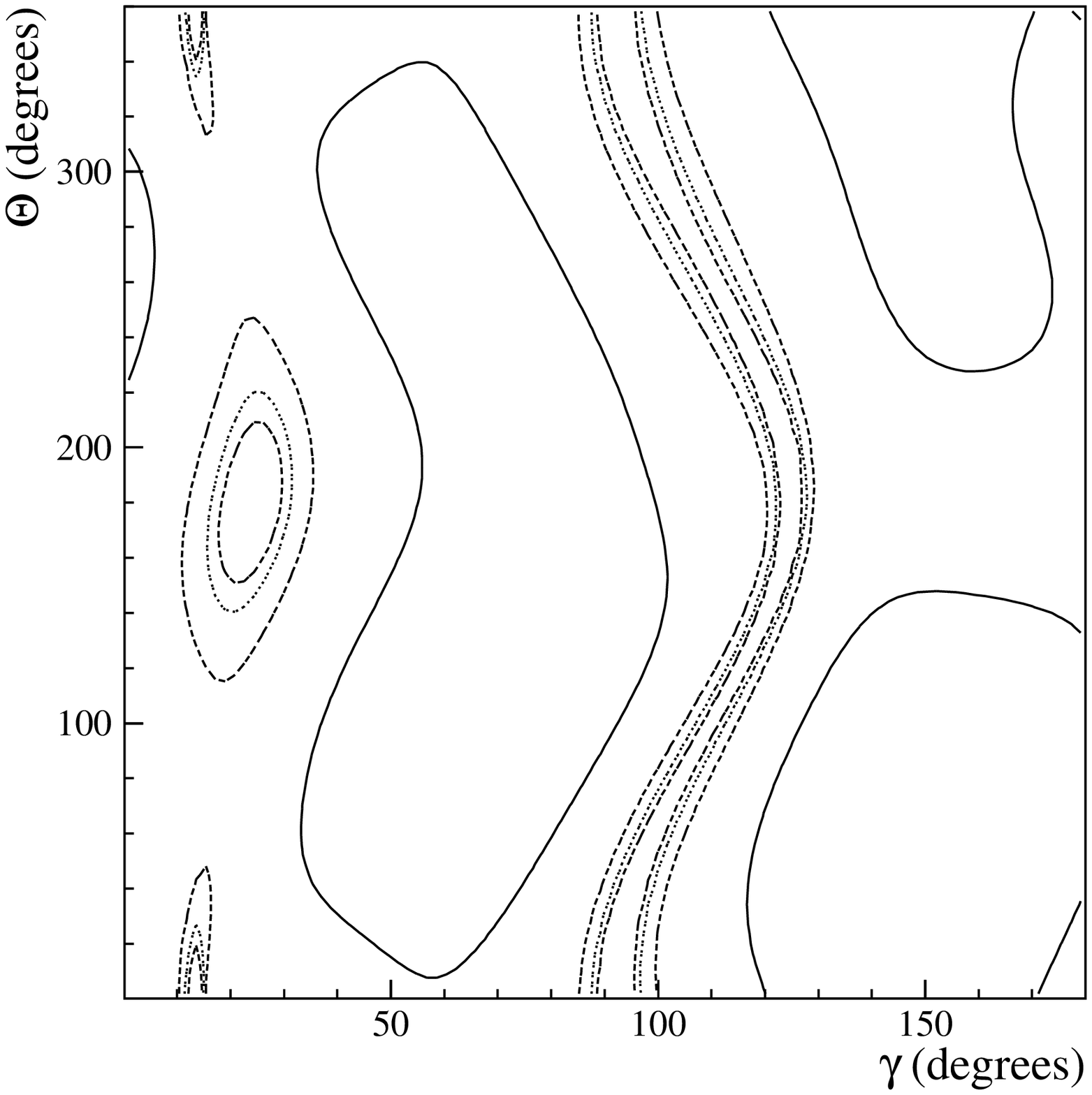,width=0.48\textwidth}}
\subfigure[ Sensitivity on $\theta-\theta'$ for the same fit as in (a)
except the relaxed $\theta=\theta'$ constraint. The contour lines
correspond to sensitivities of 
$10^\circ$ (solid), $15^\circ$ (dashed), $20^\circ$ (dotted) and
$50^\circ$ (dotted-dashed).]
{\epsfig{file=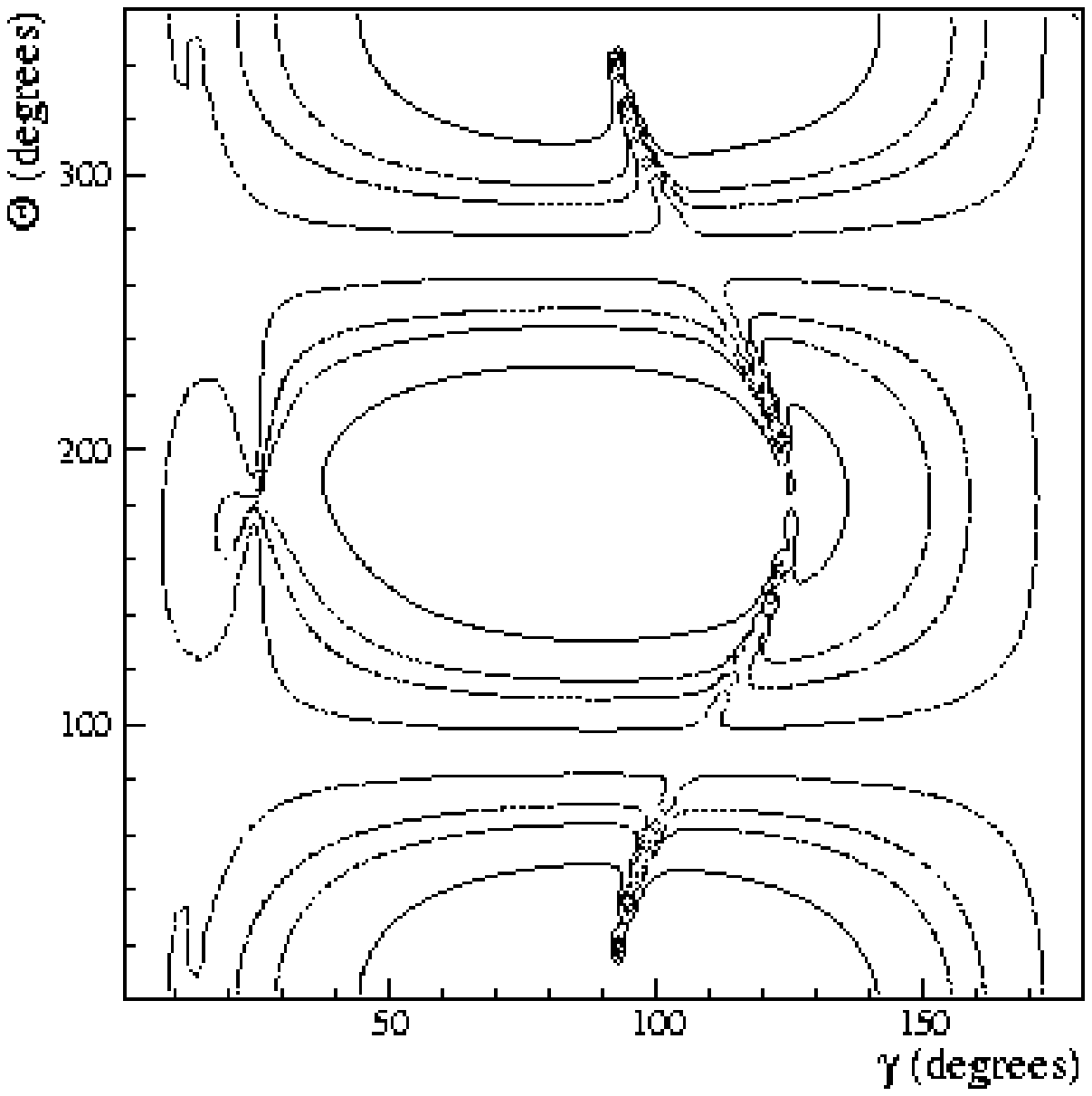,width=0.48\textwidth}}
\subfigure[Sensitivity on $d-d'$ for the same fit as in (a)
except the relaxed $d=d'$ constraint. The contour lines
correspond to sensitivities of 
$0.05$ (solid), $0.1$ (dashed), $0.2$ (dotted) and 
$0.4$ (dotted-dashed).]
{\epsfig{file=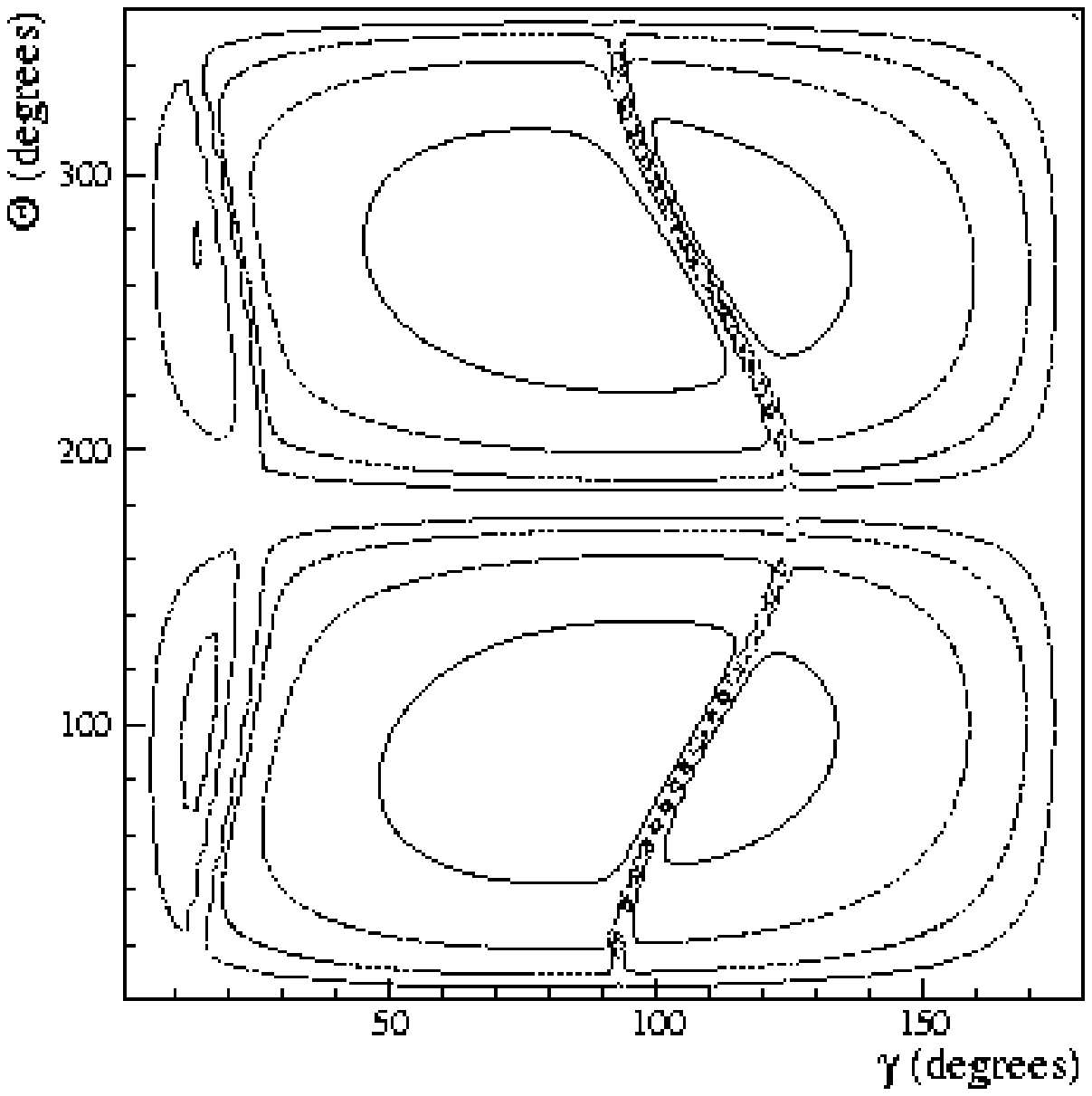,width=0.48\textwidth}}
\end{center}
\vspace*{-0.8cm}
\caption{Sensitivity on fits to the LHC combined \Bpipi\ and \BsKK\ samples.}
\label{fig_bskk_s2d}
\end{figure}

To give an indication on how the sensitivity depends on the scenario,
Fig.~\ref{fig_bskk_s2d} shows the ultimate sensitivity for 5 years 
of LHC, in the scenario given above but as a function of the true
value of $\gamma$ and $\theta=\theta'$.
For most values of $\gamma$ and $\theta$, the 
sensitivity on $\gamma$ is better than $4^\circ$, except in regions 
around $\gamma=90^\circ$ and $\gamma=20^\circ$. The sensitivity 
depends significatively of the assumed value of $d=d'$: it decreases 
(increases) by a factor of two if $d=d'\simeq 0$
($d=d'=0.5$). 

The number of degrees of freedom is such that one
of the constraint $d=d'$ or $\theta=\theta'$ can be relaxed. This
approximately doubles the uncertainty on $\gamma$, but allows U-spin flavour
symmetry relations $d=d'$ and $\theta=\theta'$ to be checked. 
Figure~\ref{fig_bskk_s2d} shows that typical precision of $15^\circ$
on $\theta-\theta'$ and 0.1 on $d-d'$ can be reached, but on regions 
that are largely disjoint in $\theta$. These numbers also indicate the 
level to which U-spin symmetry must hold in order to improve  
the estimation of $\gamma$ without biasing it.

\subsection{Conclusions}

The LHC experiments are well suited to the combined analysis of
\Bpipi\ and \BsKK,  on account of their high yield 
in two-body decays and good proper time resolution.
This analysis offers a powerful and precise way to determine the
angle $\gamma$ in a manner sensitive to new physics
contributions.  

\setcounter{equation}{0}
\def\Brhopi{$\rm B^0_d\rightarrow \rho\pi \: $}
\def\BDstarpi{$\rm  B^0_d\rightarrow \overline{D}{}^{*-} \pi^+$}
\def\Bpipi{$\rm B^0_d\rightarrow\pi^+\pi^-$}
\def\BKpi{$\rm B^0_d\rightarrow K^+\pi^-$}
\def\BsKK{$\rm B^0_s\rightarrow K^+K^-$}
\def\BsKpi{$\rm B^0_s\rightarrow K^+\pi^-$}
\def\BJK{$\rm B^0_d\rightarrow J/\psi K_S^0$}
\def\BJKbm{\rm \overline{B^0_d}\rightarrow J/\psi K_S^0}
\def\BJKm{\rm B^0_d \rightarrow J/\psi K_S^0}
\def\BsDsK{$\rm  B^0_s\rightarrow D_s K$}
\def\BJKb{$\rm \overline{B^0_d}\rightarrow J/\psi K_S^0$}
\def\BJKpm{$\rm B^\pm\rightarrow J/\psi K^\pm$}
\def\BJKst{$\rm B^0_d\rightarrow J/\psi K^{0*}$}
\def\BsDspi{$\rm  B^0_s\rightarrow D_s \pi$}
\section[SYSTEMATIC ERROR CONSIDERATIONS IN CP 
MEASUREMENTS]{SYSTEMATIC ERROR CONSIDERATIONS IN CP 
MEASUREMENTS\protect\footnote{Section coordinators: R. Fleischer and
G. Wilkinson.}}\label{sec:syst}

\subsection{Introduction}

The excellent statistical precision expected in many CP-violation
measurements at the LHC
demands that there be good control of systematic uncertainties.
The challenges posed by hadronic effects in interpreting certain observables
are discussed elsewhere in this Chapter;  here, biases
from experimental factors and initial state asymmetries will be considered,
and possible control strategies examined.

\subsection{Sources and Categories of Systematic Bias}

CP measurements require the reconstruction of a final state, and 
frequently the tagging of the initial state flavour.   Time dependent 
rates, or branching ratios,  are then combined into asymmetries from
which CKM phases can be extracted.   These measurements are inherently robust,
in that to first order experimental unknowns will cancel  or can be assumed
to be the same for all processes under consideration.  However,  certain
charge- and flavour-dependent effects may exist,  which can indeed 
bias the measurement:
\begin{itemize}

\item{{\bf Production asymmetries} \\  As explained in the Chapter on
    $b$ production \cite{bcprod}, the initial fraction of $b$ and $\bar b$ 
hadrons at the LHC
is not expected to be identical.  A production asymmetry will exist,  and
this asymmetry will vary as a function of rapidity and $p_T$, reaching
values of several percent.  
Furthermore, this asymmetry can be different for each hadron species.
In this section, the fractions of $\rm B^0_d,\,\overline{B^0_d},\, B^0_s,\,   
\overline{B^0_s}, \, B^+$ and $\rm B^-$ mesons per event are denoted 
 by $f_0, \, \overline{f_0}, \, f_s, \, \overline{f_s},
f_+$ and $f_-$.}

\item{{\bf Tagging efficiency} \\  All methods of flavour tagging rely
on measuring the charge of one or more selected tracks.   If the track
reconstruction efficiency, or particle assignment (for lepton or kaon tags),
has a charge dependence, then a difference in the tagging efficiency for
$b$ and $\bar b$ hadrons will result.   Such a dependence
is certainly possible,  for instance in LHCb where positive and negative
tracks are preferentially swept by the dipole to different areas of the 
detector.  Furthermore, an asymmetric tagging efficiency can develop
from effects such as a difference in interaction cross-sections for
$\rm K^+$ and $\rm K^-$.  The tagging efficiency for B and 
$\bar{\mbox{B}}$
mesons will be denoted by $\epsilon$ and $\overline{\epsilon}$.}

\item{{\bf Mistag rate} \\ Assuming a flavour tag has been performed, the
probability of that tag being correct can also have a flavour dependence.
For instance in a lepton tag,  different reconstructed momentum spectra
for $l^+$ and $l^-$ are conceivable.  These will result not only in 
different efficiencies, but also in different $\rm B \rightarrow l$ purities 
for the two samples.  The mistag rates for B and $\bar{\mbox{B}}$
mesons will be represented by $\omega$ and $\overline{\omega}$.}

\item{{\bf Final state acceptance} \\ 
Clearly, in any measurement where different 
final states are being compared, the
relative trigger and reconstruction efficiencies can be different.
However,  even if the asymmetry involves a single topology in the final
state, the efficiency may differ for the charge-conjugate case,
for the same charge acceptance reasons as explained above.
}

\end{itemize}
\noindent Background is obviously an additional source of possible bias,
and will require careful attention.  However,  this is a problem common
to most physics measurements, and therefore is not considered
here. 

These effects will have different consequences for each
category of measurement.  The present discussion focuses on
measurements involving decays into CP eigenstates, such as 
\BJK.  Here the observed asymmetry ${\cal A}^{\rm obs}(t)$ is constructed
from the number of  flavour-tagged $\rm B^0$ and $\rm \overline{B^0}$
decays 
into $\rm J/\psi K_S^0$, as a function of proper time.
Allowing for the factors considered above, ${\cal A}^{\rm obs}(t)$ is 
related to the true decay distributions $R^{\rm{true}}_{\rm B^0, 
\overline{B^0} \rightarrow  J/\psi K^s_0}$ as follows:
\begin{eqnarray}
{\cal A}^{\rm obs}(t) & = & 
\frac{ (1 - 2\omega) R^{\rm{true}}_{\BJKm}(t) \, - \, 
\frac{ \overline{f_0} \overline{\epsilon} } {f_0 \epsilon} (1 - 2 
\overline{\omega}) 
R^{\rm{true}}_{\BJKbm}(t)}
{R^{\rm{true}}_{\BJKm}(t) \, + \, \frac{ \overline{f_0}
    \overline{\epsilon} 
}{f_0 \epsilon}\,
R^{\rm{true}}_{\BJKbm}(t)}\,.
\end{eqnarray}
Assuming that the flavour dependent effects in tagging
and production are small,   ${\cal A}^{\rm obs}(t)$ is related to the true 
physics asymmetry 
$$A^{\rm phy}(t) = \frac{R^{\rm{true}}_{\BJKm}(t) -
R^{\rm{true}}_{\BJKbm}(t)}{R^{\rm{true}}_{\BJKm}(t) +
R^{\rm{true}}_{\BJKbm}(t)}$$
as follows:
\begin{equation}
{\cal A}^{\rm obs}(t) \approx  
(1-2\omega) \, 
\left[ A^{\rm phy}(t) \,-\, 
\frac{1}{2}\left(\frac{\overline{f_0}\overline{\epsilon}}{f_0
    \epsilon} - 1 
\right)
(1- {A^{\rm phy}(t)}^2) \,-\,
\left( \frac{\omega - \overline{\omega}}{1 - 2 \omega} \right) \, (1 -
A^{\rm 
phy}(t)) \right].
\label{eq:aobsaphy}
\end{equation}
In the absence of production or tagging asymmetries, this
reduces to the well known expression ${\cal A}^{\rm obs}(t) =
(1-2\omega) \, 
{\cal A}^{\rm phy}(t)$.
Even here, therefore,  the extraction of ${\cal A}^{\rm phy}(t)$
requires that 
the mistag rate
$\omega$ be known.   In the more general case it is also necessary to know
$\overline{f_0}/f_0$, $\overline{\epsilon}/\epsilon$ and $\omega - 
\overline{\omega}$.
Note because there is only a single final state involved, there is no
dependence on any acceptance.
In the following, we consider strategies to determine the tagging and 
production factors.

\subsection{Use of Control Channels}

\subsubsection{Introduction and Event Yields}
\begin{figure}
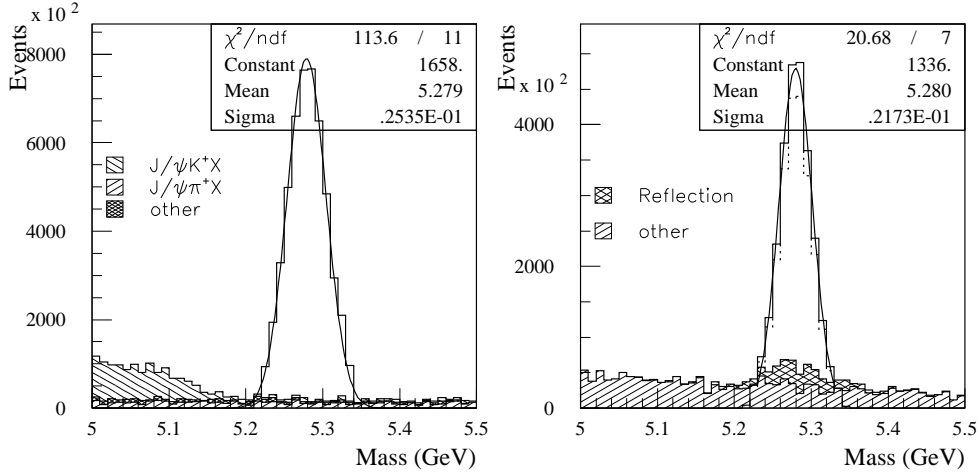

  \begin{center}
    \includegraphics[width=0.4\textwidth,
      clip]{Tbplusstat.epsi}
    \includegraphics[width=0.4\textwidth,
      ]{Tbstarstat.epsi}
 \end{center}
\vspace*{-0.8cm}
      \caption[]{Invariant mass distributions (open histograms) for \BJKpm\ 
   (left) and \BJKst\ (right) with superimposed  
   estimated background contributions (shaded histograms) at ATLAS 
after 3 years of running.}
      \label{fig:sys_ctl}
 \end{figure}                                         

Several channels are useful for controlling systematic biases of the type
considered above.   Three which are discussed here are
\BJKpm, \BJKst\ and \BsDspi.  The LHC experiments expect significant
event yields in these modes,  as is shown in Tab.~\ref{tab:yield}, with
background levels well under control. Sample invariant mass distributions
for \BJKpm and  \BJKst\ are shown in Fig.~\ref{fig:sys_ctl}.

\begin{table}
\begin{center}
\begin{tabular}{| l || c| c| c|} \hline
Channel & ATLAS & CMS & LHCb \\ \hline
\BJKpm  & 1700k & 5100k & 880k \\
\BJKst  & \phantom{k}880k & 2900k & 800k \\
\BsDspi &  \phantom{00}3.5k &  \phantom{00}4.5k & 
\phantom{0}86k \\ \hline
\end{tabular}
\end{center}
\vspace*{-0.5cm}
\caption[]{Untagged annual event yields in selected control channels.
The ATLAS numbers assume a Level-2 trigger muon threshold 
of 3$\,$GeV~\protect\cite{ATtag}.}\label{tab:yield}
\end{table}

Here an approach is  presented which shows how any  flavour
dependent tagging effects and production asymmetries may be 
determined from these channels alone.  This is to demonstrate
the power of the available constraints.  
In practice it is envisaged that a combination of these channels,
Monte Carlo,  and detailed detector cross-checks will be used.
An example of the latter is the intention of LHCb to
take data sets with swapped dipole polarity,  thereby constraining
any charge-acceptance systematics.

\subsubsection{\BJKpm}

By reconstructing and flavour tagging \BJKpm\  
decays,  the tagging efficiencies
and mistag rates $\epsilon, \, \overline{\epsilon}, \, \omega$ and 
$\overline{\omega}$
may be directly measured.    The expected event yields enable this to be 
done with annual relative precision of a few $10^{-3}$ per experiment,  which
is certainly adequate for the CP asymmetry measurements.   These
factors can be 
determined in bins of tag-method, trigger-category, $p$, $p_T$ and
rapidity,  in
order to account for correlations.

Comparing the number of untagged $\rm J/\psi K^+$ and  $\rm J/\psi K^-$ events
gives sensitivity to the $\rm B^+/B^-$ production fractions $f^+/f^-$.
However,  what is generally of interest are the $\rm B^0_d$ and $\rm B^0_s$
quantities, $f^0_d/\overline{f^0_d}$ and   $f^0_s/\overline{f^0_s}$.
More importantly, any observed asymmetry may well receive contributions from
direct CP violation and detector effects,  and the decoupling of these factors
will be very difficult.  This motivates the use of other control channels.

\subsubsection{\BJKst}

The final state of the family of modes
$\rm B^0, \, \overline{B^0} \, \rightarrow \, J/\psi K^{0*}, \, J/\psi 
\overline{K^{0*}}$
is flavour specific to the meson at decay,  therefore enabling these
events 
to be used
in a similar manner to \BJKpm.   However, the oscillation of the
mesons 
before decay
provides additional observables which may be usefully exploited.

Consider the four decay rates 
$R^{\rm B^0,\overline{B^0} \rightarrow B^0,\overline{B^0}} (t)$
of genuine $\rm B^0$ and $\rm \overline{B^0}$ mesons into reconstructed 
$\rm B^0$ and $\rm \overline{B^0}$ final states:
$$
\begin{array}{c@{~\propto~}c@{\,;\quad}c@{~\propto~}c}
R^{\rm B^0 \rightarrow B^0}(t) & f_0 \, |A|^2 \, a(t) \, 
(1 + \cos \Delta m t) \, e^{-\Gamma t} &
R^{\rm \overline{B^0} \rightarrow \overline{B^0}}(t) & 
\overline{f_0} \, |\overline{A}|^2 \, \overline{a}(t) \, (1 + \cos
\Delta m t) \, e^{-\Gamma t};\\
R^{\rm B^0 \rightarrow \overline{B^0}}(t) & f_0 \, |\overline{A}|^2 \, 
\overline{a}(t) \, (1 - \cos \Delta m t) \, e^{-\Gamma t} & 
R^{\rm \overline{B^0} \rightarrow B^0}(t) &  
\overline{f_0} \, |A|^2 \, a(t) \, (1 - \cos \Delta m t) \, e^{-\Gamma t},
\end{array}
$$
where $|A|$ and $|\overline{A}|$ represent the absolute rates of the decays,
which may be different because of direct CP violation,  and $a(t)$ and 
$\overline{a}(t)$ are acceptance factors for the two final states.
Then the observed untagged decay distribution 
into $\rm B^0$ events, $R^{\rm X \rightarrow B^0 }(t)$, is: 
\begin{eqnarray}
R^{\rm X \rightarrow B^0}(t) & = & |A|^2a(t) \, (f_0 + \overline{f_0}) \,
\left[1 \, + \, \frac{f_0 - \overline{f_0}}{f_0 + \overline{f_0}} \, 
\cos\Delta m t \right]\,e^{-\Gamma t} , 
\end{eqnarray}
with the conjugated expression for $R^{\rm X \rightarrow \overline{B^0}}(t)$.
Therefore evidence of any oscillation term in the untagged rates signifies
an initial state production asymmetry, independent of CP violation and
detector effects.  Fitting this term enables the ratio $f_0/\overline{f_0}$ to
be determined.  

Information on the flavour dependence of the tagging efficiency can also be 
obtained.  The observed decay distribution for $\rm B^0$ mesons
of initial state flavour tagged $\rm B^0$ {\it and} $\rm
\overline{B^0}$  
events
is $R^{\rm X_{tag} \rightarrow B^0 }(t)$, where:
\begin{eqnarray}
R^{\rm X_{tag} \rightarrow B^0}(t) & = & |A|^2 a(t) \, (f_0 \epsilon + 
\overline{f_0} \overline{\epsilon}) \,
\left[1 \, + \, \frac{f_0 \epsilon - \overline{f_0}
    \overline{\epsilon} 
}{f_0 \epsilon + \overline{f_0} \overline{\epsilon}} \, 
\cos\Delta m t \right]\,e^{-\Gamma t} . 
\end{eqnarray}
Thus here, and in the charge conjugated case, fitting an oscillation
 amplitude
 to
the decay distribution enables the ratio
 $\overline{f_0}\overline{\epsilon} 
\,/ \,f_0\epsilon$ to
be determined.

Finally, there are four decay distributions for initial state tagged
$\rm  B^0, \, \overline{B^0}$ mesons decaying as $\rm  B^0, \,
\overline{B^0}$, denoted by
$R^{\rm B^0,\overline{B^0}_{tag} \rightarrow B^0\overline{B^0}}(t)$ with
\begin{equation}
R^{\rm B^0_{tag} \rightarrow B^0}(t)  =  
|A|^2 a(t) \, (f_0 \epsilon (1-\omega) \, + \, \overline{f_0} 
\overline{\epsilon} \overline{\omega}) \,
\left[1 \, + \, \frac{f_0 \epsilon (1-\omega) \, - \, \overline{f_0} 
\overline{\epsilon} \overline{\omega} }
{f_0 \epsilon (1-\omega) \, + \, \overline{f_0} \overline{\epsilon} 
\overline{\omega} } \,
\cos\Delta m t \right]\,e^{-\Gamma t} . 
\end{equation}
Fitting the oscillation amplitude for $R^{\rm B^0_{tag} \rightarrow B^0}(t)$
and $R^{\rm B^0_{tag} \rightarrow \overline{B^0}}(t)$ and using the previous
results enables $\overline{\omega} / (1-\omega)$ to be determined.
The other two distributions do the same for $\omega / (1-\overline{\omega})$.
{}From these results $\omega$ and $\overline{\omega}$ can be fixed.

These expressions show how the necessary correction factors can be
extracted from data. However, the arguments presented so far do not
account for any proper time dependence in the acceptance, which is
certainly not realistic.  If the time dependence is identical for $a(t)$ and
$\overline{a}(t)$, then the extractions are still possible,  as it will cancel
in the ratios of say, $R^{\rm X \rightarrow B^0}(t)$ and
$R^{\rm X \rightarrow \overline{B^0}}(t)$.  

A still more general approach is possible, which dispenses with 
any assumption on the proper time and flavour dependence of the 
acceptance.  Consider the ratio
\begin{eqnarray}
\frac{R^{\rm B^0_{tag} \rightarrow B^0}(t) / R^{\rm
    \overline{B^0_{tag}} 
\rightarrow B^0}(t)}
{R^{\rm B^0_{tag} \rightarrow  \overline{B^0}} (t) / 
R^{\rm \overline{B^0_{tag}} \rightarrow  \overline{B^0}} (t)}
& = &
\frac{ \left[1 + \frac{1-\overline{\eta}}{1+\overline{\eta}} \cos
    \Delta mt 
\right]
       \left[1 + \frac{1-\eta}{1+\eta} \cos \Delta mt \right] }
     { \left[1 - \frac{1-\eta}{1+\eta} \cos \Delta mt \right] 
       \left[1 - \frac{1-\overline{\eta}}{1+\overline{\eta}} \cos
    \Delta mt 
\right]  }\,,
\label{eq:etaexp}
\end{eqnarray}
where $\eta$ is given as $\epsilon \omega f_0 / 
\overline{\epsilon} (1-\overline{\omega}) \overline{f_0}$,
and $\overline{\eta}$ is the conjugated expression.   These factors
may be 
simultaneously 
fitted and combined with the \BJKpm\ 
results to extract $f_0/\overline{f_0}$.  
Alternatively, they may be used directly to extract $\sin 2 \beta$ from
the 
\BJK\ decay rates.
Rather than constructing the conventional CP asymmetry, the ratio of
the 
$\rm B^0$ tagged
and $\rm \overline{B^0}$ tagged decays may be formed:
\begin{eqnarray}
\frac{R^{\rm B^0_{tag} \rightarrow J/\psi K^s_0} (t)}
{R^{\rm \overline{B^0_{tag}}\rightarrow J/\psi K^s_0} (t)} & = &
K \left[  \frac{1 - (\frac{1 -\overline{\eta}}{1 + \overline{\eta}})
    \sin 2 
\beta \sin \Delta m t } 
{1 + (\frac{1 - \eta}{1 + \eta}) \sin 2 \beta \sin \Delta m t } 
\right],
\nonumber
\end{eqnarray}
where $K$ is a normalisation factor and $\eta, \overline{\eta}$ are the
factors 
determined
from (\ref{eq:etaexp}).   With this method, $\sin 2 \beta$ can be
cleanly 
determined,
although the need to also fit $K$ reduces the statistical precision with
respect 
to the conventional 
approach.

\subsubsection{\BsDspi}

In controlling tagging systematics in $\rm B^0_s$ measurements,  the values of
$\epsilon$, $\overline{\epsilon}$, $\omega$ and $\overline{\omega}$ measured
in the  $\rm B^0_d$ channels may be used.  However, constraints are 
required on the production ratio $f_s/\overline{f_s}$.   Here it is 
impracticable
to use $\rm J/\psi K$ channels,  as these are suppressed with respect to the
$\rm B^0_d$ case.   Rather it is preferable to use the decay \BsDspi,  where
no CP violation is expected.   Attention must be given to detector acceptance
effects in the final state,  but it should prove possible to control these
to the level required by the precision of $\rm B^0_s$ measurements.

\subsection[Application to the $\rm B_d \rightarrow J/\psi K^0_S$
Sample]{Application to the \protect\boldmath $\rm B_d \rightarrow J/\psi K^0_S$
Sample}

\begin{table}
\begin{center}
\begin{tabular}{|c|c|c|}
\hline
  Measurement & \BJKpm &  \BJKst  \\
\hline
\hline
 $\delta (f_0-\overline{f_0}/f_0+\overline{f_0})$     & 0.05\%   & 0.07\%   \\
\hline
  $\delta D/D$ (Lepton Tagging)   & 0.0038  & 0.0047  \\
\hline
   $\delta D/D$ ($B$--$\pi$ Tagging) & 0.0030  & 0.0039  \\
\hline
\end{tabular}
\end{center}
\vspace*{-0.5cm}
\caption{Estimated ATLAS uncertainties on the determination of the production
asymmetry, $f_0-\overline{f_0}/f_0+\overline{f_0}$, and of the
dilution, $D\,=\,1-2\omega$, 
for lepton tagging and $B$--$\pi$ tagging 
using \BJKpm\  and   \BJKst\ control samples, after 3 years
of running.}
\label{tab:sys_tab}
\end{table}

To give a quantitative impression of the precision expected 
from the control channels, table~\ref{tab:sys_tab} shows
the results of an ATLAS study into the expected uncertainties 
after 3 years operation on the $B_d^0 - \overline{B}_d^0$ production asymmetry,
$f_0-\overline{f_0}/f_0+\overline{f_0}$, and the
tagging dilution, $D\,=\,1-2\omega$.  $D$ has been evaluated
separately for lepton tagging and  
$B$--$\pi$ correlation tagging (see Sec.~\ref{exp_ft})~\cite{ATtag}.
Uncertainties have been calculated with both the
\BJKpm\ and the \BJKst\ samples. 
The study has been done in the context of the $B^0_d \rightarrow
J/\psi K^0_s$ analysis (leading to the estimate of the systematic
uncertainty on the $\sin{2\beta}$ measurement given in
Sec.~\ref{subsec:BdpsiKS}), 
but the results are more general.  The errors are small compared
to the expected statistical uncertainty of the  $\sin{2\beta}$.

\subsection{Other Measurements and Conclusions}

The discussion so far has focused on \BJK, since this is a very
important measurement, with an excellent statistical precision expected. 
However there are other classes of measurement planned for the LHC:

\begin{itemize}
\item{{\bf Asymmetries involving decays to non-CP eigenstates} \\
Measurements such as the determination of $\gamma$ from \BDstarpi\ 
involve the comparison of four different decay rates, as explained in 
Sec.~\ref{sec:bdstarpi}.
Although there are two final states which may have different
acceptances,  due to detector-charge effects,  the asymmetries
which are formed to extract the physics unknowns do not compare these states.
Therefore charge acceptance effects will not bias the measurement.
Information on tagging factors and production asymmetries is obtained
from the usual control channels.}
\item{{\bf Branching ratio comparisons} \\
Methods such as the $\rm B^0_d \rightarrow \pi K$ strategies 
to determine $\gamma$,
described in Sec.~\ref{sec:bdpik},  rely on the comparison of 
several branching ratios.   Here it is necessary to know well
the relative reconstruction efficiencies, in particular
the contribution of the trigger.  Although challenging,
this should prove possible at a  level  which will
be adequate alongside the statistical and theoretical
uncertainties.}
\end{itemize}
It can be concluded that there is no a priori reason why
tagging related biases, production asymmetries or detector
effects should prevent the experiments from properly exploiting the
enormous B statistics at the LHC.

\setcounter{equation}{0}
\newcommand{\bb}{B--$\bar{\mbox{B}}$}

\section[B--$\bar{\mbox{B}}$ MIXING]{B--$\bar{\mbox{B}}$ 
MIXING\protect\footnote{Section coordinators:
G. Buchalla, L. Lellouch and P. Vikas with help from V. Ghete,
O. Schneider and A. Starodumov.}}\label{sec:mix}

The physics of \bb\  mixing is of prime importance for the
study of flavour dynamics.  Today, the experimental information on
$B_d$ and $B_s$ mixing, i.e.\ the mass
differences $\Delta M_d$ and $\Delta M_s$, implies already significant
constraints on the unitarity triangle.  A precise measurement of
$\Delta M_s$, for which only a lower limit exists so far, will be an
invaluable piece of information on the flavour sector of either the
SM or its possible extension. Even if $\Delta M_s$ is
measured before, LHC's B physics capabilities are likely
to remain indispensable to fully exploit the potential of \bb\ 
mixing.  In addition to $\Delta M_s$, also the lifetime difference
$\Delta\Gamma_s$ provides us with interesting opportunities.  The
measurement of this quantity is likewise very difficult and will be a
suitable goal for the LHC B physics programme.

The main theory input needed is, on the one hand, perturbative QCD
corrections and, on the other hand, hadronic matrix elements of
four-quark operators, schematically
$$
\langle B_q \mid (\bar q \Gamma b)(\bar q \Gamma' b) \mid \bar
B_q\rangle,
$$
where $\Gamma$, $\Gamma'$ stand for the relevant combinations of Dirac
matrices and $q\in \{s,d\}$.  Whereas the perturbative terms are known
to NLO in QCD \cite{BJW,DGamma-cal1}, 
hadronic matrix elements can be obtained
from first principles using lattice QCD and we start this section by
an overview of the relevant lattice results.  We then discuss
specifically the mass and width difference $\Delta M_q$ and
$\Delta\Gamma_q$ of the $B_q$ system and give
predictions for the expected ranges of $\Delta M_s$ and
$\Delta\Gamma_s$ in the SM.  The section concludes with experimental
considerations on the measurement of $B_s^0$ oscillations at the LHC.

The numerical results presented in this section are obtained using the
following input parameters:
\begin{equation}
m_b=4.8\,\mbox{GeV},\quad \bar m_b(m_b)=4.4\,\mbox{GeV},\quad \bar
m_s(m_b)=0.1\,\mbox{GeV},
\quad \bar m_t(m_t) = 167\,\mbox{GeV},
\label{inparams}
\end{equation}
$$
M_B=5.28\,\mbox{GeV},\quad M_{B_s}=5.37\,\mbox{GeV},\quad B(B_s\to Xe\nu)=0.104
\ ,
$$
and the two-loop expression for $\alpha_s$ with 
$\Lambda^{(5)}_{\overline{\mbox{\scriptsize 
MS}}}=225\,\mbox{MeV}$. Above, $m_b$ is the pole mass and
the barred masses refer to the $\overline{\mbox{MS}}$ scheme.

\subsection{Hadronic Matrix Elements from Lattice
  Calculations}\label{sub:lattice} 

The matrix elements relevant for B mixing are
\begin{eqnarray}\label{mebbq}
\langle B_q\mid(\bar qb)_{V-A}(\bar qb)_{V-A}\mid\bar B_q\rangle
&\equiv& \frac{8}{3} B_{B_q}(\mu) f^2_{B_q} M^2_{B_q}\,,\\
\langle B_s\mid(\bar qb)_{S+P}(\bar qb)_{S+P}\mid\bar B_s\rangle &=&
-\frac{5}{3}\, \frac{M^2_{B_s}B_S(\mu)}{(\bar m_b(\mu)+\bar m_s(\mu))^2}\,
f^2_{B_s}M^2_{B_s},\label{bbs}\\
\langle 0 \mid \bar q \gamma_\mu \gamma_5 b \mid 
\bar B_q\rangle & = & i f_{B_q} p_\mu,
\end{eqnarray}
which are parametrized in terms of the leptonic decay constants
$f_{B_q}$ and the B-parameters $B_{(B_q,S)}(\mu)$. Instead of the
scale- and scheme-dependent parameter $B_{B_q}$, one usually
introduces the renormalization-group invariant parameter
$\hat{B}_{B_q}$, which to NLO in QCD is given by \cite{BJW,BBL-rev}
\begin{equation}\label{bbqinv}
{\hat B}^{\rm nlo}_{B_q}=B_{B_q}(\mu)[\alpha_s(\mu)]^{-6/23}
\left[ 1+\frac{\alpha_s(\mu)}{4\pi}J_5\right],\quad 
J_5 =\frac{5165}{3174}\quad\mbox{(NDR scheme)}.
\end{equation}
While the matrix elements (\ref{mebbq}) and (\ref{bbs}) can be
determined as such on the lattice, the dimensionless quantities
$B_{B_q}$ and $M^2_{B_s}B_S/(\bar m_b+\bar m_s)^2$ are obtained from
ratios of Euclidean correlation functions in which many statistical
and systematic uncertainties are expected to cancel. Thus, it is
advantageous to get the matrix elements from an independent
determination of the above quantities and $f_{B_q}$, combined with the
experimental value of $M_{B_q}$.

Because the $b$ quark with mass $m_b\sim 5\,\mbox{GeV}$ has a compton
wavelength that is not large compared to typical (quenched) lattice
spacings, $a\sim (2-4)\,\mbox{GeV}^{-1}$, it cannot be simulated
directly as a relativistic quark on present day lattices. This has led
to a variety of approaches for studying hadrons composed of a heavy
quark and light degrees of freedom. In the {\em relativistic}
approach, calculations are performed with a discretisation of the
relativistic Dirac action, for  heavy quarks with masses around that
of the
charm and extrapolated in mass up to $m_b$, using heavy
quark effective theory as a guide. There are also effective theory
approaches, in which QCD is expanded in inverse powers of the
b quark mass. Of these, there is the {\em static-quark} approach, in
which the heavy quark is treated as an infinite-mass, spin-1/2, static
source of colour; a variant of this approach, in which a number of
leading $1/m_b$-corrections to the static limit are included in the
action, goes under the name of {\em non-relativistic QCD} or {\em
NRQCD}. Finally, there is a {\em hybrid} approach in which results,
calculated at $m_b$ with a relativistic action, are given a
non-relativistic interpretation. While we favour the {\em relativistic}
approach, which does not suffer from the typical ills of effective
theories (operator proliferation and power divergences when
higher-order corrections are taken into account), the different
approaches should be viewed as complementary and any significant
disagreement amongst them should be understood.

An important source of uncertainty in many present day lattice
calculations is the quenched approximation ($N_f=0$), in which the
feedback of quarks on the gauge fields is neglected. More and more,
though, groups are doing away with this approximation and are
performing full QCD calculations with 2 flavours of sea quarks
($N_f=2$), usually with masses around that of the strange quark. Even
then, there is some way to go to reach our physical world where there
are $N_f=3$ light sea quarks: the two very light up and down quarks,
and the more massive strange quark.

Because this is not the place for a full fledged review, we will only very
rarely quote individual results and rather give summary numbers, which
are meant to reflect the present state of lattice calculations. The
results taken into account are those obtained as of January 2000, most
of which are referenced in one of the reviews 
in Ref.~\cite{Aoki:1999ue}.

\subsubsection{Leptonic Decay Constants}

Lattice calculations of the leptonic decay constants $f_{B_q}$ have a
long history and results obtained in the quenched approximation with
the different approaches to heavy quarks described above are gradually
converging. The dominant systematic errors (quenching aside)
depend on the approach used, but they are typically of the order of
10\%.

In the past year or two, a number of groups have begun studying the
effect of unquenching on decay constants by performing $N_f=2$
calculations with a variety of approaches to heavy quarks. While these
calculations are still in rather early stages, and should therefore be
given time to mature, they nevertheless suggest an ${\mathcal
O}(10{-}20\%)$ increase in $f_{B_q}$. $f_{B_s}/f_B$, however, appears
to change very little, indicating that theoretical uncertainties,
including the effects of quenching, cancel in such SU(3)-breaking
ratios.
Because systematic errors depend on the approach and parameters used,
it is difficult to combine systematically results from different
groups. We therefore choose to give, in Tab.~\ref{fbtab}, summary
numbers for the quenched and unquenched decay constants which are
meant to reflect the present situation. 
\begin{table}
\begin{center}
\begin{tabular}{|c||c|c|}
\hline
Quantity & $N_f=0$ & $N_f=2$\\
\hline
$f_B\,\,(\mbox{MeV})$ & $175\pm 20$ & $200\pm 30$\\
$f_{B_s}\,\,(\mbox{MeV})$ & $200\pm 20$ & $230\pm 30$\\
$f_{B_s}/f_B$ & $1.14\pm 0.05$ & $1.15\pm 0.07$ \\
\hline
\end{tabular}
\end{center}
\vspace*{-0.5cm}
\caption[]{Summary of the results for leptonic decay constants of B
mesons from lattice QCD in the quenched ($N_f=0$) approximation and with two
flavours of sea quarks ($N_f=2$). It is evident that the values for
$f_{B_q}$ are sensitive to quenching effects, whereas their ratio is 
not.}\label{fbtab}
\end{table}

Because a final number is needed for phenomenological purposes, 
we provide the following summary of the summaries, taking into
account the fact that the unquenched results are still rather
preliminary and correspond to $N_f=2$:
\begin{equation}
f_B=(200\pm 40)\,\mbox{MeV},\quad f_{B_s}=(230\pm 40)\,\mbox{MeV}\quad
\mbox{and}\quad
\frac{f_{B_s}}{f_B}=1.15\pm 0.07
\ .\label{fbsum}
\end{equation}
These are the values of the decay constants to be used for numerical
estimates in the subsequent subsections.  The errors will certainly
come down significantly once the unquenched calculations mature.

\subsubsection{B-Parameters for $\Delta M$}

The lattice calculation of these B-parameters is less mature than
that of leptonic decay constants. None\-the\-less, there have been a
number of calculations over the years. 

Agreement amongst calculations using the relativistic approach is
good, and recent work at different values of the lattice spacing
\cite{Bernard:1998dg,lat1} indicates that discretization errors are
small in this approach. Agreement with the NRQCD calculation of
Ref.~\cite{Hashimoto:1999ck} is less good. However, in matching
the lattice results to $\overline{\mbox{MS}}$, the authors use the
one-loop static instead of NRQCD coefficients, thereby inducing large
systematic uncertainties.  Thus, until the NRQCD results are
finalised, we choose to use the relativistic results to establish our
summary numbers for B-parameters. In any case, all methods predict that
$B_{B_s}/B_B$ is very close to one.

An effect that has not yet been addressed in B-parameter
calculations is the error associated with the quenched approximation:
there exist no unquenched calculations of $B_{B_q}$ to date. However,
because these parameters correspond to ratios of rather similar matrix
elements, their errors are expected to be smaller than those
of decay constants. 

Compiling the relativistic results, we give for the
B-parameters:
\begin{equation}
B_{B_q}(m_b)=0.91\pm0.06,\quad\hat B_{B_q}^{\mathrm{nlo}}=1.40\pm0.09
\quad\quad\mbox{and}\quad\quad\frac{B_{B_s}}{B_B}=1.00(3)
\ ,
\label{bparsum}
\end{equation}
where we do not distinguish $q=d$ from $q=s$. The
renormalization group invariant
parameter $\hat B_{B_q}^{\mathrm{nlo}}$ is obtained from
$B_{B_q}(m_b)$ using (\ref{bbqinv}) with the input parameters of
(\ref{inparams}).

The theoretical determination $\Delta M_s/\Delta M_d$ requires calculation
of the non-perturbative parameter $R_{sd}$ (or $\xi$), defined through:
\begin{equation}
\frac{\Delta M_s}{\Delta M_d}=\left|\frac{V_{ts}}{V_{td}}\right|^2 
R_{sd}
=\left|\frac{V_{ts}}{V_{td}}\right|^2 \left(\frac{M_{B_s}}{M_{B_d}}\right)\xi^2
\label{rsddef}
\ .\end{equation}
While there are at least two possible ways of obtaining $R_{sd}$
from the lattice, the most accurate and most reliable, at present, 
is via:
\begin{equation}
R_{sd}\equiv \left(\frac{M_{B_s}}{M_B}\right)\left(
\frac{f_{B_s}}{f_{B}}\right)^2
\left(\frac{B_{B_s}}{B_{B}}\right)
\ ,
\end{equation}
with $\left(f_{B_s}/f_{B}\right)$ and $\left(B_{B_s}/B_{B}\right)$ 
determined on the
lattice and $\left(M_{B_s}/M_B\right)$ measured experimentally.  The
 different approaches have been explored using relativistic quarks by
 two groups
\cite{Bernard:1998dg,lat1}.

Because the results obtained by these groups are fully compatible with
the value of $R_{sd}$ obtained using the results (\ref{fbsum}) and
(\ref{bparsum}), we quote the latter value as our summary number:
\begin{equation}\label{rsdnum}
R_{sd} = 1.35(17)\quad\quad\mbox{or}\quad\quad \xi\equiv \sqrt{R_{sd}
\left(\frac{M_B}{M_{B_s}}\right)}=1.15(7)
\ .\end{equation}

\subsubsection{B-Parameter for $\Delta\Gamma_s$}

No complete calculation of $m^2_{B_s}B_S/(\bar m_b+\bar m_s)^2$ in
(\ref{bbs}) exists to date. There has been one calculation performed
within the relativistic approach, but with only a single heavy quark
whose mass is close to that of the charm \cite{Gupta:1997yt}. There is
also an NRQCD calculation, but where the matching of the lattice to
$\overline{\mbox{MS}}$ is performed using the one-loop static instead
of NRQCD coefficients \cite{DGamma-cal2}. Both are quenched.

The two results are, respectively:
\begin{equation}
\frac{M^2_{B_s}B_S(m_b)}{(\bar m_b(m_b)+\bar m_s(m_b))^2}=1.07(1)
\quad\mbox{and}\quad 1.54(3)(24)
\ ,\end{equation}
where the first was obtained from \cite{Gupta:1997yt} using the
conversion of \cite{DGamma-cal1} and the masses in (\ref{inparams}).
Both these numbers should be considered preliminary, though the second
does include an estimate of systematic errors. So, for the moment, we take
\begin{equation} \frac{M^2_{B_s}B_S(m_b)}{(\bar m_b(m_b)+\bar
    m_s(m_b))^2}= 1.4(4). \label{bssum}\end{equation}
The near future, however, should bring new results.

\subsection[The Mass Difference 
$\Delta M$]{The Mass Difference \protect\boldmath $\Delta M$}\label{sec:dMS}

In the SM the $B_q$ mass difference,
calculated from box diagrams with virtual top exchange, is given
by
\begin{equation}\label{delmq}
\Delta M_q =\frac{G^2_F M^2_W}{6\pi^2}\, \eta_B S_0(x_t)\,
M_{B_q} {\hat B}_{B_q} f^2_{B_q}\, |V_{tq}|^2.
\end{equation}
Here $S_0(x_t)$, where $x_t=\bar{m}^2_t/M^2_W$, is the top-quark mass
dependent Inami-Lim function for \bb\ mixing. 
To an accuracy of better than 1\%, 
$S_0(x_t)\simeq 0.784 x_t^{0.76}$.
$\eta_B$ is a correction factor describing short-distance 
QCD effects. It has been calculated at next-to-leading order
in \cite{BJW}. With the definition of ${\hat B}_{B_q}$ in (\ref{bbqinv}),
and employing the running mass $\bar m_t(m_t)$ in $S_0(x_t)$,
the numerical value is $\eta_B=0.55$ (with negligible uncertainty).
Note that $\eta_B$, being a short-distance quantity, is independent
of the flavour content of the B meson: it is identical for
$B_d$ and $B_s$. The dependence on the light-quark flavour
$q=d$, $s$ belongs to the non-perturbative, long-distance effects, which are
isolated in the matrix element (\ref{mebbq}) \cite{BJW,BBL-rev}.

Experimentally, $\Delta M_q$ can be measured from flavour oscillations
of neutral $B_q$ mesons. The current world average is given by
\cite{LEPBOSC}
\begin{equation}\label{dmd}
\Delta M_d = (0.476\pm 0.016)\, \mbox{ps}^{-1},\quad
\Delta M_s > 14.3\, \mbox{ps}^{-1} \,\, @\, 95\%\, \mbox{CL}.
\end{equation}
The measurement of $\Delta M_d$ can be used to
constrain $|V_{td}|$ via (\ref{delmq}). While the short-distance
quantity $\eta_B S_0(x_t)$ is known very precisely, large uncertainties
are still present in the hadronic matrix element $B_{B_q} f^2_{B_q}$.
Numerically,
\begin{equation}\label{mdsum}
|V_{td}|=7.36\times 10^{-3}
\left[\frac{167\,{\rm GeV}}{\bar m_t(m_t)}\right]^{0.76}
\left[\frac{237\,{\rm MeV}}{f_{B_d}\sqrt{\hat B^{\rm nlo}_{B_d}}}\right]
\left[\frac{\Delta M_d}{0.476\,{\rm ps}^{-1}}\right]^{0.5}.
\end{equation}
The theoretical uncertainties are reduced considerably in the
ratio $\Delta M_s/\Delta M_d$, as given in (\ref{rsddef}).
With the results (\ref{dmd}), an upper limit
on $|V_{td}/V_{ts}|$ can be inferred from (\ref{rsddef}). This
limit already represents a very interesting CKM constraint, which 
disfavours negative values of the Wolfenstein parameter $\varrho$.
A future precision measurement of $\Delta M_s$ will be a crucial
input for the phenomenology of quark mixing.
Using $|V_{td}/V_{ts}| > 0.17$ \cite{CP-revs2} and
Eqs.\ (\ref{rsddef}), (\ref{rsdnum}), (\ref{dmd}), we
find a SM prediction of
\begin{equation}\label{smdms}
\Delta M_s=(14.3\, \mbox{--}\, 26)\,\mbox{ps}^{-1}.
\end{equation}

\subsection[The Width Difference $\Delta\Gamma$]{The Width Difference 
\protect\boldmath $\Delta\Gamma$}

$(\Delta\Gamma/\Gamma)_{B_s}$ is expected to be one
of the largest rate differences in the $b$ hadron sector,\footnote{The width
difference in the $B_d$ system is Cabibbo suppressed. We thus only
consider the $B_s$ sector.}
with typical size of (10--20)\% \cite{BIG1,BENEKE}.
The measurement of a substantial $(\Delta\Gamma/\Gamma)_{B_s}$
would open new possibilities for CP violation studies with
untagged $B_s$ mesons \cite{dun,FD2,FD1}. 
Numerically, one has, using NLO coefficients \cite{DGamma-cal1}:
\begin{equation}\label{dggnum}
\left(\frac{|\Delta\Gamma|}{\Gamma}\right)_{B_s}=
\left(\frac{f_{B_s}}{230~{\rm MeV}}\right)^2
\left[0.007\, B(m_b)
 +0.132\, \frac{M^2_{B_s} B_S(m_b)}{(\bar m_b(m_b)+ \bar m_s(m_b))^2}
 - 0.078\right]
=  0.11(7)
\end{equation}
with the B-parameters as discussed in Sec.~\ref{sub:lattice}.
Note that the B-parameters are to be
taken in the NDR scheme as defined in \cite{DGamma-cal1}.
The last term in (\ref{dggnum}), --0.078, represents
$1/m_b$ corrections \cite{BENEKE} and has a relative 
uncertainty of at least 20\%. An additional 30\% scale-ambiguity
from perturbation theory has not been displayed in (\ref{dggnum}).

\subsection[Measurement of $B_s^0$ 
Oscillations]{Measurement of \protect\boldmath $B_s^0$ Oscillations}

The probability density to observe an initial $B_s^0$ meson decaying
as a $\overline{B_s^0}$ meson at time $t$ after its creation is given by:
\begin{equation}\label{P_osc}
P_{B_s^0 \rightarrow \overline{B_s^0}}(t) = 
                 \frac{\Gamma_s^2-(\Delta\Gamma_s/2)^2}{2\Gamma_s}
  e^{-\Gamma_s t}\left[\cosh\frac{\Delta\Gamma_s t}{2}+
\mu\cos (\Delta M_s t)\right],
\end{equation}
where $\mu = -1$, $\Delta\Gamma_s = \Gamma_H - \Gamma_L$ and 
$\Gamma_s=(\Gamma_H + \Gamma_L)/2$. If the initial $B_s^0$ meson decays as
a $B_s^0$ at time $t$, the probability density 
${P_{B_s^0 \rightarrow B_s^0}}$ is given
by the above expression with $\mu = +1$.  
Experimentally, $\Delta M_s$ can be determined by measuring the following
time-dependent asymmetry:
\begin{equation} \label{exasy}
A(t) = \frac{P_{B_s^0 \rightarrow B_s^0\vphantom{\overline{B_s^0}
}}(t)-P_{B_s^0 \rightarrow \overline{B_s^0}}(t)}
            {P_{\vphantom{\overline{B_s^0}}
B_s^0 \rightarrow B_s^0}(t)+P_{B_s^0 \rightarrow \overline{B_s^0}}(t)}
=
       \frac{\cos (\Delta M_s t)}
            {\cosh \frac{\Delta \Gamma_s t}{2}}.
\end{equation}
The mass difference $\Delta M_s$ is $2\pi$ times the
oscillation frequency. Within the SM, one
has, using the formulae of \cite{DGamma-cal1} and the matrix elements of
Sec.~\ref{sub:lattice},\footnote{Note that according to the 
sign convention
  used in this report, (\ref{delm}), $\Delta\Gamma_s$ is negative in
  the SM.} suppressing a 30\% renormalization scale-uncertainty,
\begin{equation}\label{eq:7.20}
\frac{|\Delta \Gamma_s|}{\Delta M_s} = (4.3\pm2.0)\times 10^{-3},
\end{equation}
which is independent of uncertainties due to CKM matrix elements. 
It has mainly hadronic uncertainties which are expected to decrease in the
future. Therefore, within the SM, $\Delta M_s$ can in principle be
inferred from a direct measurement of $\Delta\Gamma_s$, although with a large
error. Small values of $\Delta \Gamma_s$ and large values of $\Delta M_s$
are difficult to measure. However, Eq.~(\ref{eq:7.20}) implies that the smaller
$\Delta \Gamma_s$ is, the easier it should be to measure $\Delta M_s$, and,
inversely, the larger $\Delta M_s$ is, the easier it should be to measure
$\Delta \Gamma_s$.
 
\begin{table}[tb]
\renewcommand{\arraystretch}{0.9}
\begin{center}
\begin{tabular}{|r||c|c|c|}
\hline
& ATLAS & CMS & LHCb \\
\hline\hline
\multicolumn{4}{l}{Channels used:} \\
$B_s^0$ decay channels          &$D_s^- \pi^+$ &$D_s^- \pi^+$ &$D_s^- \pi^+$\\ 
                             &$D_s^- a_1^+$ &              &            \\
$D_s^-$ decay channels       &$\phi\pi^-$
                             & $\phi\pi^-$  &$\phi\pi^-$                \\ 
                             &              & $K^{*0} K^-$ & (see text) \\ 
$\phi$ decay channel         &$ K^+K^-$     & $K^+K^-$     &$ K^+K^-$   \\ 
$a_1^+$ decay channel        &$\rho^0 \pi^+$&              &            \\ 
$K^{*0}$ decay channel       &              & $K^+ \pi^-$  &            \\
\hline
\multicolumn{4}{l}{Assumptions:} \\
$B(\bar{b}\to B^0_s$)     & 0.105      & 0.105       & 0.12    \\
$B(B^0_s \to D^-_s \pi^+)$ & 3.0$\,\times\,$10$^{-3}$ 
                                 & 3.0$\,\times\,$10$^{-3}$            
                                 & 3.0$\,\times\,$10$^{-3}$ \\
$B(B^0_s \to D^-_s a^+_1)$ & 6.0$\,\times\,$10$^{-3}$    & -- & -- \\
$B(D^-_s \to \phi \pi^-)$ & 0.036      & 0.036      &      \\
$B(D^-_s \to K^{*0} K^-)$ & --         & 0.033 & --      \\
$B(D^-_s \to K^+ K^-\pi^-)$ & --         & -- & 0.04      \\
$B_s^0$ lifetime                    & 1.54 ps    & 1.61 ps    & 1.57 ps \\
\hline
\multicolumn{4}{l}{Analysis performance:} \\
Reconstructed signal events per year     & 3457 & 4500 & 86000 \\
Rec.\ and tagged signal events per year  & 3457 & 4500 & 34500 \\
$B_s^0$ purity of tagged sample             & 0.38 & 0.5  & 0.95 \\
Wrong tag probability                    & 0.22 & 0.22 & 0.30 \\
Proper time resolution(Gaussian function(s))& 50 fs (60.5\%) & 65 fs & 43 fs \\
                                         & 93 fs (39.5\%) &       &       \\
\hline
\multicolumn{4}{l}{$\Delta M_s$ reach after one year of running:} \\
Measurable values of $\Delta M_s$ 
 up to & 30$\,{\rm \mbox{ps}}^{-1}$ & 
26$\,\mbox{ps}^{-1}$ & 48$\,\mbox{ps}^{-1}$ \\
95\% CL excl.\ of $\Delta M_s$ 
values up to    & --        & 29$\,\mbox{ps}^{-1}$ & 58$\,\mbox{ps}^{-1}$ \\
$\sigma(\Delta m_s)$ for $\Delta m_s=20\,$ps$^{-1}$ & 0.11 
                            & -- & 0.011 \\ 
\hline
\multicolumn{4}{l}{$x_s$ reach after one year of running:} \\
Measurable values of $x_s$ up to & 46  & 42 & 75 \\
95\% CL excl.\ of $x_s$ values    up to & --  & 47 & 91 \\
\hline
\end{tabular}
\end{center}
\renewcommand{\arraystretch}{1}
\vspace*{-0.5cm}
\caption[]{Summary of the analyses and results for $B_s^0$ oscillation 
         frequency measurements
         by the LHC experiments.}\label{tab:exp_bmix}
\end{table}

\begin{figure}
   \vspace{0cm}
\epsfxsize=7cm
   \centerline{\epsffile{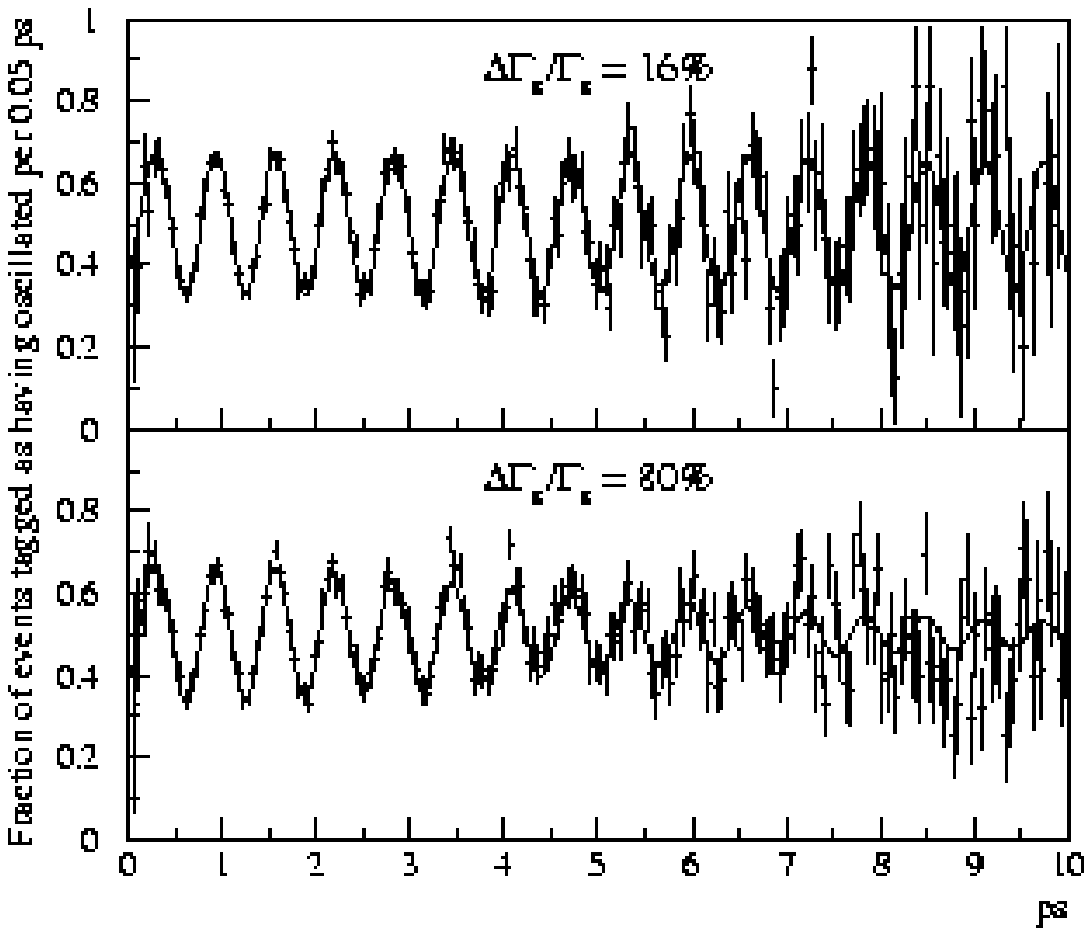}}
   \vspace*{-0.6cm}
\caption[]{Fraction of events tagged as having oscillated as a function
         of proper time for two different values of 
         $\Delta \Gamma_s /\Gamma_s$, for 
         $\Delta M_s =10$ ps$^{-1}$~\cite{LHCbTP}.
         The curves display the result of the maximum likelihood fit 
         to the total sample.}\label{fig:lhcb_1}

~\\[-0.3cm]

\begin{minipage}[t]{0.47\textwidth}
   \epsfxsize=0.9\textwidth
   \centerline{\epsffile{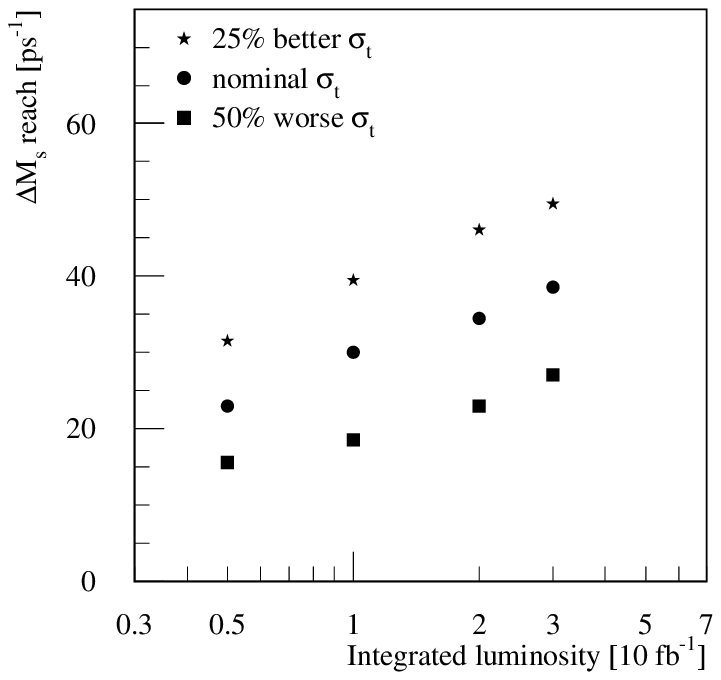}}
\vspace*{-0.8cm}
\caption[]{$\Delta M_s$ reach of ATLAS as a function of the integrated 
           luminosity for various
           proper time resolutions $\sigma_t$.}\label{fig:atlasbosc1}
\end{minipage}
\hspace*{0.5cm}
\begin{minipage}[t]{0.47\textwidth}
\epsfxsize=0.9\textwidth
   \centerline{\epsffile{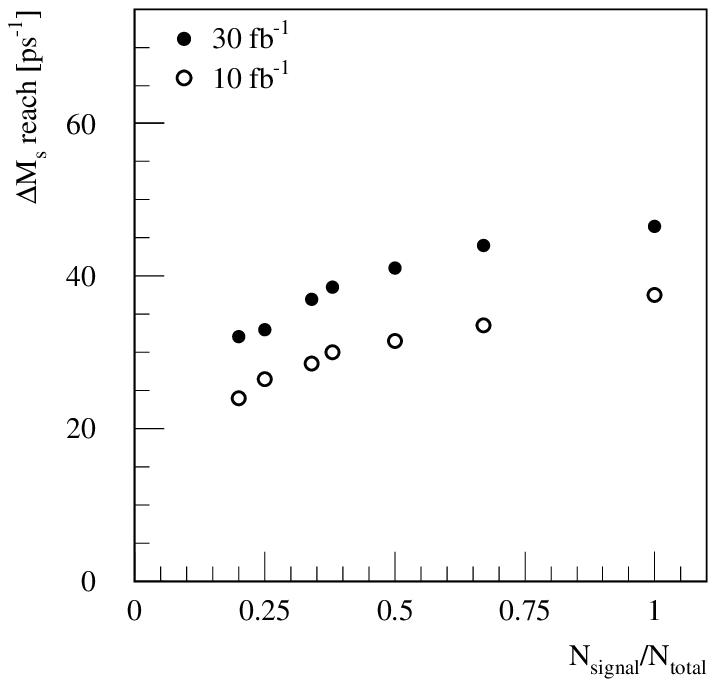}}
   \vspace*{-0.8cm}
\caption[]{$\Delta M_s$ reach of ATLAS as a function of the signal 
           content of the sample for
           nominal proper time resolution and integrated luminosities of
           10 fb$^{-1}$ and 30 fb$^{-1}$.}\label{fig:atlasbosc2}
\end{minipage}

\begin{minipage}[t]{0.47\textwidth}
   \epsfxsize=0.9\textwidth
   \centerline{\epsffile{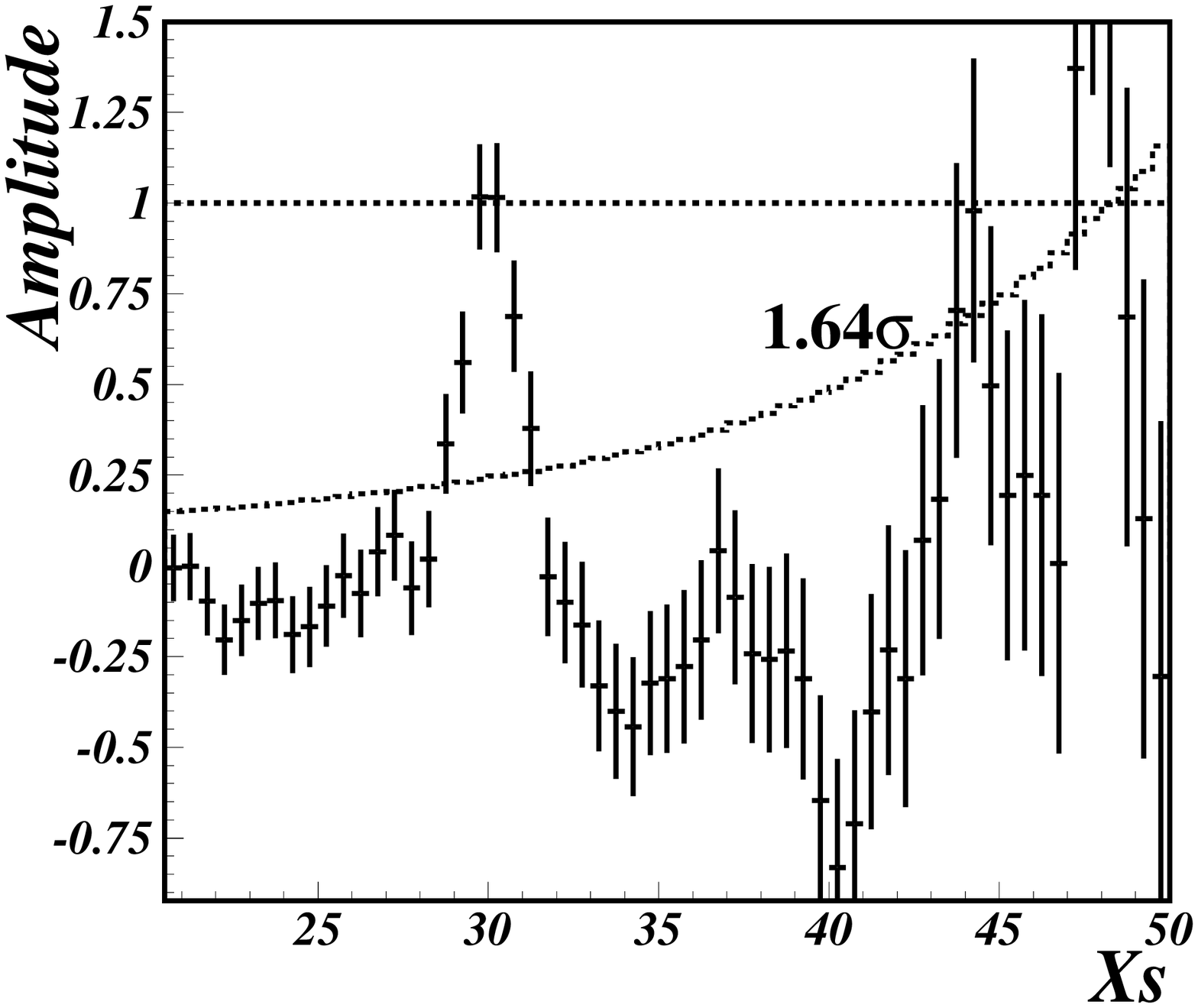}}
\vspace*{-0.5cm}
\caption[]{The amplitude with its error, $\sigma_A$, for an input
           value of $x_s=30$ from CMS.}\label{fig:cmsbosc}
\end{minipage}
\hspace*{0.5cm}
\begin{minipage}[t]{0.47\textwidth}
\epsfxsize=0.9\textwidth
   \centerline{\epsffile{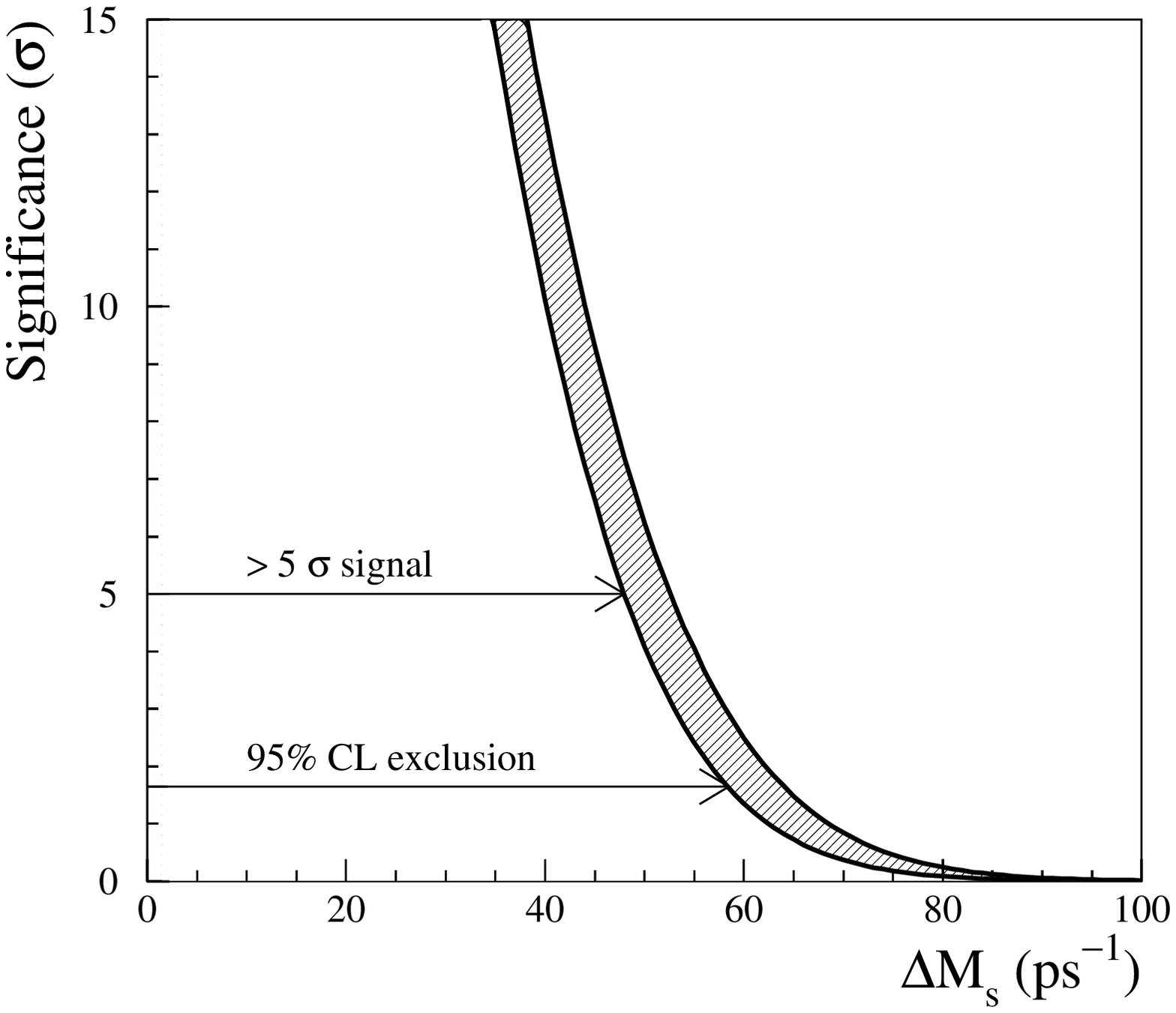}}
   \vspace*{-0.5cm}
\caption[]{Statistical significance of the $B_s^0$ oscillation signal
           as a function of $\Delta M_s$. The band delimited by the two
           curves reflects the $1\sigma$ statistical uncertainty on the
         proper-time resolution of $\sigma_t =
           (\mbox{43}\pm\mbox{2})$fs.}\label{fig:lhcb_2}
\end{minipage}
\end{figure}

The effect of $\Delta \Gamma_s$ being non-zero is to damp the
$B_s^0$ oscillations with a time-dependent factor. Figure~\ref{fig:lhcb_1} 
 shows the proper time
distributions of $B_s^0\rightarrow D_s^- \pi^+$ candidates generated with
two different values of $\Delta\Gamma_s$~\cite{LHCbTP}. The curves display 
the result of a
maximum likelihood fit to the total sample. The damping of the $B_s^0$ 
oscillations due to $\Delta\Gamma_s /\Gamma_s$ is not significant at the
expected value of 16\%, but could be important  
if $\Delta\Gamma_s$ turns out to be unexpectedly large. The $B_s^0$
decay-width difference can be obtained by fitting proper time
distributions of untagged samples of events simultaneously for the mean
$B_s^0$ lifetime $\tau_{B_s} = 1/\Gamma_s$ and $\Delta\Gamma_s/\Gamma_s$.
All three experiments will use their $B_s^0\rightarrow J/\psi \phi$ events 
for this measurement as described in Sec.~\ref{sec:4.2}. In addition,
LHCb will have an untagged sample of $B_s^0\rightarrow D_s^- X$ events thanks 
to their low-level hadronic triggers. LHCb expect to directly
observe and measure $\Delta \Gamma_s/\Gamma_s$ after one year
of data-taking
with their untagged $B_s\rightarrow D_s^- \pi^+$ sample,
if $\Delta \Gamma_s/\Gamma_s$ is at least 20\%~\cite{LHCbTP}.  

The B meson flavour at production and decay time and the $B_s^0$ proper time 
with good resolution are the ingredients needed to measure 
$\Delta M_s$. 
The best channels to make this measurement are 
$B_s^0$ decays to exclusive, flavour specific states 
like $B_s^0 \rightarrow D_s^- \pi^+$. The flavour of the $B_s^0$ at its
decay is unambiguously tagged by the sign of the $D_s^-$. The  
$B_s^0$ flavour at production can be determined from the sign of 
the decay product(s) of the other $b$ hadron in the event. 
The factors which affect the sensitivity of an experiment 
for measuring $\Delta M_s$ are the wrong tag fraction, 
$\omega_{tag}$, the presence of background and the proper time
resolution, $\sigma_t$. The corresponding dilution factors for the time
dependent asymmetry are $D_{tag}=1-2\omega_{tag}$, 
$D_{bkg}\approx N_{signal}/(N_{signal}+N_{bkg})$ and 
$D_{time}\approx\mbox{exp}\, (-(\Delta M_s \sigma_t)^2/2)$. Here, $N_{signal}$
and $N_{bkg}$ are the number of signal and background events, respectively. 
The measured asymmetry is given by
\begin{equation}
A_{meas}(t)= A(t) \cdot D_{tag}\cdot D_{bkg}\cdot D_{time}.
\end{equation}
 The amplitude fit method \cite{BOSC_FIT} has been used to determine 
the experimental reach for a $\Delta M_s$ measurement from the 
time-dependent asymmetry. 
In this method, $\mbox{cos}\,(\Delta{M_s}t)$ 
is multiplied by an amplitude parameter
$A$. The value of the parameter and its error $\sigma_A$ are determined
for each $\Delta M_s$ value by a maximum likelihood fit.
For a measurement of $\Delta M_s$ in a
region well inside the sensitivity of an experiment, the standard
maximum likelihood method is foreseen.

ATLAS~\cite{ATLASPTDR}, CMS~\cite{CMSNOTE1,CMSNOTE2} and 
LHCb~\cite{LHCbTP} have 
determined their sensitivities to 
$\Delta M_s$ using events generated by Pythia~\cite{PYTHIA} and 
then passed
through detailed detector simulation. 
Table~\ref{tab:exp_bmix} summarizes the channels used, assumptions,
performance and results of the three analyses. 
All three have used 
$B_s^0 \rightarrow D_s^- \pi^+$
and ATLAS has also used $B_s^0 \rightarrow D_s^- a^+_1$ followed by
$a_1^+\rightarrow\rho^0\pi^+$. The $D_s^-$ is reconstructed via its decay
into $\phi\pi^-$ followed by $\phi\rightarrow K^+ K^-$ by all three
experiments
and also $D_s^-\rightarrow K^{*0} K^-$ followed by $K^+ \pi^-$ by CMS.
CMS has assumed a 50\% efficiency of the higher level triggers
for calculating the final yield of reconstructed $B_s^0$ mesons.
ATLAS also reconstructed $D_s^-\rightarrow K^{*0} K^-$, but did not include it
in their final analysis since after applying the cuts needed to obtain
a reasonable rate of the level 2 trigger, the addition of this mode 
did not improve their limit.
$D_s^-$ decay modes other than $\phi\pi^-$ contributing to the
$K^+ K^- \pi^-$ final state will also be reconstructed by LHCb; for the 
yield presented in Tab.~\ref{tab:exp_bmix}, an effective 
$D_s^- \rightarrow K^+ K^- \pi^-$ branching ratio of 4\% is assumed, with
the same efficiency and purity as for $D_s^-\rightarrow \phi\pi^-$.
For flavour tagging
at production, ATLAS and CMS have used the trigger muon, which primarily
comes from the semileptonic decay of the other $b$ hadron in the event.
LHCb use identified muons, electrons and kaons from the
decay of the other $b$ hadron. Other tagging techniques will be developed
in the future.
 
Figures~\ref{fig:atlasbosc1} and \ref{fig:atlasbosc2} from ATLAS illustrate
the sensitivity for $\Delta M_s$ measurements as a function of the integrated 
luminosity and the signal content of the sample. 1000
experiments were performed at each $\Delta M_s$ point and a 
$\Delta M_s$ value was
considered ``reachable'' if 95\% of the experiments gave a value
within 2$\sigma$ of the input value. 
CMS and LHCb have defined two kinds of reaches --- one for a 
measurement and the other one for 95\% CL exclusion. Figure~\ref{fig:cmsbosc} 
shows the result for $\Delta M_s$ reach from CMS using the amplitude method.
The amplitude, $A$, together with its error, $\sigma_A$, is shown 
for different $x_s$, where $x_s\equiv\Delta M_s/\Gamma_s$. 
$x_s$ values below the intersection point of the 1.645$\,\sigma_A$
curve and the line $A=1$ are excluded at 95\% CL. 
CMS determined their reach by a method similar to 
ATLAS, but an experiment 
was considered ``successful'' if the  $x_s$ value corresponded to the
highest peak in the amplitude spectrum and was in the vicinity of 
$x_s^{\mbox{\scriptsize true}}$ within the natural width ($\pm\,$1.5
in $x_s$) of the amplitude distribution. The two methods yielded the 
same results. 
Figure~\ref{fig:lhcb_2} shows the statistical significance $S=1/\sigma_A$ 
of the 
$B_s^0$ oscillation signal as a function of $\Delta M_s$ from LHCb. 
The LHCb reach for $\Delta M_s$ quoted in
Tab.~\ref{tab:exp_bmix} is for $S=5$ ($5\sigma$ measurement) and  
$S=1.645$ (95\% CL exclusion).  
According to these studies, $\Delta M_s$ can be measured up to
30$\,$ps$^{-1}$ (ATLAS),
26$\,$ps$^{-1}$ (CMS) and 48$\,$ps$^{-1}$ (LHCb) with one year of data. The 
addition of more 
channels is likely to improve the reach. Thus, each of the three
experiments will be able to fully explore the $\Delta M_s$ range 
allowed in the SM, Eq.~(\ref{smdms}), after one year of data-taking.
In addition, the likely precision on $\Delta M_s$ will be such that the
extraction of $|V_{ts}/V_{td}|^2$ will be limited
by the theoretical uncertainty on $R_{sd}$ (see expressions
(\ref{rsddef}) to (\ref{rsdnum})).

\setcounter{equation}{0}
\section[RARE DECAYS]{RARE DECAYS\protect\footnote{Section 
coordinators: P. Ball
and F. Rizatdinova with help from
  P. Bartalini, P. Koppenburg, M. Misiak, A.~Nikitenko, N.~Nikitin
  and E. Polycarpo.}}\label{sec:rare}

Flavour-changing neutral current decays involving $b\to s$ or $b\to d$
transitions occur only at loop-level in the SM, come with small
exclusive branching ratios $\sim O(10^{-5})$ or smaller and thus
provide an excellent probe of indirect effects of new physics and
information on the masses and couplings of the virtual SM or
beyond-the-SM particles participating. Within the SM, these decays are
sensitive to the CKM matrix elements $|V_{ts}|$ and $|V_{td}|$,
respectively; a measurement of these parameters or their ratio would
be complementary to their determination from B mixing, discussed in
Sec.~\ref{sec:mix}.

The effective field theory
for $b\to s(d)$ transitions is universal for all the channels
discussed here. Due to space-restrictions, we cannot review all
important features of that effective theory; for a quick overview we
refer to Chapter~9 of the BaBar Physics Book \cite{BaBar}, where also
references to more detailed reviews can be found. Here we simply state
that the effective Hamiltonian governing rare decays can be obtained
from the SM Hamiltonian by performing an operator product expansion
yielding
\begin{equation}\label{eq:hammel}
{\cal H}^q_{\mbox{\scriptsize eff}} = -\frac{4 G_F}{\sqrt{2}}\, 
V_{tb}^{\vphantom{*}}
V_{tq}^* \,\sum\limits_{i=1}^{11} C_i(\mu) {\cal O}^q_i(\mu),
\end{equation}
where the ${\cal O}^q_i$ are local renormalized operators and
$V_{tb}^{\vphantom{*}}V_{tq}^*$ are CKM matrix elements with $q=s,d$.
The Wilson-coefficients 
$C_i$ can be calculated in perturbation theory and encode the
relevant short-distance physics, in particular any potential
new-physics effects. The renormalization-scale $\mu$ can be viewed as 
separating the
long- and short-distance regimes. For calculating decay rates with
the help of (\ref{eq:hammel}), the value of $\mu$ has to be chosen as
$\mu\sim m_b$ in a truncated perturbative expansion.
The Hamiltonian (\ref{eq:hammel}) is suitable to describe physics in the
SM as well as in a number of its extensions, for instance the minimal
supersymmetric model. The operator basis in (\ref{eq:hammel})
is, however, not always complete, and in some models, for instance
those exhibiting left-right symmetry, new physics also shows up in the
form of new operators. This proviso should be kept in mind when
analysing rare B decays for new-physics effects by measuring
Wilson-coefficients.

At present, the following channels have been evaluated for LHCb,
CMS and ATLAS:
\begin{itemize}
\item purely muonic decays  $B^0_{d,s}\to \mu^+\mu^-$ (all experiments);
\item the radiative decay $B^0_d \to {K^*}^0 \gamma $ (LHCb only);
\item semimuonic decays 
      $B^0_d\to \rho^0 \mu^+\mu^-$,
      $B^0_d \to {K^*}^0 \mu^+\mu^-$,
      $B^0_s\to \phi^0 \mu^+\mu^-$ (all experiments). 
\end{itemize}
As a reflection of this rather preliminary status of rare B
decay studies for the LHC, we confine this section's discussion to
channels most of which are in principle  accessible at $e^+
e^-$ B factories and can also be studied at the
Tevatron. This applies in particular to the
radiative decay $B\to K^*\gamma$ that has already been measured at
CLEO~\cite{CLEO_Brad} and for which at the time of the first
physics runs at the LHC rather accurate measurements should be
available. The situation is different for $B\to\mu^+\mu^-$, which will
be seen before the start of the LHC only if it is enhanced drastically, i.e.\
by orders of magnitude, by new-physics effects. Also the
measurement of the spectra of $B\to K^*\mu^+\mu^-$ will be reserved to
the LHC, although the decay itself should be seen at the B factories before.
In general, and in contrast to the exploration of CP
violation, the main impact of the LHC on the study of rare decays will be to
provide radically increased statistics rather than opening new,
alternative channels.

\subsection[$B^0\to\mu^+\mu^-$]{\protect\boldmath $B^0\to\mu^+\mu^-$}

This decay is an experimental favourite thanks to its unique signature
and at the same time a challenge, as its SM branching ratio is of
order $10^{-9}$. The motivation for measuring this decay lies mainly
in its r\^{o}le as indicator for possible new physics which might 
significantly 
enhance the branching ratio. The present experimental bounds from
Tevatron are in the $10^{-6}$ range.

\subsubsection{Theoretical Framework}

The purely muonic neutral B decays are described by only three
operators \cite{GLN}: 
$$
O^q_P    = (\bar q \gamma_5 b) (\bar \mu \gamma_5\mu), \quad 
~~O^{q'}_P = (\bar q \gamma_5 b) (\bar \mu \mu), \quad 
~~O^q_A    = (\bar q \gamma^\alpha\gamma_5 b) 
(\bar \mu \gamma_\alpha\gamma_5\mu),
$$
with $q=s,d$. In the SM, these transitions proceed through
electroweak penguin diagrams with $Z$ and $H^0$ exchange as well as
$W$ box diagrams. Introducing dimensionless Wilson-coefficients
$C_{P,A}^{q,q'}$, the branching ratio is given by
\begin{equation}
B(B_q\to \mu^+\mu^-) = \frac{G_F^2}{8\pi}\, \tau_B f_B^2 m_B^3
\sqrt{1-\frac{4 m_\mu^2}{m_B^2}}\left\{ \left| C_P^q -
    \frac{2m_\mu}{m_B}\, C_A^q \right|^2 + \left( 1 - \frac{4
      m_\mu^2}{m_B^2} \right) \left| C_P^{q'} \right|^2 \right\}.
\end{equation}
In the SM, the coefficients $C_P$ arise from penguin diagrams with
physical and unphysical neutral scalar exchange and are suppressed by
a factor $(m_b/m_W)^2$ \cite{SK}. The decay rate is then determined
solely by the coefficient 
\begin{equation}
C_{A,SM}^q = \frac{\alpha V_{tb}^{\vphantom{*}}
  V_{tq}^*}{\sqrt{8}\pi\sin^2\theta_w}\, Y(x_t),
\end{equation}
where $x_t\equiv m_t^2/m_W^2$, $\sin^2\theta_w$ is the weak mixing
angle and the function $Y(x)$ is at leading order in QCD given by \cite{IL}
\begin{equation}
Y(x) = \frac{x}{8} \left[ \frac{x-4}{x-1} + \frac{3x}{(x-1)^2}\, \ln x
\right].
\end{equation}
The SM branching fractions are then given by (with $f_{B_q}$ from
(\ref{fbsum}),  $|V_{td}|$
 from (\ref{mdsum}) and $m_t = 167\,$GeV)
\begin{eqnarray}
B(B_d\to \mu^+\mu^-) & = & (1.0\pm 0.5)\times 10^{-10} \left[
  \frac{f_{B_d}}{200\,{\rm MeV}}\right]^2 \left[
  \frac{\overline{m}_t(\overline{m}_t)}{167\,{\rm GeV}}\right]^{3.12} 
\left[\frac{|V_{td}|}{0.0074}\right]^2
  \left(\frac{\tau_{B_d}}{1.56\,{\rm ps}}\right),\label{eq:Bdmumu}\\
B(B_s\to \mu^+\mu^-) & = & (3.7\pm 1.0) \times 10^{-9} \left[
  \frac{f_{B_s}}{230\,{\rm MeV}}\right]^2 \left[
  \frac{\overline{m}_t(\overline{m}_t)}{167\,{\rm GeV}}\right]^{3.12} 
\left[\frac{|V_{ts}|}{0.040}\right]^2
  \left(\frac{\tau_{B_s}}{1.54\,{\rm ps}}\right).
\end{eqnarray}
Due to these tiny SM branching ratios and the favourable experimental
signature, these decay processes are ideal candidates for new physics
to be observed, for example flavour-changing neutral Higgses. 
New-physics scenarios have been investigated e.g.\ in
Refs.~\cite{SK,laura}.

\subsubsection{Experimental Considerations}

Purely muonic B decays, so-called "self-triggering" channels,
have a clear signature that can be used at level-1 trigger  in all LHC
experiments. Only muon identification is necessary. 
The expected numbers of events quoted in the following refer to the
SM branching ratios $B(B^0_s \to \mu^+ \mu^-)=(3.5 \pm 1.0)\times
10^{-9}$ and $B(B^0_d \to \mu^+ \mu^-)=1.5\times 10^{-10}$, i.e.\ the
``optimistic'' end of the theory prediction (\ref{eq:Bdmumu}).

The CMS collaboration has performed a detailed study of the
observability of $B_s^0\to\mu^+\mu^-$ \cite{CMSNote} at both low and
high luminosity, implementing the
complete pattern recognition and track reconstruction procedure. Both
the gluon fusion and the gluon splitting production mechanism are
included and yield comparable contributions. CMS has tuned the 
experimental selection criteria to optimize the signal-to-background
ratio as follows:
\begin{itemize}
\item[1.] Only muon pairs satisfying the requirement 
$0.4 < \Delta R_{\mu \mu}< 1.2$ were considered as candidates for 
$B^0_s\to \mu^+\mu^-$; the transverse momentum of the muon pair 
$p_T^{\mu \mu}$ must be larger than  12$\,$GeV and $p_{T}$ of either
muon be larger than 4.3$\,$GeV.
\item[2.] The effective mass of the dimuon pair was required to be
  within a 80$\,$MeV mass
window around the nominal $B^0_s$ mass. Only 1.1\% of background combinations
are retained after this mass cut.
\item[3.] The third set of cuts is based on the secondary vertex
reconstruction: the distance between $B^0_s$ and primary vertex in the
transverse plane is required to be larger than 12$\sigma_{\rm vtx}$, 
about 820 $\mu$m, where $\sigma_{\rm vtx}$
is the vertex resolution. The angle $\alpha$ between the line joining primary
and secondary
vertex and transverse momentum vector was required to satisfy
$\cos\alpha> 0.9997$. The absolute error of the secondary vertex 
reconstruction was required to be less than 80 $\mu$m.
The distance between the two muons, $d_2$, had to be smaller than
50$\,\mu$m and the ratio $d_2/\sigma(d_2)$ smaller than 2.
\item[4.] Isolation of the dimuon pair in the tracker was required, i.e.\ 
no charged particles with $p_T >\,$0.9$\,$GeV must be found in the cone 
$R< 0.5 \times \Delta R_{\mu\mu} + 0.4$ around the dimuon momentum
direction. The isolation requirement is important for suppressing the
background induced by gluon-splitting. 
About 50\% of the signal and 3\% of the background events passed through
the isolation cut in the tracker. An additional factor 2.3 of background
suppression was obtained by  requiring isolation of the
dimuon pair in the calorimeters, i.e.\ the transverse energy in the 
electromagnetic and hadron calorimeters was required to be less than 
4$\,$GeV in the same tracker cone.
\end{itemize}
After applying these cuts, the number of expected events detected by CMS
after 3 years running at 
low luminosity is 21 with less than 3
 background events at 90\% C.L., assuming the SM 
branching ratio. CMS will observe this channel even after
1 year running at low luminosity. Taking into account the production
ratio $B_d^0/B_s^0=0.40/0.11$ and the expected SM branching ratio
(\ref{eq:Bdmumu}), 
CMS also expects, for three years running at low luminosity, to find 
2.2$\pm$1.1 $B_d\to\mu^+\mu^-$ events with again essentially no background.

LHCb's sensitivity to the decay $B_s\to\mu^+\mu^-$ has been studied 
using fully GEANT generated samples of both signal and background events.
Good quality tracks are combined into a vertex if they are identified as
muon tracks with high confidence level and are within 50$\,\mu$m
in space.
The secondary vertex must also satisfy quality criteria and be well
displaced from the primary vertex. The impact parameter of the
reconstructed $B_s$ candidate is required to be smaller than
35$\,\mu$m and a mass window of 20~MeV around the nominal $B_s$ mass is
applied. After all those selection cuts 11 signal events per year are
expected.
Since the initial background sample was very small compared to the number
 of events in one year of LHCb operation, pions which are a direct product
 of B decays were allowed to make pairs with muons, ``faking'' the
 background signature, in order to increase the statistics of the sample.
 Using this procedure, it was possible to estimate the rejection power of
 the cuts in the impact parameter and the mass of the $B_s$ candidate,
 assuming they are uncorrelated and that the mass distribution in
a mass window of 200~MeV around the nominal value is flat. The expected
 background yield in one year is 3.3~events. 
 Studies with high statistics samples of full GEANT simulation are under
way, in order to make the background estimate more precise. Hence LHCb 
will observe the decay $B_s\to\mu^+\mu^-$ within 1~year of running.

The ATLAS collaboration has made a detailed study of the decay mode
$B^0_s\to \mu^+\mu^-$, using fully simulated samples \cite{ATLASPTDR}. 
To suppress the
combinatorial background, cuts on the quality of vertex
reconstruction and on the decay length of the reconstructed B meson were
applied. Further background reduction was obtained by imposing
cuts on the angle  between the line joining primary and secondary
vertex and the transverse momentum vector and on the isolation of the
dimuon pair formed in the decay of the B meson. The mass resolution 
obtained after all selection cuts is $\sigma(M)=68\,$MeV. The mass
window ${}^{+2\sigma}_{-1\sigma}$ was taken for estimating the number of
signal and background events.
 After applying cuts, the number of expected events detected by ATLAS
 after 3 years running at 
low luminosity, assuming the SM branching fractions, 
is 27 with 93 background events.
For $B^0_d \to \mu^+\mu^-$, one can expect 
4 signal events with 93 background events. 

Hence, all three experiments will be able to measure the SM branching fraction
of $B^0_s\to \mu^+\mu^-$. The
numbers of events expected by the three
collaborations after 3 years' data collection are given in Tab.~\ref{LLSUM}.

Both ATLAS and CMS are planning to continue the study of purely muonic
decays at high luminosity $10^{34}{\rm cm}^{-2}{\rm s}^{-1}$. 
This is made possible by the low dimuon trigger 
rate which is expected to be around 30 Hz in ATLAS. In both
experiments, the number of minimum bias 
events accepted together with the triggered events is expected to be
10 times larger than at the LHC run at low luminosity. 
The CMS collaboration  estimated the
possibility to detect the purely muonic decay  using a high
luminosity pixel configuration that leads to degradation of the 
vertex resolution.
The ATLAS collaboration assumed that the geometry of the Inner
Detector will be the
same as at low luminosity (no degradation in vertex and $p_T$
resolution is expected 
compared to low luminosity results). 
The same analysis-cuts as at low luminosity were applied to the signal
and background events by both collaborations. The resulting
numbers of events expected by the CMS and
ATLAS collaborations  after 1 year  running at high luminosity
are given in Tab.~\ref{HL}, assuming the SM branching fraction.
\begin{table}
\begin{minipage}[b]{0.48\textwidth}
\begin{center}
\begin{tabular}{|l||c|c|c|}  \hline 
Experiment & ATLAS & CMS & LHCb\\ \hline
Signal & 27 & 21 & 33\\
Background & 93 & \phantom{2}3 & 10\\ \hline
\end{tabular}
\end{center}
\vspace*{-0.5cm}
\caption[]{Expected signal and background events for
  $B_s\to\mu^+\mu^-$ after 3
  years running at low luminosity.}\label{LLSUM}
\end{minipage}
\hspace*{10pt}
\begin{minipage}[b]{0.48\textwidth}
\begin{center}
\begin{tabular}{|c||c|r|}  \hline
Experiment &  ATLAS & \multicolumn{1}{c|}{CMS}\\ \hline
$B^0_s \to \mu^+\mu^-$ & \phantom{6}92 & 26\\
$B^0_d \to \mu^+\mu^-$ & \phantom{6}14 & 4.1\\
Background &  660 & $<\,$6.4\\ \hline
\end{tabular}
\end{center}
\vspace*{-0.5cm}
\caption[]{The expected statistics for purely muonic decays after one
  year running at high luminosity.}\label{HL}
\end{minipage}
\end{table}
The decay $B^0_s\to \mu^+\mu^-$ can clearly  be observed after 1
 year running at high luminosity by both collaborations. Concerning  
$B^0_d\to \mu^+\mu^-$, the sensitivity of
ATLAS to the branching ratio will be at the level of $3\times 10^{-10}$, i.e.\
 roughly a factor 3 above the SM prediction. 
High luminosity measurements of 
the purely muonic decays would significantly improve the data to be obtained 
at low luminosity.

\subsection[$B\to K^* \gamma$]{\protect\boldmath $B\to K^* \gamma$}

In this subsection we discuss the specifics of the radiative FCNC
transition $B\to K^*\gamma$ relevant for the LHC, concentrating on
non-perturbative QCD effects.  For the treatment
of perturbative issues, in particular the reduction of
renormalization-scale dependence and remaining uncertainties, we refer
to \cite{tobias,mikolaj}.

\subsubsection{Theoretical Framework}

The theoretical description of the $B\to K^*\gamma$ decay is quite
involved with regard to both long- and short-distance contributions.
In terms of the effective Hamiltonian (\ref{eq:hammel}), the decay
amplitude can be written as
\begin{equation}\label{eq:bsg}
{\cal A}(\bar{B}\to \bar{K}^*\gamma) = -\frac{4 G_F}{\sqrt{2}}\,
V_{tb}^{\vphantom{*}} V^*_{ts}
\,\langle \bar{K}^* \gamma | C_7 O_7 + i\epsilon^\mu \sum_{i\neq 7}
C_i \int d^4x
e^{iqx} T \{ j_\mu^{em}(x)O_i(0)\} | \bar{B}\rangle\,,
\end{equation}
where $j_\mu^{em}$ is the electromagnetic current and $\epsilon_\mu$ the
polarization vector of the photon. $O_7$ is the only operator 
containing the photon field at tree-level:
\begin{equation}
O_7 = \frac{e}{16\pi^2}\, m_b \bar s \sigma_{\mu\nu} R b F^{\mu\nu}
\end{equation}
with $R=(1+\gamma_5)/2$. Other operators, the
second term in (\ref{eq:bsg}), contribute mainly closed fermion loops.  The
first complication is now that the first term in (\ref{eq:bsg})
depends on the regularization- and renormalization-scheme. For this
reason, one usually introduces a scheme-independent linear combination
of coefficients, called ``effective coefficient'' (see \cite{mikolaj}
and references therein):
\begin{equation}
C_7^{\mbox{\scriptsize eff}}(\mu) = C_7(\mu) + \sum\limits_{i=3}^6 y_i 
C_i(\mu),
\end{equation}
where the numerical coefficients $y_i$ are given in
\cite{mikolaj}.

The current-current operators 
\begin{equation}\label{eq:O12}
O_1 = (\bar s\gamma^\mu L b)(\bar c\gamma_\mu L c), 
\qquad O_2 = (\bar s\gamma_\mu L c)(\bar c\gamma^\mu L b) 
\end{equation}
give vanishing contribution to the perturbative $b \to s \gamma$
amplitude at one loop.  Thus, to leading logarithmic accuracy (LLA) in
QCD and neglecting long-distance contributions from $O_{1,2}$ to the
$b\bar{s}\gamma X$ Green's functions, the $\bar{B}\to \bar{K}^*\gamma$
amplitude is given by
\begin{equation}
{\cal A}_{O_7}^{\rm LLA}(\bar{B}\to \bar{K}^*\gamma) = 
-\,\frac{4G_F}{\sqrt{2}}\, V_{tb}^{\vphantom{*}}
 V_{ts}^* C_7^{(0){\mbox{\scriptsize eff}}}
\langle \bar{K}^*\gamma | O_7 | \bar{B}\rangle.
\end{equation}
Here, $C_7^{(0){\mbox{\scriptsize eff}}}$ denotes the leading logarithmic
approximation to $C_7^{\mbox{\scriptsize eff}}$. The above expression
is, however,
not the end of the story, as the second term in (\ref{eq:bsg}) also
contains long-distance contributions. Some of them can be viewed as
the effect of virtual intermediate resonances $\bar{B}\to \bar{K}^*
V^*\to \bar{K}^*\gamma$.  The main effect comes from $c\bar c$
resonances and is contributed by the operators $O_1$ and $O_2$ in
(\ref{eq:bsg}). It is governed by the virtuality of $V^*$, which, for
a real photon, is just $-1/m_{V^*}^2\sim -1/4 m_c^2$. The presence of
such power-suppressed terms $\sim 1/m_c^2$ has first been derived for
inclusive decays in Ref.~\cite{voloshin} in a framework based on
operator product expansion. 
The first, and to date only, study for exclusive decays was done in
\cite{stoll}.  Technically, one performs an operator product expansion
of the correlation function in (\ref{eq:bsg}), with a soft
non-perturbative gluon being attached to the charm loop, resulting in
terms being parametrically suppressed by inverse powers of the charm
quark mass.  As pointed out in \cite{mp}, although the power increases
for additional soft gluons, it is possible that contributions of
additional external hard gluons could remove the power-suppression.
This question is also relevant for inclusive decays and deserves
further study.

After inclusion of the power-suppressed terms $\sim 1/m_c^2$, the 
$\bar{B}\to \bar{K}^*\gamma$ amplitude reads
\begin{equation} \label{eq:with.OF}
{\cal A}^{\rm LLA}(B\to \bar{K}^*\gamma) = -\,\frac{4G_F}{\sqrt{2}}\, 
V_{tb} V_{ts}^*
\langle \bar{K}^*\gamma \mid C_7^{(0){\mbox{\scriptsize eff}}} O_7 + 
\frac{1}{4m_c^2}\, C_2^{(0)} O_F \mid \bar{B}\rangle.
\end{equation}
Here, $O_F$ is the effective quark-quark-gluon operator obtained in
\cite{stoll}, which describes the leading non-perturbative
corrections. The two hadronic matrix elements can be described in
terms of three form factors, $T_1$, $L$ and $\tilde{L}$:
\begin{eqnarray}
\langle \bar{K}^*(p)\gamma|\bar s \sigma_{\mu\nu} q^\nu b 
|\bar{B}(p_B)\rangle & = & i\epsilon_{\mu\nu\rho\sigma}\epsilon^{*\mu}_\gamma
\epsilon^{*\nu}_{K^*} p_B^\rho p^\sigma 2 T_1(0),\nonumber\\
\langle \bar{K}^*(p)\gamma|O_F|\bar{B}(p_B)\rangle & = &
\frac{e}{36\pi^2} \left[
  L(0) \epsilon_{\mu\nu\rho\sigma} \epsilon^{*\mu}_\gamma \epsilon^{*\nu}_{K^*}
  p_B^\rho p^\sigma\right.\nonumber\\
& & \left.{} + i \tilde{L}(0) \left\{ (\epsilon^*_{K^*} p_B) 
(\epsilon^*_{\gamma} p_B) - \frac{1}{2}\, (\epsilon^*_{K^*} 
\epsilon^*_{\gamma}) (m_B^2 -
m_{K^*}^2) \right\} \right].
\end{eqnarray}
The calculation of the above form factors requires genuinely
non-perturbative input. Available methods include, but do not exhaust,
lattice calculations and QCD sum rules. Again, a discussion of the
respective strengths and weaknesses of these approaches is beyond the
scope of this report. Let it suffice to say that --- at least at
present --- lattice cannot reach the point $(p_B-p)^2=0$ relevant for $B\to
K^*\gamma$, and that QCD sum rules on the light-cone predict \cite{BB}
\begin{equation}
T_1(0) = 0.38 \pm 20\%
\end{equation}
at the renormalization scale $\mu = 4.8\,$GeV. For the other two form
factors, QCD sum rules predict \cite{stoll}
\begin{equation}
L(0) = (0.55\pm 0.10)\,{\rm GeV}^3,\qquad 
\tilde{L}(0) = (0.7\pm 0.1)\,{\rm GeV}^3.
\end{equation}
Numerically, these corrections increase the decay rate by about 5 to 10\%.
After their inclusion, one obtains
\begin{eqnarray}
B(B\to K^*\gamma) & = & \frac{\alpha}{32\pi^4}\, G_F^2 |V_{tb}
V_{ts}^*|^2 \left|C_7^{(0){\mbox{\scriptsize eff}}}\right|^2 m_b^2\,
\frac{(m_B^2-m_{K^*}^2)^3}{m_B^3} \left| T_1(0) \right|^2\nonumber\\
& & \times \left( 1 -
  \frac{1}{18m_c^2} \, \frac{C_2^{(0)}}{C_7^{(0){\mbox{\scriptsize
          eff}}}}\, \frac{1}{m_b}\,
\frac{L(0)+\tilde{L}(0)}{T_1(0)} \right)\nonumber\\
& = & 4.4\times 10^{-5} (1+8\%)
\end{eqnarray}
for the central values of the QCD sum rule results, which agrees with
the experimental measurement.

Let us close this subsection with a few remarks on the decay
$B\to\rho\gamma$. Although at first glance it might seem that its
structure is the same as that of $B\to K^*\gamma$, this is actually
not the case. There are additional long-distance 
contributions to $B\to\rho\gamma$,
which are CKM-suppressed for $B\to K^*\gamma$ and have been neglected
in the previous discussion; these contributions comprise
\begin{itemize}
\item weak annihilation mediated by $O^{u}_{1,2}$ with non-perturbative
  photon emission from light quarks; these contributions are
  discussed in \cite{Brhog} and found to be of order 10\% at the
  amplitude level;
\item effects of virtual $u\bar u$ resonances ($\rho$,
  $\omega$,\dots); they are often said to be small, but actually have
  not been studied yet in a genuinely
  non-perturbative framework, so that statements about their smallness
  lack proper justification.
\end{itemize}
For the above reasons it is, at present, premature to aim at an
accurate determination of $|V_{ts}|/|V_{td}|$ from a measurement of
$B(B\to\rho\gamma)$ and $B(B\to K^*\gamma)$. A very recent discussion
of long-distance effects in $B\to V\gamma$ decays can also be found in
Ref.~\cite{newshit}. 

\subsubsection{Experimental Considerations}

The radiative decay 
$B^0_d \to {K^*}^0 \gamma $ has been studied by the LHCb collaboration
at both the particle and  the full-simulation level \cite{LHCb_BKgamma}.
The event selection and reconstruction can be summarized as follows:
\begin{itemize}
\item selection: $X^+X^-\gamma$ combinations;
tracks are consistent with $K^-$ and $\pi^+$ hypotheses;\\ 
$|M(K^-\pi^+)$--$M(K^{*0})| < 55$ MeV;
cluster in the electromagnetic calorimeter with $E_T > 4\,$GeV;
\item geometrical cuts: $\chi^2<9$ of secondary vertex fit;
 $|\Delta(Z) | >\,$1.5$\,$mm between primary and secondary vertex;
impact parameters of both tracks $>400 \mu$m; 
the angle between the momentum vector and the line joining primary and
secondary vertex smaller than 0.1 rad; the angle $\theta$ between $B^0_d$
and $K^-$ in the $K^{*0}$ rest frame $|\cos\theta| < 0.6 $;
\item $p_T > 4\,$GeV of reconstructed $B^0_d$.   
\end{itemize}
The mass resolution obtained at the particle-level study is 67 MeV.
The mass window taken for estimates is 200 MeV around the nominal
$B^0_d$ mass. 
Assuming $B(B^0_d \to {K^*}^0 \gamma)=(4.9 \pm 2.0)\times 10^{-5}$, 
the expected number of
signal events after 1 year  running is 26000, with $S/B \sim 1$.
This will be sufficient to measure the branching fraction with
high accuracy. The expected accuracy in the CP asymmetry measurement is 
$\delta_{CP} = 0.01$. The SM predicts a CP asymmetry of order 1\%.

\subsection[$B\to K^*\mu^+\mu^-$]{\protect\boldmath $B\to K^*\mu^+\mu^-$} 

Like with $B\to K^*\gamma$, we can only review the essentials and put
emphasis on recent developments in theory and the specifics for the
LHC experiments. A slightly more detailed discussion and relevant
references can be found in the BaBar physics book \cite{BaBar}. The
current state-of-the-art of perturbation theory is summarized in
Ref.~\cite{mikolaj2}. The
motivation for studying this decay is either, assuming the SM to be
correct, the measurement of the CKM matrix element $|V_{ts}|$, or the
search for manifestations of new physics in non-standard values of the
Wilson-coefficients. A very suitable observable for the latter purpose is the
forward-backward asymmetry which is independent of CKM matrix elements
and, due to extremely small event numbers, only accessible at the
LHC. Of all the rare decay channels discussed in this section, $B\to
K^*\mu^+\mu^-$ is definitely the one whose detailed study is only
possible at the LHC and which has the potential for high impact both
on SM physics and beyond.

\subsubsection{Theoretical Framework}

The presentation in this section follows closely
Ref.~\cite{ABHH}; for other relevant recent papers treating $B\to
K^*\mu^+\mu^-$, see \cite{aliev_gang}.

At the quark-level, the effective Hamiltonian (\ref{eq:hammel}) leads to the
following decay amplitude:
\begin{eqnarray}
        {\cal A}(b\to s\mu^+\mu^-) & = & \frac{G_F \alpha}{\sqrt{2}  \pi} 
                V_{t s}^\ast V_{tb}  \left\{
                  C_9^{\mbox{\scriptsize eff}}(s) \left[ \bar{s}
                \gamma_\alpha  L  b \right] 
                          \left[ \bar\mu  \gamma^\alpha  \mu \right]
                + C_{10}  \left[ \bar{s}  \gamma_\alpha  L  b \right] 
                         \left[ \bar\mu  \gamma^\alpha  \gamma_5  \mu \right]
                \right. \nonumber \\
        & &  \left.
                - 2 m_b  C_7^{\mbox{\scriptsize eff}}  \left[ \bar{s}
                i  \sigma_{\alpha \nu}
                        \frac{q^{\nu}}{s}  R  b \right]
                        \left[ \bar\mu  \gamma^\alpha  \mu \right]
                \right\}  .
        \label{eq:m}
\end{eqnarray}
Here, $L/R = (1\mp\gamma_5)/2$, $s=q^2$, $q=p_++p_-$, where $p_\pm$
are the four-momenta of the leptons. We neglect the
strange quark mass, but keep the leptons massive.  Already the free
quark decay amplitude ${\cal A}(b\to s\mu^+\mu^-)$ contains certain
long-distance effects which usually are absorbed into a redefinition
of the Wilson-coefficient $C_9$. To be specific, we define, for exclusive
decays, the {\em momentum-dependent} effective coefficient of the
operator ${\cal O}_9 = e^2/(16\pi^2) (\bar s \gamma_\alpha Lb)(\bar\mu
\gamma^\alpha \mu)$ as
\begin{equation}
C_9^{\mbox{\scriptsize eff}}(s) = C_9 + Y(s),
\end{equation}
where $Y(s)$ stands for matrix elements of
four-quark operators. Formulas can be found in \cite{AH}. The
prominent contribution to $Y(s)$ comes from the $c\bar c$ resonances
$J/\psi$, $\psi'$, $\psi''$ which show up as peaks in the dimuon
spectrum, but are irrelevant for the short-distance physics one is
interested in. Note that
the effective coefficient depends on the process being considered and
is, in particular, not the same for exclusive and inclusive decays: in the
latter ones, also virtual and bremsstrahlung corrections to $\langle 
\mu^+\mu^- s| {\cal O}_9 | b\rangle$, usually denoted by $\omega(s)$,
 are included, whereas for
exclusive decays, they are contained in the  hadronic
matrix elements to be defined below. 
For $s$ far below the $c\bar c$ threshold, perturbation theory,
augmented by non-perturbative 
power-corrections in $1/m_c^2$, is expected to yield a
reliable estimate for long-distance effects in $C_9^{\mbox{\scriptsize
    eff}}$. In
contrast to inclusive decays, however, the corresponding $1/m_c^2$
terms have not yet been worked out for exclusive decays. 
To date, one has to rely on
phenomenological prescriptions for incorporating non-perturbative
contributions to $Y(s)$ \cite{c9eff}. The resulting
uncertainties on $C_9^{\mbox{\scriptsize eff}}$ and on various
distributions in inclusive
decays have been worked out in Refs.~\cite{AH,mikolaj2} to which we refer
for a detailed discussion. 

Other long-distance corrections, specific for the exclusive decay
$B\to K^*\mu^+\mu^-$, are described in terms of
 matrix elements of the quark operators in (\ref{eq:m}) between
meson states and can be parametrized in terms of form
factors. Denoting by $\epsilon_\mu$ the polarization vector of the
$K^*$ vector meson, we define
\begin{eqnarray}
\langle K^*(p) | (V-A)_\mu | B(p_B)\rangle & = & -i \epsilon^*_\mu
(m_B+m_{K^*}) A_1(s) + i (p_B+p)_\mu (\epsilon^* p_B)\,
\frac{A_2(s)}{m_B+m_{K^*}}\nonumber\\
\lefteqn{+ i q_\mu (\epsilon^* p_B) \,\frac{2m_{K^*}}{s}\,
\left(A_3(s)-A_0(s)\right) +
\epsilon_{\mu\nu\rho\sigma}\epsilon^{*\nu} p_B^\rho p^\sigma\,
\frac{2V(s)}{m_B+m_{K^*}}\,,}\hspace*{1cm}\nonumber\\
\langle {K^*}(p) | \bar s \sigma_{\mu\nu} q^\nu (1+\gamma_5) b |
B(p_B)\rangle & = & i\epsilon_{\mu\nu\rho\sigma} \epsilon^{*\nu}
p_B^\rho p^\sigma \, 2 T_1(s)\nonumber\\
\lefteqn{+ T_2(s) \left\{ \epsilon^*_\mu
  (m_B^2-m_{{K^*}}^2) - (\epsilon^* p_B) \,(p_B+p)_\mu \right\}  + T_3(s)
(\epsilon^* p_B) \left\{ q_\mu - \frac{s}{m_B^2-m_{{K^*}}^2}\, (p_B+p)_\mu
\right\}}\hspace*{4.5cm}\nonumber\\\label{eq:FFrare}
\end{eqnarray}
with 
\begin{eqnarray*}
 A_3(s) & = & \frac{m_B+m_{K^*}}{2m_{K^*}}\, A_1(s) -
\frac{m_B-m_{K^*}}{2m_{K^*}}\, A_2(s),\qquad
\vphantom{\frac{m_B-m_{K^*}}{2m_{K^*}}}A_0(0) = A_3(0), \qquad
 T_1(0) =  T_2(0).
\end{eqnarray*}
The form factors $T_i$ are renormalization-scale dependent.
All signs are defined in such a way as to render the form factors
positive. The physical range in $s$ extends from $s_{min}=0$ to
$s_{max}=(m_B-m_{K^*})^2$. As described in the last subsection for the
$B\to K^*\gamma$ form factor $T_1$, the above form factors are
essentially non-perturbative. 
Lacking results from lattice
calculations, we quote the form factors as calculated from QCD sum rules
on the light-cone \cite{ball98,BB}, in the parametrization
suggested in \cite{ABHH}, where also a discussion of the theoretical
uncertainties can be found. The form factors can be parametrized as
\begin{equation}
F(s) = F(0) \exp( c_1 \hat{s} + c_2 \hat{s}^2)\label{eq:para}
\end{equation}
with $\hat{s} = s/m_B^2$. The central values of the parameters $c_i$
are given in Tab.~\ref{tab:7.1}.
\begin{table}
\addtolength{\arraycolsep}{3pt}
\renewcommand{\arraystretch}{1.1}
$$
\begin{array}{|l||llll|lll|}
\hline
 & A_1 & A_2 & A_0 & V & T_1 & T_2 & T_3\\ \hline
F(0) & 0.337 & 0.282 & 0.471 &
0.457 & 0.379 & 0.379 & 0.260\\
c_1 & 0.602 & 1.172 & 1.505 & 1.482
& 1.519 & 0.517 & 1.129\\
c_2  & 0.258 & 0.567 & 0.710 & 1.015 & 1.030 &
0.426 & 1.128\\\hline
\end{array}
$$
\vspace*{-0.5cm}
\caption[]{Central values of parameters for the parametrization
  (\protect{\ref{eq:para}}) of
  the $B\to K^*$ form factors. Renormalization scale for
   $T_i$ is $\mu = m_b$.}\label{tab:7.1}
\addtolength{\arraycolsep}{-3pt}
\renewcommand{\arraystretch}{1}
\end{table}

\begin{figure}
\begin{minipage}[t]{5cm}
\epsfxsize=5cm
\centerline{\epsffile{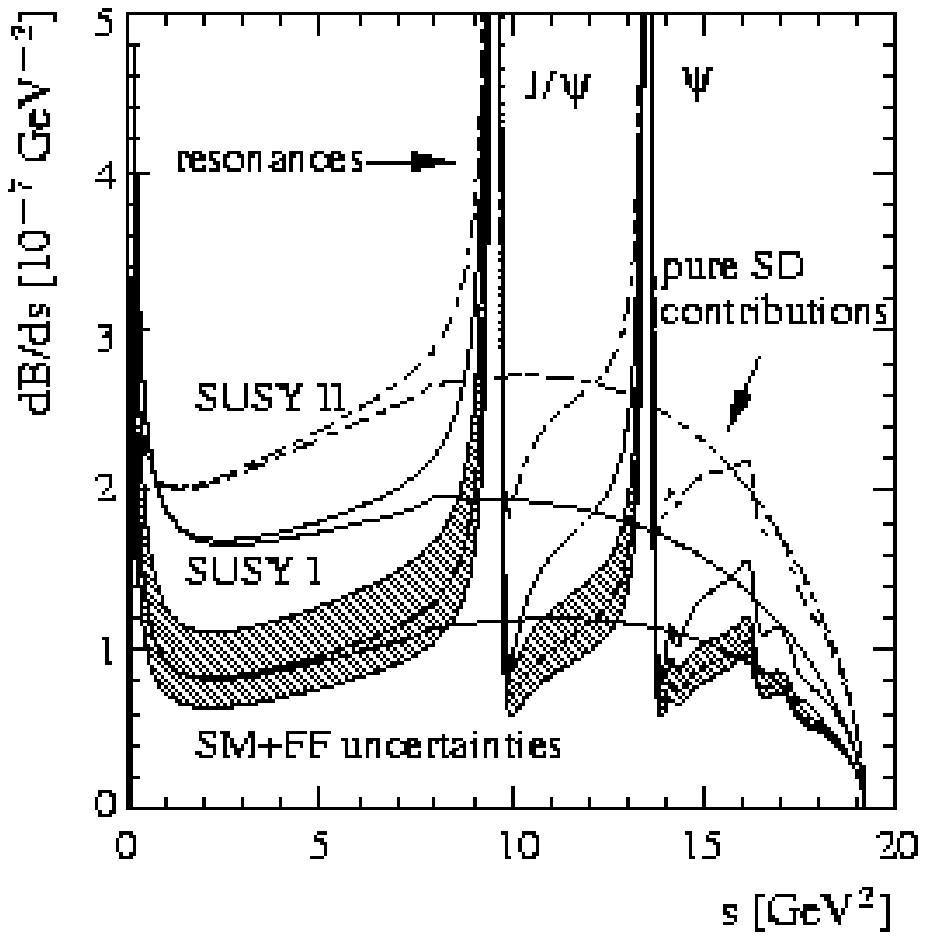}}
\vspace*{-0.5cm}
\caption[]{Dimuon-mass spectrum of $B\to K^*\mu^+\mu^-$ in the SM and
  two SUSY models}\label{fig:7.1}
\end{minipage}\hspace*{0.4cm}
\begin{minipage}[t]{5cm}
\epsfysize=5.1cm
\centerline{\epsffile{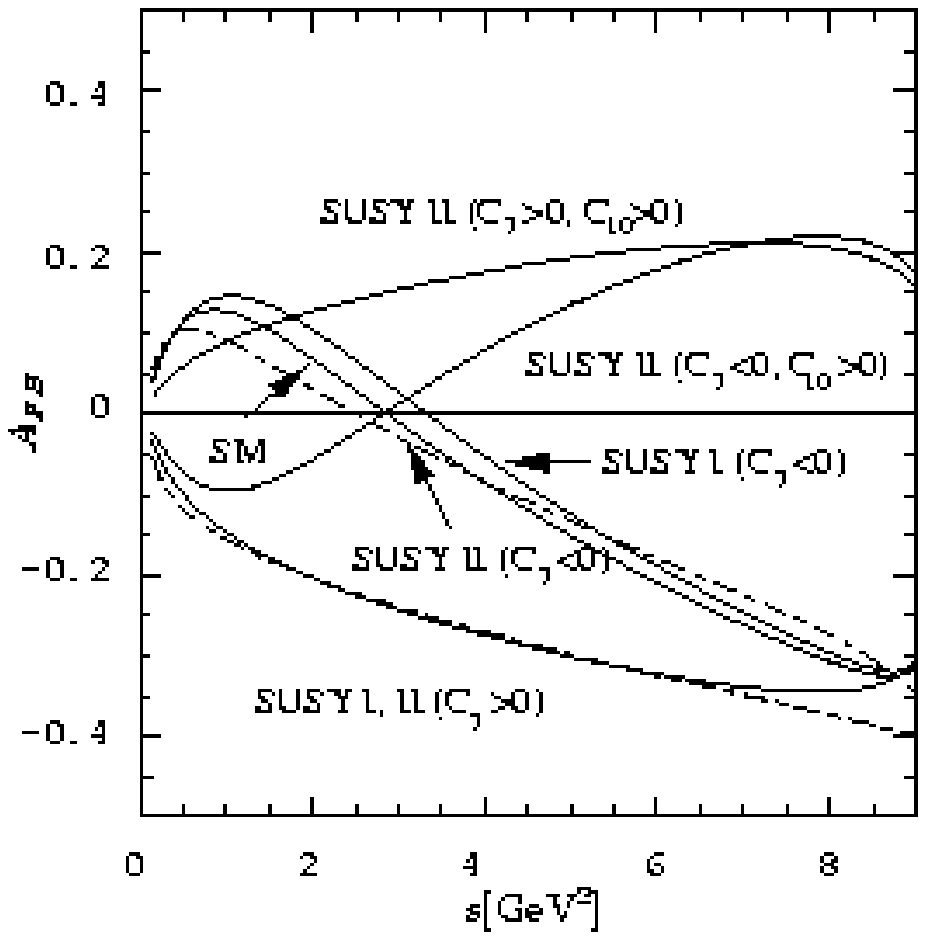}}
\vspace*{-0.5cm}
\caption[]{Forward-backward asymmetry of $B\to K^*\mu^+\mu^-$ in the SM and
  two SUSY models.}\label{fig:7.2}
\end{minipage}\hspace*{0.4cm}
\begin{minipage}[t]{5cm}
\epsfxsize=5cm
\centerline{\epsffile{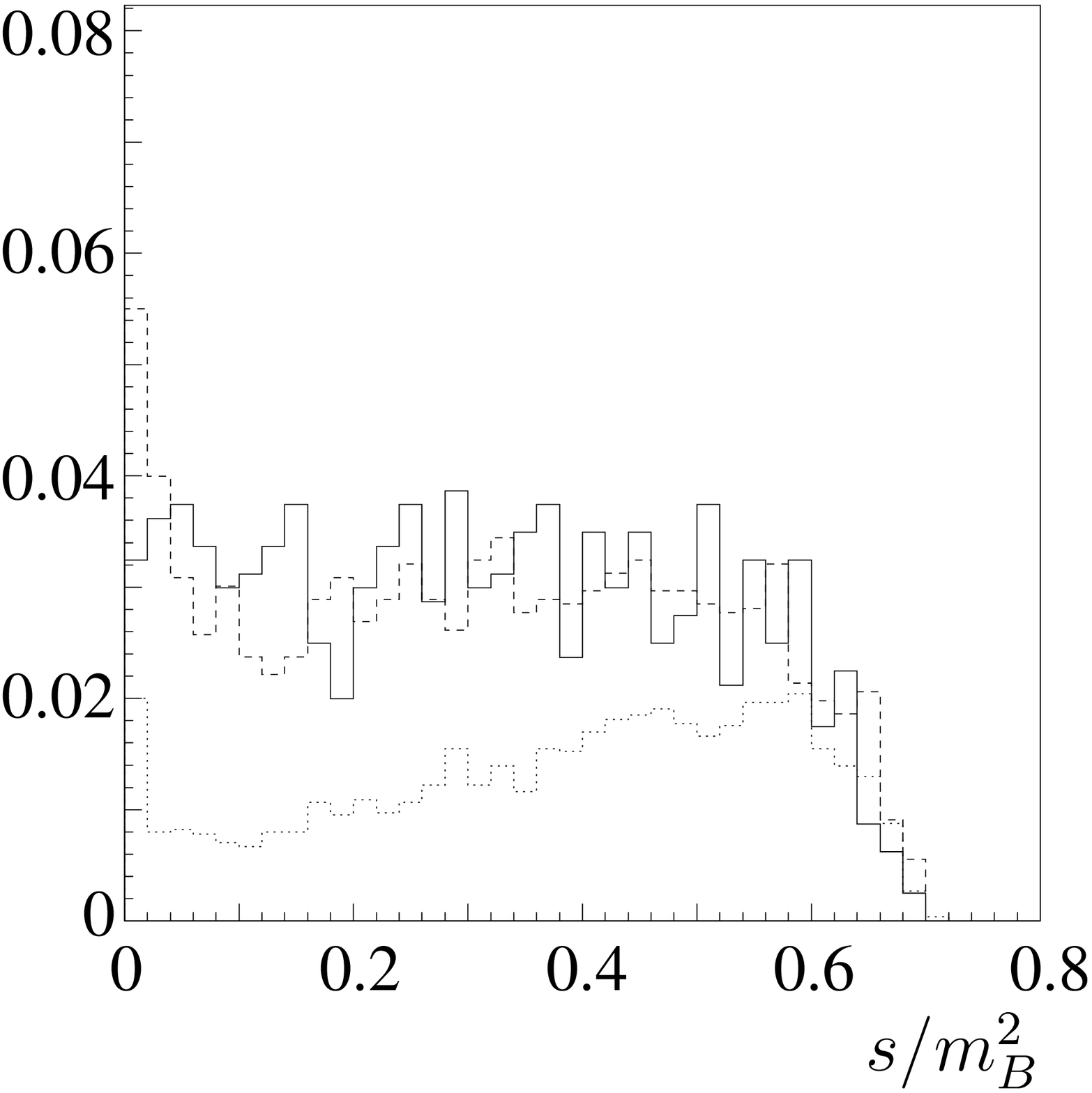}}
\vspace*{-0.5cm}
\caption[]{ATLAS' dilepton-mass distribution for 3
 data sets: solid line: PYTHIA, dashed: GI, dotted: 
ISGW2.}\label{matrixel}
\end{minipage}
\end{figure}

Let us now turn to the various decay distributions relevant for the
phenomenological analysis. For lack of space, we cannot give
detailed expressions for decay amplitudes and spectra in terms of the
hadronic matrix elements (\ref{eq:FFrare}); they can
be found in \cite{ABHH}. Besides the
total branching fraction and the spectrum in the dimuon mass, it is in
particular the forward-backward asymmetry that is interesting
for phenomenology. It is defined as
\begin{equation}
A_{FB}(s) = \frac{1}{d\Gamma/ds}\left( \int\limits_0^1
  d(\cos\theta) \frac{d^2\Gamma}{dsd\cos\theta} - \int\limits_{-1}^0
  d(\cos\theta) \frac{d^2\Gamma}{dsd\cos\theta}\right),\label{eq:defAFB}
\end{equation}
where $\theta$ is the angle between the momenta of the $B$ meson and
the $\mu^+$ in the dilepton CMS. The asymmetry is governed by
\begin{equation}
A_{FB}\propto  C_{10} \left[ {\rm Re}\left(
    C_9^{\mbox{\scriptsize eff}}\right) V(s) A_1(s)
+ \frac{\hat{m}_b}{\hat{s}} \,
    C_7^{\mbox{\scriptsize eff}} \left\{ V(s) T_2(s) \left( 1 - 
\hat{m}_{K^*}\right) +
    A_1(s) T_1(s) \left( 1 + \hat{m}_{K^*}\right)\right\}\right].
\end{equation}
In the SM, $A_{FB}$ exhibits a zero at $s=s_0$, given by
\begin{equation}
{\rm Re}\left(C_9^{\mbox{\scriptsize eff}}(s_0)\right) = 
-\frac{\hat{m}_b}{\hat{s}} \,
    C_7^{\mbox{\scriptsize eff}} \left\{\frac{T_2(s_0)}{A_1(s_0)}\, 
(1 - \hat{m}_{K^*}) + 
\frac{T_1(s_0)}{V(s_0)}\, (1 + \hat{m}_{K^*})\right\}.\label{eq:fbzero}
\end{equation}
The forward-backward asymmetry has a zero if and only if 
\begin{equation}\label{eq:condition}
{\rm sign} (C_7^{\mbox{\scriptsize eff}} {\rm
  Re}\,C_9^{\mbox{\scriptsize eff}}) = -1.
\end{equation}
It is interesting to observe that in  the Large Energy Effective
 Theory (LEET) \cite{LEET}, both ratios of the form factors appearing in
Eq.~(\ref{eq:fbzero}) have essentially no hadronic uncertainty,
 i.e.\ all dependence on
intrinsically non-perturbative quantities cancels, and one has simply
\begin{equation}
\frac{T_2(s)}{A_1(s)}= \frac{1+\hat{m}_{K^*}}{1+\hat{m}_{K^*}^2-\hat{s}}
\left(1-\frac{\hat{s}}{1-\hat{m}_{K^*}^2}\right), \qquad
\frac{T_1(s)}{V(s)}   =  \frac{1}{1+\hat{m}_{K^*}} \; .\label{eq:FBA}
\end{equation}
These relations are fulfilled by QCD sum rules on the light-cone to
2\% accuracy, which indicates that corrections to the LEET limit are
extremely small.
In that limit, one thus has a particularly simple form for the equation
determining $s_0$, namely
\begin{equation}
{\rm Re}(C_9^{\mbox{\scriptsize eff}}(s_0)) =- 2\,
\frac{\hat{m}_b}{\hat{s}_0}\, 
C_7^{\mbox{\scriptsize eff}}
\frac{1-\hat{s}_0}{1+\hat{m}_{K^*}^2 -\hat{s}_0} \; .
\label{eq:fbzeroleet}
\end{equation}
Thus, the precision of the zero of the forward-backward asymmetry in $B \to K^*
\mu^+ \mu^-$ is determined essentially
by the precision of the ratio of the effective coefficients and $m_b$
and is {\it largely independent of hadronic uncertainties}.
The insensitivity of $s_0$ to the decay form factors in $B \to K^*
\mu^+ \mu^-$ is a remarkable result, which  has also been discussed in
\cite{burdman}. However, the LEET-based
result in Eq.~(\ref{eq:FBA}) stands theoretically on more rigorous
grounds than the arguments based on scanning over a number of form factor
models as done in \cite{burdman}.
In the SM, one finds $s_0\approx 2.9\, \mbox{GeV}^2$ at $\mu = 4.8\,$GeV.
{}From Eq.~(\ref{eq:fbzero}) it also
follows that there is no zero below the
$c \bar{c}$ resonances if both $C_9^{\mbox{\scriptsize eff}}$ and 
$C_7^{\mbox{\scriptsize eff}}$ have
the same sign as
predicted in some beyond-the-SM models.
Thus, condition (\ref{eq:condition}) 
provides a discrimination between the SM and certain models
with new physics. 
Due to space limitations we cannot discuss in detail the possible
impact of particular beyond-the-SM scenarios on the decay
distributions introduced above. To illustrate the fact that large
effects are indeed possible, we show, in Figs.~\ref{fig:7.1} and
\ref{fig:7.2}, the results for the dimuon spectrum and the
forward-backward asymmetry obtained in \cite{ABHH} for several
SUSY-extensions of the SM. 

Note that the above formulas and considerations cannot immediately
be applied to the decay $B\to\rho\mu^+\mu^-$, whose measurement could,
in principle, together with that of $B\to K^*\mu^+\mu^-$, be used to 
determine the
ratio of CKM matrix elements $|V_{ts}/V_{td}|$, as an alternative
to the determination from B mixing. The problem lies in new
contributions to $C_9^{\mbox{\scriptsize eff}}$ originating from light
quark loops and
associated with the presence of low-lying resonances, for instance $\rho$
and $\omega$, in the dimuon spectrum. These contributions are
CKM-suppressed in $B\to K^*\mu^+\mu^-$, so that the corresponding uncertainties
can be neglected, but they are unsuppressed in
$B\to\rho\mu^+\mu^-$ decays. The problematic part in that is that the
theory tools that allow one to treat $c\bar c$ resonance contributions to 
$B\to K^*\mu^+\mu^-$ are not applicable anymore: perturbation theory does only
work in the unphysical region $s<0$, and an operator-product
expansion which would indicate potential power-suppressed terms also
fails. No satisfactory solution to that problem is presently available.

Finally, we note that the analysis of $B_s\to\phi \mu^+\mu^-$
parallels exactly that of $B_d\to K^*\mu^+\mu^-$; the corresponding
form factors can be found in Ref.~\cite{BB}. Also semimuonic decays
with a pseudoscalar meson in the final state, e.g.\ $B_d\to
K\mu^+\mu^-$ and $B_d\to\pi\mu^+\mu^-$, are, from a theoretical point
of view, viable sources for information on short-distance physics and
CKM matrix elements. Their experimental detection is, however,
extremely difficult and no experimental feasibility studies exist to date.

\subsubsection{Experimental Considerations}

As with $B\to\mu^+\mu^-$, the 
semimuonic decays $B^0_{d}\to K^* \mu^+ \mu^-$ are "self-triggering" 
channels thanks to the presence of two muons with high $p_T$ in the
final state. Particle identification helps decisively in 
separating the final-state hadrons.
All three experiments assume the branching ratio 
$B(B^0_d\to {K^*}^0 \mu^+\mu^-) =
1.5\times 10^{-6}$ for estimating the number of events to be observed. 
 
ATLAS have investigated form factor effects on the
detection of $B^0_d\to K^{*0} \mu^+\mu^-$; 
details of the analysis can be found in \cite{ATLASXmumu}.
Two different parametrizations of the hadronic matrix
 elements (\ref{eq:FFrare}), GI and ISGW2, were implemented
 into PYTHIA and the final numbers 
of expected events after trigger cuts were evaluated for these two samples of 
signal events. The dimuon mass distribution is shown in 
Fig.~\ref{matrixel}  for the case of the 
phase-space decay, GI and ISGW2 parametrizations. 
It was found that the matrix elements practically do not change the
 inclusive parameters of the muons and the $K^{*0}$ meson, which is important
 for triggering these events. They do, however, strongly influence the spectrum
 in the dimuon mass and the forward-backward asymmetry. Although quark
 model calculations of form factors like GI and ISGW2 may serve as
 rough guidelines for first estimates, they do not reflect the modern
 state-of-the-art of theoretical calculations. For this
 reason, it is important to extend existing studies, taking
 advantage of the recent developments
 in the theoretical calculation of hadronic matrix elements as discussed
 in the last subsection, and in particular to use only such model
 calculations that reproduce the model-independent results
 for certain form factor ratios like (\ref{eq:FBA}). 

\begin{figure}
\addtolength{\arraycolsep}{-1pt}
$$
\begin{array}{ccc}
\epsfxsize=5cm
\epsffile{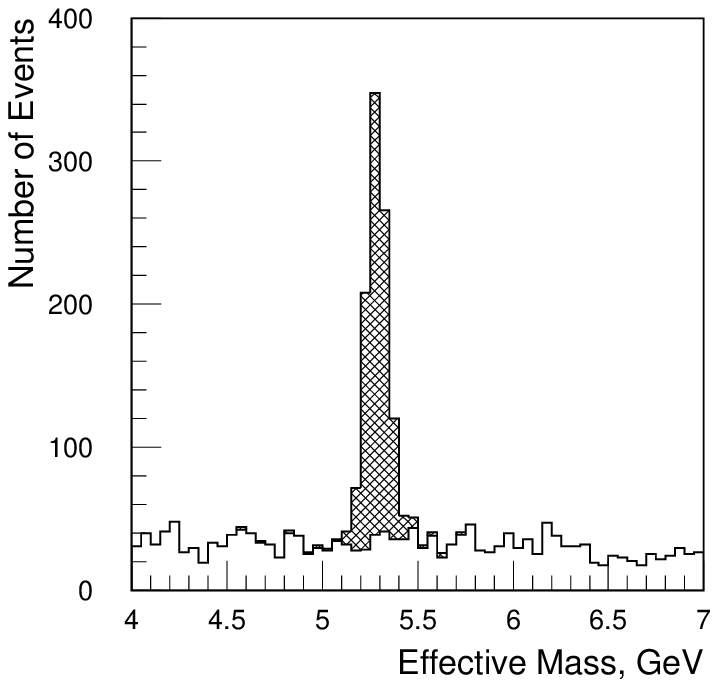}  &
\epsfxsize=5cm
\epsffile{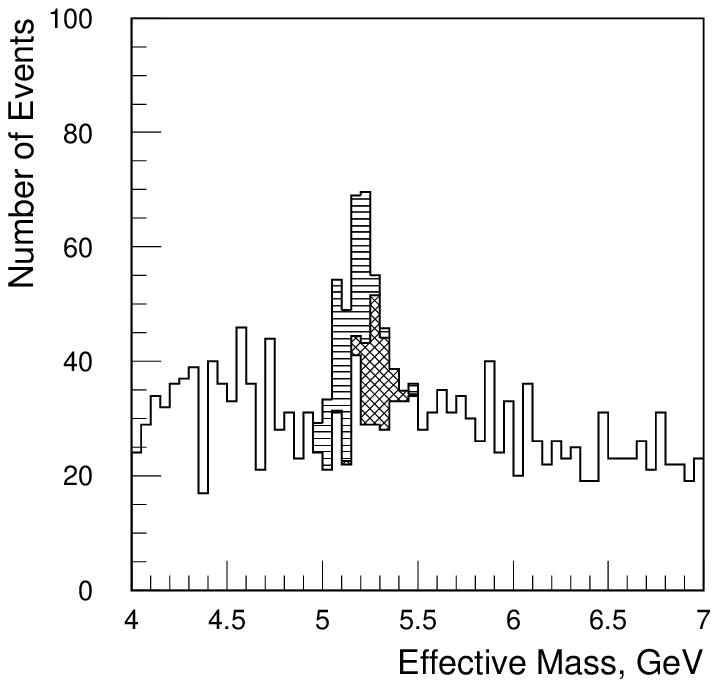} & 
\epsfxsize=5cm
\epsffile{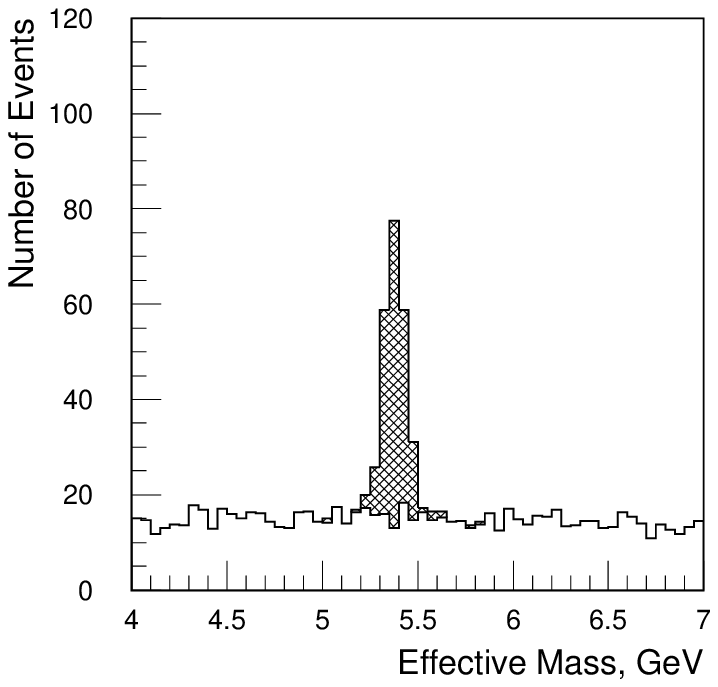}  
\end{array}
$$
\addtolength{\arraycolsep}{1pt}
\vspace*{-1.2cm}
\caption[]{$B^0_d\to {K^*}^0 \mu^+\mu^-$ (left), 
$B^0_d\to \rho^0 \mu^+\mu^-$ (centre) and 
$B^0_s\to \phi^0 \mu^+\mu^-$ (right) 
signals with background as simulated by ATLAS. The cross-hatched
histogram shows the $B^0_d\to \rho^0 \mu^+\mu^-$ 
signal, and the horizontally hatched one the reflection 
of $B^0_d\to {K^*}^0 \mu^+\mu^-$ to  $B^0_d\to \rho^0 \mu^+\mu^-$. 
}\label{signalplusbg}
\end{figure}

The ATLAS collaboration has studied the decays $B^0_d\to \rho^0 \mu^+\mu^-$, 
$B^0_d\to {K^*}^0 \mu^+\mu^-$ and $B^0_s\to \phi^0 \mu^+\mu^-$. All
these channels 
were fully simulated and reconstructed in the Inner Detector. As possible
background, the following reactions have been considered:
$B^0_d$ meson decays to $J/\psi K^0_s$,
$\omega^0 \mu^+\mu^-$, reflection of $B^0_d\to \rho^0 \mu^+\mu^-$ and
$B^0_d\to {K^*}^0 \mu^+\mu^-$ to other signal channels;
$B^0_s$ meson decays to  $K^{*0} (\phi)  \mu^+\mu^-$,
semimuonic decays of one of the $b$ quarks and 
semimuonic decays of both $b$ quarks. 
An additional minimum bias of 2.4 events in the precision tracker and
3.2 events in the transition radiation tracker were taken into account
when studying the signal and background. The expected results for
observing these three channels are shown
in  Fig.~\ref{signalplusbg}. 

Assuming the SM to be valid, 
the measurement of the branching fractions of the decays 
$B^0_d\to \rho^0 \mu^+\mu^-$ and $B^0_d\to {K^*}^0 \mu^+\mu^-$ gives,
in principle, the
possibility to extract the ratio of the CKM elements 
$|V_{td}|/| V_{ts}|$ using the following equation:  
\begin{equation}
\frac{N(B^0_d\to \rho^0 \mu^+\mu^-)}{N(B^0_d\to {K^*}^0 \mu^+\mu^-)} =
 k_d\,\frac{|V_{td} |^2}{|V_{ts}|^2}\,.
\end{equation}
The quantity $k_d$ depends on form factors and Wilson-coefficients
and also on the experimental cuts. Although there exist claims in the
literature that, with proper cuts, $k_d$ may be calculated with small
hadronic uncertainties, see e.g.\ \cite{bloedsinn}, these papers tend
to underestimate the uncertainty associated with the impact on $c\bar
c$ resonances on the spectrum (for $B^0_d\to \rho^0 \mu^+\mu^-$, there
are also $u\bar u$ resonances whose contributions are often completely
ignored). Our present knowledge of these long-distance
effects in $C_9^{\mbox{\scriptsize eff}}$ is, as has also been discussed in the
theory subsection, unsatisfactory and calls for improved theory
studies.

ATLAS also studied the prospects for measuring 
the forward-backward (FB) asymmetry $A_{FB}$, defined in
(\ref{eq:defAFB}). Experimentally, the following quantity will be measured: 
\begin{equation}
\langle A_{FB}\rangle _{[s_1, s_2]}=
\frac{\langle N_F\rangle _{[s_1, s_2]}-\langle N_B
\rangle _{[s_1, s_2]}} {\langle N_F\rangle _{[s_1, 
  s_2]}+\langle N_B\rangle _{[s_1, s_2]}}\,,
\end{equation}
where $\langle N_F\rangle _{[s_1, s_2]}$ and $\langle N_B
\rangle_{[s_1, s_2]}$ are 
the numbers of positive leptons (including background ones) 
moving in the forward and backward directions of
the B meson, respectively, in the range of the squared 
dimuon mass $s \in [s_1, s_2]$. 
In Fig.~\ref{fig:7.2}, we show the SM prediction for $A_{FB}$ 
together with predictions in several supersymmetric extensions of the
SM, which are characterized by the possibility that the
Wilson-coefficients $C_7^{\mbox{\scriptsize eff}}$ and/or 
$C_9^{\mbox{\scriptsize eff}}$ can change 
sign with respect  to the SM. As discussed in the previous subsection,
the behavior of the asymmetry with 
$s$ depends crucially on these signs. For example, if 
the asymmetry turns out to be negative at 
small $s$, then this  means that there is new physics beyond the SM.

The precision for asymmetry measurements in three different $s$ intervals 
was estimated by ATLAS. The data are presented in
Tab.~\ref{asymmetrymeas}, together with asymmetry values in the SM and one
exemplary supersymmetric model, integrated over the corresponding
intervals in $\hat s=s/m_B^2$. 
\begin{table}
\begin{center}
\begin{tabular}{|c|c|c|c|}
\hline 
Interval & ${\hat s}_{min} \div 0.14 $ &$0.14 \div 0.33 $ & $0.55 \div 
{\hat s}_{max}$   \\ \hline
ATLAS $\delta A_{FB}$  (3 years)  & 5 \% & 4.5 \% & 6.5 \% \\ \hline
LHCb $\delta A_{FB}$ (1 year)  & 2.4 \% & 2.4 \% & 5.8 \% \\ \hline
SM $A_{FB}$ & 10\% & $-$14\% & -29 \%   \\ \hline
MSSM $A_{FB}$ & $(-17 \div 0.5)$\% &$(-35 \div -13)$\%  & $(-33 \div
-29)$ \% \\ \hline
\end{tabular}
\end{center}
\vspace*{-0.5cm}
\caption[]{Expected precision for asymmetry measurements at ATLAS and
  LHCb, for 3 and 1 years running, respectively, at low luminosity
 and assuming SM 
branching ratios; the experimental numbers rely on \protect\cite{flora} and
the theoretical predictions on the form factors
in the GI parametrization and MSSM parameters as
discussed in \protect\cite{missing}. 
The kinematic limits are given by ${\hat
  s}_{min}=4m^2_\mu/m^2_B$ and 
${\hat s}_{max}=(m_B-m_{K^*})^2/m^2_B$.}\label{asymmetrymeas}
\end{table}
The expected accuracy of the asymmetry measurement with the ATLAS
detector will be
sufficient to distinguish between the SM and some of its extensions. It
should, however, be stressed that new-physics effects {\it not}
yielding sign-flips of the Wilson-coefficients do not change
$A_{FB}$ dramatically as compared to the SM.

\begin{figure}
\begin{center}
\epsfxsize=6cm
\epsffile{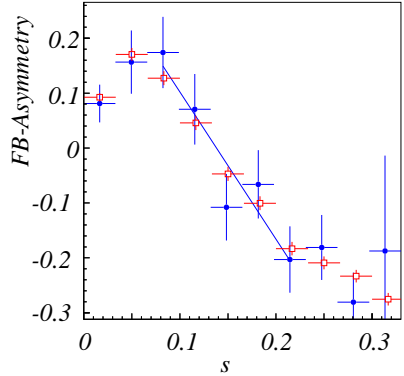}  
\end{center}
\vspace*{-0.8cm}
\caption[]{LHCb fit of the FB-asymmetry $A_{FB}$ for 
$B\to K^{*}\mu^{+}\mu^{-}$ around the zero $\hat{s}_0$ with
 $A_{FB}(\hat{s}_0)=0$. Squares denote generated
 and dots reconstructed data (one year statistics). The linear fit of 
reconstructed data intersects at $0.138 \pm 0.035$ (only statistical 
error).}\label{FBFIT}
\end{figure}

LHCb has also performed an analysis of $B^0_d\to
 K^{*0} \mu^+\mu^-$.  The matrix elements reproducing 
the correct dimuon mass distribution were implemented into PYTHIA. 
The detector response for both signal and background events was
 simulated and the charged particles were 
reconstructed in the detector. LHCb expects to observe 4500
 $B^0_d\to K^{*0} \mu^+\mu^-$ events per year.
For background studies, the following reactions were simulated with PYTHIA:
$B^0_d\to {K^*}^0 \mu^+\mu^-$, 
$B^0_d\to J/\psi (K^{*0},K^0_s,\phi,K^+)$, with the subsequent
decay of $J/\psi$ into two muons, inclusive  $B\to 4 \pi$,
$b\to \mu X, \overline{b} \to \mu X $ and
$B \to  \mu D(\mu X) X $. The total number of background events is
expected to be 280. The large signal statistics with very low
 background gives a
nice possibility to study this channel in detail. LHCb also evaluated the 
sensitivity of $A_{FB}$ measurements. The results are shown in the
 Tab.~\ref{asymmetrymeas}. 
Promising results
were obtained by LHCb for measuring the position of the zero 
of $A_{FB}$, $\hat{s}_0$ with $A_{FB}(\hat{s}_0)=0$. As
 discussed in the previous subsection, the 
position of the zero is proportional to the ratio of two 
Wilson-coefficients, $C_9^{\mbox{\scriptsize
 eff}}/C_7^{\mbox{\scriptsize eff}}$, with only small
 hadronic uncertainties from form factors. Note, however,
 that it is the {\it effective} Wilson-coefficients that determine
 $\hat{s}_0$ and that these coefficients encode both short-distance
 SM and -- potentially -- new physics effects and long-distance QCD
 effects, which latter ones {\it do} come with a certain hadronic
 uncertainty that to date has not been investigated in sufficient detail.
LHCb simulated the expected measurements of the asymmetry, see 
Fig.~\ref{FBFIT},   and made a
linear fit of the "experimental points".  It is shown that  $\hat s_0$ can be
measured with 25\% 
accuracy, which  leads to a 4\% 
error in extracting the ratio 
$C_9^{\mbox{\scriptsize eff}}/C_7^{\mbox{\scriptsize eff}}$.

The CMS collaboration studied three rare  B meson decay channels,
 $B^0_d\to {K^*}^0 \mu^+\mu^-$, 
 $B^0_d\to \rho^0 \mu^+\mu^-$,
 $B^0_s\to \phi^0 \mu^+\mu^-$, 
at the particle level. No full simulation of the signal and background
 in 
the CMS detector has been performed yet. Secondary vertex
 reconstruction was however studied in detail. The main source of
 uncertainty in the CMS evaluation is the efficiency of higher-level
 triggering of dimuons with continuum mass distribution. The complete
 event reconstruction, using object-oriented techniques, the
 implementation of various higher trigger level strategies and the
 evaluation of triggering efficiencies, is now under way in CMS. 
The sources  of background  considered are $B^0\to J/\psi (\mu^+\mu^-) X$,
$B^0\to \mu Y \to \mu \mu +X$, reflection of $B^0_s\to \phi^0 
 \mu^+\mu^-$ and  $B^0_d\to K^{*0} \mu^+\mu^-$ to other signal
 channels and semimuonic decays of both $b$ quarks.

The numbers of signal and background 
events  expected by ATLAS, CMS and LHCb are given in
Tab.~\ref{expectedsignalxll}.

\begin{table}
\renewcommand{\baselinestretch}{1.4}
\begin{center}
\begin{tabular}{|l|r||c|c||c||c|c|}  \hline
& & \multicolumn{2}{c||}{ATLAS} & \multicolumn{1}{c||}{CMS} &
    \multicolumn{2}{c|}{LHCb}\\\cline{3-7}
\multicolumn{1}{|c|}{Channel} & \multicolumn{1}{c||}{$B$} & 
\multicolumn{1}{c|}{Signal} & \multicolumn{1}{c||}{BG} &
    \multicolumn{1}{c||}{Signal} &
    \multicolumn{1}{c|}{Signal}& \multicolumn{1}{c|}{BG}\\\hline
$B^0_d\to \rho^0 \mu^+\mu^-$  & 10$^{-7}$ & 222 & 950  & 1050  &
    \multicolumn{2}{c|}{not yet estimated}\\ \cline{6-7}
$B^0_d\to K^{*0} \mu^+\mu^-$  & 1.5$\,\times\,$10$^{-6}$ & 1995 & 290 &
    12600 & 22350 & 1400 ($<$ 4300 \@ 95\% C.L.)\\\cline{6-7}
$B^0_s\to \phi^0 \mu^+\mu^-$  & 10$^{-6}$ & 411 & 140 & 3600  &
    \multicolumn{2}{c|}{not yet estimated}\\ \hline
\end{tabular}
\end{center}
\renewcommand{\baselinestretch}{1}
\vspace*{-0.5cm}
\caption[]{Expected signal and background statistics for rare
  semimuonic decays, for 3 years' running of ATLAS
  and CMS at low luminosity and 5 years' of LHCb. The CMS simulation
  was done at the particle level only.}\label{expectedsignalxll}
\end{table}

\subsection{Inclusive Decays}

The inclusive decay mode $B\to X_s\gamma$ has received much attention
in connection with its measurement at CLEO, $B(B\to X_s \gamma) = (3.15
\pm 0.35 \pm 0.32 \pm 0.26) \times 10^{-4}$ \cite{CLEO99}, which
should become much more accurate with data being taken at the $e^+e^-$
B factories. A state-of-the-art review on inclusive decays can be
found in the corresponding Chapter of the BaBar physics book
\cite{BaBar}.  The experimental environment of a hadronic machine
makes it very hard to measure inclusive decays. Nevertheless, the
D$\emptyset$ collaboration at Fermilab was able to set a 90\%~CL bound
$B( B\to X_s \mu^+ \mu^-) < 3.2 \times 10^{-4}$ \cite{D0}, which
should be compared to the corresponding CLEO \cite{CLEO98} result of
$5.8 \times 10^{-5}$ and the SM expectation of $6 \times 10^{-6}$.  In
the D$\emptyset$ analysis, no displaced vertex was required for the
muon pair, contrary to the CDF analysis of the exclusive $B\to
K^*\mu^+\mu^-$ mode \cite{CDF-rare}, where a sensitivity of order $10^{-6}$
has been reached.  It is an interesting question to ask
whether LHC could improve the
D$\emptyset$ result (e.g. by requiring a displaced vertex) and whether
it could possibly reach the SM sensitivity for $B\to X_s\gamma$ or
$B\to X_s \mu^+ \mu^-$.

The theoretical advantage of the inclusive decays $B\to X_s\gamma$ and
$B\to X_s \mu^+ \mu^-$ over particular exclusive channels lies
in the fact that non-perturbative contributions to the inclusive modes
can be calculated in a model-independent way with the help of the Operator
Product Expansion (OPE) within the Heavy Quark Effective Theory (HQET)
\cite{FLS94}.  Actually, this statement is true only at the leading
order in hard strong interactions (i.e. in $\alpha_s(m_b)/\pi$) and
only after imposing certain kinematic cuts (see e.g.
\cite{mikolaj2,mp}). Even with these restrictions, the accuracy of
theoretical predictions for the inclusive branching ratios is expected
to be better than in the exclusive case.

The theoretical analysis of $\bar{B} \to X_s \gamma$ proceeds along
the same lines as in the $\bar{B} \to \bar{K}^* \gamma$ case, up to
Eq.~(\ref{eq:with.OF}), where $\bar{K}^*$ has to be replaced by
any $S=-1$ hadronic state $X_s$. Then, the modulus squared of the
amplitude is taken, and a sum over all the states $X_s$ is performed.
The obtained sum can be related via optical theorem to the imaginary
part of the $\bar{B}\gamma \to \bar{B}\gamma$ elastic scattering
amplitude, analogously to what is done in the analysis of $\bar{B} \to
X_{u,c} e \bar{\nu}$ \cite{CGG90}. After OPE and
calculating matrix elements of several local operators between
$\bar{B}$ meson states at rest, one finds that the ``subtracted''
branching ratio
\begin{equation}
B( \bar{B} \to X_s \gamma)^{\mbox{\scriptsize 
subtracted}\;\psi}_{E_{\gamma} > E_{\rm cut}}
\equiv B( \bar{B} \to X_s \gamma)_{E_{\gamma} > E_{\rm cut}}
-      B(\bar{B} \to X^{(1)}_{\mbox{\scriptsize no charm}} \psi  )
\times B( \psi   \to X^{(2)}_{\mbox{\scriptsize no charm}} \gamma)
\end{equation}
is given in terms of the purely perturbative $b$ quark decay width, up to
small non-perturbative corrections
\begin{eqnarray}
\frac{ \Gamma( \bar{B} \to X_s \gamma)^{\mbox{\scriptsize subtracted}
\;\psi}_{E_{\gamma} > 
E_{\rm cut}}}
{ \Gamma( \bar{B} \to X_c e \bar{\nu}_e)} &\simeq&
\frac{ \Gamma( b \to X_s \gamma)^{\mbox{\scriptsize 
perturbative NLO}}_{E_{\gamma} > 
E_{\rm cut}}}
{ \Gamma( b \to X_c e \bar{\nu}_e)^{\mbox{\scriptsize perturbative
      NLO}}} \times
\nonumber \\ \label{eq:bratio}
&\times& \left[ 1 + ({\cal O}(\Lambda^2/m_b^2)\simeq 1\%)
                + ({\cal O}(\Lambda^2/m_c^2)\simeq 3\%) \right].
\end{eqnarray}
The normalization to the semileptonic rate has been used here to
cancel uncertainties due to $m_b^5$, CKM-angles and some of the
non-perturbative corrections. One has to keep in mind that
(\ref{eq:bratio}) becomes a bad approximation for $E_{\gamma}^{\rm
  cut} \ll 1$~GeV, and that non-perturbative corrections grow
dramatically when $E_{\gamma}^{\rm cut} > 2$~GeV. Moreover,
non-perturbative effects arising at ${\cal O}(\alpha_s(m_b))$ are not
included in (\ref{eq:bratio}). Estimating the size of such
non-perturbative effects requires further study, see Ref.~\cite{mp}.  
For $E_{\rm  cut} = 1$~GeV, Eq.~(\ref{eq:bratio}) gives
\begin{equation}
\label{nbratio}
B( \bar{B} \to X_s \gamma)^{\mbox{\scriptsize
subtracted}\;\psi}_{E_{\gamma} > E_{\rm cut}} = (3.29 \pm 0.33) \times 10^{-4},
\end{equation}
where the dominant uncertainties originate from the uncalculated
${\cal O}(\alpha_s^2)$ effects and from the ratio $m_c/m_b$ in the
semileptonic decay (around 7\% each).

The calculation of $\bar{B} \to X_s \mu^+ \mu^-$ for small dimuon
invariant mass is conceptually analogous to $\bar{B} \to X_s \gamma$,
but technically more complicated, because more operators become
important. Here, we shall quote only the numerical estimate
\cite{mikolaj2}
\begin{equation}
B( B \to X_s \mu^+ \mu^-)_{s \in [0.05 m_b^2, 0.25 m_b^2]} 
= (1.46 \pm 0.19) \times 10^{-6},
\end{equation}
where only the error from $\mu$-dependence of the perturbative
amplitude is included.

\subsection{Conclusions}

The LHC experiments will be able to make precise measurements of
rare radiative, 
semimuonic and muonic B decays.  ATLAS and CMS will measure 
rare decays in the central $\eta$ region, which will be
complementary to the 
data to be taken by LHCb. A first assessment of LHC's
potential to measure rare B decays, presented in this report, 
has shown that it will be possible to
\begin{itemize}
\item observe $B^0_s\to \mu^+\mu^-$, measure its branching ratio,
  which is of order $10^{-9}$ in the SM, and
 perform a high sensitivity search for $B^0_d\to \mu^+\mu^-$;
\item measure the branching ratio and decay characteristics
  of  $B^0_d\to K^{*0} \gamma$ at LHCb;
\item   measure the branching ratios of 
 $B^0_s\to \phi^0 \mu^+\mu^-$, $B^0_d\to \rho^0 \mu^+\mu^-$
and $B^0_d\to K^{*0} \mu^+\mu^- $ and study the dynamics of
these decays;
\item  measure the FB-asymmetry in 
 $B^0_d\to K^{*0} \mu^+\mu^- $, which allows the distinction between
 the SM and a large class of SUSY models.
\end{itemize}
Studying rare muonic decays at high luminosity with the ATLAS and
CMS detectors would 
significantly improve the results that can be obtained at low luminosity. 

Open questions to be discussed in the future:
\begin{itemize}
\item assessment of the combined performance of LHC experiments
on rare muonic and semimuonic decays;
\item studies of CP asymmetries in rare semileptonic
B decays at LHC;
\item evaluation of the potential of ATLAS, CMS and LHCb to
measure inclusive $B^0_{d,s}\to \rm{X} \mu^+\mu^-$  branching ratios;
\item detection of rare 
decays with a $\tau$ in the final state;
\item feasibility study for measuring semimuonic decays with a
  pseudoscalar meson in the final state, e.g.\ $B_d\to \pi\mu^+\mu^-$,
  $B_d\to K\mu^+\mu^-$.
\end{itemize}

{}From the theory point of view, the most urgent question left open is
the precise assessment of long-distance effects both in the radiative
B decays $B\to(K^*,\rho)\gamma$ and in the semimuonic ones, encoded in
the effective
Wilson-coefficient $C_9^{\mbox{\scriptsize eff}}$; the lack of 
knowledge of these effects
limits the precision with which CKM matrix elements and
short-distance coefficients can be extracted from semimuonic decays.
Other tasks remaining are the improvement of form factor calculations,
for instance from lattice, and the parametrization of form factors in
a form that includes as much known information on the positions of
poles and cuts as possible. Also, the possible size of CP asymmetries
in semimuonic decays deserves further study; only few papers treat
that subject, see e.g.\ \cite{semimuonicCP}.

\setcounter{equation}{0}
\section[THEORETICAL DESCRIPTION OF NONLEPTONIC DECAYS]{THEORETICAL 
DESCRIPTION OF NONLEPTONIC DECAYS\protect\footnote{Section coordinator: 
P. Ball, with help from M. Beneke, G. Buchalla, I. Caprini and 
A. Khodjamirian.}}
\label{sec:nonlept}

Exclusive nonleptonic B decays form an important
part of LHC's B physics programme and at the same time pose a big
challenge for theory. In the standard approach using an effective weak
Hamiltonian, nonleptonic decay amplitudes are reduced to products of
short-distance Wilson-coefficients and hadronic matrix elements. The
calculation of the latter ones requires genuine knowledge of
nonperturbative QCD and is often done in the so-called factorization
approximation, where a matrix element over typically a four-quark
operator is ``factorized'' into a product of matrix elements over
current operators, which are much easier to calculate:
$$
\langle J/\psi K_S \mid (\bar c \gamma_\mu c) (\bar s \gamma^\mu b)
\mid B\rangle \to \langle J/\psi  \mid (\bar c \gamma_\mu c) \mid 0
\rangle \times \langle K_S \mid  (\bar s \gamma^\mu b) \mid B\rangle\,.
$$
The factorization approximation is, of course, not exact and the
assessment of ``nonfactorizable contributions'', including
final-state-interaction phases, is a fundamental
problem of strong interactions, which affects both the extraction of
weak phases from CP asymmetries, like ${\cal A}_{\rm CP}(B\to\pi\pi)$,
and the determination of CKM angles or new physics from rare
decays. Whereas in Secs.~3 to 5 a pragmatic approach has been
presented which aims at constraining strong-interaction effects from
experiment, it remains a big challenge for theory to
{\it predict} these effects from first principles. 
For this reason we devote a separate section to review
several ans\"{a}tze for solving or rather approaching the problem,
although it is to be admitted that a complete solution is still far
beyond our power. In three subsections we discuss the calculation of
nonfactorizable contributions to $B\to J/\psi K^{(*)}$ from QCD sum
rules on the light-cone \cite{KR,KR2}, a method to obtain information
on the strong phase in $B\to\pi\pi$ from dispersion relations
\cite{CaMi} and, finally, an approach that applies the methods
developed for hard exclusive QCD reactions to certain B decays in the
heavy quark limit $m_b\to \infty$ \cite{BBNS}. 
We would like to stress, however, that the problem of how to calculate 
non-factorizable contributions and, in particular,
final-state-interaction phases, is very challenging indeed and that a
lot of theory work remains to be done. We thus can present, 
instead of a coherent picture, only facettes, albeit scintillating 
ones. 

\subsection[Nonfactorizable Contributions to 
$B\to J/\psi K^{(*)}$]{Nonfactorizable Contributions to \protect\boldmath 
$B\to J/\psi K^{(*)}$}\label{sec:nonlept1}

The nonfactorizable contributions 
to the amplitudes of $B\to J/\psi K^{(*)}$ decays have recently been  
estimated \cite{KR,KR2}  using operator product expansion (OPE) and 
QCD light-cone sum rules. In this subsection, we outline the main results 
of this study. 

With the effective Hamiltonian (\ref{eq:hammel}), 
the matrix element of $B \rightarrow J/\psi K^{(*)} $  
has the following form:
\begin{eqnarray}
\langle K^{(*)} J/\psi \mid {\cal H}^s_{\mbox{\scriptsize eff}}  \mid B\rangle
 &= &4\,\frac{G_F}{\sqrt{2}}V_{cb}V_{cs}^*
\Bigg [\left(C_1(\mu)+\frac{C_2(\mu)}3\right)
\langle K^{(*)} J/\psi \mid O^s_1(\mu) \mid B\rangle \nonumber\\
& & {}+\frac{1}{2}\,C_2(\mu)
\langle K^{(*)} J/\psi \mid \widetilde{O}^s_1(\mu) \mid B\rangle \Bigg].
\label{ampl}
\end{eqnarray}
The explicit form of the four-quark operators $O^s_{1,2}$ is given in
(\ref{eq:O12}). The operator 
$$
\widetilde{O}^s_1=(\bar{c}\Gamma^\rho \frac{\lambda^a}{2}c)
(\bar{s}\Gamma_\rho \frac{\lambda^a}{2}b)
$$
with $\Gamma_\rho=\gamma_\rho(1-\gamma_5)$ originates from the Fierz 
rearrangement of $O^s_2$.
In the  factorization approximation, the matrix 
elements of $\widetilde{O}^s_1$  vanish and the matrix elements of 
$O^s_1$ are split into the  product
\begin{equation}
\langle
K^{(*)} J/\psi \mid O^s_1(\mu)\mid B\rangle
 =  \frac{1}{4}\,
\langle J/\psi \mid \bar{c}\Gamma^\rho c  \mid 0 \rangle
\langle K^{(*)}  \mid \bar{s}\Gamma_\rho b \mid B\rangle ~,
\label{fact}
\end{equation}
involving simpler matrix elements of quark currents:
$
\langle 0 \mid \bar{c}\Gamma^\rho c \mid J/\psi(p)\rangle= 
f_\psi m_{\psi}\epsilon_\psi^\rho 
$
and 
\begin{equation}
\langle K(p) \mid \bar{s}\Gamma_\rho b \mid B(p_B)\rangle
 = f_+(s)(p_{B\rho}+p_\rho) + f_-(s) q_\rho.
\label{formf}
\end{equation}
The form factor decomposition of the matrix element
$
\langle K^{(*)}(p) \mid \bar{s}\Gamma^\rho b \mid B(p_B)\rangle
$
can be found in (\ref{eq:FFrare}).
In the above, $q=p_B-p$, $s=q^2$, 
$f_\psi$ is the $J/\psi$ decay constant, 
$\epsilon_\psi$ , $\epsilon_{K^*}$ are the polarization vectors
of $J/\psi$ and $K^*$, respectively, and $f_{\pm}$ are the form 
factors for $B\to K$. 
For the numerical analysis we use the form factors as calculated from
QCD sum rules on the light-cone \cite{BKR,BB,ball98,ABHH}. 

The short-distance coefficients $C_{1,2}(\mu)$ 
and the matrix elements entering (\ref{ampl}) are scale-dependent, 
whereas the decay constants and form factors determining the
right-hand side 
of (\ref{fact}) are physical scale-independent quantities. Therefore,   
factorization can at best be an approximation valid 
at one particular scale.  
In fact, in both $B\to J/\psi K$  and $B\to J/\psi K^*$, 
factorization does not work at $\mu = {\cal O}(m_b)$ and is unable 
to reproduce both partial widths and their ratio 
as can be seen from Tab.~\ref{tab:alex}. 
\begin{table}
\renewcommand{\arraystretch}{1.2}
\begin{center}
\begin{tabular}{|c||c|c|l|}
\hline
Decay Parameter& (a) 
& (b)
& \multicolumn{1}{|c|}{Experiment}\\
\hline
$\Gamma(B \to J/\psi K)$ ( in 10$^8$ sec$^{-1}$) & 
1.0 $\div$ 1.5 &  0.15 $\div$ 0.2  & \parbox[c]{3.3cm}{5.8 $\pm$ 0.8
  ($B^0$)\cite{PDG} \\6.1 $\pm$ 0.6 ($B^\pm$)\cite{PDG}} \\
\hline
$\Gamma(B \to J/\psi K^*)$ ( in 10$^8$ sec$^{-1}$) & 3.9  $\div$ 6.0
& 0.6 $\div$ 0.9 &\parbox[c]{3.3cm}{9.7 $\pm$ 1.1 ($B^{0}$)\cite{PDG}\\
9.0 $\pm$ 1.6 ($B^{\pm}$)\cite{PDG}}\\
\hline
$\Gamma(B \to J/\psi K^*)/\Gamma(B \to J/\psi K)$ 
& 
\multicolumn{2}{|c|} 
{2.6 $\div$ 6.2 }
& 1.45 $\pm$ 0.26 \cite{CLEO}\\
$P_L= \Gamma_L/\Gamma$ & \multicolumn{2}{|c|}{0.475 $\div$ 0.465} &
\parbox[c]{3.3cm}{0.52 $\pm$ 0.08\cite{CLEO}\\0.65 $\pm$ 
0.11\cite{CDF-schmidt,CLEO}} \\
\hline
$\displaystyle\left|a_2^{B\psi K}\right|$ & 0.14 & 0.055 & 0.31 $\pm$
0.02 $\pm$ 0.03\\ 
$\displaystyle\left|a_{2,1}^{B\psi K^*}\right|$& 0.14 & 0.055 &  
0.18$^{+0.03}_{-0.04} \pm\,$0.02 \\
$\displaystyle\left|a_{2,2}^{B\psi K^*}\right|$
& 0.14 & 0.055 &\parbox[c]{3.3cm}{0.13$^{+0.09}_{-0.10} \pm\,$0.01\\ 
0.69$^{+0.07}_{-0.08} \pm\,$0.05} \\
$|a_{2,V}^{B\psi K^*}|$&  0.14&  0.055 &
0.16$^{+0.04}_{-0.05} \pm\,$0.02\\
\hline
\end{tabular}
\renewcommand{\arraystretch}{1}
\end{center}
\vspace*{-0.5cm}
\caption[]{ $ B \to J/\psi K^{(*)}$ decay characteristics 
calculated in naive factorization 
approximation, neglecting nonfactorizable contributions and taking 
$C_{1,2}(\mu)$ from \cite{BaBar} in NLO at (a) $\mu=m_b$, (b) $\mu= m_b/2$ 
and compared with experiment. The intervals of 
theoretical predictions reflect the uncertainties in the 
$B\to K$ and $B \to K^*$ form factors taken from \cite{ABHH}.}\label{tab:alex}
\end{table}
Factorization in these channels 
has to be generalized 
by replacing the short-distance coefficient $C_1(\mu)+C_2(\mu)/3$ 
by effective coefficients $a_2$ which are supposed 
to be scale-independent and incorporate possible nonfactorizable 
effects. The most general decomposition of the matrix elements in 
(\ref{ampl}) includes one  effective coefficient
for $B\to J/\psi K$  and three for $B\to J/\psi K^{(*)}$ (one for 
each partial wave):
\begin{eqnarray}
\langle K(p) J/\psi(q) \mid {\cal H}^s_{\mbox{\scriptsize eff}}  
\mid B(p_B)\rangle &=& \sqrt{2}G_F V_{cb}V_{cs}^*a_2^{B\psi K}f_\psi
f_+m_{\psi}( \epsilon^*_\psi \cdot p)~,
\label{a2K}\\
\langle
K^*(p) J/\psi(q) \mid {\cal H}^s_{\mbox{\scriptsize eff}} \mid B(p_B) \rangle
&=&\frac{G_F}{\sqrt{2}}V_{cb}V^*_{cs}m_\psi f_\psi \epsilon_{\psi}^{*\rho}
\Big[-i(m_B+m_{K^*})\epsilon^{*}_{K^*\rho} 
a_{2,1}^{B \psi K^*}A_1(s)\nonumber\\
\lefteqn{+i\frac{(\epsilon^*_{K^*}\cdot q)(p_B+p)_\rho }{m_B+m_{K^*}}
a_{2,2}^{B \psi K^*} A_2(s)
+2\frac{\epsilon_{\rho\nu\alpha\beta}\epsilon_{K^*}^{*\nu}
q^\alpha p^\beta}{m_B+m_{K^*}} a_{2,V}^{B \psi K^*}
V(s)\Big].}\hspace*{1cm}
\label{a2Kst}
\end{eqnarray}
{}From experimental data we obtain  estimates 
for these coefficients as displayed in Tab.~\ref{tab:alex}.
Only the absolute values of $a_{2}^{B\psi K}$ and $a_{2,i}^{B\psi K^*}$ 
$(i=1,2,V)$ can be extracted, whereas the relative sign of 
$a_{2,2}^{B\psi K^*}$ and $a_{2,1}^{B\psi K^*}$ turns out to be  positive
with a twofold ambiguity for the coefficient $a_{2,2}^{B\psi K^*}$. 
It is important to notice that experimental data themselves
signal non-universality of the $a_2$-coefficients, 
although the accuracy still has to be improved. 

Now we turn to describing how these coefficients can be 
estimated theoretically. 
The main nonfactorizable contributions to $a_2$ come from the 
matrix elements of $\widetilde{O}^s_1$, which are
parametrized in the form \cite{KR}
\begin{eqnarray}
\langle K J/\psi \mid\widetilde{O}^s_1(\mu) \mid B \rangle
&=& 2f_\psi m_{\psi}\tilde{f}(\mu)(\epsilon^*_\psi \cdot p)\,,
\label{def}\\
\langle
K^* J/\psi \mid\widetilde{O}^s_1(\mu) \mid B \rangle
&=&m_\psi f_\psi \epsilon_{\psi}^{*\rho}
\Big[-i(m_B+m_{K^*})\epsilon^{*}_{K^*\rho} 
\widetilde{A}_1(s)\nonumber\\
&&{}+i\,\frac{(\epsilon^{*}_{K^*}\cdot q)(p_B+p)_\rho }{m_B+m_{K^*}}
\widetilde{A}_2(s)
+2\,\frac{\epsilon_{\rho\nu\alpha\beta}\epsilon_{K^*}^{*\nu}
q^\alpha p^\beta}{m_B+m_{K^*}} \widetilde{V}(s)\Big].
\label{defst}
\end{eqnarray}
$a_2^{B\psi K}$ can be expressed as
\begin{equation}
a_2^{B\psi K}=C_1(\mu)+\frac{C_2(\mu)}3 + 
2C_2(\mu)\,\frac{\tilde{f}(\mu)}{f_+(m_\psi^2)} + ...~
\label{a2}
\end{equation}
and similar expressions for $a_{2,1}^{B\psi K^*}$, 
$a_{2,2}^{B\psi K^*}$ and $a_{2,V}^{B\psi K^*}$ 
with the ratios  $\widetilde{A}_1/A_1(m_\psi^2)$,
$\widetilde{A}_2/A_2(m_\psi^2)$ and $\widetilde{V}/V(m_\psi^2)$,
respectively, replacing $\tilde{f}(\mu)/f_+(m_\psi^2)$.
In the above, the ellipses denote neglected nonfactorizable contributions 
of $O^s_2$, which are supposed to be subdominant.
The nonfactorizable amplitudes $\tilde{f}_{B\psi K}$, $\widetilde{A}_{1,2}$
and $\widetilde{V}$ have been 
estimated  in Ref.~\cite{KR2} following the approach suggested in 
Ref.~\cite{BS93} 
and using OPE. In this report we do not have the space to explain the
method in
detail, but simply state the results. At the current level of accuracy, 
one predicts  the following ranges of nonfactorizable 
amplitudes:
\begin{eqnarray}
\tilde{f}(\mu_0) & = & -(0.06 \pm 0.02)\,,\\
\widetilde{A}_{1}(\mu_0) = 0.0050 \pm 0.0025,~~
\widetilde{A}_{2}(\mu_0) & = & -(0.002 \pm 0.001),~~
\widetilde{V}(\mu_0) = -(0.09 \pm 0.04)~.
\label{numberKst}
\end{eqnarray}
These estimates reveal substantial non-universality in  
absolute values and difference in signs of the nonfactorizable amplitudes. 
Although the ratios of these amplitudes 
to form factors, e.g. $\tilde{f}(\mu_0)/f_+(m_\psi^2)$ $\simeq 0.1$  
are small,  they have a strong impact on the coefficients $a_2$ because 
of a strong cancellation  in  $C_1(\mu_0)+C_2(\mu_0)/3 \simeq 0.055$,
$\mu_0=2m_c=2.6\,$GeV (which is numerically close to $m_b/2$) 
being the relevant scale in the process. 
{}From (\ref{a2}) and the corresponding relations for the other $a_2$, 
we obtain:
\begin{equation}
a_2^{B\psi K} = -(0.09 \div 0.23),~~ 
a_{2,1}^{B\psi K^*}= 0.07 \div 0.09,~~
a_{2,2}^{B\psi K^*}= 0.04 \div 0.05,~~ 
a_{2,V}^{B\psi K^*}= -(0.05 \div 0.26)\,,
\label{estst}
\end{equation}
where an additional $\pm (10 \div 20) \%$ uncertainty from the 
form factors should be added. 
Although in comparison with the experimental numbers 
for $|a_2^{B\psi K}|$ and $|a_{2,1}^{B\psi K^*}|$, 
the estimates (\ref{estst}) fall somewhat short, 
the gap between naive factorization at $\mu=m_b/2$ and 
experiment is narrowed considerably. Note also that the 
sum rule estimates for 
$a_2^{B\psi K}$ and $a_{2,V}^{B\psi K^*}$ yield negative 
sign for these two coefficients in contradiction to the 
global fit of the factorized decay amplitudes to the data 
\cite{neu-ste}, yielding a universal positive $a_2$. 
For $a_{2,2}^{B\psi K^*}$ and $a_{2,V}^{B\psi K^*}$,
the estimates in (\ref{estst}) are not very conclusive 
in view of the large experimental uncertainties of these two 
coefficients. Clearly, further improvements 
in the sum rules are needed to achieve more accurate estimates.
Nevertheless, the above calculation has demonstrated
 that for future theoretical 
studies of exclusive nonleptonic decays of heavy mesons 
QCD sum rule techniques provide new ways to 
go beyond factorization. 

\subsection[Dispersion Relations for B Nonleptonic  Decays into Light 
Pseudoscalar Mesons]{Dispersion Relations for B Nonleptonic  Decays into Light 
Pseudoscalar Mesons}

Rescattering effects in
nonleptonic  B decays into light pseudoscalar mesons were
investigated in \cite{CaMi} by the method of 
dispersion relations in terms of the external masses.
Defining the  weak decay amplitude $A_{B\to P_1 P_2}=A(m_B^2, m_1^2, m_2^2)$,
where $P_1$, $P_2$ are pseudoscalar mesons, 
one can show \cite{CaMi} that the weak amplitude satisfies the
following dispersion representation:
\begin{equation}\label{direl}
A(m_B^2, m_1^2, m_2^2)= A^{(0)}(m_B^2, m_1^2, m_2^2)\,+\,
{1\over \pi} \int\limits_0^{(m_B-m_2)^2}{\rm d}z\,
{\mbox{Disc}\, A(m_B^2, z, m_2^2)\over z-m_1^2-i\epsilon}\,.
\end{equation}
The first term in this representation is the amplitude in the factorization
limit, while in the second term the dispersion variable is the mass
squared of the
meson which does not contain the spectator quark. The representation
(\ref{direl}) allows one to recover the amplitude in the factorization
approximation  when the strong rescattering is switched off, which is a
reasonable consistency condition. As shown in \cite{CaMi}, in the
two-particle approximation
(\ref{direl}) can be written as
\begin{equation}\label{sumint1}
 A_{B\to P_1P_2}
= A^{(0)}_{B\to P_1P_2}\,+\,{1\over 2}\sum _{\{P_3P_4\}} \Gamma_{P_3P_4;P_1P_2}
\bar A_{B\to P_3 P_4}+ {1\over 2}\sum_{\{P_3P_4\}}
\overline\Gamma_{P_3P_4;P_1P_2}
A_{B\to P_3 P_4}\,.
\end{equation}
In this relation $A^{(0)}_{B\to P_1P_2}$ is the amplitude in the
factorization limit,
$\bar A_{B\to P_3 P_4}$ is obtained from $A_{B\to P_3 P_4}$ by changing the
 sign of the strong phases,
the coefficients $\Gamma_{P_3 P_4; P_1 P_2}$ are computed as
dispersive integrals
\begin{equation}\label{gama}
\Gamma_{P_3 P_4; P_1 P_2}={1\over \pi}
\int\limits_0^{(m_B-m_2)^2}{\mathrm d}z\,{C_{ P_3P_4;
P_1P_2}(z)\over z-m_1^2-i
\epsilon}\,,
\end{equation} and  $\overline\Gamma_{P_3 P_4;P_1 P_2}$ are defined as in 
(\ref{gama}), with the numerator $C$ replaced by $C^*$, where
\begin{equation}\label{coef}
 C_{P_3P_4;P_1P_2}(z)
={1\over 2}\int
{{\mathrm d}^3{\mathbf k}_3\over(2\pi)^{3} 2\omega_3}{{\mathrm d}^3{\mathbf
k}_4\over(2\pi)^{3}2\omega_4}
(2\pi)^4\delta^{(4)}(p-k_3-k_4) {\mathcal M}_{P_3 P_4 \to P_1 P_2}(s,t)\,.
\end{equation}
The strong amplitudes ${\mathcal M}_{P_3 P_4;P_1 P_2}(s,t)$
 entering this expression are
evaluated for an off-shell meson $P_1$ of mass squared equal to $z$,
 at the 
c.m.\ energy squared $s=m_B^2$, which is high enough to justify the
application of  Regge-theory. A detailed calculation \cite{CaMi} takes
 into account both the $t$-channel trajectories describing the
 scattering at small angles and the $u$-channel trajectories
describing the scattering at large angles.

\begin{figure}
$$
\begin{array}{c@{\qquad}c}
\epsfxsize=4.5cm\epsfysize=6cm\epsfbox{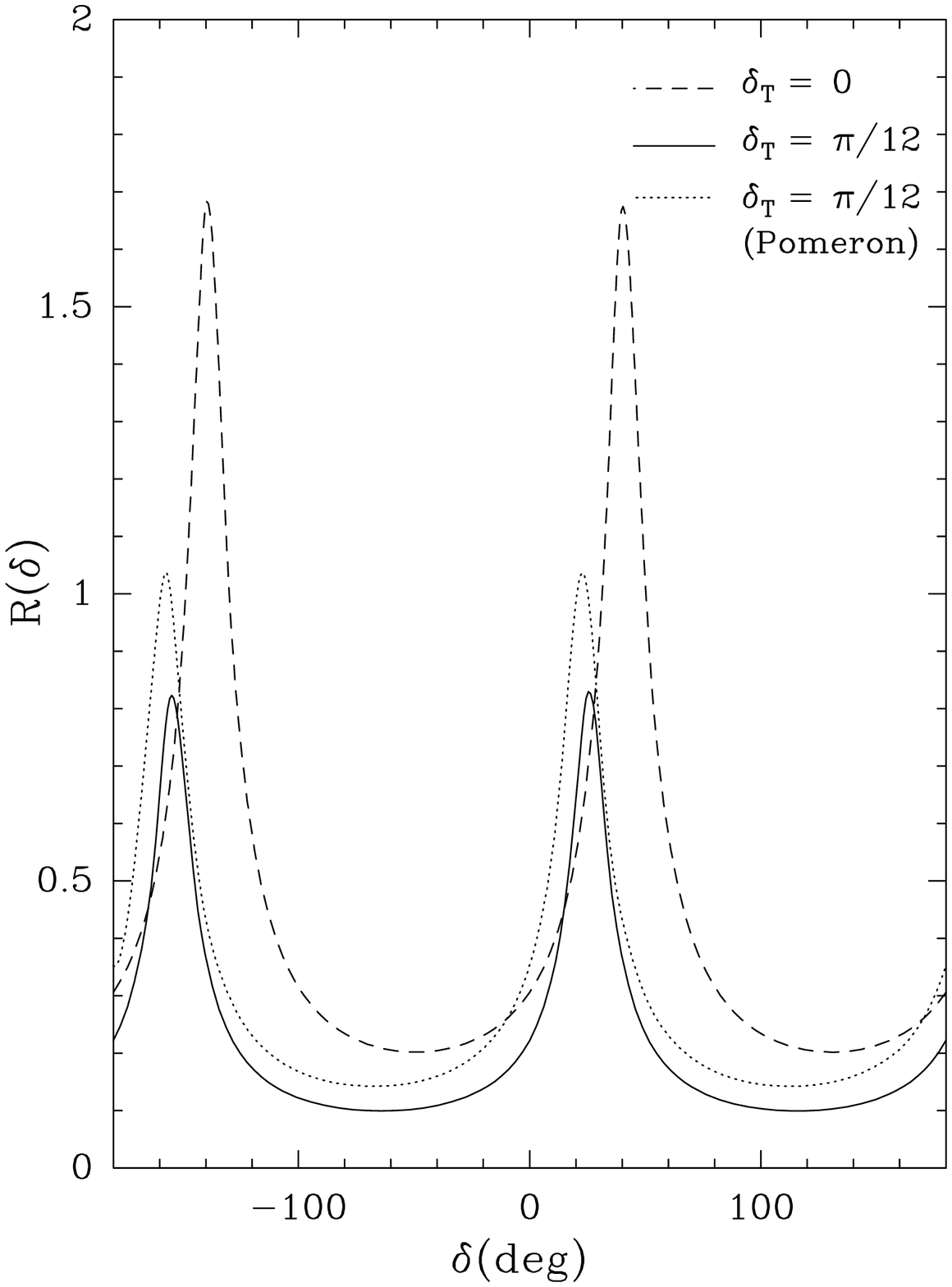} &
\epsfxsize=4.5cm\epsfysize=6cm\epsfbox{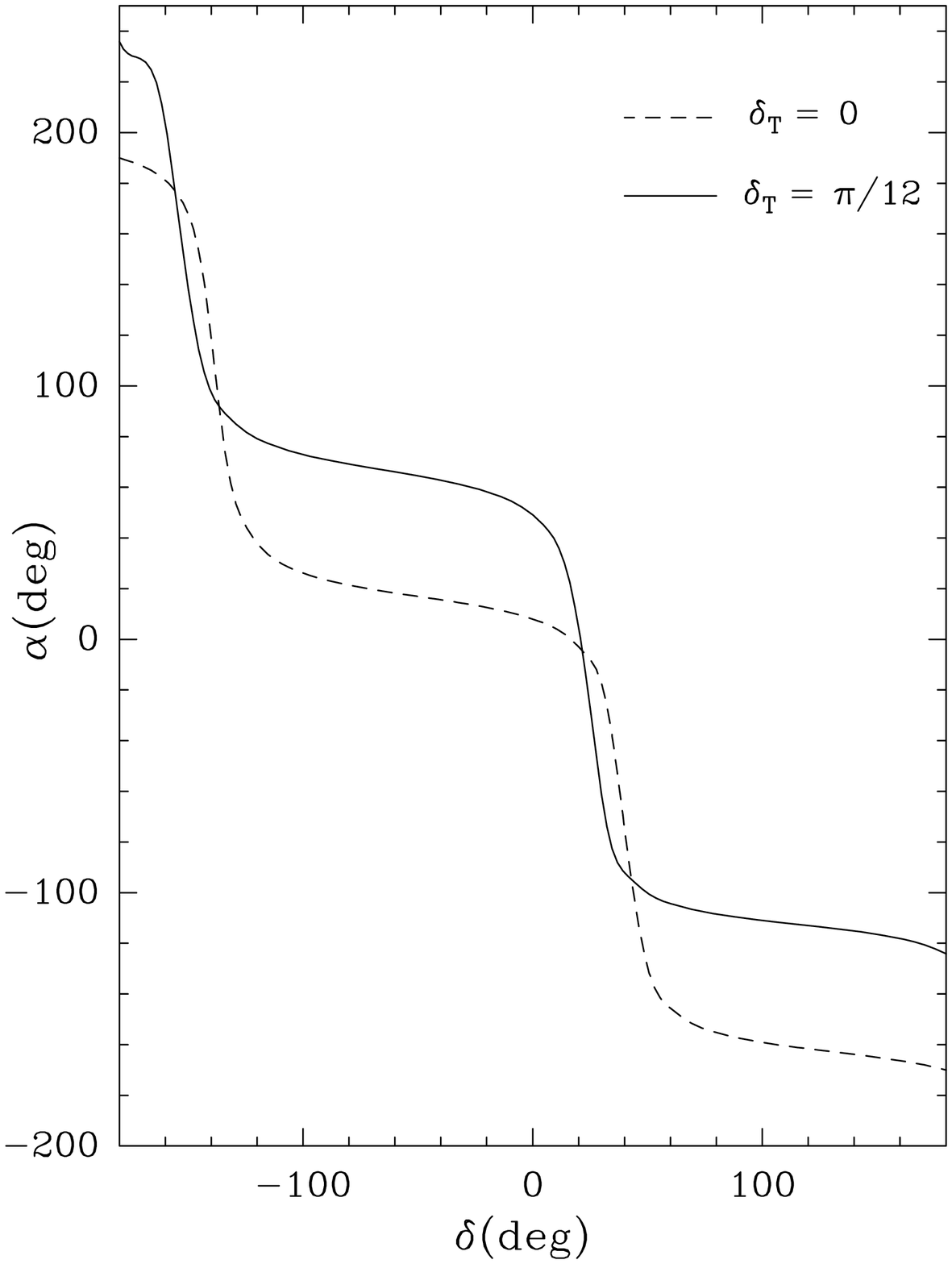}
\end{array}
$$
\vspace*{-0.7cm}
\caption[]{The ratio $R=\vert A_P/ A_T\vert$ (left) and the weak phase 
$\alpha$ (right), as functions of the strong phase difference
$\delta$, solid curve $\delta_T=\pi/12$, dashed curve
$\delta_T=0$.}\label{fig:cap1}
\end{figure}

Let us apply the dispersive formalism to the decay $B^0\to
\pi^+\pi^-$, 
taking as intermediate states in  the dispersion relation (\ref{sumint1})
 the pseudoscalar mesons  $\pi^+\pi^-$,
$\pi^0\pi^0$, $ K^+ K^-$, $ K^0\bar K^0$,
$\eta_8 \eta_8$,  $\eta_1 \eta_1$ and  $\eta_1 \eta_8$.  Assuming 
SU(3) flavour symmetry and keeping only the contribution of the 
dominant quark topologies, the dispersion relation  (\ref{sumint1}) becomes an
algebraic equation involving tree and penguin amplitudes, 
$A_T$ and $A_P$. With the Regge parameters discussed in \cite{CaMi}, 
this relation can be written as
\addtolength{\arraycolsep}{-2pt}
\begin{eqnarray}
A(B^0\to\pi^+\pi^-)/A_T &=&  {\mathrm e}^{i\gamma}+
 R {\mathrm e}^{i\delta}{\mathrm e}^{-i\beta}\nonumber\\
& = &  
{{\mathrm e}^{-i\delta_T}\over A_T} 
\left[A_T^{(0)} {\mathrm e}^{i\gamma}+A_P^{(0)}
{\mathrm e}^{-i\beta}\right]
-\left[(0.01+1.27\,i) +(0.75 - 1.01 \,i)\, {\mathrm
e}^{-2i\delta_T}\right]\, {\mathrm e}^{i\gamma}\nonumber\\
& &+R\left[ -(1.97+2.64\,i)\, {\mathrm e}^{i\delta} 
-(1.79-2.00 i)\, {\mathrm e}^{-i\delta} 
{\mathrm e}^{-2i\delta_T}\right]\,{\mathrm e}^{-i\beta}\, .\label{drpi}
\end{eqnarray} 
\addtolength{\arraycolsep}{2pt}
Here $A_T^{(0)}$ and  $A_P^{(0)}$ are the amplitudes in the
factorization approximation,
 $R = \vert A_P/ A_T \vert$ and  $\delta=\delta_P-\delta_T$,
$\delta_T$ ($\delta_P$) being the strong phase of  $A_T$ ($A_P$), respectively.
 It is seen that the weak angles appear in the combination
$\gamma+\beta=\pi-\alpha$. Solving  the complex equation (\ref{drpi})
 for  
$R$ and  $\alpha$,  we derive their expressions as functions of
 $\delta_T$ 
and $\delta$.
The evaluation of these expressions requires also
the knowledge of the ratios $A_P^{(0)}/A_T^{(0)}$ and $A_T^{(0)}/A_T$.
 In Fig.~\ref{fig:cap1}
we represent  $R$ and  $\alpha$ as functions of the phase
difference $\delta$, for two values of $\delta_T$, using as input 
$A_P^{(0)}/A_T^{(0)}=0.08$  and  $A_T^{(0)}/A_T\approx 0.9$
 \cite{BBNS}. Values of the
ratio $R$ less than one are obtained for both $\delta_T=0$ and 
$\delta_T=\pi/12$. The dominant contribution is given by the
elastic channel, more precisely by the Pomeron, as is seen in 
Fig.~\ref{fig:cap1},
where the dotted curve shows the ratio $R$ for  $\delta_T=\pi/12$, 
keeping only the contribution of the Pomeron in the Regge amplitudes.

The above results show that the dispersive formalism  is
consistent with the treatment
based on factorization and perturbative QCD in the heavy quark limit
presented in Ref.~\cite{BBNS}, 
supporting therefore the physical idea of parton-hadron duality.
{}From a practical point of view, the dispersion
representations in the external mass provide a set of algebraic
equations for on-shell decay amplitudes, leading to nontrivial
constraints on the hadronic parameters. 

\subsection[QCD Factorization for Exclusive Nonleptonic B
Decays]{QCD Factorization for Exclusive Nonleptonic B
Decays}

\begin{figure}[htb]
\epsfxsize=0.5\textwidth
   \centerline{\epsffile{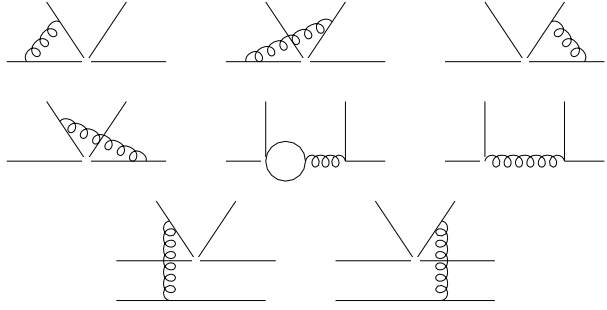}}
\vspace*{-0.2cm}
\caption[dummy]{\small Order $\alpha_s$ corrections to the hard 
scattering kernels $T^I_i$ (first two rows) and $T^{II}_i$ 
(last row). In the case of $T^I_i$, the spectator quark does 
not participate in the hard interaction and is not drawn. 
The two lines directed upwards represent the two quarks forming
the emitted pion. $T^{I}_i$ starts 
at ${\cal O}(\alpha^0_s)$, $T^{II}_i$ at ${\cal O}(\alpha^1_s)$.
\label{figBB2}}
\end{figure}

The theory of hadronic B decay matrix elements is
a crucial basis for precision flavour physics with
nonleptonic modes, which is one of the central goals
of the B physics programme at the LHC.
A new, systematic approach towards this problem, 
going beyond previous attempts, was recently proposed in 
\cite{BBNS}.
It solves the problem of how to calculate nonfactorizable
contributions, and in particular final state interactions,
in the heavy quark limit and constitutes a promising approach, 
complementary to 
the one discussed in the preceding sections. In this approach,
the statement of QCD factorization in the case of
$B\to\pi\pi$, for instance, can be schematically written as
\begin{equation}\label{qfac}
A(B\to\pi\pi)=\langle\pi|j_1|B\rangle\, \langle\pi|j_2|0\rangle\cdot
\left[1+{\cal O}(\alpha_s)+
{\cal O}\left(\frac{\Lambda_{\rm QCD}}{m_B}\right)\right]\, .
\end{equation}
Up to corrections suppressed by $\Lambda_{\rm QCD}/m_B$ the amplitude is
calculable in terms of simpler hadronic objects: It factorizes,
to lowest order in $\alpha_s$, into matrix elements of bilinear
quark currents ($j_{1,2}$).
To higher order in $\alpha_s$, but still to leading order in
$\Lambda_{\rm QCD}/m_B$, there are `nonfactorizable' corrections, which are
however governed by hard gluon exchange. They are therefore again
calculable in terms of few universal hadronic quantities.
More explicitly, the matrix elements of four-quark operators
$Q_i$ are expressed by the factorization formula
\begin{equation}\label{fform}
\langle\pi(p')\pi(q)|Q_i|\bar B(p)\rangle =
f^{B\to\pi}(q^2)\int^1_0 dx\, T^{I}_i(x)\Phi_\pi(x)
 +\int^1_0 d\xi dx dy\, T^{II}_i(\xi,x,y)\Phi_B(\xi)
\Phi_\pi(x) \Phi_\pi(y),
\end{equation}
which is valid up to corrections of relative order 
$\Lambda_{\rm QCD}/m_b$.
Here $f^{B\to\pi}(q^2)$ is a $B\to\pi$ form factor \cite{KR,ball98}
evaluated at $q^2=m^2_\pi\approx 0$, and $\Phi_\pi$ ($\Phi_B$) are 
leading-twist light-cone
distribution amplitudes of the pion (B meson).
The $T^{I,II}_i$ denote hard-scattering kernels, 
which are calculable in perturbation theory. The corresponding
diagrams are shown in Fig.~\ref{figBB2}.

This treatment of hadronic B decays is based on the analysis of
Feynman diagrams in the heavy quark limit, utilizing consistent
power counting to identify the leading contributions. The framework
is very similar in spirit to more conventional applications  of
perturbative QCD in exclusive hadronic processes with a large 
momentum transfer, as the pion electromagnetic form factor 
\cite{FJ,ER,LB}.
It may be viewed as a consistent formalization of 
Bjorken's colour transparency argument \cite{bjorken}. In addition
the method includes, for $B\to\pi\pi$, the hard
nonfactorizable spectator interactions, penguin contributions
and rescattering effects. 
As a corollary, one finds that strong
rescattering phases are either of ${\cal O}(\alpha_s)$, and
calculable, or power suppressed. In any case they vanish therefore
in the heavy quark limit.
QCD factorization is valid for cases where the emitted particle
(the meson created from the vacuum in the weak process, as opposed
to the one that absorbs the $b$ quark spectator) is a small size
colour-singlet object, e.g.\ either a fast light meson
($\pi$, $\varrho$, $K$, $K^*$) or a $J/\psi$. 
For the special case of the ratio
$\Gamma(B\to D^{*}\pi)/\Gamma(B\to D\pi)$ the perturbative
corrections to naive factorization have been evaluated in \cite{PW}
using a formalism similar to the one described above.
Note that factorization
cannot be justified in this way if the emitted particle is a
heavy-light meson ($D^{(*)}$), which is not a compact object
and has strong overlap with the remaining hadronic environment.

\subsubsection{Final State Interactions}

A general issue in hadronic B decays, with important implications
for CP violation, is the question of final state interactions.
When discussing this problem, we may choose a 
partonic or a hadronic language. The partonic language can be justified 
by the dominance of hard rescattering in the heavy quark limit. In 
this limit the number of physical intermediate states is arbitrarily 
large. We may then argue on the grounds of parton-hadron duality 
that their average is described well enough (say, up to $\Lambda_{\rm QCD}/
m_b$ corrections) by a partonic calculation. This is the picture 
implied by (\ref{fform}). The hadronic language is in principle 
exact. However, the large number of intermediate states makes 
it almost impossible to observe systematic cancellations, which 
usually occur in an inclusive sum of intermediate states.

Consider again the decay of a B meson into two pions. 
Unitarity implies $\mbox{Im}\,A(B\to \pi\pi)$ 
$\sim$ $\sum_n A(B\to n) A^*(n\to \pi\pi)$.
The elastic rescattering contribution ($n=\pi\pi$)
is related to the $\pi\pi$ 
scattering amplitude, which exhibits Regge behaviour in the high-energy 
($m_b\to\infty$) limit. Hence the soft, elastic rescattering 
phase increases slowly in the heavy quark limit \cite{DGPS96}. On 
general grounds, it is rather improbable that elastic rescattering 
gives an appropriate description at large $m_b$. This 
expectation is also borne out in the framework of Regge behaviour, see  
\cite{DGPS96}, where the importance of inelastic rescattering is 
emphasized. However, the approach pursued in \cite{DGPS96} leaves 
open the possibility of soft rescattering phases that do not vanish in 
the heavy quark limit, as well as the possibility of systematic 
cancellations, for which the Regge language does not provide an 
appropriate theoretical framework.

Eq.~(\ref{fform}) implies that such systematic cancellations 
do occur in the sum over all intermediate states $n$. It is worth 
recalling that such cancellations are not uncommon for hard 
processes. Consider the example of $e^+ e^-\to\,$hadrons at large 
energy $q$. While the production of any hadronic final state 
occurs on a time scale of order $1/\Lambda_{\rm QCD}$ (and would 
lead to infrared divergences if we attempted to describe it in 
perturbation theory), the inclusive cross section given by the sum 
over all hadronic final states is described very well by a 
$q\bar{q}$ pair that lives over a short time scale of order $1/q$. In 
close analogy, while each particular hadronic intermediate state $n$ 
cannot be described partonically, the 
sum over all intermediate states is accurately represented by 
a $q\bar{q}$ fluctuation of small transverse size of order $1/m_b$, 
which therefore interacts little with its environment. Note that 
precisely because the $q\bar{q}$ pair is small, the physical picture of 
rescattering is very different from elastic $\pi\pi$ scattering --  
hence the Regge picture is difficult to justify in the heavy quark 
limit. 

As is clear from the discussion, parton-hadron duality is crucial for 
the validity of (\ref{fform}) beyond perturbative factorization. 
A quantitative proof of how accurately duality holds is a yet 
unsolved problem in QCD. Short of a solution, it is worth noting that the 
same (often implicit) assumption is fundamental to 
many successful QCD predictions in 
jet and hadron-hadron physics 
or heavy quark decays.

\subsubsection{QCD Factorization in $B\to\pi\pi$}

Let us finally illustrate one phenomenological application
of QCD factorization in the heavy quark limit for $\bar
B\to\pi^+\pi^-$ \cite{BBNS}.
The $\bar B_d\to\pi^+\pi^-$ decay amplitude $A$ reads
\begin{equation}\label{abpm}
A = i\frac{G_F}{\sqrt{2}}m^2_B f_+(0)f_\pi
|\lambda_c|\cdot \bigl[R_b e^{-i\gamma}
\left(a^u_1(\pi\pi)+a^u_4(\pi\pi)+a^u_6(\pi\pi)r_\chi\right)
-\left(a^c_4(\pi\pi)+a^c_6(\pi\pi)r_\chi\right)\bigr].
\end{equation}
Here $R_b$ is the ratio of CMK matrix elements defined in
(\ref{Rb-intro}), $\gamma$ is the phase of $V^*_{ub}$,
and we will use $|V_{cb}|=0.039\pm 0.002$, $|V_{ub}/V_{cb}|= 
0.085\pm 0.020$. 
We also take $f_\pi=131\,$MeV, $f_B=(180\pm 20)\,$MeV, 
$f_+(0)=0.275\pm0.025$, and $\tau(B_d)=1.56\,$ps;
$\lambda_c\equiv V^*_{cd}V_{cb}$.
The contribution of $a_6^p(\pi\pi)$ is multiplied by
$r_\chi=2 m_\pi^2/(m_b (m_u+m_d))\sim\Lambda_{\rm QCD}/m_b$.
It is thus formally power suppressed, but numerically
relevant since $r_\chi\approx 1$.
The coefficients $a_i$ are estimated in Tab.~\ref{tab:ai}.
\begin{table}
{\small
\begin{center}
\begin{tabular}{|c|c|c|c|}
\hline
$a^u_1(\pi\pi)$ & $a^u_4(\pi\pi)$ & $a^c_4(\pi\pi)$ & 
$a^{p}_6(\pi\pi)r_\chi$ \\ \hline
$1.038+0.018i$ & $-0.029-0.015i$ & $-0.034-0.008i$ & -- \\
$(1.020)$ & $(-0.020)$ & $(-0.020)$ & $(-0.030)$ \\
\hline
\end{tabular}
\end{center}}
\vspace*{-0.5cm}
\caption[]{ 
QCD coefficients $a^p_i(\pi\pi)$ for $\bar B\to\pi^+\pi^-$
at NLO (renormalization scale $\mu=m_b$). 
Leading order values are shown in parenthesis for comparison.}
\label{tab:ai}
\end{table}
We then find 
for the branching fraction 
\begin{equation}\label{brpi}
B(\bar B_d\to\pi^+\pi^-) = 
6.5\,[6.1]\cdot 10^{-6}
\left|\,e^{-i\gamma} + 0.09\,[0.18] \,e^{i\cdot 12.7\,[6.7]^\circ}\,
\right|^{\,2}\!\!.
\end{equation}
The default values correspond to $a_6^p(\pi\pi)=0$,  
the values in brackets use $a_6^p(\pi\pi)$ at leading order. The 
predictions for the $\pi^+\pi^-$ final state are 
relatively robust, with errors of the order of $\pm\,$30\% due to the 
input parameters. 
The direct CP asymmetry in the $\pi^+\pi^-$ mode is
approximately $4\%\times\sin\gamma$.

\begin{figure}
   \epsfxsize=7cm
   \centerline{\epsffile{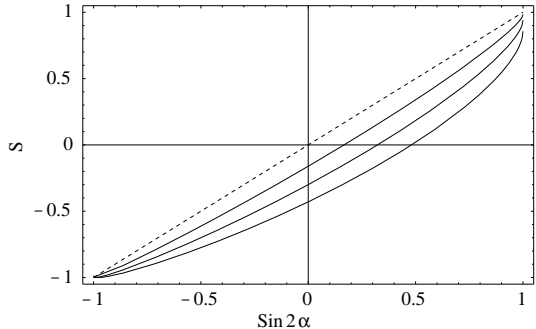}}
   \vspace*{-0.2cm}
\caption[dummy]{\small Coefficient of $-\sin(\Delta M_{B_d} t)$ vs. 
$\sin 2\alpha$. $\sin 2\beta=\,$0.7 has been assumed. 
See text for explanation.
\label{alpha}}
\end{figure}

As a further example we
use the factorization formula to compute the time-dependent,  
mixing-induced asymmetry in $B_d\to \pi^+\pi^-$ decay, 
\begin{equation}
{\cal A}(t) = -S\cdot \sin(\Delta M_{B_d}t)+C\cdot\cos(\Delta M_{B_d} t).
\end{equation}
In the absence of a penguin contribution (defined as the contribution 
to the amplitude which does not carry the weak phase $\gamma$ in 
standard phase conventions), $S=\sin 2\alpha$ (where $\alpha$ refers to 
one of the angles of the CKM unitarity triangle) and $C=0$.  
Fig.~\ref{alpha} shows $S$ as a function of $\sin 2\alpha$ with the 
amplitudes computed according to (\ref{fform}) and (\ref{abpm}). 
The central of the 
solid lines refers to the heavy quark limit including $\alpha_s$ 
corrections to naive factorization and including the power-suppressed 
term $a_6 r_\chi$ that is usually 
also kept in naive factorization. The other two solid lines correspond to 
dropping this term or multiplying it by a factor of 2. This exercise 
shows that formally power-suppressed terms can be non-negligible, 
but it also shows that a measurement of $S$ can be 
converted into a range for $\sin 2\alpha$ which may already provide 
a very useful constraint on CP violation.

More work remains to be done. The proof of factorization 
has to be completed. 
Power corrections are an important issue, as $m_b$ is not arbitrarily 
large. There exist `chirally enhanced' corrections
$\sim r_\chi$. All such terms can be identified, but they involve 
nonfactorizable soft gluons. 
The size of these terms has to be estimated to arrive 
at a realistic phenomenology. 
If this can be done, one may expect promising 
constraints and predictions for a large number of nonleptonic two-body 
final states.
We emphasize in particular the experimentally attractive
possibility to determine $\sin 2\alpha$ from $B\to\pi^+\pi^-$ decays
alone.

\section[\protect\boldmath $B_c$ PHYSICS]{\protect\boldmath $B_c$ 
PHYSICS\protect\footnote{Section coordinators:
G. Buchalla, P. Colangelo and
  F. De Fazio, with acknowledgements for V.V.\ Kiselev and A. Likhoded
  for useful discussions.}}\label{sec:Bc}

The $B^+_c$ meson is the lowest lying bound state of two
heavy quarks, $\bar b$ and $c$. The QCD dynamics of this state
is therefore similar to that of quarkonium systems, such as the
$\bar bb$ or $\bar cc$ families, which are approximately
nonrelativistic. In contrast to the common quarkonia, however,
$B_c$ carries open flavour and the ground state is stable under
strong interactions. In fact, $B_c$ is the only hadron combining
these features and forming a flavoured, weakly decaying quarkonium.
Since the complicated interplay of strong and weak
forces is the key problem in the theoretical 
analysis of weak decays of hadrons,
the quarkonium-like $B_c$  provides us with a very interesting
special case to study such a general question.
Calculational tools, as for example heavy quark expansions, 
nonrelativistic QCD (NRQCD), factorization, which are important
in many areas of heavy flavour physics, can be tested in a
complementary setting. 

The observation of the $B_c$ meson by the CDF Collaboration
in the channel $B_c\to J/\psi l\nu$, with measured
mass and lifetime \cite{cdf}
\begin{equation}
M_{B_c}=6.40 \pm 0.39\, \mbox{(stat)} \pm 0.13\,
\mbox{(syst)}\,\mbox{GeV},
\quad
\tau_{B_c}=0.46^{+0.18}_{-0.16}\,\mbox{(stat)} \pm 0.03\, 
\mbox{(syst)}\, \mbox{ps}, \label{cdfbc}
\end{equation}
opened up the experimental investigations of the 
$\bar b c$ hadronic system. 
$B_c$ physics can also be pursued at the LHC, where a copious 
production of the $B_c$ meson and of its radial and orbital excitations is 
expected \cite{bcprod} (see also \cite{GERS} for a recent review).
No full experimental studies have been performed yet, and thus we
concentrate on a brief summary of $B_c$
decay properties, collecting useful information and illustrating a 
range of opportunities that may be pursued in this field.
We also shortly summarize the present status of experimental studies.

\subsection[$B_c$ Lifetime and Inclusive Decays]{\protect\boldmath 
$B_c$ Lifetime and Inclusive Decays}

The total decay rate of the $B_c$ can be computed starting from
a heavy-quark expansion of the transition operator, supplemented by
NRQCD. This framework is familiar from the study of
ordinary, heavy-light $b$ hadron lifetimes \cite{BIG1},
with the basic difference that in the heavy-light case the r\^{o}le
of NRQCD is played by heavy-quark effective theory (HQET).
For $B_c$ the characteristic features of NRQCD result in a
particularly intuitive expression for the total rate:
\begin{equation}\label{gammabc}
\Gamma_{B_c}=\Gamma_b\left(1-\frac{v^2_b}{2}\right)+
\Gamma_c\left(1-\frac{v^2_c}{2}\right)+
\Delta\Gamma_{PI}+\Delta\Gamma_{WA}+{\cal O}(v^4)\, .
\end{equation}
Eq. (\ref{gammabc}) is written as an expansion in the heavy
quark velocities $v$, complete through order $v^3$.
To lowest order, $v^0$, we have $\Gamma_{B_c}=\Gamma_b+\Gamma_c$,
the sum of the free decay rates of the heavy quark constituents
$\bar b$ and $c$. The first bound state corrections arise at 
${\cal O}(v^2)$
only and are equivalent to time dilatation.
The effect of binding compels the heavy quarks to move around
each other, thus retarding their decay.
At ${\cal O}(v^3)$ there are two terms. First,
a correction from Pauli interference (PI) of the
two $c$ quarks in the final state of
$(c)\bar b\to (c)\bar c c\bar s$ decay. Second, a contribution
from the weak annihilation of the constituents $\bar bc$, either into
hadrons or into leptons, the latter dominated by $B_c\to\tau\nu$.
A numerical analysis of (\ref{gammabc}) gives the estimate
\cite{BBBC}
\begin{equation}\label{taubc}
\tau_{B_c}\equiv \frac{1}{\Gamma_{B_c}}=(0.4 \,\mbox{--}\, 0.7)\,{\rm ps}
\, ,
\end{equation} 
with a large uncertainty from the heavy quark masses ($m_c$),
but in agreement with the measurement (\ref{cdfbc}).
The same framework can be used to calculate other inclusive
decay properties of $B_c$, for instance the semileptonic branching
fraction $B(B_c\to Xe\nu)$, which turns out  to be
$\sim\,$12\%. More
details and references can be found in \cite{BBBC,CQ,BIGBC}.

\subsection[Leptonic and Radiative Leptonic $B_c$ 
Decays]{Leptonic and Radiative Leptonic \protect\boldmath $B_c$ Decays}

The purely muonic $B_c$ branching ratio is determined by the decay
constant $f_{B_c}$:
\begin{eqnarray}
B(B_c \to \mu \nu)&=& \tau_{B_c}{G_F^2 \over 8 \pi} |V_{cb}|^2
f_{B_c}^2 M_{B_c}^3
\Bigg( {m_\mu \over M_{B_c}} \Bigg)^2  \Bigg( 1-{m_\mu^2 \over
M_{B_c}^2}\Bigg)^2 \hskip 3 pt  \nonumber\\
&=&6.8\times 10^{-5}\frac{M_{B_c}}{6.28\,{\rm GeV}}
\left(\frac{|V_{cb}|}{0.04}\right)^2\left(
\frac{f_{B_c}}{400\,{\rm MeV}}\right)^2
\frac{\tau_{B_c}}{0.46\,{\rm ps}}\, .\label{brmu}
\end{eqnarray}
The value of $f_{B_c}$ has been computed by lattice NRQCD:
$f_{B_c}=(420 \pm 13)\,$MeV \cite{NRlat}, 
QCD Sum Rules:  $f_{B_c}=(360 \pm 60)\,$MeV \cite{qcdsr0,qcdsr1}, 
and various quark models with predictions in the range 
$f_{B_c}=[430-570]\,$MeV \cite{qmod0}.

The photon emission in the radiative muonic decay 
$B_c \to \mu \nu \gamma$
removes the helicity suppression of the purely muonic mode. 
In the nonrelativistic limit, the radiative muonic
decay width is also determined by $f_{B_c}$ \cite{CFP}.
In this limit one obtains the ratio
$\Gamma(B_c\to\mu\nu\gamma)/\Gamma(B_c\to\mu\nu)\simeq 0.8$.
Corrections to this result within a relativistic quark model
have been discussed in \cite{bcmng}.

\subsection[Semileptonic $B_c$ Decay Modes]{
Semileptonic\protect\boldmath  $B_c$ Decay Modes}

The calculation of the matrix elements governing the exclusive semileptonic
$B_c$ decay modes has been carried out using  
QCD sum rules \cite{qcdsr1,qcdsr2} and  quark models
\cite{qmod,qmod1}. The  predictions for the various exclusive decay rates
are reported
(in rather conservative ranges) in Tab.~\ref{t:tab1},
with the conclusion that the semileptonic $B_c$ decay width is dominated by
the modes induced by the charm decay.

\begin{table}[ht]
\begin{center}
\vspace*{0.5cm}
\begin{tabular}{|l||c|r|} \hline 
~~Channel &$\Gamma   (10^{-15}$ GeV) & \multicolumn{1}{|c|}{$B$} \\ \hline
$B_c^+ \to B_s e^+ \nu   $&$11 - 61 $&$[8 - 42] \times 10^{-3}$\\
$B_c^+ \to B^*_s e^+ \nu $&$30 - 79 $&$[21 - 55] \times 10^{-3}$\\ \hline
$B_c^+ \to B_d e^+ \nu   $&$1 - 4   $&$[7 - 28]\times 10^{-4}$\\
$B_c^+ \to B^*_d e^+ \nu $&$2 - 6   $&$[14 - 42]\times 10^{-4}$\\ \hline
$B_c^+ \to \eta_c e^+ \nu$&$2 - 14$&$[14 - 98]  \times 10^{-4}$\\
$B_c^+ \to J/\psi e^+ \nu$&$22 - 35$&$[15 - 24]\times 10^{-3}$\\ \hline
$B_c^+ \to \eta_c^\prime e^+ \nu$&$0.3  - 0.7 $&$[2 - 5]\times 10^{-4}$\\
$B_c^+ \to \psi^\prime e^+ \nu    $&$1  -  2 $&$[7 - 14 ]\times 10^{-4}$\\
\hline
$B_c^+ \to D^0 e^+ \nu   $&$ 0.01 - 0.09 $&$[7 - 63]\times 10^{-6}$\\
$B_c^+ \to D^{*0} e^+ \nu$&$0.1 - 0. 3    $&$[7 - 21] \times 10^{-5}$ \\ \hline
\end{tabular}
\end{center}
\vspace*{-0.5cm}
\caption[]{Semileptonic $B_c^+$ decay widths and branching fractions
($\tau_{B_c}=0.46$ ps).}
\label{t:tab1}
\end{table}

The calculation of the semileptonic matrix elements
can be put on a firmer theoretical ground taking into account
the decoupling of the spin of the heavy quarks of the $B_c$
meson, as well as of the meson produced in the semileptonic decays, i.e.
mesons belonging to the $\bar c c$ family ($\eta_c, J/\psi$, etc.) and
mesons containing a single heavy quark
($B_s^{(*)}$, $B_d^{(*)}$, $D^{(*)}$).
The decoupling occurs in the heavy quark  limit
($m_b, m_c \gg \Lambda_{QCD}$); it produces
a symmetry, the heavy quark spin symmetry, allowing to relate, 
near the zero-recoil point,  the form
factors governing the $B_c$ decays into a $0^-$ and $1^-$ final meson
to a few invariant functions \cite{jenkins}. 
Examples are the processes
$B_c \to (B_s, B_s^*)\mu \nu$ and 
$B_c \to (B_d, B^*_d) \mu \nu $, where
the energy released to the final hadronic 
system  is much less than
$m_b$, thus leaving the $b$ quark almost unaffected.
The final $B_a$ meson ($a=d,s$) keeps 
the same $B_c$ four-velocity $v$, apart
from a small residual momentum $q$.
Defining
$p_{B_c}=M_{B_c}v$ and $p_{B_a}=M_{B_a}v+q$, 
one has:
\begin{eqnarray}
\langle B_a,v,q|V_\mu|B_c,v\rangle  &=& 
\sqrt{2M_{B_c}2M_{B_a}}[\Omega_1^a~ v_\mu +a_0~
\Omega_2^a~ q_\mu], \nonumber \\
\langle B_a^*,v,q,\epsilon|V_\mu|B_c,v\rangle &=& -i \; 
\sqrt{2M_{B_c}2M_{B^*_a}}~a_0~
\Omega_2^a~ \epsilon_{\mu \nu \alpha \beta} \epsilon^{*\nu} q^\alpha
v^\beta,
\nonumber \\
\langle B_a^*,v,q,\epsilon|A_\mu|B_c,v\rangle &=& 
\sqrt{2M_{B_c}2M_{B^*_a}}[\Omega_1^a~\epsilon^*_\mu 
+a_0~\Omega_2^a~ \epsilon^* \cdot q ~v_\mu],
\end{eqnarray}
\noindent i.e.\ only two form factors are needed to describe the previous
transitions.
The scale parameter $a_0$ is related to the $B_c$ Bohr radius \cite{jenkins}.
For the $B_c$ transitions into a $\bar c c$ meson,
$B_c \to (\eta_c, J/\psi)\mu \nu$, spin symmetry implies that the
semileptonic matrix elements can be expressed, 
near the zero-recoil point, in terms of a single 
form factor:
\begin{equation}
\langle \eta_c,v,q|V_\mu|B_c,v\rangle = \sqrt{2M_{B_c}2M_{\eta_c}} 
~\Delta ~v_\mu\, ,\quad
\langle J/\psi,v,q,\epsilon|A_\mu|B_c,v\rangle = 
\sqrt{2M_{B_c}2M_{J/\psi}} ~\Delta
~\epsilon^*_\mu \, . \label{qqbar}
\end{equation}
Model-independent results exist in the heavy-quark limit for
$\Delta$ and $\Omega_1^a$ at the zero-recoil point \cite{jenkins}.
Additional
information on the form factors $\Delta$ and $\Omega_i^a$ is available from
quark models \cite{qmod1, sanchiz} and NRQCD sum rules \cite{kiselev}. 
The related predictions are included in the ranges  
reported in Tab.~\ref{t:tab1}.
Moreover, spin symmetry implies
relations between $B_c$ decays to pseudoscalar and vector states,
near the non-recoil point,
that can be experimentally tested at the LHC \cite{sanchiz}.

\subsection{Nonleptonic Decay Modes}

Two-body nonleptonic decays are of prime importance for the measurement of 
the $B_c$ mass. In particular, decay modes having a $J/\psi$ in the 
final state are suitable for an efficient background rejection.

The nonleptonic $B_c$ decay rates have been computed in
the factorization  approximation, 
using various parametrizations of the semileptonic form factors
and different prescriptions for the 
parameters $a_1$ and $a_2$ appearing
in the factorized matrix elements  
\cite{qmod,qmod1,sanchiz}.
Predictions for the various decay modes, 
induced by the beauty and charm quark transitions, are reported
in Tabs.~\ref{t:tab2} and \ref{t:tab3}, respectively,
using $M_{B_c}=6.28$ GeV. For several modes,
ranges of values for the branching fractions are reported;
they are obtained considering the spread of predictions  
by  different approaches, and suggest
the size of the theoretical uncertainty for each decay mode.

\begin{table}[ht]
\begin{center}
\begin{tabular}{|l|r||l|r|} \hline
\multicolumn{1}{|c|}{Channel} &\multicolumn{1}{|c||}{$B $} & 
\multicolumn{1}{|c|}{Channel} &\multicolumn{1}{|c|}{$B $} \\
\hline
$ B_c^+ \to \eta_c \pi^+  $& $[3 - 25] \times 10^{-4}$
&$ B_c^+ \to\eta_c K^+    $&$[2-17] \times 10^{-5}$\\
$B_c^+ \to \eta_c \rho^+ $& $[7 - 60] \times 10^{-4}$
&$B_c^+ \to \eta_c K^{*+} $&$[4 - 31] \times 10^{-5}$\\
$B_c^+ \to \eta_c a_1^+  $&$9 \times 10^{-4}$
&$B_c^+ \to \eta_c K_1^+  $&$5 \times 10^{-5}$\\\hline
$B_c^+ \to J/\psi \pi^+  $& $[1 - 2] \times 10^{-3}$
&$ B_c^+ \to J/\psi K^+$&$[7-17] \times 10^{-5}$\\
$ B_c^+ \to J/\psi \rho^+  $& $[4-7] \times 10^{-3}$
&$B_c^+ \to J/\psi K^{*+}$&$[2-4] \times 10^{-4}$\\
$B_c^+ \to J/\psi a_1^+  $&$5 \times 10^{-3}$
 &$B_c^+ \to J/\psi K_1^+$&$3 \times 10^{-4}$\\\hline
$B_c^+ \to \psi^\prime \pi^+  $& $[2-3] \times 10^{-4}$
&$B_c^+ \to \psi^\prime K^+$&$[1-2] \times 10^{-5}$\\
$B_c^+ \to \psi^\prime \rho^+  $&$[5-8] \times 10^{-4}$
&$B_c^+ \to \psi^\prime   K^{*+}$&$[3-4] \times 10^{-5}$\\
$B_c^+ \to \psi^\prime a_1^+  $& $6 \times 10^{-4}$
&$B_c^+ \to \psi^\prime K_1^+  $&$3 \times 10^{-5}$\\\hline
$B_c^+ \to D^+ {\bar D}^0$ & $[1-12] \times 10^{-5}$
& $B_c^+ \to D^+_s {\bar D}^0$ &$[6-62] \times 10^{-7}$\\
$B_c^+ \to D^+ {\bar D}^{*0}$ & $[1-12] \times 10^{-5}$
& $B_c^+ \to D^+_s {\bar D}^{*0}$ &$[6-62] \times 10^{-7}$\\
$B_c^+ \to D^{*+} {\bar D}^0$ &$[8-10] \times 10^{-5}$ 
& $B_c^+ \to D^{*+}_s {\bar D}^0$ &$[5-6] \times 10^{-6}$\\
$B_c^+ \to D^{*+} {\bar D}^{*0}$ & $[1-2] \times 10^{-4}$
& $B_c^+ \to D^{*+}_s {\bar D}^{*0}$ &$[8-11] \times 10^{-6}$\\\hline
$B_c^+ \to \eta_c D_s $ &$[5-7] \times 10^{-3}$ &
$B_c^+ \to \eta_c D^+ $ &$[5-8] \times 10^{-5}$ \\
$B_c^+ \to \eta_c D_s^*$&$[4-6] \times 10^{-4}$&
$B_c^+ \to \eta_c D^{*+} $ &$[2-6] \times 10^{-5}$ \\\hline
$B_c^+ \to J/\psi D_s $ &$[2-3] \times 10^{-3}$ &
$B_c^+ \to J/\psi D^+ $ &$[5-13] \times 10^{-5}$ \\
$B_c^+ \to J/\psi D_s^* $ & $[6-12] \times 10^{-3}$&
$B_c^+ \to J/\psi D^{*+} $ & $[2-4] \times 10^{-4}$\\ \hline
\end{tabular}
\end{center}
\vspace*{-0.5cm}
\caption[]{Branching fractions of $B_c^+$ nonleptonic decays induced
  by 
$b \to c,u$ transitions.}
\label{t:tab2}
\end{table}

\begin{table}[ht]
\begin{center}
\begin{tabular}{|l|r||l|r|}\hline
\multicolumn{1}{|c|}{Channel}& \multicolumn{1}{|c||}{$B$} &
\multicolumn{1}{|c|}{Channel}& \multicolumn{1}{|c|}{$B$} \\ \hline
$B_c^+ \to B_s \pi^+$ &$[4 - 17] \times 10^{-2}$ & 
$B_c^+ \to B_s K^+$   &$[3 - 12] \times 10^{-3}$ \\ 
$B_c^+ \to B_s \rho^+$&$[2 - 7] \times 10^{-2}$ & 
$B_c^+ \to B_s K^{*+}$&$[5 - 9] \times 10^{-5}$\\ 
$B_c^+ \to B_s^* \pi^+$&$[3 - 7] \times 10^{-2}$ & 
$B_c^+ \to B_s^* K^+$&$[2 - 5] \times 10^{-3}$\\ 
$B_c^+ \to B_s^* \rho^+$ &$[14 - 19] \times 10^{-2}$&  &\\ \hline
$B_c^+ \to B_d \pi^+$ &  $[2 - 4] \times 10^{-3}$ & 
$B_c^+ \to B_d K^+$   &  $[2 - 3] \times 10^{-4}$ \\ 
$B_c^+ \to B_d \rho^+$&  $[2 - 7] \times 10^{-3}$ & 
$B_c^+ \to B_d K^{*+}$ &  $[4 - 20] \times 10^{-5}$ \\ 
$B_c^+ \to B_d^* \pi^+$&  $[2 - 4] \times 10^{-3}$ & 
$B_c^+ \to B_d^* K^+$  &  $[1 - 3] \times 10^{-4}$ \\ 
$B_c^+ \to B_d^* \rho^+$& $[1 - 2] \times 10^{-2}$ & 
$B_c^+ \to B_d^* K^{*+}$&$[4 - 6] \times 10^{-4}$ \\ \hline
\end{tabular}
\end{center}
\vspace*{-0.5cm}
\caption[]{Branching fractions of 
$B_c^+$ decays induced by $c \to s,d$ transitions.}
\label{t:tab3}
\end{table}

\subsection[$B_c$ Decays induced by FCNC
Transitions]{\protect\boldmath $B_c$ Decays induced by FCNC Transitions}

Among the rare $B_c$ decay processes that have been discussed
in the literature are
the radiative decays $B_c \to B^*_u \gamma$ and 
$B_c \to D^*_s \gamma$, induced at the quark level by the $c \to u \gamma$
and $b \to s \gamma$ transitions, respectively \cite{fajfer}.
The interest for the former decay mode is related to the possibility
of studying the $c \to u$ electromagnetic penguin transition, 
which in the charm mesons is overwhelmed by long-distance 
contributions. In the case of $B_c$, long-  and short-distance 
contributions have been estimated to be of comparable size, 
and the branching fraction $B(B_c \to B^*_u \gamma)$ is predicted, in the
SM, at the level of $10^{-8}$.

\subsection[CP Violation in $B_c$ Decays]{CP 
Violation in \protect\boldmath $B_c$ Decays}

$B_c$ decays can  give  information about CP violation and 
the weak CKM phases. 
Promising channels are $B_c^{\pm} \to ({\bar c}c) D^{\pm}$,
in particular the one where the charmonium state is a $J/\psi$, whose
decay mode to $\mu^+ \mu^-$ can be easily identified. In this case, CP
violation is due to the difference between the weak phases of the tree and
penguin diagrams contributing to the decay. The CP asymmetry 
${\cal A}(B_c^\pm \to J/\psi D^\pm)$ has been estimated:
${\cal A}(B_c^\pm \to J/\psi D^\pm)\simeq 4 \times 10^{-3}$
\cite{masetti}.
Interesting channels are also those having a light meson in the final
state, e.g. $B_c \to D \rho$ and $B_c \to D \pi$. However, 
in this case the
sizeable r\^{o}le played by the annihilation mechanism makes it difficult
to predict the decay rates and the CP asymmetries.
Decay modes such as $B_c \to D^0 D_s$ can also be considered, although
considerable difficulties would be met in the
experimental detection of $D_s$ and  in the removal of the background 
from $B_u$ decays.

Finally,
the decay $B^+_c\to B^{(*)}_s l^+\nu$ has been proposed as
an interesting source of flavour-tagged $B_s$ mesons for the
study of mixing and CP violation in the $B_s$ sector \cite{CQ}.

\subsection{Experimental Considerations}

 ATLAS have studied the reconstruction of $B_c$ mesons using the decays
 $B_c\to J/\psi\pi$ and $B_c\to J/\psi \ell \nu$, with $J/\psi\to \mu^+
 \mu^-$ (see~\cite{ATLASPTDR}). For this
 study, the following branching ratios have been assumed: $B(B_c \to J/\psi
 \pi)=0.2\times 10^{-2}$ and $B(B_c\to J/\psi \mu \nu)=2\times
 10^{-2}$. It is estimated that after 3 years of running at low
 luminosity, it will be possible to fully reconstruct 12000 $B_c\to J/\psi \pi$
 events and 3$\,\times\,$10$^6$ events in the $B_c\to J/\psi \mu \nu$
 channel. The statistics would allow a
 very precise determination of the $B_c$ mass and lifetime.

\subsection{Concluding Remarks}

The $B_c$ meson is of particular interest as a unique case to
study the impact of QCD dynamics on weak decays. Applications
in flavour physics (CKM parameters, rare decays, $B_s$ flavour tagging)
have also been considered in the literature.
Important theoretical questions that need further attention are
the issues of quark-hadron duality for inclusive decays and, for exclusive
modes, the importance of corrections to the heavy-quark and
nonrelativistic limits, as well as corrections to the 
factorization approximation.
The experimental feasibility for various observables needs
likewise to be assessed in more detail.
The aim of the present section has been to give a flavour of the
special opportunities that exist, from a theoretical perspective,
in studying the physics of the $B_c$. 
Some of these are realizable at the LHC, where it will be possible 
to investigate also the production, spectrum, lifetimes and decays of baryons
containing two heavy quarks \cite{baryons}. It is to be hoped that the
results summarized in this section will trigger 
more detailed 
experimental studies.

\setcounter{equation}{0}
\section[CONCLUSIONS]{CONCLUSIONS\protect\footnote{Section 
coordinators: P. Ball, R. Fleischer, G.F.\ Tartarelli, P. Vikas and
G. Wilkinson.}}\label{concl}

The studies presented at and initiated by the workshop have clearly
shown that the LHC is very well equipped and prepared to pursue a
rigorous $b$ physics programme. The main emphasis in the studies
presented here has been on exploring LHC's potential for measuring CP
violating phenomena and, on the theory side, a meaningful
extraction of
information on the underlying mechanism on CP violation in the
SM. Most of the presented ``strategies'' aim at
extracting the three angles of the unitarity triangle, $\alpha$,
$\beta$ and $\gamma$, as well as $\delta\gamma$, in as many different
ways as possible; any significant discrepancy between the extracted
values or with
the known lengths of the sides of the triangle would constitute
evidence for new physics. Apart from detailed studies of the $e^+$--$\,e^-$ B
factory ``benchmark modes'', also the hadron collider ``gold-plated''
mode $B_s\to J/\psi \phi$ has been studied, and new strategies for
measuring $\beta$ and $\gamma$ from $B_s$ decays, which cannot be
accessed at $e^+$--$\,e^-$ B factories, have been developed. We conclude
that the three experiments are well prepared to solve the ``mystery of
CP violation'' (p.~\pageref{p:mystery}). 

Another important goal to be pursued is the measurement of B mixing
     parameters, and the studies summarized im this report make clear that all
     3 LHC experiments have excellent potential.    There is sensitivity
     in one year's operation to a mass difference in the $B_s$ system 
     far beyond the SM expectation,  and similarly good prospects for 
     a rapid measurement of the width difference.

The second focus of the workshop was the assessment of LHC's reach in
rare decays. The discussion centered on decay modes with the
favourable experimental signature of two muons or
one photon in the final state. It has been demonstrated that the decay
$B_s\to \mu^+\mu^-$ with a SM branching ratio of $\sim 10^{-9}$ can
be seen within one year's running. It has also been shown that decay
spectra of semimuonic rare decays like $B\to K^*\mu^+\mu^-$ are
accessible, which opens the possibility to extract information on
short-distance (new) physics in a theoretically controlled way. LHC's
full potential for rare decays has, however, not yet been fully
plumbed, and further studies, in particular about the feasibility of
inclusive measurements, are ongoing.

Of the many other possible $b$-physics topics, only a few could be
marked out, and we have reported some recent developments in the
theoretical description of nonleptonic decays and discussed a few
issues in $B_c$ physics. The exploration of other exciting topics,
such as physics with $b$ baryons or (non-rare) semileptonic decays, to
name only a few, has to await a second round of workshops.

\newcommand{\prd}[3]{Phys.\ Rev.\ {\bf D{#1}}, {#2} ({#3})}
\newcommand{\npb}[3]{Nucl.\ Phys.\ {\bf B{#1}}, {#2} ({#3})}
\newcommand{\plb}[3]{Phys.\ Lett.\ {\bf B{#1}}, {#2} ({#3})}
\newcommand{\zpc}[3]{Z. Phys.\ {\bf C{#1}}, {#2} ({#3})}
\newcommand{\prl}[3]{Phys.\ Rev.\ Lett.\ {\bf {#1}}, {#2} ({#3})}
\newcommand{\mpl}[3]{Mod.\ Phys.\ Lett.\ {\bf A{#1}}, {#2} ({#3})}
\newcommand{\ptp}[3]{Prog.\ Th.\ Phys.\  {\bf {#1}}, {#2} ({#3})}
\newcommand{\epj}[3]{Eur.\ Phys.\ J.  {\bf C{#1}}, {#2} ({#3})}
\newcommand{\jhep}[3]{JHEP {\bf {#1}}, {#2} ({#3})}
\newcommand{\hep}[1]{hep--ph/{#1}}
\newcommand{\yad}[3]{Yad.\ Fiz.\ {\bf{#1}}, {#2} ({#3})}
\newcommand{\pan}[3]{Phys.\ Atom.\ Nucl.\ {\bf{#1}}, {#2} ({#3})}
\newcommand{\sjn}[3]{Sov.\ J. Nucl.\ Phys.\ {\bf{#1}}, {#2} ({#3})}
\newcommand{\nca}[3]{Nuovo Cim.\ {\bf{#1}A}, {#2} ({#3})}
\newcommand{\jpg}[3]{J.\ Phys.\ G {\bf{#1}}, {#2} ({#3})}
\newcommand{\hip}[3]{Heavy Ion Phys.\ {\bf{#1}}, {#2} ({#3})}
\newcommand{\nim}[3]{Nucl.\ Instrum.\ Meth.\ {\bf A{#1}}, {#2} ({#3})}
\newcommand{\cpc}[3]{Comp.\ Phys.\ Comm.\  {\bf{#1}}, {#2} ({#3})}
\newcommand{\npps}[3]{Nucl.\ Phys.\ Proc.\ Suppl.\ {\bf{#1}}, {#2}
  ({#3})}
\newcommand{\lat}[1]{hep--lat/{#1}}
\newcommand{\ex}[1]{hep--ex/{#1}}
\newcommand{\ar}[3]{Annu.\ Rev.\ Nucl.\ Part.\ Sci.\ {\bf{#1}}, {#2}
    ({#3})}
\newcommand{\ijmp}[3]{Int.\ J. Mod.\ Phys. {\bf A{#1}}, {#2} ({#3})}
\newcommand{\rmp}[3]{Rev.\ Mod.\ Phys.\ {\bf{#1}}, {#2} ({#3})}
\newcommand{\pr}[3]{Phys.\ Rept.\ {\bf{#1}}, {#2} ({#3})}


\begin{thebibliography}{999}

\parskip 0pt                  
\itemsep=2pt


\bibitem{CP-discovery}
J.H.\ Christenson et al., \prl{13}{138}{1964}.

\bibitem{paula} P. Eerola, \ex{9910067}.

\bibitem{book} G.C.\ Branco, L. Lavoura and J.P.\ Silva, {\em CP
    Violation}, Clarendon Press, Oxford, UK, 1999.

\bibitem{CP-revs1}For reviews, see e.g.\ 
Y. Nir, \hep{9911321}; M. Gronau, \hep{9908343};
R. Fleischer, \hep{9908340}.

\bibitem{CP-revs2} A.J.\ Buras, \hep{9905437}.

\bibitem{BaBar}``The BaBar Physics Book'', eds.\ P.F. Harrison and H.R. Quinn,
SLAC Report 504 (1998).

\bibitem{cab}N. Cabibbo, \prl{10}{531}{1963}.

\bibitem{KM}M. Kobayashi and T. Maskawa, \ptp{49}{652}{1973}.

\bibitem{jarlskog}C. Jarlskog, \prl{55}{1039}{1985};
  \zpc{29}{491}{1985}.

\bibitem{new-phys}For reviews, see, for instance, Y. Grossman, Y. Nir and 
R. Rattazzi, \hep{9701231};\\ 
M. Gronau and D. London, \prd{55}{2845}{1997};\\
Y. Nir and H.R. Quinn, \ar{42}{211}{1992};\\
R. Fleischer, \hep{9709291};\\
L. Wolfenstein, \prd{57}{6857}{1998}.

\bibitem{wolf}L. Wolfenstein, \prl{51}{1945}{1983}.

\bibitem{BLO} A.J.\ Buras, M.E.\ Lautenbacher and G. Ostermaier,
  \prd{50}{3433}{1994}.

\bibitem{AKL}R. Aleksan, B. Kayser and D. London, \prl{73}{18}{1994}.

\bibitem{ut}L.L.\ Chau and W.-Y.\ Keung, \prl{53}{1802}{1984};\\
C. Jarlskog and R. Stora, \plb{208}{268}{1988}.

\bibitem{RF-EWP1}R. Fleischer, \zpc{62}{81}{1994}.

\bibitem{RF-EWP2} R. Fleischer, \plb{321}{259}{1994}; \plb{332}{419}{1994}.

\bibitem{RF-rev}R. Fleischer, \ijmp{12}{2459}{1997}.

\bibitem{BBL-rev}G. Buchalla, A.J.\ Buras and M.E.\ Lautenbacher, 
\rmp{68}{1125}{1996}.

\bibitem{BIG1} I. Bigi et al., \hep{9401298}.

\bibitem{BSZ} B. Blok, M. Shifman and D.-X.\ Zhang,
  \prd{57}{2691}{1998}; (E) \prd{59}{019901}{1999}.

\bibitem{BIG2} I. Bigi et al., \prd{59}{054011}{1999};\\
I. Bigi and N. Uraltsev, \prd{60}{114034}{1999};
\plb{457}{163}{1999}.

\bibitem{GL}
B. Grinstein and R. Lebed, \prd{57}{1366}{1998}; \prd{59}{054022}{1999}.

\bibitem{DGamma-cal1} M. Beneke et al., \plb{459}{631}{1999}.

\bibitem{DGamma-cal2}  S. Hashimoto et al., \hep{9912318}.

\bibitem{dun}I. Dunietz, \prd{52}{3048}{1995}.

\bibitem{FD1}R. Fleischer and I. Dunietz, \prd{55}{259}{1997}.

\bibitem{FD2}R. Fleischer and I. Dunietz, \plb{387}{361}{1996}.

\bibitem{RF-Bs}R. Fleischer, \prd{58}{093001}{1998}.

\bibitem{epsprime}A. Alavi-Harati et al.\ (KTeV Coll.), \prl{83}{22}{1999};\\ 
V. Fanti et al.\ (NA48 Coll.), \plb{465}{335}{1999}.

\bibitem{gw}M. Gronau and D. Wyler, \plb{265}{172}{1991}.

\bibitem{GRL}M. Gronau, J.L.\ Rosner and D. London, \prl{73}{21}{1994}.

\bibitem{GHLR}O.F.\ Hern\'andez et al., \plb{333}{500}{1994}; 
\prd{50}{4529}{1994}.

\bibitem{ads}D. Atwood, I. Dunietz and A. Soni, \prl{78}{3257}{1997}.

\bibitem{new-over}For an overview, see R. Fleischer, \hep{9908341}.

\bibitem{BB}P. Ball and V.M. Braun, \prd{58}{094016}{1998}.

\bibitem{ATLASTP}
  The ATLAS Coll., {\em ATLAS Technical Proposal,}
  CERN/LHCC/94-43. 

\bibitem{ATLASPTDR}
  The ATLAS Coll., {\em ATLAS Detector and Physics Performance
   Technical Design Report,} CERN/LHCC/99-14 and CERN/LHCC/99-15. In
 particular, the {\em Inner Detector} is described in Vol.~I,
 Chapter 3, pp.~53--98 and the B-physics performance in Vol.~II,
 Chapter 17, pp.~561--618.

\bibitem{CMSPTDR}
  The CMS Coll.,
  {\em CMS Technical Proposal,}
  CERN/LHCC/94-38.

\bibitem{LHCbTP}
 The LHCb Coll., {\em LHCb Technical Proposal}, CERN/LHCC/98-4.
  
%
%
\bibitem{ATtrk}
  The ATLAS Coll.,
  {\em Inner Detector Technical Design Report,} 
  CERN/LHCC/97-16 and CERN/LHCC/97-17;
   {\em Pixel Detector Technical Design Report,}
   CERN/LHCC/98-13.
\bibitem{CMtrk}
  The CMS Coll.,
  {\em The Tracker System Project Technical Design Report,} 
  CERN/LHCC/98-6.
%
%


\bibitem{ATcalo}
  The ATLAS Coll.,
  {\em Calorimeter Performance Technical Design Report,}
  CERN/LHCC/96-40;
  {\em Liquid Argon Calorimeter Technical
  Design Report,}
  CERN/LHCC/96-41;
  {\em Tile Calorimeter Technical Design Report,}
  CERN/LHCC/96-42.
\bibitem{CMcalo}
  The CMS Coll.,
  {\em The Hadronic Calorimeter Technical Design Report,}
  CERN/LHCC/97-31;
  {\em The Electromagnetic Calorimeter Technical Design Report,}  
  CERN/LHCC/97-33.
%
%
\bibitem{ATmu}
  The ATLAS Coll.,
  {\em Muon Spectrometer Technical Design Report,}
  CERN/LHCC/97-22.
\bibitem{CMmu}
  The CMS Coll.,
  {\em Muon Technical Design Report,}
  CERN/LHCC/97-32. 
%
\bibitem{PYTHIA}
  T. Sj\"ostrand, \cpc{82}{74}{1994}.

\bibitem{bcprod}
See the Chapter on B production in this Book, \hep{0003142}.

\bibitem{CTEQ}
  H.L.\ Lai et al.\ (CTEQ Coll.), \prd{51}{4763}{1995}.

\bibitem{GEANT}
  GEANT 3.21, CERN Program Library Long Write-up W5013.

%
%

\bibitem{CMSpid} S.\ Banerjee et. al., CMS Note CMS-NOTE-1999-056.


%
%
\bibitem{atl0}
  The ATLAS Coll.,
  {\em Trigger Performance Status Report,}
  CERN/LHCC/98-15.
\bibitem{cms0}
 G. Wrochna, CMS Note CMS-CR-1996-002.
%
%
\bibitem{atl1}
  The ATLAS Coll.,
  {\em Level-1 Trigger Technical Design Report,}
  CERN/LHCC/98-14.
\bibitem{cms1}
 J. Pilszka and G. Wrochna, CMS Note CMS-NOTE-1998-075.
%
\bibitem{atl2}
  The ATLAS Coll.,
  {\em DAQ, Event Filter, Level-2 and DCS Technical Progress Report,} 
  CERN/LHCC/98-16.
%
\bibitem{ATtag} Y. Coadou et al., ATLAS Note ATL-PHYS-99-022.

\bibitem{CMtag} Y. Lemoigne and V. Roinishvili, CMS Note
     CMS-NOTE-2000-018.

\bibitem{GroNiRo} M. Gronau, A. Nippe and J.L.\ Rosner,
   \prd{47}{1988}{1993}.

\bibitem{Bstarstar} The ALEPH Coll., \plb{425}{215}{1998};\\
The DELPHI Coll., \plb{345}{598}{1995};\\ The OPAL Coll.,
\zpc{66}{19}{1995}.


%


\bibitem{bisa}A.B.\ Carter and A.I.\ Sanda, \prl{45}{952}{1980};
\prd{23}{1567}{1981};\\
I.I.\ Bigi and A.I.\ Sanda, \npb{193}{85}{1981}.

\bibitem{RF-BdsPsiK}R. Fleischer, \epj{10}{299}{1999}.

\bibitem{nir-sil}Y. Nir and D. Silverman, \npb{345}{301}{1990}.

\bibitem{sin2b-exp}K. Ackerstaff et al.\ (OPAL Coll.),
 \epj{5}{1998}{379};\\
F. Abe et al.\ (CDF Coll.), \prl{81}{1998}{5513} (for an updated analysis,
see Preprint CDF/PUB/BOTTOM/CDF/4855);\\
R. Forty et al.\ (ALEPH Coll.), Preprint ALEPH 99-099.

\bibitem{PDG} 
C. Caso et al.\ (PDG), \epj{3}{1}{1998}.


\bibitem{alpha-uncert}M. Gronau, \plb{300}{163}{1993};\\
J.P.\ Silva and L. Wolfenstein, \prd{49}{R1151}{1994};\\
R. Aleksan et al., \plb{356}{95}{1995};\\
A.J.\ Buras and R. Fleischer, \plb{360}{138}{1995};\\
F. DeJongh and P. Sphicas, \prd{53}{4930}{1996};\\
M. Ciuchini et al., \npb{501}{271}{1997};\\
P.S.\ Marrocchesi and N. Paver, \ijmp{13}{251}{1998};\\
A. Ali, G. Kramer and C.D.\ L\"u, \prd{59}{014005}{1999}.

\bibitem{charles} J. Charles, \prd{59}{054007}{1999}.

\bibitem{gl}M. Gronau and D. London, \prl{65}{3381}{1990}.

\bibitem{BF}A.J.\ Buras and R. Fleischer, \epj{11}{93}{1999}.

\bibitem{GPY}M. Gronau, D. Pirjol and T.M.\ Yan, \prd{60}{034021}{1999}.

\bibitem{gq-alpha}Y. Grossman and H.R.\ Quinn, \prd{58}{017504}{1998}.

\bibitem{Brhopi}H. Lipkin et al., \prd{44}{1454}{1991}.

\bibitem{SQ}A. Snyder and H.R.\ Quinn, \prd{48}{2139}{1993}.

\bibitem{cleo-Bpipi}D.E.\ Jaffe (CLEO Coll.), \ex{9910055}.

\bibitem{RF-Bpipi} R. Fleischer, \hep{0001253}.

\bibitem{FM-Bpipi}R. Fleischer and T. Mannel, \plb{397}{269}{1997}.

\bibitem{BBNS}M. Beneke et al., \prl{83}{1914}{1999}.



\bibitem{j_alek} R. Aleksan et al., \npb{361}{141}{1991}.

\bibitem{j_sv} S. Versill\'e, PhD thesis, Universit\'e de Paris-Sud
(April 1999). Available (in French) at the URL 
\texttt{http://www-lpnhep.in2p3.fr/babar/public/versille/Thesis/} .

\bibitem{jc_phd} J. Charles, PhD thesis, Universit\'e de Paris-Sud
(April 1999). Available (in French) at the URL\\
\texttt{http://qcd.th.u-psud.fr/preprints\_labo/physique\_particule/\\
art1999/art1999.html}.

\bibitem{j_4} S. Versill\'e, J. Charles, R.N.\ Cahn and F. le
  Diberder, in preparation.


\bibitem{BDpi}R.G.\ Sachs, Preprint EFI-85-22 (1985) (unpublished);\\
I. Dunietz and R.G.\ Sachs, \prd{37}{3186}{1988}; (E)
\prd{39}{3515}{1989};\\
I. Dunietz, \plb{427}{179}{1998}.


\bibitem{adk}R. Aleksan, I. Dunietz and B. Kayser, \zpc{54}{653}{1992}.

\bibitem{GroLoBs}M. Gronau and D. London, \plb{253}{483}{1991}.


\bibitem{cleo-BDK}M. Athanas et al.\ (CLEO Coll.),
  \prl{80}{5493}{1998}.

\bibitem{dun-BDK}I. Dunietz, \plb{270}{75}{1991}.


\bibitem{CDF-schmidt}M.P. Schmidt (CDF Coll.), \ex{9906029}.

\bibitem{DDLR}A.S.\ Dighe et al., \plb{369}{144}{1996}.

\bibitem{ddf1}A.S.\ Dighe, I. Dunietz and R. Fleischer, \epj{6}{647}{1999}.

\bibitem{DFN}I. Dunietz, R. Fleischer and U. Nierste, in preparation.

\bibitem{sil}D. Silverman, \prd{58}{095006}{1998}.

\bibitem{bfNP}P. Ball and R. Fleischer, \plb{475}{111}{2000} [\hep{9912319}].

\bibitem{pol}J.L.\ Rosner, \prd{42}{3732}{1990}.

\bibitem{RF-ang}R. Fleischer, \prd{60}{073008}{1999}.

\bibitem{ddf2}A.S.\ Dighe, I. Dunietz and R. Fleischer,
  \plb{433}{147}{1998}.

\bibitem{ambig} Y. Grossman and H.R.\ Quinn, \prd{56}{7259}{1997};\\
J. Charles et al., \plb{425}{375}{1998};\\
B. Kayser and D. London, \hep{9909560}.

\bibitem{LR-refs} See, e.g.\ D. Chang, \npb{214}{435}{1983};\\
  G. Ecker and W. Grimus, \npb{258}{328}{1985}; \zpc{30}{293}{1986}.

\bibitem{JMF} J.-M.\ Fr\`{e}re et al., \prd{46}{337}{1992}.

\bibitem{my}P. Ball, J.-M. Fr\`ere and J. Matias, \hep{9910211}.

\bibitem{AL}A. Ali and D. London, \epj{9}{687}{1999}.

\bibitem{quim} G. Barenboim et al., \prd{60}{016003}{1999}.

\bibitem{grossman}Y. Grossman, \plb{380}{99}{1996}.

\bibitem{LHCbJpsiphi} P. Kooijman and N. Zaitsev, LHCb NOTE 98-067, 
and N. Zaitsev, private communication.

\bibitem{ATLASJpsiphi} 
 M.Smizanska for the ATLAS Coll., Talk given at the  6th 
 Conference on B Physics with Hadron Machines, June 1999, Bled.
 To be published in Nucl. Instrum. Meth. A.

\bibitem{CMSJpsiphi} P. Galumian, private communication. 

\bibitem{DIGHESEN} A. Dighe and S. Sen, \prd{59}{074002}{1999}.

\bibitem{BSW}M. Bauer, B. Stech and M. Wirbel, \zpc{29}{637}{1985}; 
\zpc{34}{103}{1987}.

\bibitem{CLEO}
C.P. Jessop et al.\ (CLEO Coll.), \prl{79}{4533}{1997}.

\bibitem{BENEKE} M. Beneke, G.Buchalla and I.Dunietz, \prd{54}{4419}{1996}.





\bibitem{BpiK-revs}For overviews, see R. Fleischer, \hep{9904313}; 
M. Neubert, \hep{9909564}.

\bibitem{RF-BsKK}R. Fleischer, \plb{459}{306}{1999}.

\bibitem{dun-snowmass}I. Dunietz, Proceedings of the Workshop on
B Physics at Hadron Accelerators, Snowmass (CO), USA, eds.\ P. McBride 
and C. Shekhar Mishra, p.\ 83.

\bibitem{lipkin}H.J. Lipkin, \plb{415}{186}{1997}.

\bibitem{bfm}A.J.\ Buras, R. Fleischer and T. Mannel, \npb{533}{3}{1998}.

\bibitem{defan}R. Fleischer, \epj{6}{451}{1999}.

\bibitem{BKK}R. Fleischer, \plb{435}{221}{1998}.

\bibitem{pirjol}D. Pirjol, \prd{60}{054020}{1999}.

\bibitem{cleo-BpiK99}Y. Kwon et al.\ (CLEO Coll.), \ex{9908029}; \ex{9908039}.

\bibitem{FMbound}R. Fleischer and T. Mannel, \prd{57}{2752}{1998}.

\bibitem{NRbound}M. Neubert and J.L.\ Rosner, \plb{441}{403}{1998}.

\bibitem{PAPIII}R. Fleischer, \plb{365}{399}{1996}.

\bibitem{GroRo}M. Gronau and J.L.\ Rosner, \prd{57}{6843}{1998}.

\bibitem{NR}M. Neubert and J.L.\ Rosner, \prl{81}{5076}{1998}.

\bibitem{PAPI}A.J.\ Buras and R. Fleischer, \plb{365}{398}{1996}.

\bibitem{facto}J. Schwinger, \prl{12}{630}{1964};\\
R.P.\ Feynman, in ``Symmetries in Particle Physics'', ed.\ A. Zichichi 
(Acad.\ Press 1965);\\
O. Haan and B. Stech, \npb{22}{448}{1970};\\
D. Fakirov and B. Stech, \npb{133}{315}{1978};\\
L.L.\ Chau, \pr{95}{1}{1983}.

\bibitem{bpiff} A. Khodjamirian, R. R\"uckl and C.W.\ Winhart,
  \prd{58}{054013}{1998};\\
E. Bagan, P. Ball and V.M.\ Braun,
  \plb{417}{154}{1998}.

\bibitem{ball98} P. Ball, \jhep{9809}{005}{1998}.

\bibitem{bjorken}J.D. Bjorken, \npps{11}{325}{1989};
Proceedings of the SLAC Summer Institute 1990, p.\ 167.

\bibitem{FSI}L. Wolfenstein, \prd{52}{537}{1995};\\
J.-M. G\'erard and J. Weyers, \epj{7}{1}{1999};\\
A.F. Falk et al., \prd{57}{4290}{1998};\\
D. Atwood and A. Soni, \prd{58}{036005}{1998}.

\bibitem{neubert}M. Neubert, \plb{424}{152}{1998}.

\bibitem{groro-FSI}M. Gronau and J.L.\ Rosner, \prd{58}{113005}{1998}.

\bibitem{Neubert-BpiK}M. Neubert, \jhep{9902}{014}{1999}.

\bibitem{dh}N.G.\ Deshpande and X.G.\ He, \prl{74}{26}{1995}; 
(E) \prl{74}{4099}{1995}.

\bibitem{GHLR-EWP}M. Gronau et al., \prd{52}{6374}{1995}.

\bibitem{FSI-recent}M. Gronau and D. Pirjol, \plb{449}{321}{1999};
  \prd{61}{013005}{2000};\\
K. Agashe and N.G.\ Deshpande, \plb{451}{215}{1999}; \plb{454}{359}{1999}.

\bibitem{BpiK-NP}R. Fleischer and T. Mannel, \hep{9706261};\\
R. Fleischer, in Ref.~\cite{new-phys}; M. Neubert, in 
Ref.~\cite{Neubert-BpiK};\\
D. Choudhury, B. Dutta and A. Kundu, \plb{456}{185}{1999};\\
X.G.\ He, C.L.\ Hsueh and J.Q.\ Shi, \prl{84}{18}{2000}.

\bibitem{GNK}Y. Grossman, M. Neubert and A.L.\ Kagan, \hep{9909297}.

\bibitem{FMat}R. Fleischer and J. Matias, \prd{61}{074004}{2000}.




\bibitem{neu-ste}M. Neubert and B. Stech, \hep{9705292}.

\bibitem{HMChiPT1}E. Jenkins and M.J. Savage, \plb{281}{331}{1992}.

\bibitem{HMChiPT2}B. Grinstein et al., \npb{380}{369}{1992}.

\bibitem{lat1} L. Lellouch and C.-J.D.\ Lin
 (UKQCD Coll.), \npps{73}{357}{1999} [\lat{9809018}]. 


\bibitem{pen-calc}R. Fleischer, \zpc{58}{483}{1993};\\
G. Kramer, W.F.\ Palmer and H. Simma, \zpc{66}{429}{1995}.

\bibitem{bss}M. Bander, D. Silverman and A. Soni, \prl{43}{242}{1979}.


\bibitem{BJW}
A.J. Buras, M. Jamin and P.H. Weisz, \npb{347}{491}{1990}.

\bibitem{Aoki:1999ue}
S. Aoki, \hep{9912288}; S. Hashimoto, \hep{9909136}; 
L. Lellouch, \hep{9906497};
J.M.\ Flynn and C.T.\ Sachrajda, \lat{9710057}.

\bibitem{Bernard:1998dg}
C. Bernard, T. Blum and A. Soni, \prd{58}{014501}{1998};\\
L. Lellouch and C.J.\ Lin (UKCD Coll.), \hep{9912322}.

\bibitem{Hashimoto:1999ck}
S. Hashimoto et al., \prd{60}{094503}{1999}.

\bibitem{Gupta:1997yt}
R. Gupta, T. Bhattacharya and S. Sharpe, \prd{55}{4036}{1997}.


\bibitem{LEPBOSC}
G. Blaylock, \ex{9912038};\\ see also {\tt
  http://www.cern.ch/LEPBOSC/combined\_results/summer\_1999/}.

\bibitem{BOSC_FIT}
H.-G. Moser and A. Roussarie, \nim{384}{491}{1997}.

\bibitem{CMSNOTE1} A. Starodumov and Z. Xie, CMS NOTE 1999/006.

\bibitem{CMSNOTE2} Z. Xie, F. Palla and A. Starodumov, CMS NOTE in 
preparation.


\bibitem{CLEO_Brad} R. Ammar et al.\ (CLEO Coll.), \prl{71}{674}{1993}.

\bibitem{GLN} Y. Grossman, Z. Ligeti and E. Nardi, \prd{55}{2768}{1997}.

\bibitem{SK} W. Skiba and J. Kalinowski, \npb{404}{3}{1993}.

\bibitem{IL} T. Inami and C.S.\ Lim, \ptp{65}{297}{1981}; (E) 1772.

\bibitem{laura} D. Atwood, L. Reina and A. Soni, \prd{55}{3156}{1997};\\
K.S.\ Babu and C. Kolda, \prl{84}{228}{2000}.

\bibitem{CMSNote} A. Nikitenko, A. Starodumov and N. Stepanov,
  CMS-NOTE-1999-039 (\hep{9907256}).


\bibitem{tobias} C. Greub, T. Hurth and D. Wyler,
  \prd{54}{3350}{1996};\\ C. Greub and T. Hurth, \prd{56}{2934}{1997}.

\bibitem{mikolaj}
K. Chetyrkin, M. Misiak and M. M\"{u}nz,
  \plb{400}{206}{1997}; (E) {\bf B425}, 414 (1998).

\bibitem{voloshin} M.B.\ Voloshin, \plb{397}{275}{1997}.

\bibitem{stoll} A. Khodjamirian et al., \plb{402}{167}{1997}.

\bibitem{mp} M. Misiak, \hep{0002007}.

\bibitem{Brhog} A. Khodjamirian, G. Stoll and D. Wyler,
 \plb{358}{129}{1995};\\
A. Ali and V.M.\ Braun, \plb{359}{223}{1995}.

\bibitem{newshit} B. Grinstein and D. Pirjol, \hep{0002216}.

\bibitem{LHCb_BKgamma} G. Kostina and P.
Pakhlov, LHCb Note LHCB/97--022\\ ({\tt 
http://lhcb.cern.ch/notes/postscript/97notes/97-022.ps}).

\bibitem{mikolaj2} C. Bobeth, M. Misiak and J. Urban, \hep{9910220}.

\bibitem{ABHH} A. Ali et al., \prd{61}{074024}{2000}.

\bibitem{aliev_gang} 
D. Melikhov, N. Nikitin and S. Simula, \prd{57}{6814}{1998}; 
\plb{430}{332}{1998}; \plb{442}{381}{1998};\\
T.M.\ Aliev, C.S.\ Kim and Y.G.\ Kim, \hep{9910501};\\ 
C.S.\ Kim, Y.G.\ Kim and C.D.\ Lu, \hep{0001151}.

\bibitem{AH} A. Ali and G. Hiller, \epj{8}{619}{1999}; \prd{60}{034017}{1999}.

\bibitem{c9eff}
A. Ali, T. Mannel and T. Morozumi, \plb{273}{505}{1991};\\
F. Kr\"uger and L.M.\ Sehgal, \plb{380}{199}{1996};\\
Z. Ligeti, I.W.\ Stewart and M.B.\ Wise, \plb{420}{359}{1998}.

\bibitem{LEET} J. Charles et al., \prd{60}{014001}{1999}.

\bibitem{burdman} G. Burdman, \prd{57}{4254}{1998}.


\bibitem{ATLASXmumu} D. Melikhov et al., ATLAS Note ATL--PHYS--98--123.

\bibitem{bloedsinn} T.M.\ Aliev, C.S.\ Kim and M. Savci,
  \plb{441}{410}{1998}.

\bibitem{flora} N. Nikitin, F. Rizatdinova and L. Smirnova,
  \pan{62}{1697}{1999} [\yad{62}{1823}{1999}].

\bibitem{missing} P. Cho, M. Misiak and D. Wyler, \prd{54}{3329}{1996}.

\bibitem{CLEO99} S. Ahmed et al.\ (CLEO Coll.), \ex{9908022}.

\bibitem{D0} B. Abbott et al.\ (D0 Coll.), 
\plb{423}{419}{1998}.

\bibitem{CLEO98} S. Glenn et al.\ (CLEO Coll.), \prl{80}{2289}{1998}.

\bibitem{CDF-rare} T. Affolder et al.\ (CDF Coll.), \prl{83}{3378}{1999}.

\bibitem{FLS94} A.F.\ Falk, M. Luke and M.J.\ Savage, \prd{49}{3367}{1994}.

\bibitem{CGG90} J. Chay, H. Georgi and B. Grinstein, \plb{247}{399}{1990}.

\bibitem{semimuonicCP} T.M.\ Aliev, D.A.\ Demir and M. Savci, \hep{9912525}.


\bibitem{KR}
A. Khodjamirian and R. R\"uckl, \hep{9801443}. 

\bibitem{KR2}
A. Khodjamirian and R. R\"uckl, \hep{9807495};\\ 
A. Khodjamirian, R. R\"uckl and C.W.\ Winhart, in preparation.

\bibitem{CaMi} I. Caprini, L. Micu and C. Bourrely,
  \prd{60}{074016}{1999}; \hep{9910297}.

\bibitem{BKR}
V.M.\ Belyaev, A. Khodjamirian and R. R\"uckl, \zpc{60}{349}{1993}.

\bibitem{BS93} 
B. Blok and M. Shifman, \npb{389}{534}{1993}.

\bibitem{FJ}
G.R.\ Farrar and D.R.\ Jackson, \prl{43}{246}{1979}.

\bibitem{ER}
A.V.\ Efremov and A.V.\ Radyushkin, \plb{94}{245}{1980}.

\bibitem{LB}
G.P.\ Lepage and S.J.\ Brodsky, \prd{22}{2157}{1980}.

\bibitem{PW}
H.D.\ Politzer and M.B.\ Wise, \plb{257}{399}{1991}.

\bibitem{DGPS96}
J.F. Donoghue et al., \prl{77}{2178}{1996}.


\bibitem{cdf}
F. Abe  et al.\ (CDF Coll.), \prl{81}{2432}{1998}; \prd{58}{112004}{1998}.

\bibitem{GERS}
S.S.\ Gershtein et al., \hep{9803433}.

\bibitem{BBBC}
M. Beneke and G. Buchalla, \prd{53}{4991}{1996}.

\bibitem{CQ}
C. Quigg, in Proc.\ of the Workshop on B Physics at Hadron
Accelerators, Snowmass (CO), USA, 1993, eds.\ P. McBride and
C.S.\ Mishra.

\bibitem{BIGBC}
I.I. Bigi, \plb{371}{105}{1996}.

\bibitem{NRlat}
B.D.\ Jones and R.M.\ Woloshyn, \prd{60}{014502}{1999}.

\bibitem{qcdsr0}
V.V.\ Kiselev and A. Tkabladze, \sjn{50}{1063}{1989};\\
T.M.\ Aliev and O. Yilmaz, \nca{105}{827}{1992};\\
M. Chabab, \plb{325}{205}{1994}.

\bibitem{qcdsr1}
P. Colangelo, G. Nardulli and N. Paver, \zpc{57}{43}{1993};\\
E. Bagan et al., \zpc{64}{57}{1994}.

\bibitem{qmod0}
S. Godfrey and N. Isgur, \prd{32}{189}{1985};\\
P. Colangelo, G. Nardulli and M. Pietroni, \prd{43}{3002}{1991};\\
E. Eichten and C. Quigg, \prd{49}{5845}{1994};\\
S.S.\ Gershtein et al., \prd{51}{3613}{1995};\\
L.P. Fulcher, \prd{60}{074006}{1999}.

\bibitem{CFP}
G. Chiladze, A.F.\ Falk and A.A.\ Petrov, \prd{60}{034011}{1999}.

\bibitem{bcmng}
P. Colangelo and F. De Fazio, \mpl{33}{2303}{1999}.

\bibitem{qcdsr2}
V.V.\ Kiselev and A. Tkabladze, \prd{48}{5208}{1993}.

\bibitem{qmod}
D. Du and Z. Wang, \prd{39}{1342}{1989};\\
M. Lusignoli and M. Masetti, \zpc{51}{549}{1991}.

\bibitem{qmod1}
C.H.\ Chang and Y.Q.\ Chen, \prd{49}{3399}{1994};\\
J.F.\ Liu and K.T.\ Chao, \prd{56}{4133}{1997}.

\bibitem{jenkins}
E. Jenkins et al., \npb{390}{463}{1993}.

\bibitem{sanchiz}
M.A.\ Sanchiz-Lozano,\npb{440}{251}{1995};\\
M. Galdon and M.A.\ Sanchiz-Lozano, \zpc{71}{277}{1996};\\
P. Colangelo and F. De Fazio, \prd{61}{034012}{2000}.

\bibitem{kiselev}
V.V.\ Kiselev, A.K.\ Likhoded and A.I.\ Onishchenko, \hep{9905359}.

\bibitem{fajfer}
D. Du, X. Li and Y. Yang, \plb{380}{193}{1996};\\
S. Fajfer, D. Prelovsek and P. Singer, \prd{59}{114003}{1999};\\
T.M. Aliev and M. Savci, \hep{9908203}.

\bibitem{masetti}
M. Masetti, \plb{286}{160}{1992};\\
Y.S.\ Dai and D.S.\ Du, \epj{9}{557}{1999}.

\bibitem{baryons}
E. Bagan et al., \plb{287}{176}{1992}; \plb{301}{243}{1993};\\
D.B.\ Lichtenberg, R. Roncaglia and E. Predazzi,\prd{52}{1722}{1995};\\
S.S.\ Gershtein et al., \hip{9}{133}{1999};\\
V.V.\ Kiselev, A.K.\ Likhoded and A.I.\ Onishchenko, \hep{9901224}.

\end{thebibliography}
\end{document}